%% file: thesis.tex
\documentclass[12pt,oneside,letterpaper]{report}

\usepackage{amsmath}
\newcommand{\draftfinal}[2]{\ifdefined\draftversion#1\else#2\fi}

\newcommand{\finalonly}[1]{\draftfinal{}{#1}}

\newcommand{\thesistitle}{Dualities and Symmetries of Quantum Field Theories from Brane Engineering}
\newcommand{\thesisauthor}{Xingyang Yu}
\newcommand{\thesisadvisor}{Prof.  Massimo Porrati}
\newcommand{\secondadvisor}{Prof. Sebasti\'{a}n Franco}

\newcommand{\thesisdept}{Physics}
\newcommand{\gradmonth}{September}
\newcommand{\gradyear}{2023}

\RequirePackage[margin=1in, includefoot, letterpaper]{geometry}

\RequirePackage[prologue]{xcolor}
\definecolor[named]{ThesisBlue}{cmyk}{1,0.1,0,0.1}
\definecolor[named]{ThesisYellow}{cmyk}{0,0.16,1,0}
\definecolor[named]{ThesisOrange}{cmyk}{0,0.42,1,0.01}
\definecolor[named]{ThesisRed}{cmyk}{0,0.90,0.86,0}
\definecolor[named]{ThesisLightBlue}{cmyk}{0.49,0.01,0,0}
\definecolor[named]{ThesisGreen}{cmyk}{0.20,0,1,0.19}
\definecolor[named]{ThesisPurple}{cmyk}{0.55,1,0,0.15}
\definecolor[named]{ThesisDarkBlue}{cmyk}{1,0.58,0,0.21}

\definecolor{SchoolColor}{rgb}{0.3412, 0.0235, 0.5490} 
\definecolor{chaptergrey}{rgb}{0.2600, 0.0200, 0.4600} 
\definecolor{midgrey}{rgb}{0.4, 0.4, 0.4}

\usepackage{hhline}
\usepackage[utf8]{inputenc}
\usepackage{float}
\usepackage{mathtools}
\usepackage{dsfont}
\usepackage{cancel}

\usepackage{amsfonts}
\usepackage{amssymb}
\usepackage{ytableau}
\usepackage{cellspace}
\usepackage{array}
\usepackage{makecell}
\usepackage{pgfplots}
\usepackage{tikz}
\usepackage{tikz-cd}
\usetikzlibrary{shapes.misc,shadows,fit,decorations.pathmorphing,positioning,trees,decorations.markings,decorations.pathreplacing,calc,shapes,patterns,arrows}
\usepackage{chemformula}
\usepackage{array}

\RequirePackage[labelfont={bf,sf,small,singlespacing},
                textfont={sf,small,singlespacing},
                margin=0pt,
                figurewithin=chapter,
                tablewithin=chapter]{caption}

\usepackage{fix-cm}
\RequirePackage[raggedright,sc]{titlesec}
\definecolor{gray75}{gray}{0.75}
\newcommand{\hsp}{\hspace{20pt}}

\titleformat{\chapter}[hang]
{\Huge\sc}
{\textcolor{SchoolColor}{\thechapter}\hsp\textcolor{gray75}{|}\hsp}
{0pt}{\Huge\sc\raggedright}

\usepackage{setspace}

\finalonly{
  \doublespacing 
}

\usepackage{lipsum}

\input{defs}

\usepackage{comment}

\usepackage{hyperref}
\hypersetup{colorlinks,
  linkcolor=ThesisDarkBlue,
  citecolor=ThesisPurple,
  urlcolor=ThesisDarkBlue,
  filecolor=ThesisDarkBlue}

\usepackage[capitalize]{cleveref} 
\usepackage[
    backend=biber,
    style=numeric-comp,
    natbib=true,
    url=false, 
    doi=true,
    eprint=false,
    maxbibnames=99,
    sorting=none
]{biblatex}
 \usepackage{lscape}
 
\begin{document}

\pagenumbering{roman}
%
\thispagestyle{empty}
%

\vspace*{25pt}
\begin{center}

  {\Large
    \begin{doublespace}
      {\textcolor{SchoolColor}{\textsc{\thesistitle}}}
    \end{doublespace}
  }
  \vspace{.7in}

  by
  \vspace{.7in}

  \thesisauthor
  \vfill

  \begin{doublespace}
    \textsc{
    A dissertation submitted in partial fulfillment\\
    of the requirements for the degree of\\
    Doctor of Philosophy\\
    Department of \thesisdept\\
    New York University\\
    \gradmonth, \gradyear}
  \end{doublespace}
\end{center}
\vfill

\noindent\makebox[\textwidth]{\hfill\makebox[2.5in]{\hrulefill}}\\
\makebox[\textwidth]{\hfill\makebox[2.5in]{\hfill\thesisadvisor}}
\noindent\makebox[\textwidth]{\hfill\makebox[2.5in]{\hrulefill}}\\
\makebox[\textwidth]{\hfill\makebox[2.5in]{\hfill\secondadvisor}}

\newpage

\thispagestyle{empty}
\vspace*{25pt}
\begin{center}
  \scshape \noindent \small \copyright \  \small  \thesisauthor \\
  All rights reserved, \gradyear
\end{center}
\vspace*{0in}
\newpage

\cleardoublepage
\phantomsection
\chapter*{Dedication}
\addcontentsline{toc}{chapter}{Dedication}
To my father, Zhenguo Yu, and my mother, Hongyue Zhang, for teaching me to have a dream and pursue it. To my wife, Fang-Yi Chu, and my son, Isaac Daohuan Yu, for being the love of my life.
\vfill
\newpage

\chapter*{Acknowledgements}
\addcontentsline{toc}{chapter}{Acknowledgments}

 \input{acknowledge}


\chapter*{Abstract}
\addcontentsline{toc}{chapter}{Abstract}

\input{abstract}

\newpage

\tableofcontents

\cleardoublepage
\phantomsection
\addcontentsline{toc}{chapter}{List of Figures}
\listoffigures
\newpage

\listoftables

\pagenumbering{arabic} 




\chapter{Introduction} \input{chapters/1}\label{ch1}

\chapter{$2d$ $\mathcal{N}=(0,1)$ gauge theories and Spin(7) Orientifolds}\label{cha:ch2}
This chapter is adapted from \cite{Franco:2021ixh}.

\input{chapters/2}

\chapter{Spin(7) Orientifolds and $2d$ $\mathcal{N}=(0,1)$ Trialities}\label{cha:ch3} 
This chapter is adapted from \cite{Franco:2021vxq}.
\input{chapters/3}

\chapter{Top Down Approach to Topological Duality Defects}\label{cha:ch4}
This chapter is adapted from \cite{Heckman:2022xgu}.
\input{chapters/4}

\appendix

\chapter{Details on $\mathbb{C}^4/\mathbb{Z}_2\times \mathbb{Z}_2$ and SPP$\times \mathbb{C}$ Theories} \label{app:C4Z2Z2SPPC}

\input{App/A}

\chapter{Details on $Q^{1,1,1}/\ZZ_2$ Theories}\label{app:Q111Z2-details}
\input{App/B}

\chapter{More on Duality Defects}\label{app:C}
\input{App/C}








\cleardoublepage
\phantomsection

\addcontentsline{toc}{chapter}{Bibliography}


\printbibliography
\end{document}


%% file: defs.tex
\usepackage{multicol}
\usepackage{amsfonts}
\usepackage{amssymb}
\usepackage{amsmath}
\usepackage{setspace}
\usepackage{verbatim}
\usepackage{color}
\usepackage{longtable}
\usepackage{standalone}
\usepackage{float}
\usepackage{subcaption}
\usepackage{epsfig}
\usepackage{epstopdf}
\usepackage{adjustbox}
\usepackage{tikz}
\usepackage[margin=1in]{geometry}
\usepackage{titletoc}
\usepackage{hyperref}%
\usepackage{braket}
\usepackage{mathtools}
\usepackage{bbm}
\usepackage{graphicx} 

 \usepackage{indentfirst}

\usepackage{amsthm} 

\usepackage{amsfonts}
\usepackage{amssymb}

\RequirePackage[T1]{fontenc}
\ifxetex
  \RequirePackage[tt=false]{libertine}
\else
  \RequirePackage[tt=false, type1=true]{libertine}

\RequirePackage[varqu]{zi4}
\RequirePackage[libertine]{newtxmath}

\usepackage{mathpartir}

\usepackage{bm}

\usepackage{booktabs}   
\usepackage[labelformat=simple]{subcaption} 

\usepackage[T1]{fontenc} 
\usepackage[utf8]{inputenc} 

\usepackage{amssymb}
\usepackage{longtable}
\usepackage[final]{listings}
\usepackage{tikz}
\usetikzlibrary{shapes.misc,shadows,fit,decorations.pathmorphing,positioning,trees,decorations.markings,decorations.pathreplacing,calc,shapes,patterns,arrows}
\usepackage[shortlabels, inline]{enumitem}
\makeatletter
\newcommand{\myitem}[1][]{
  \protected@edef\@currentlabel{#1}%
\item[#1]
}
\makeatother

\usepackage{mathtools}

\usepackage{etoolbox}

\usepackage{array}
\newcolumntype{L}{>$l<$}

\usepackage{stmaryrd}

\usepackage{dashbox}

\usepackage{scalerel}

\usepackage{mathpartir}

\usepackage{xspace}
\usepackage{multirow}

\usepackage{breakcites}

\theoremstyle{plain}

\theoremstyle{definition}

\theoremstyle{plain}

\makeatletter
\def\squarecorner#1{
	\pgf@x=\the\wd\pgfnodeparttextbox%
	\pgfmathsetlength\pgf@xc{\pgfkeysvalueof{/pgf/inner xsep}}%
	\advance\pgf@x by 2\pgf@xc%
	\pgfmathsetlength\pgf@xb{\pgfkeysvalueof{/pgf/minimum width}}%
	\ifdim\pgf@x<\pgf@xb%
	\pgf@x=\pgf@xb%
	\fi%
	\pgf@y=\ht\pgfnodeparttextbox%
	\advance\pgf@y by\dp\pgfnodeparttextbox%
	\pgfmathsetlength\pgf@yc{\pgfkeysvalueof{/pgf/inner ysep}}%
	\advance\pgf@y by 2\pgf@yc%
	\pgfmathsetlength\pgf@yb{\pgfkeysvalueof{/pgf/minimum height}}%
	\ifdim\pgf@y<\pgf@yb%
	\pgf@y=\pgf@yb%
	\fi%
	\ifdim\pgf@x<\pgf@y%
	\pgf@x=\pgf@y%
	\else
	\pgf@y=\pgf@x%
	\fi
	\pgf@x=#1.5\pgf@x%
	\advance\pgf@x by.5\wd\pgfnodeparttextbox%
	\pgfmathsetlength\pgf@xa{\pgfkeysvalueof{/pgf/outer xsep}}%
	\advance\pgf@x by#1\pgf@xa%
	\pgf@y=#1.5\pgf@y%
	\advance\pgf@y by-.5\dp\pgfnodeparttextbox%
	\advance\pgf@y by.5\ht\pgfnodeparttextbox%
	\pgfmathsetlength\pgf@ya{\pgfkeysvalueof{/pgf/outer ysep}}%
	\advance\pgf@y by#1\pgf@ya%
}
\makeatother

\pgfdeclareshape{square}{
	\savedanchor\northeast{\squarecorner{}}
	\savedanchor\southwest{\squarecorner{-}}
	
	\foreach \x in {east,west} \foreach \y in {north,mid,base,south} {
		\inheritanchor[from=rectangle]{\y\space\x}
	}
	\foreach \x in {east,west,north,mid,base,south,center,text} {
		\inheritanchor[from=rectangle]{\x}
	}
	\inheritanchorborder[from=rectangle]
	\inheritbackgroundpath[from=rectangle]
}

\renewcommand{\ge}{\geqslant}

\newcommand{\R}{\ensuremath{\mathbb{R}}}

\newcommand{\CC}{\mathbb{C}}

\newcommand{\ZZ}{\mathbb{Z}}
\newcommand{\ID}{\mathds{1}}
\newcommand{\IM}{\mathbf{M}}

\newcommand{\coma}{\, , \,}
\newcommand{\fstop}{\, .}
\newcommand{\hc}{\text{ h.c.}}

\def\quadro{\rotatebox[origin=c]{45}{$\blacksquare$}}

\newcommand{\drawsquare}[2]{\hbox{%
\rule{#2pt}{#1pt}\hskip-#2pt
\rule{#1pt}{#2pt}\hskip-#1pt
\rule[#1pt]{#1pt}{#2pt}}\rule[#1pt]{#2pt}{#2pt}\hskip-#2pt
\rule{#2pt}{#1pt}}

\newcommand{\fund}{~\raisebox{-.5pt}{\drawsquare{6.5}{0.4}}~}
\newcommand{\antifund}{~\overline{\raisebox{-.5pt}{\drawsquare{6.5}{0.4}}}~}

\newcommand*\centermathcell[1]{\omit\hfil$\displaystyle#1$\hfil\ignorespaces}

\definecolor{redX}{RGB}{218,59,38}
\definecolor{blueX}{RGB}{71,159,248}
\definecolor{yellowX}{RGB}{239,189,64}
\definecolor{greenX}{HTML}{54ae32}

\allowdisplaybreaks[1] 

\DeclareMathOperator{\Res}{Res}
\DeclareMathOperator{\U}{U}
\DeclareMathOperator{\SU}{SU}
\DeclareMathOperator{\SO}{SO}

\DeclareMathOperator{\USp}{USp}
\DeclareMathOperator{\Spin}{Spin}
\DeclareMathOperator{\SPP}{SPP}

\DeclareMathOperator{\PE}{PE}
\DeclareMathOperator{\PL}{PL}
\DeclareMathOperator{\HS}{HS}

\def\quadro{\rotatebox[origin=c]{45}{$\blacksquare$}}

\definecolor{cyanX}{HTML}{90e0ef}

\providecommand{\U}[1]{\protect\rule{.1in}{.1in}}
\pdfoutput=1
\newsavebox{\mysavebox}

\hypersetup{colorlinks,citecolor=black,filecolor=black,linkcolor=black,urlcolor=black}
\numberwithin{equation}{section}

\newcommand{\ba}{\begin{eqnarray}}
\newcommand{\ea}{\end{eqnarray}}

\newcommand{\be}{\begin{equation}}
\newcommand{\ee}{\end{equation}}

\usepackage{scalerel}

\usepackage{tikz}
\usetikzlibrary{arrows}
\usetikzlibrary{arrows.meta}
\usetikzlibrary{shapes.geometric,calc,arrows, positioning,shapes.misc,decorations.markings}
\tikzset{
  big arrow/.style={
    decoration={markings,mark=at position 1 with {\arrow[scale=2,#1]{>}}},
    postaction={decorate},
    shorten >=0.4pt},
  big arrow/.default=black}

\pgfdeclarelayer{edgelayer}
\pgfdeclarelayer{nodelayer}
\pgfsetlayers{edgelayer,nodelayer,main}
\tikzstyle{none}=[inner sep=0pt]

\tikzstyle{NodeCross}=[draw, shape=circle, cross out, inner sep=0pt, minimum size=6pt,line width=0.25mm]
\tikzstyle{Circle}=[draw, shape=circle, black, fill=black, inner sep=0pt, minimum size=6pt]
\tikzstyle{Star}=[draw, shape=star, fill=black, star points=8, inner sep=0pt, minimum size=8pt]
\tikzstyle{CircleRed}=[draw, shape=circle, black, fill=red, inner sep=0pt, minimum size=4pt]
\tikzstyle{StarP}=[draw={rgb,255: red,128; green,0; blue,128}, shape=star, fill={rgb,256: red,128; green,0; blue,128}, star points=8, inner sep=0pt, minimum size=12pt]

\tikzstyle{DashedLine}=[-, densely dashed, line width=0.25mm]
\tikzstyle{DottedLine}=[-, dotted, line width=0.25mm]
\tikzstyle{ThickLine}=[-, line width=0.25mm]
\tikzstyle{ArrowLineRight}=[-, -{Stealth[scale=1.75]}, line width=0.1mm, scale=5]
\tikzstyle{ArrowLineRed}=[-, draw={rgb,255: red,191; green,0; blue,0}, -{Stealth[scale=1.75]}, line width=0.1mm, scale=5]
\tikzstyle{RedLine}=[-, draw={rgb,255: red,191; green,0; blue,0}, fill=none, line width=0.25mm]
\tikzstyle{DashedLineThin}=[-, densely dashed, line width=0.125mm, fill=none, draw=black]
\tikzstyle{DottedRed}=[-, dotted, draw={rgb,255: red,191; green,0; blue,0}, fill=none, line width=0.25mm]
\tikzstyle{DashedRed}=[-, densely dashed, draw={rgb,255: red,191; green,0; blue,0}, fill=none, line width=0.25mm]
\tikzstyle{BlueLine}=[-, draw={rgb,255: red,0; green,0; blue,191}, fill=none, line width=0.25mm]
\tikzstyle{ArrowLineBlue}=[-, draw={rgb,255: red,0; green,0; blue,191}, -{Stealth[scale=1.75]}, line width=0.1mm, scale=5]
\tikzstyle{GreenDoubleArrow}=[<->, draw={rgb,155: red,0; green,255; blue,0},  line width= 0.5mm, scale=5]
\tikzstyle{RedDoubleArrow}=[<->, draw={rgb,255: red,255; green,0; blue,0},  line width= 0.5mm, scale=5]


%% file: acknowledge.tex
I would like to express my sincere gratitude to the following individuals for their invaluable support and assistance throughout the completion of my Ph.D.:

First and foremost, I would like to thank my thesis advisors, Massimo Porrati and Sebasti\'{a}n Franco for their unwavering guidance, patience, and encouragement. Over the years, I have benefited greatly from Massimo's profound knowledge and work ethic. He has taught me a lot about independent thinking, asking good questions, and digging into details no matter how obvious they seemingly are.  
As for Sebasti\'{a}n, I have learnt a lot from his work enthusiasm and dedication. He has taught me to go down to earth to implement ideas and isolate the difficulties, no matter how fancy or crazy they look. Massimo and Sebasti\'{a}n set up role models for theoretical physicists and perfect mentors which I wish I could inherit a bit.

I would also like to extend my appreciation to my thesis committee members, Sergei Dubovsky, Glennys Farrar, and Matthew Kleban, for their insightful suggestions and valuable help. 

I would like to express special gratitude to Jonathan J. Heckman for his kindly host at Upenn particle theory group. Jonathan has been a reliable collaborator and a perfect mentor. I have benefited a lot from his broad knowledge during our countless enlightening conversations. It is really amazing to have the opportunity to explore so many aspects of quantum field theories and string theory with him. I also want to extend my thanks to the whole Upenn particle theory group for the nicely host during the last two years of my graduate studies.

I would also like to thank my excellent collaborators: Azeem Hasan for fruitful discussions at the early stage of my graduate studies; Max Hubner for always generously sharing his ideas and teaching me so many things about geometric aspects of string theory; Alessandro Mininno for being an enthusiastic workmate; Ethan Torres for sharing his thoughts and insightful conversations in various topics; Andrew Turner for helping me understand and collaborating on holography and quantum information topics, as well as his help during my postdoc application; \'{A}ngel M. Uranga for teaching me various solid concepts in string theory; Hao Y. Zhang, for sharing ideas and knowledge in F-theory and for being a perfect coworker.

During my graduate studies, I have been fortunate to interact with many great physicists and mathematicians, including Vijay Balasubramanian, Florent Baume, Nathan Benjamin, T Daniel Brennan, Sergei Gukov, Amihay Hanany, Monica J.Kang, Ho Tat Lam, Craig Lawrie, Si Li, Qiuyue Liang, Henry Maxfield, Gregg Musiker, Hirosi Ooguri, Monica Pate, Du Pei, Martin Roček, Victor Alonso Rodriguez, Mauricio Romo, Vladimir Rosenhaus, Savdeep Sethi, Shu-Heng Shao, Marcus Sperling, Jonathan Sorce, Zimo Sun, Cumrun Vafa, Yifan Wang, Fei Yan, Zhenbin Yang, and Shing-Tung Yau. Discussions with them have become an essential part of shaping me as a physicist. 

Among all these great scientists, I would like to express my special thanks to Ho Tat Lam, Du Pei, Shu-Heng Shao, and Yifan Wang. I want to thank Ho Tat for always answering my naive questions, discussing my naively proposed idea, and teaching me many exciting aspects of global symmetries. I want to thank Du for sharing his fruitful knowledge on mathematical physics, for his patient explanation during our collaborations, and also for being a perfect mentor during my hard time. I would like to thank Yifan for always resolving my confusions quickly and for all our insightful discussions, from which I indeed benifited a lot, and for him giving honest career advice to me. Last but not least, I would like to strongly express my gratitude to Shu-Heng for him training me 2D CFT technique and collaboration at the early stage of my graduate studies, for him being patient no matter how annoying I am when I ask questions, and for always explaining nicely the spirit of his latest works to me.


I would like to thank all my friends in CCPP and NYU physics department for their support and for being friendly workmates together. I would also like to thank all my friends in the US and in China for coloring my life, cheering me up, and making the six years of my doctoral study possible.

I want to thank all my family. In particular, I am very grateful to my parents, Zhenguo Yu and Hongyue Zhang, for showing faith in me, giving me
the liberty to choose what I desire, and encouraging me to pursue it. I appreciate my parents-in-law, Yi-Ming Chu and Wen-Wan Hsueh,
for their support and for providing care to my son.

And most of all, I owe my deepest gratitude to my wife, Fang-Yi Chu, for her affection,
encouragement, and understanding. It is super hard for two Ph.D. students to raise a baby, but we
made it! My thesis acknowledgment would be incomplete without thanking my little boy,
Isaac Daohuan Yu. Having him midway through my Ph.D. was undoubtedly difficult for us, but he made
my life wonderful. My parents, wife, and son form the backbone of my success, and I
dedicate my dissertation to them.

%% file: abstract.tex
This dissertation presents a study of dualities and generalized global symmetries in quantum field theories (QFTs) from the string theory perspective. Chapter 2 is based on the work with Sebasti\'{a}n Franco, Alessandro Minino and \'{A}ngel M. Uranga \cite{Franco:2021ixh}. It introduces a new class of string theory backgrounds, Spin(7) orientifolds, allowing for the engineering of 2$d$ $\mathcal{N}=(0,1)$ gauge theories on D1-branes and illustrating the perspective on 2$d$ $\mathcal{N}=(0,1)$ theories as real slices of 2$d$ $\mathcal{N}=(0,2)$ ones. Chapter 3 is based on the work with Sebasti\'{a}n Franco, Alessandro Minino and \'{A}ngel M. Uranga \cite{Franco:2021vxq}. It presents a new, geometric perspective on the triality of 2$d$ $\mathcal{N}=(0,1)$ gauge theories, based on their brane engineering introduced in Chapter 2. It also shows that general Spin(7) orientifolds extend triality to theories consisting of coupled 2$d$ $\mathcal{N}=(0,2)$ and $\mathcal{N}=(0,1)$ sectors, leading to extensions of triality that interpolate between pure 2$d$ $\mathcal{N}=(0,2)$ and $\mathcal{N}=(0,1)$ cases. Chapter 4 is based on the work with Jonathan J. Heckman, Max Hubner, Ethan Torres and Hao Y. Zhang \cite{Heckman:2022xgu}. It presents a top-down construction of non-invertible duality symmetries in 4$d$ QFTs. The realization of QFTs is through D3-branes probing a Calabi-Yau threefold with an isolated singularity. The non-invertible duality defect then arises from configurations of 7-branes "at infinity". The study shows that different field-theoretic realizations of duality defects simply amount to distinct choices of where to place 7-brane branch cuts in the 5D bulk.

%% file: chapters/1.tex
Quantum Field Theory (QFT) is an active area of research in theoretical physics, with a wide range of applications from high-energy physics to condensed matter physics. Despite its great success so far, there are still open questions in QFT at the fundamental level. In particular, investigating non-perturbative aspects in QFT can be challenging. This is where the powerful concepts of duality and symmetry come in.

Duality refers to the idea that two apparently different physical theories can be equivalent in some sense. In the context of QFT, duality has proven to be an incredibly powerful tool for understanding non-perturbative effects, since one strongly coupled theory can be dual to a weakly coupled one. Two seemingly unrelated theories can actually describe the same physics, giving us insight into the underlying structure of the theory.

Symmetry is another powerful concept in QFT. It refers to the idea that a physical system remains unchanged under certain transformations. In the context of QFT, symmetry can help  simplify calculations and constrain the possible interactions, sometimes even without the need for a Lagrangian or a Hamiltonian. In particular, the concept of generalized global symmetries has been thoroughly investigated in recent years and provides many insights into QFTs in diverse dimensions.

A celebrated approach to studying QFTs is brane engineering from string theory. This approach involves constructing QFTs living on the worldvolume of branes (extended objects in higher-dimensional spaces) and studying their properties within non-trivial string theory backgrounds. Brane engineering is particularly useful because it naturally gives geometric perspectives of dualities and symmetries of QFTs. Furthermore, it can also provide evidences for new dualities and symmetries, which are not apparent from the field-theoretic perspective. 

This introduction presents the basic concepts necessary for understanding this dissertation's main content. We begin by introducing the basic setup of brane engineering in string theory in Section \ref{sec:brane engineering}. Then in Section \ref{sec:dualities from brane engineering}, we discuss generically how QFT dualities can be viewed from the brane engineering perspective. In Section \ref{sec:Generalized Global Symmetries and Branes}, we give a lighting review of generalized global symmetries and how to engineer them using branes. Section \ref{sec:overview and summary} presents an overview and summary of this dissertation.

\section{Brane Engineering}\label{sec:brane engineering}

In string theory, branes are fundamental objects that play a central role in many aspects of the theory. They are extended objects in spacetime, and can be thought of as hypersurfaces that are embedded in a higher-dimensional space. Branes exist in different dimensions, with the most common examples being D-branes, which can serve as boundaries for open strings and carry charges for certain Ramond-Ramond (RR) gauge fields in the closed-string sector.

Branes can be used as a powerful tool for studying QFTs in the context of string theory. One of the most widely studied setups is $N$ D3-branes in the flat space in IIB string theory. The low-energy effective theory on the worldvolume of these coincident D3-branes is the celebrated $SU(N)$ Yang-Mills theory with $\mathcal{N}=4$ supersymmetry. Many new results have been discovered based on this brane engineering, including the famous AdS/CFT correspondence \cite{Maldacena:1997re}. 

\begin{figure}
    \centering
    \includegraphics[width=7cm]{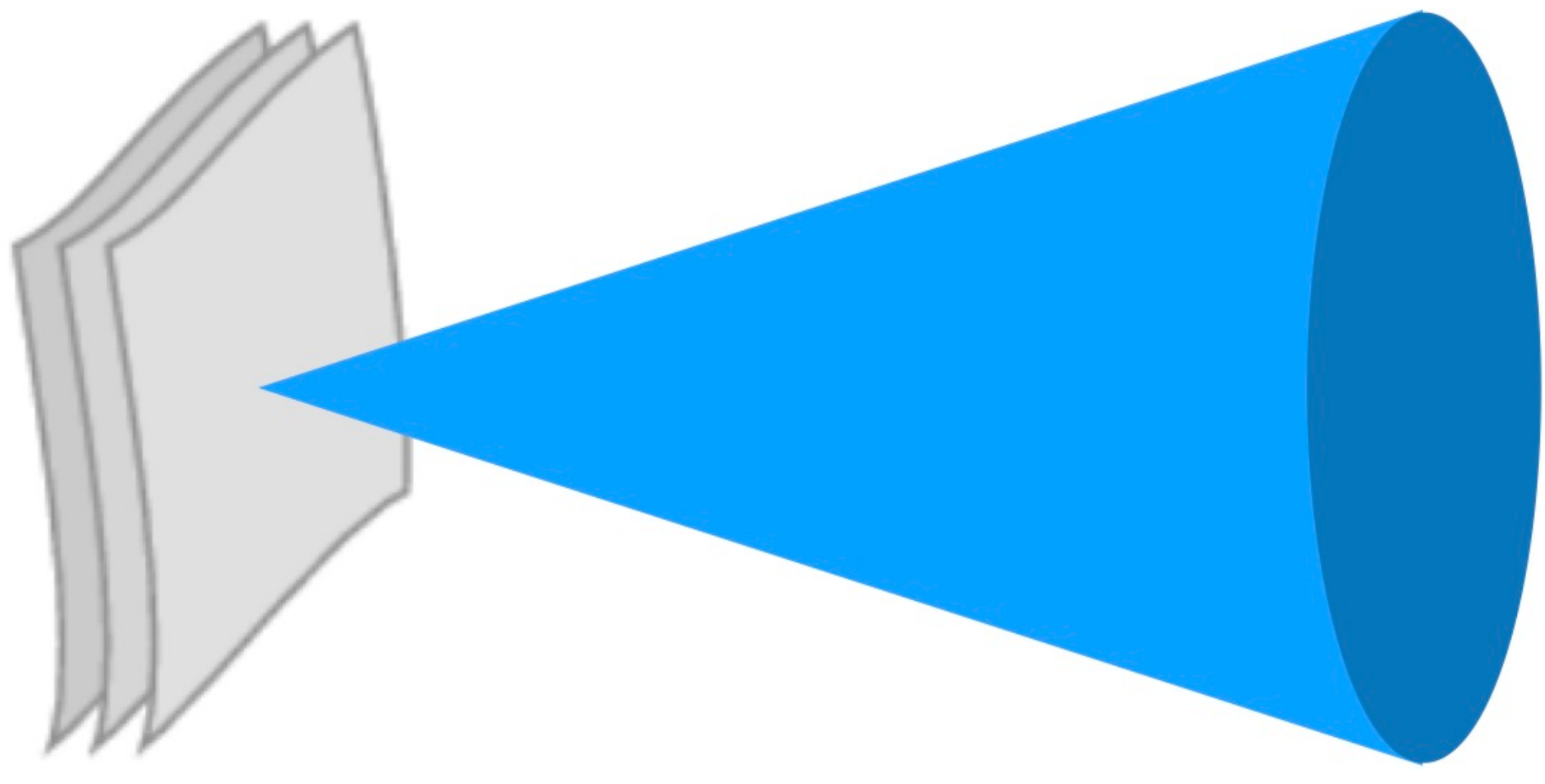}
    \caption{Branes probe a noncompact singular geometry. QFTs decoupled from gravity are determined by the degrees of freedom localized on branes based on the property of the singularity.}
    \label{fig:brane probing singularity}
\end{figure}

In addition to considering a single stack of branes sitting in the flat spacetime, there are various ways to generalize the configuration of the brane engineering. In this dissertation, we focus on branes probing singularities. In this setup, the branes are placed at a singularity in the spacetime geometry.  We further require the singular geometry to be locally a noncompact space; thus, gravity is decoupled due to the infinite volume of the geometry probed by branes. See Figure \ref{fig:brane probing singularity} for an illustration. Therefore, one is allowed to investigate QFTs with degrees of freedom solely localized on branes. This setup translates the study of QFTs to the study of the behavior of the singularity probed by branes. 

\begin{figure}
    \centering
    \includegraphics[width=4cm]{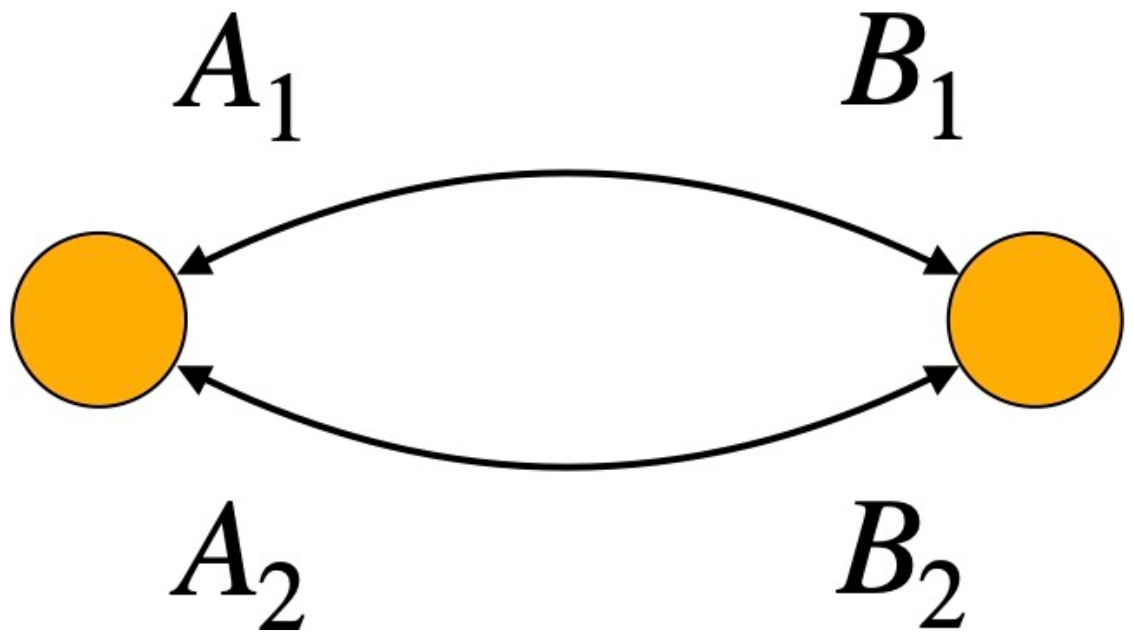}
    \caption{The quiver shows field content of the Klebanov-Witten Theory. Each round node denotes a $SU(N)$ gauge factor. Four arrows denote bifundamental chiral superfields, named as $A_{1,2}$ and $B_{1,2}$.}
    \label{fig:KW quiver}
\end{figure}

For example, consider IIB string theory with background $\mathbb{R}^{1,3}\times X_6$ where $\mathbb{R}^{1,3}$ is the $4d$ spacetime and $X_6$ is 6-dimensional geometry defined by equation $xy-z^2=0$. This geometry, known as the conifold, has a singularity at $x=y=z=0$. Now introduce $N$ D3-branes extending the 4$d$ spacetime and placed at the singularity. The resulting QFT living on the D3-branes is a 4$d$ $SU(N)\times SU(N)$ gauge theory with $\mathcal{N}=1$ supersymmetry, known as the Klebanov-Witten theory \cite{Klebanov:1998hh}. Its matter content consists of four chiral superfiels under the bifundamental representation of gauge symmetry. The matter content can be summarized in the quiver shown in Figure \ref{fig:KW quiver}.
There are also interactions between matter fields, which can be expressed in the following superpotential.
\begin{equation}
    W=\text{Tr}\left[ A_1B_1A_2B_2-A_1B_2A_2B_1 \right].
\end{equation}

The brane setup for the Klebanov-Witten theory can be regarded as a generalization of the engineering of $\mathcal{N}=4$ Super Yang-Mills theory obtained by substituting the conifold geometry $xy=z^2$ to the flat space $\mathbb{C}^3$. In fact, the conifold belongs to an infinite family of 6-dimensional geometries known as Calabi-Yau 3-folds. This class of geometries has a special holonomy group $SU(3)$, and the D3-brane probe of these geometries gives rise to an infinite family of  $\mathcal{N}=1$ gauge theories (see, e.g., \cite{Franco:2005rj}). This is our main stage in Chapter \ref{cha:ch4}. 

More generically, one can consider the D-dimensional string/M-theory background $\mathbb{R}^{1,p}\times X_{D-1-p}$, with D=10 for string theory and D=11 for M-theory. Introduce branes, with $p$ spatial dimensions, extending in $\mathbb{R}^{1,p}$ and probing the singularity of $X_{D-1-p}$. Certain classes of $X_{D-1-9}$ will engineer interesting and tractable QFTs on the worldvolume of branes. In Chapter \ref{cha:ch2} and \ref{cha:ch3}, we study the case $D=10, p=1$, namely D1-branes probing 8-dimensional geometry.

\section{QFT Dualities from Brane Engineering}
\label{sec:dualities from brane engineering}
QFTs can enjoy dualities - a phenomenon implying that different theories may describe the same physical system. One of the simplest examples is the Kramers-Wannier duality, which relates the free energy of a $2d$ Ising model at a low temperature to another Ising model at a high temperature. Various dualities have been found in diverse dimensions, especially for QFTs with supersymmetry. 

\begin{figure}
    \centering
    \includegraphics[width=10cm]{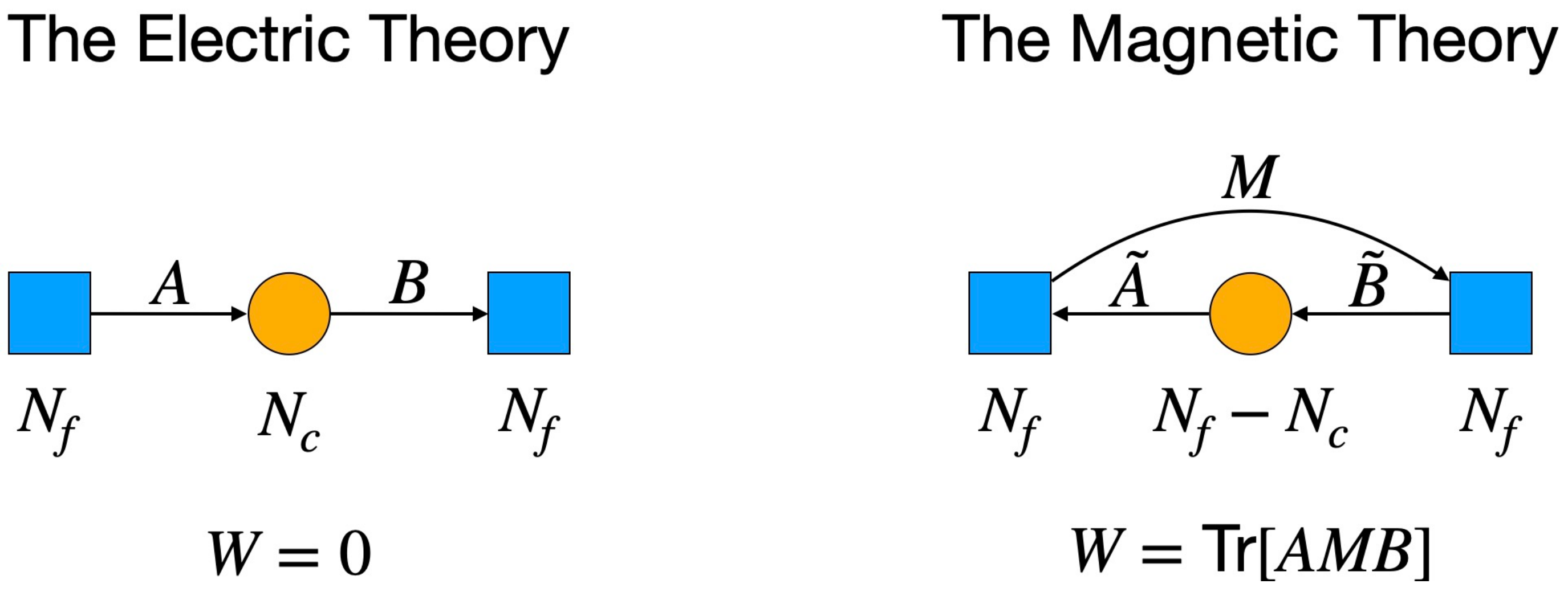}
    \caption{The electric and magnetic theories under Seiberg duality. They have different gauge symmetries, field contents, and interactions but have the same flavor symmetry.}
    \label{fig:Seiberg duality}
\end{figure}

A celebrated example is the Seiberg duality for $\mathcal{N}=1$ supersymmetric gauge theories. This duality is found in \cite{Seiberg:1994pq} by considering two simple $\mathcal{N}=1$ gauge theories. The electric theory is a $SU(N_c)$ gauge theory with $2N_f$ flavors. This theory can be regarded as a close cousin of QCD with $N_c$ colors and $N_f$ flavors. Seiberg found that there exists a magnetic theory dual to the electric one; namely, they both flow to the same infrared physics. The magnetic theory has quite a different description from the electric one: It has $SU(N_f-N_c)$ gauge symmetry, $N_f$ dual flavors, and $N_f^2$ gauge-invariant mesons with interactions between flavors and mesons. This duality is summarized in Figure \ref{fig:Seiberg duality}

\begin{figure}[H]
    \centering
    \includegraphics[width=10cm]{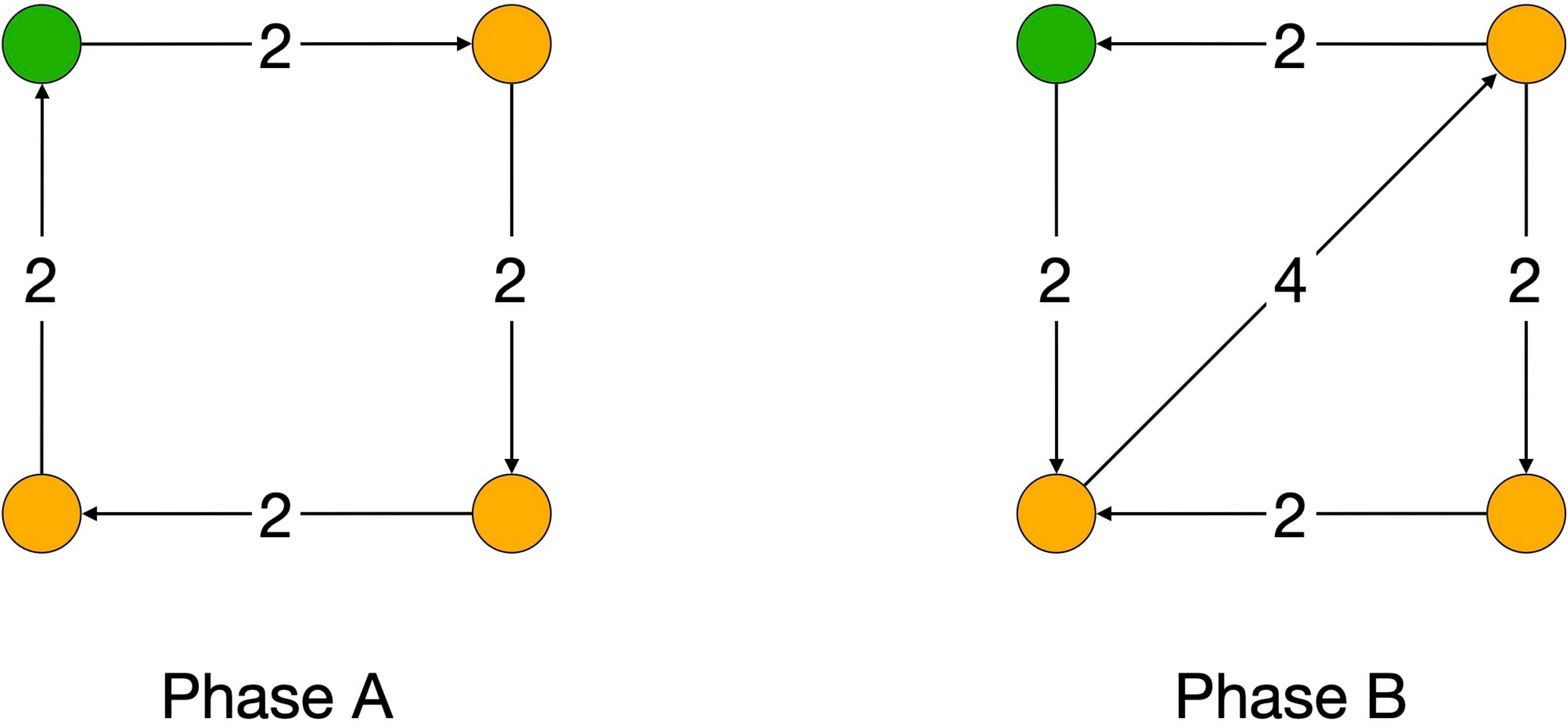}
    \caption{From the string theory perspective, one gets two gauge theories constructed from D3-branes probing the complex cone over $\mathbb{F}_0$. From the field theory perspective, they are dual theories under Seiberg duality on the upper left gauge node.}
    \label{fig:phasesF0}
\end{figure}

As we discussed in Section \ref{sec:brane engineering}, an infinite family of $4d$ $\mathcal{N}=1$ gauge theories can be engineered on D3-branes probing Calabi-Yau 3-fold singularities; it is natural to ask whether this brane engineering provides a geometric interpretation of the Seiberg duality. This is indeed the case. For example, consider a Calabi-Yau 3-fold $X$ as the complex cone over the Zeroth Hirzebruch surface $\mathbb{F}_0\simeq \mathbb{P}^1\times \mathbb{P}^1$. Computing the effective QFT living on the D3-brane probe, one finds the result is not unique. There are two theories referred to as phase A and B, both corresponding to this geometry. They are shown in Figure \ref{fig:phasesF0}.
These two gauge theories describe the low energy physics of the same system, namely D3-branes probing the complex cone over $\mathbb{F}_0$, so they can be regarded as dual theories. Interestingly, the duality connecting these two theories is exactly the Seiberg duality \cite{Franco:2002mu}. To see this, one can consider, for instance, the upper left gauge node in \ref{fig:phasesF0} as the gauge group and all other nodes as flavor groups, then phases A and B are dual to each other following exactly the same transformation rule as the original Seiberg duality shown in \ref{fig:Seiberg duality}. 

The above example falls in a generic approach to the geometric interpretation of QFT dualities from brane engineering: for branes probing a certain geometry background in string theory or M-theory, the non-uniqueness of the resulting QFTs computed from the geometric data can give rise to field-theoretic dualities. This idea provides new results for existing QFT dualities in diverse dimensions with various amounts of supersymmetry (see, e.g., \cite{Franco:2020ijt}). Furthermore, brane engineering also shows evidence for new QFT dualities, which may not be fully recognized from the field-theoretic point of view. In Chapter \ref{cha:ch3}, we will see a new type of dualities shows up in $2d$ gauge theories via brane engineering.

\section{Generalized Global Symmetries and Branes}\label{sec:Generalized Global Symmetries and Branes}

One of the recent significant advances in the study of QFT is the realization that generalized symmetries can be better understood through corresponding topological operators \cite{Gaiotto:2014kfa}. 

The simplest example to start with is an ordinary global symmetry. Consider a $d$-dimensional QFT with a continuous internal global symmetry $G$. The corresponding conserved current $J^\mu$ satisfies $\partial_\mu J^\mu=0$ implied by Noether’s theorem. Define a $(d-1)$-form $J\equiv \epsilon_{\mu_1\cdots \mu_d}J^{\mu_1}dx^{\mu_2}\cdots dx^{\mu_d}$ and then translate the conserved current equation as the condition that $J$ is a closed form $dJ=0$. The conserved charge operator is then supported on closed $(d-1)$-manifolds $M_{d-1}$:
\begin{equation}
    Q(M_{d-1})=\oint_{M_{d-1}}J.
\end{equation}
Exponentiating charge operators, we derive symmetry operators $U_g(M_{d-1})$ labeled by group elements $g\in G$. Since $J$ is a closed form, the symmetry operator is unchanged under small deformation of the manifold $M_{d-1}$. Therefore, $U_g(M_{d-1})$ is topological operators. The action of topological symmetry operators act on a local charged operator $O_i(x)$ in a representation $R$ as
\begin{equation}
    U_g(M_{d-1})O_i(x)=R^j_i(g)O_j(x),
\end{equation}
where $M_{d-1}$ surrounds the point $x$, and $R^j_i$ is the representation of $g$. The fusion of symmetry operators is given by the group multiplication law as 
\begin{equation}
    U_{g_1}(M_{d-1})U_{g_2}(M_{d-1})=U_{g_1g_2}(M_{d-1}).
\end{equation}

The idea of generalized global symmetries is realized by changing the properties of the topological symmetry operators. Generalize the supporting manifold $M_{d-1}$ to have lower dimensions $M_{d-p-1}$, we get higher $p$-form symmetries. In this sense, the ordinary symmetry is 0-form symmetry \cite{Gaiotto:2014kfa}. Interestingly, the fact that generalized symmetry operators are topological also suggests that the “worldvolume” itself may support a non-trivial topological field theory (TFT). One striking consequence of this fact is that the product of two symmetry operators may produce a sum of symmetry operators 
\begin{equation}\label{eq:non-invertible fusion rule}
    U_i(M_{d-p-1})U_j(M_{d-p-1})=\sum_k c_{ijk}U_k(M_{d-p-1}),
\end{equation}
i.e., there can be a non-trivial fusion rule instead of the group multiplication law. Note that the right-hand side of (\ref{eq:non-invertible fusion rule}) is not a single operator but a sum of operators. This implies in general there is no inverse for the operator under this fusion rule. This is closely tied to the appearance of “non-invertible” symmetry operators. 

Given the fact that many QFTs have elegant string theory realization, it is natural to ask whether these topological symmetry operators can be directly realized in terms of objects in string theory. In particular, one might hope that performing this analysis could provide additional insight into the associated worldvolume topological field theories for non-invertible symmetry operators and provide a systematic method for extracting the corresponding fusion rules for these generalized symmetry operators.

Along these lines, it was recently shown in \cite{Apruzzi:2022rei, GarciaEtxebarria:2022vzq,Heckman:2022muc} that for QFTs constructed via geometric and brane engineering, generalized symmetry operators are obtained from “branes at infinity.” The resulting defects are topological in the sense that they do not contribute to the stress-energy tensor of the QFT localized around the singularity. Starting from the topological terms of a brane “at infinity,” one can then extract the resulting TFT on its worldvolume, and consequently extract the resulting fusion rules for the associated generalized symmetry operators. In Chapter \ref{cha:ch4}, we will follow this idea and engineer non-invertible duality operators \cite{Choi:2021kmx} in 4D QFTs via non-perturbative 7-branes.

\section{Overview and Summary}\label{sec:overview and summary}
This dissertation's overall theme is exploring non-perturbative dualities and global symmetries in QFTs, and their brane engineering perspective. We first explore the brane engineering of a new class of $2d$ minimally supersymmetric gauge theories. We then move on to study the dualities of these $2d$ theories from a string theory perspective. Finally, we discuss a brane realization of non-invertible duality symmetry defects in $4d$ supersymmetric gauge theories. The content of each chapter is summarized below.

In Chapter \ref{cha:ch2} based on \cite{Franco:2021ixh}, we initiate the geometric engineering of 2d $\mathcal{N}=(0,1)$ gauge theories on D1-branes probing singularities. To do so, we introduce a new class of backgrounds obtained as quotients of Calabi-Yau 4-folds by a combination of an anti-holomorphic involution leading to a Spin(7) cone and worldsheet parity. We refer to such constructions as {\it Spin(7) orientifolds}. Spin(7) orientifolds explicitly realize 2d $\mathcal{N}=(0,1)$  theories as real slices of $\mathcal{N}=(0,2)$ ones. Remarkably, this projection is geometrically realized as Joyce’s construction of Spin(7) manifolds via quotients of Calabi-Yau 4-folds by anti-holomorphic involutions \cite{Joyce:1999nk}. We illustrate this construction in numerous examples with both orbifold and non-orbifold parent singularities, discuss the role of the choice of vector structure in the orientifold quotient, and study partial resolutions.

In Chapter \ref{cha:ch3} based on \cite{Franco:2021vxq}, we present a new, geometric perspective on the recently proposed triality of 2d $\mathcal{N}=(0,1)$ gauge theories, based on its engineering in terms of D1-branes probing Spin(7) orientifolds. In this context, triality translates into the fact that multiple gauge theories correspond to the same underlying orientifold. We show how Spin(7) orientifolds based on a particular involution, which we call the universal involution, gives rise to precisely the original version of $\mathcal{N}=(0,1)$ triality. Interestingly, our work also shows that the space of possibilities is significantly richer. Indeed, general Spin(7) orientifolds extend triality to theories that can be regarded as consisting of coupled $\mathcal{N}=(0,2)$ and $(0,1)$ sectors. The engineering of 2d gauge theories in terms of D1-branes at singularities therefore leads to extensions of triality that interpolate between the pure $\mathcal{N}=(0,2)$ and $(0,1)$ cases.

In Chapter \ref{cha:ch4} based on \cite{Heckman:2022xgu}, we present a top-down construction of non-invertible duality defects, which exist in QFTs with lower supersymmetry, where 0-form symmetries are often present. 
We realize the QFTs of interest via D3-branes probing $X$, which is a Calabi-Yau threefold cone with an isolated 
singularity at the tip of the cone. The IIB duality group descends to dualities of the 4D worldvolume theory. Non-trivial codimension-one topological interfaces arise from configurations of 7-branes ``at infinity'' that implement a suitable $SL(2,\mathbb{Z})$ transformation when the defect crosses 7-branes. Dimensional reduction of the topological term in IIB string theory on the boundary $\partial X$ results in a 5D TFT, known as the symmetry TFT \cite{Kaidi:2022cpf}. Different realizations of duality defects, such as the gauging of 1-form symmetries with certain mixed anomalies 
and half-space gauging constructions, which will be defined in Chapter \ref{cha:ch4}, simply amount to distinct choices of where to place the branch cuts in the 5D bulk.

Appendix A, based on \cite{Franco:2021ixh}, serves as a supplement to Chapter \ref{cha:ch2}. It presents details of $2d$ theories on D1-branes probing $\mathbb{C}^4/\mathbb{Z}_2\times \mathbb{Z}_2$ and SPP$\times \mathbb{C}$ geometries, as well as their Spin(7) orientifolds. Appendix B, based on \cite{Franco:2021vxq}, serves as a supplement to Chapter \ref{cha:ch3}. It presents details of $2d$ theories on D1-branes probing $Q^{1,1,1}/\mathbb{Z}_2$ geometry, as well as its Spin(7) orientifolds. Appendix C, based on \cite{Heckman:2022xgu}, serves as a supplement to Chapter \ref{cha:ch4}. It presents how the various defects in  Chapter \ref{cha:ch4} are implemented
in other top down constructions. It also provides a proposal for reducing the relevant topological
terms of a non-perturbative 7-brane to a suitable 3D TFT, as well as details on the $4d$ theory of D3-branes probing $\mathbb{C}^3/\mathbb{Z}_3$.

%% file: chapters/2.tex
\section{Introduction}
Engineering gauge theories in string or M-theory provides alternative perspectives, often geometric, on their dynamics. Such realizations typically lead to a deeper understanding of the theories at hand, suggest natural generalizations, and even contribute to the discovery of new results.

Our understanding of $2$d $\mathcal{N}=(0,2)$ gauge theories has significantly progressed in recent years. The new results include $c$-extremization \cite{Benini:2012cz,Benini:2013cda}, $\mathcal{N}=(0,2)$ triality \cite{Gadde:2013lxa} and connections to gauge theories in higher dimensions \cite{Benini:2013cda,Gadde:2013sca,Kutasov:2013ffl,Kutasov:2014hha,Benini:2015bwz}. These discoveries have fueled a renewed interest in the stringy engineering of such theories. A possible scenario involves realizing them on the world volume of D1-brane probing singular Calabi-Yau (CY) 4-folds.\footnote{For alternative setups leading to $2$d $\mathcal{N}=(0,2)$ gauge theories, see e.g. \cite{Benini:2013cda,Gadde:2013sca,Tatar:2015sga, Schafer-Nameki:2016cfr,Benini:2015bwz}.} Following the pioneering work of \cite{Garcia-Compean:1998sla}, a new class of brane configurations, denoted {\it  brane brick models}, was introduced in \cite{Franco:2015tya}. Brane brick models fully encode the $2$d $\mathcal{N}=(0,2)$ gauge theories probing toric CY 4-folds, to which they are connected by T-duality. Furthermore, they have significantly simplified the map between geometry and the corresponding gauge theories (see \cite{Franco:2016nwv,Franco:2016qxh,Franco:2016fxm,Franco:2017cjj,Franco:2018qsc,Franco:2020avj,Franco:2021elb} for further developments).

As usual, it is desirable to investigate theories with less supersymmetry. The next step corresponds to $2$d $\mathcal{N}=(0,1)$, namely minimally supersymmetric, theories. Such models are particularly interesting because while they are still supersymmetric, they no longer have holomorphy. While considerably less is known about them, new results about their dynamics have appeared in \cite{Gukov:2019lzi}, including the proposal of a new $2$d $\mathcal{N}=(0,1)$ triality. Once again, this raises the question of how to engineer these theories in string theory. In \cite{Gukov:2019lzi}, it was noted that the theories participating in $\mathcal{N}=(0,1)$ triality are, in a sense, ``real slices” of their ``complex” $\mathcal{N}=(0, 2)$ counterparts, both at the level of gauge theory description and effective non-linear sigma model. A more general formulation of such $\mathcal{N}=(0,2)/(0,1)$ correspondence was left as an open question. 

With these motivations in mind, in this chapter, we introduce \emph{Spin(7) orientifolds}, a new class of backgrounds that combine Joyce’s construction of Spin(7) manifolds via the quotient of CY 4-folds by anti-holomorphic involutions with worldsheet parity, and construct $2$d $\mathcal{N}=(0,1)$ gauge theories on D1-branes probing them. Closely related ideas were presented in the insightful paper \cite{Forcella:2009jj,Amariti:2014ewa}, whose goal was to engineer 3d $\mathcal{N}=1$ theories on M2-branes.\footnote{For applications to F-theory of Spin(7) holonomy manifolds from CY 4-folds quotients see, e.g.,~\cite{Bonetti:2013fma,Bonetti:2013nka}.}

This chapter is organized as follows. Section~\ref{sec:2dN01QFT} discusses the general structure and properties of $2$d $\mathcal{N}=(0,1)$ field theories. Section~\ref{sec:N02inN01Form} presents the decomposition of $\mathcal{N}=(0,2)$ supermultiplets in $\mathcal{N}=(0,1)$ language. Section~\ref{sec:2dN01orientifolds} explains the construction of Spin(7) cones and Spin(7) orientifolds starting from CY $4$-folds. Sections~\ref{sec:orientN01gaugeth} and \ref{sec:N01theorfromorienquot} discuss the field theory implementation of Spin(7) orientifolds. The connection between the anti-holomorphic involutions of the CY$_4$ and the gauge theory is studied in Section~\ref{sec:HSreview}. Section~\ref{sec:examplesenginN=01} considers Spin(7) orientifolds of $\mathbb{C}^4$ and its orbifolds. In Section~\ref{sec:vectorstructure} we decribe how the choice of vector structure can lead to different  gauge theories associated to the same geometric involution. Section~\ref{sec:BeyondOrbif} presents Spin(7) orientifolds of generic, non-orbifold, parent CY$_4$’s. Finally, Section~\ref{sec:HiggsingPartResol} investigates the interplay between partial resolution and higgsing. Section~\ref{section_conclusions} collects our conclusions and outlook. Appendix~\ref{app:C4Z2Z2SPPC} contains additional examples that are used in Section~\ref{sec:HiggsingPartResol}.

\section{2d $\mathcal{N}=(0,1)$ Field Theories}
\label{sec:2dN01QFT}

In this section, we briefly review the general structure of $2$d $\mathcal{N}=(0,1)$ field theories. Instead of discussing all terms in the Lagrangian, we will focus on the main facts we will use in following sections. We refer the reader to \cite{Sakamoto:1984zk,Hull:1985jv,Brooks:1986uh,Brooks:1986gd,Gukov:2019lzi} for a more detailed presentation.

\subsection{Constructing 2d $\mathcal{N}=(0,1)$ gauge Theories}
\label{sec:2dN01construT}

We describe these theories in terms of $2$d $\mathcal{N}=(0,1)$ superspace $\left(x^0,x^1,\theta^+\right)$. 
There are three types of supermultiplets as elementary building blocks:
\begin{itemize}
\item Vector multiplet:
\begin{equation}\label{(0,1) vector}
\begin{split}
		V_+&=\theta^+(A_0(x)+A_1(x)),\\
		V_-&=A_0(x)-A_1(x)+\theta^+\lambda_-(x).
\end{split}
\end{equation}
It contains a gauge boson $A_{\pm}$ and a left-moving Majorana-Weyl fermion $\lambda_-$ in the adjoint representation. 

\item Scalar multiplet:
\begin{equation}
\Phi(x,\theta)=\phi(x)+\theta^+\psi_+(x).
\end{equation}
It has a real scalar field $\phi$ and a right-moving Majorana-Weyl fermion $\psi_+$. 

\item Fermi multiplet:
\begin{equation}\label{(0,1) Fermi}
\Lambda(x,\theta)=\psi_-(x)+\theta^+F(x).
\end{equation}
It has a left-moving Majorana-Weyl spinor as its only on-shell degree of freedom. Here $F$ is an auxiliary field. 
\end{itemize}

As usual, the kinetic terms for matter fields and their gauge couplings are given by
\begin{equation}
	\begin{split}
\mathcal{L}_s+\mathcal{L}_F=\int d\theta^+~ \left(\frac{i}{2}\sum_i(\mathcal{D}_+\Phi_i\mathcal{D}_-\Phi_i)-\frac{1}{2}\sum_a(\Lambda_a\mathcal{D}_+\Lambda_a) \right),
\end{split}
\label{eq:kingauge01mattfields}
\end{equation}
where $\mathcal{D}_{\pm}$ are super-covariant derivatives~\cite{Gukov:2019lzi}. 

These theories admit another interaction, which is an $\mathcal{N}=(0,1)$ analog of the $\mathcal{N}=(0,2)$ $J$-term interaction, or $\mathcal{N}=1$ superpotential:
\begin{equation}
	\mathcal{L}_J\equiv \int d \theta^+W^{(0,1)}=\int d\theta^+\sum_a (\Lambda_aJ^a(\Phi_i)),
	\label{eq:W01super}
\end{equation}
where $J^a(\Phi_i)$ are real functions of scalar fields. Both the quiver and $W^{(0,1)}$ are necessary for fully specifying any of the $\mathcal{N}=(0,1)$ gauge theories considered in this chapter. From now on, we will refer to $W^{(0,1)}$ as the \emph{superpotential} for convenience.

After integrating out the auxiliary fields $F_a$, $\mathcal{L}_J$ produces various interactions, including Yukawa-like couplings 
\begin{equation}
	\sum_a \lambda_{-a}\frac{\partial J^a}{\partial \phi_i}\psi_{+i},
\end{equation}
as well as a scalar potential 
\begin{equation}
	\frac{1}{2}\sum_a(J^a(\phi_i))^2.
\end{equation}

\subsection{Anomalies}
\label{sec:Anomalies}

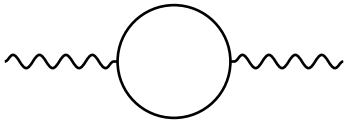
\begin{figure}[!htp]
	\centering
	\begin{tikzpicture}[scale=0.75]
	\draw[line width=1pt] (0,0) circle (1);
	\draw[line width=1pt,decoration=snake,decorate] (-3,0) -- (-1,0);
	\draw[line width=1pt,decoration=snake,decorate] (3,0) -- (1,0);
	\end{tikzpicture}
	\caption{Generic $1$-loop diagram associated with $2$d anomalies.}
	\label{fig:anom1loop}
\end{figure}

In $2$d, anomalies are given by $1$-loop diagrams of the generic form shown in Figure~\ref{fig:anom1loop}, where left- and right-moving fermions running in the loop contribute oppositely. 

Since $2$d $\mathcal{N}=(0,1)$ theories are chiral, left- and right-moving fermions are not necessarily paired up, and anomalies do not cancel automatically. For a given symmetry group, anomalies depend on the types and the representations of the fields transforming under it. Below, we focus on those groups and representations appearing in the $2$d $\mathcal{N}=(0,1)$ theories engineered in this chapter.

\paragraph{Non-Abelian Anomalies}\mbox{}

\smallskip

Let us first consider pure non-Abelian $G^2$ gauge or global anomalies, where $G$ can be $SU(N)$, $SO(N)$ or $USp(N)$ group.\footnote{In our convention $USp(2)\simeq SU(2)$, so $USp(N)$ makes sense only if $N$ is even.} 
The corresponding anomaly is given by
\begin{equation}
	\text{Tr}[\gamma^3J_GJ_G],
\end{equation}
where $\gamma^3$ is the chirality matrix in $2$d and $J_G$ is the current associated to $G$. The resulting anomaly from a field in representation $\rho$ of $G$ can be computed in terms of the Dynkin index $T(\rho)$: 
\begin{equation}
T(\rho)=C_2(\rho)\frac{d(\rho)}{d(\text{adjoint})},
\end{equation}
where $C_2(\rho)$ is the quadratic Casimir for representation $\rho$.

\begin{table}[!htp]
	\centering
\begin{tabular}{|Sc|Sc|Sc|Sc|Sc|}
	\hline
	$SU(N)$	& fundamental & adjoint & antisymmetric & symmetric\\
	\hhline{|=|=|=|=|=|} 
	vector multiplet & $\times$ & $-N$ & $\times$ &$\times$ \\
	\hline
	Fermi multiplet & $-\dfrac{1}{2}$ & $-N$ & $\dfrac{-N+2}{2}$ & $\dfrac{-N-2}{2}$\\
	\hline
	scalar multiplet & $\dfrac{1}{2}$ & $N$ &$\dfrac{N-2}{2}$ & $\dfrac{N+2}{2}$\\
	\hline
\end{tabular}
\caption{Anomaly contributions of the $2$d $\mathcal{N}=(0,1)$ multiplets in various representations of $SU(N)$. Since anomalies are quadratic in 2d, the same contributions apply for the conjugate representations.}
\label{tab:SUanomaly}
\end{table}

\begin{table}[!htp]
	\centering
\begin{tabular}{|Sc|Sc|Sc|Sc|}
	\hline
	$SO(N)$	& fundamental & antisymmetric (adjoint) & symmetric\\
	\hhline{|=|=|=|=|} 
	vector multiplet & $\times$ & $-N+2$ & $\times$ \\
	\hline
	Fermi multiplet & $-1$ & $-N+2$ & $-N-2$\\
	\hline
	scalar multiplet & $1$ & $N-2$ &$N+2$\\
	\hhline{|=|=|=|=|} 
	$USp(N)$	& fundamental & antisymmetric  & symmetric (adjoint)\\
	\hhline{|=|=|=|=|} 
	vector multiplet & $\times$ & $\times$ & $-N-2$ \\
	\hline
	Fermi multiplet & $-1$ & $-N+2$ & $-N-2$\\
	\hline
	scalar multiplet & $1$ & $N-2$ &$N+2$\\
	\hline
\end{tabular}
\caption{Anomaly contributions of the $2$d $\mathcal{N}=(0,1)$ multiplets in various representations of $SO(N)$ and $USp(N)$.}
\label{tab:SOUSpanomaly}
\end{table}

In Table~\ref{tab:SUanomaly} we present anomaly contributions for superfields in the most common representations of $SU(N)$. In Table~\ref{tab:SOUSpanomaly}, we present anomaly contributions of different types of superfields carrying various representations of $SO(N)$ and $USp(N)$ groups, computed using Dynkin indices listed in \cite{Yamatsu:2015npn}. 

In the case of gauge groups, anomalies must vanish for consistency of the theory at the quantum level. This leads to important constraints in our construction of $2$d $\mathcal{N}=(0,1)$ theories, that may require the introduction of extra flavors to cancel  anomalies. We will illustrate this with concrete examples in following sections.

Unlike gauge symmetries, global symmetries may indeed be anomalous. One important property of global anomalies is that they are preserved along the Renormalization Group (RG) flow. Therefore, they can be used to check dualities between two or more theories, namely whether these UV-different theories are IR-equivalent. Examples of using global anomalies to check dualities in $2$d $\mathcal{N}=(0,1)$ theories can be found in \cite{Gukov:2019lzi} and also in our upcoming work \cite{Franco:2021vxq}.

\paragraph{Abelian Anomalies}\mbox{}

\smallskip

For $U(N)$ groups of the worldvolume theories on D-brane probes, in addition to non-Abelian anomalies, the $U(1)$ factors can generically have $U(1)_i^2$ and mixed $U(1)_iU(1)_j$ Abelian anomalies. As before, the $U(1)$ groups can be either gauged or global.

The theories studied in this chapter generically have non-vanishing Abelian gauge anomalies. Similarly to the discussion in \cite{Franco:2015tna, Franco:2017cjj}, we expect that such anomalies are canceled by the bulk fields in the closed string sector via a generalized Green-Schwarz (GS) mechanism (see \cite{Ibanez:1998qp,Mohri:1997ef} for derivations in 4d ${\cal N}=1$ and $2$d ${\cal N}=(0,2)$ theories realized at orbifolds/orientifold singularities).

\subsection{Triality}
\label{sec:triality}

Recently, an IR triality between 2d $\mathcal{N}=(0,1)$ theories with $SO$ and $USp$ gauge groups was proposed in \cite{Gukov:2019lzi}. Evidence for the proposal includes matching of anomalies and elliptic genera. This new triality can be regarded as a cousin of the 2d $\mathcal{N}=(0,2)$ triality introduced in \cite{Gadde:2013lxa}. Interestingly, 2d $\mathcal{N}=(0,2)$ triality, together with Seiberg duality for 4d gauge theories \cite{Seiberg:1994pq}, extend to an infinite family of order $(m+1)$ dualities of $m$-graded quiver theories \cite{Franco:2016tcm,Franco:2017lpa,Closset:2018axq}.

It is natural to ask whether, within the context of gauge theories on the worldvolume of D-branes probing singularities, the $\mathcal{N}=(0,1)$ triality admits a geometric explanation. The similarity between the theories in \cite{Gukov:2019lzi} and the ones constructed in this chapter hints that this is the case. This question will be addressed in Chapter \ref{cha:ch3}, where we will show that $\mathcal{N}=(0,1)$ triality follows from the non-uniqueness of the map between Spin(7) cones and 2d $\mathcal{N}=(0,1)$ gauge theories.

\section{$\mathcal{N}=(0,2)$ Field Theories in $\mathcal{N}=(0,1)$ Formalism}
\label{sec:N02inN01Form}

In this section, we will construct $2$d $\mathcal{N}=(0,1)$ theories from $2$d $\mathcal{N}=(0,2)$ theories via orientifold quotients. Therefore, it is useful to decompose $\mathcal{N}=(0,2)$ theories in terms of the $\mathcal{N}=(0,1)$ formalism.

$\mathcal{N}=(0,2)$ theories can be expressed in superspace $\left(x^0,x^1, \theta^+, \bar{\theta}^+\right)$ and have three types of multiplets: vector, chiral and Fermi. These multiplets and the Lagrangian can be further expressed in $\mathcal{N}=(0,1)$ language using the superspace $\left(x^0, x^1, \theta^+\right)$.

\paragraph{$\mathcal{N}=(0,2)$ vector multiplet}\mbox{}

\smallskip

The $\mathcal{N}=(0,2)$ vector multiplet $V^{(0,2)}$ contains a gauge boson, a left-moving chiral fermion and an auxiliary field. It decomposes into $\mathcal{N}=(0,1)$ multiplets as follows:
\begin{equation}
\renewcommand{\arraystretch}{1.1}
    \begin{array}{c}
           \mathcal{N}=(0,2) \text{ vector multiplet } V_i^{(0,2)}  \\
         \swarrow \qquad \searrow  \\
    \mathcal{N}=(0,1) \text{ vector multiplet } V_i  \quad \oplus\quad  \mathcal{N}=(0,1) \text{ Fermi multiplet } \Lambda_{i}^R.
    \end{array}
\end{equation}
The chiral fermion in $V^{(0,2)}_i$ is separated into two Majorana-Weyl fermions, one of which is included in $V_i$ and the other is in $\Lambda_{i}^R$. The auxiliary field in $V^{(0,2)}_i$ becomes the one in $\Lambda_{i}^R$.

The kinetic term of $V_i^{(0,2)}$ in the Lagrangian can be expressed in $\mathcal{N}=(0,1)$ superspace as kinetic terms of $V_i$ and $\Lambda_{i}^R$:
\begin{equation}
\mathcal{L}_{\text{gauge}}^{(0,2)}\rightarrow \mathcal{L}_{\text{gauge}}-\frac{1}{2}\int d \theta^+\sum_a(\Lambda_{i}^R\mathcal{D}_+\Lambda_{i}^R),
\end{equation}
where $\mathcal{L}_{\text{gauge}}$ is the kinetic term of an $\mathcal{N}=(0,1)$ vector multiplet.

\paragraph{$\mathcal{N}={(0,2)}$ chiral multiplet}\mbox{}

\smallskip

The $\mathcal{N}=(0,2)$ chiral superfield contains a complex scalar $\phi^c$ and a right-moving chiral fermion $\psi^c_+$. Its expansion is
\begin{equation}
	\Phi^{(0,2)}_m=\phi^c_m +\theta^+\psi_{+m}^c-i\theta^+ \bar{\theta}^+D_+\phi^c_m.
\end{equation}
It decomposes into $\mathcal{N}=(0,1)$ multiplets as follows:
\begin{equation}
\renewcommand{\arraystretch}{1.1}
    \begin{array}{c}
           \mathcal{N}=(0,2) \text{ chiral multiplet } \Phi^{(0,2)}_m  \\
         \swarrow \qquad \searrow  \\
    \mathcal{N}=(0,1) \text{ scalar multiplet } \Phi^{1}_m \quad \oplus \quad \mathcal{N}=(0,1) \text{ scalar multiplet } \Phi_m^2.
    \end{array}
\end{equation}
The two $\mathcal{N}=(0,1)$ scalar multiplets $\Phi_m^{1,2}$ can be further combined into an $\mathcal{N}=(0,1)$ \emph{complex scalar multiplet}, so that the above decomposition is rewritten as \begin{equation}\label{(0,2) chiral in (0,1) complex scalar}
\renewcommand{\arraystretch}{1.1}
    \begin{array}{c}
           \mathcal{N}=(0,2) \text{ chiral multiplet } \Phi^{(0,2)}_m  \\
         \downarrow  \\
     \mathcal{N}=(0,1) \text{ complex scalar multiplet } \Phi_m.
    \end{array}
\end{equation}
The kinetic terms of the matter fields in $\Phi^{(0,2)}_m$ and their gauge couplings are included in the term $\mathcal{L}^{(0,2)}_{\text{chiral}}$ in $\mathcal{N}=(0,2)$ superspace. As an example, let us consider a chiral multiplet $\Phi^{(0,2)}_m$ transforming under a $U(1)$ gauge group. In this case, $\mathcal{L}_{\text{chiral}}^{(0,2)}$ reads: 
\begin{equation}
\begin{split}
		\mathcal{L}^{(0,2)}_{\text{chiral}}&=-\frac{i}{2}\int d\theta^+ d\bar{\theta}^+ (\Phi^{(0,2)}_m)^\dagger\mathcal{D}_-^{(0,2)}\Phi^{(0,2)}_m,
\end{split}
\end{equation}
where with $\dagger$ we mean the Hermitian conjugate\footnote{I.e. complex conjugate and transposition,  $\left(\bar{\Phi}^{(0,2)}\right)^T$.} of $\Phi^{(0,2)}_m$. The above Lagrangian can be regarded as a combination of two parts:
\begin{equation}\label{two parts of (0,2) chiral kinetic terms}
	\mathcal{L}^{(0,2)}_{\text{chiral}}=\text{Kinetic terms of $\Phi^{(0,2)}_m$}+\text{Interaction terms between $V^{(0,2)}$ and $\Phi^{(0,2)}_m$,}
\end{equation}
which can be further expressed in terms of $\mathcal{N}=(0,1)$ multiplets as 
\begin{equation}\label{(0,2) chiral lagrangian in (0,1) multiplets}
\begin{split}
	\mathcal{L}^{(0,2)}_{\text{chiral}}\rightarrow &~~\text{Kinetic terms of $\mathcal{N}=(0,1)$  complex scalar $\Phi_m$}\\
	&+\text{Interaction terms between $V$ and $\Phi_m$}\\
	&+\text{Interaction terms between $\Lambda^R$ and $\Phi_m$ .}
\end{split}
\end{equation}
$V$ and $\Lambda^R$ here are $\mathcal{N}=(0,1)$ vector and adjoint Fermi multiplets coming from the decomposition of $V^{(0,2)}$. From now on, the superscript $R$ is used to emphasize that a superfield is real.

$\mathcal{L}^{(0,2)}_{\text{chiral}}$ can be expressed in $\mathcal{N}=(0,1)$ superspace using~(\ref{eq:kingauge01mattfields}) and (\ref{eq:W01super}). It becomes
\begin{equation}
\begin{split}
\mathcal{L}^{(0,2)}_{\text{chiral}}&\rightarrow \mathcal{L}_{s}+\int d \theta^+ W^{(0,1)}\\
	&=-\frac{i}{4}\int d \theta^+[\mathcal{D}_+\Phi^\dagger_m\mathcal{D}_-\Phi_m+\mathcal{D}_+\Phi_m\mathcal{D}_-\Phi^\dagger_m]+\int d\theta^+\Lambda^R \Phi_m^{\dagger}\Phi_m.
\end{split}
\end{equation}

\paragraph{$\mathcal{N}=(0,2)$ Fermi Multiplet}\mbox{}

\smallskip

The $\mathcal{N}=(0,2)$ Fermi multiplet contains a left-moving chiral fermion $\lambda_{-a}^c$ and an auxiliary field $G_a$. It can be expanded as
\begin{equation}\label{(0,2) Fermi}
\Lambda^{(0,2)}_a=\lambda^c_{-a}-\theta^+G_a-i\theta^+\bar{\theta}^+D_+\lambda^c_{-a}-\bar{\theta}^+E_a^{(0,2)}(\Phi^{(0,2)}_m),
\end{equation}
where $E_a^{(0,2)}\left(\Phi^{(0,2)}_m\right)$ is a holomorphic function of chiral multiplets, called $E$-term. The decomposition of an $\mathcal{N}=(0,2)$ Fermi multiplet into $\mathcal{N}=(0,1)$ multiplets is 
\begin{equation}
\renewcommand{\arraystretch}{1.1}
    \begin{array}{c}
           \mathcal{N}=(0,2) \text{ Fermi multiplet } \Lambda^{(0,2)}_a   \\
         \swarrow \qquad \searrow  \\
    \mathcal{N}=(0,1) \text{ Fermi multiplet } \Lambda_a^1 \quad  \oplus\quad  \mathcal{N}=(0,1) \text{ Fermi multiplet } \Lambda_a^2.
    \end{array}
\end{equation}
The two $\mathcal{N}=(0,1)$ Fermi multiplets can be further combined into an $\mathcal{N}=(0,1)$ \emph{complex Fermi multiplet}. The decomposition of $\mathcal{N}=(0,2)$ Fermi multiplet is then 
\begin{equation}
\renewcommand{\arraystretch}{1.1}
    \begin{array}{c}
           \mathcal{N}=(0,2) \text{ Fermi multiplet } \Lambda^{(0,2)}_a   \\
         \downarrow  \\
     \mathcal{N}=(0,1) \text{ complex Fermi multiplet } \Lambda_a.
    \end{array}
\end{equation}

In $\mathcal{N}=(0,2)$ theories, in addition to the $E$-term, there is another holomorphic function $J^{(0,2)a}(\Phi_m)$ of chiral fields associated to the Fermi multiplet $\Lambda^{(0,2)}_a$. The kinetic terms for the Fermi multiplet and its couplings to chiral multiplets are 
\begin{equation}\label{kinetic terms of (0,2) Fermi}
\begin{split}
\mathcal{L}_{\text{Fermi}}^{(0,2)}+\mathcal{L}^{(0,2)}_J&=-\frac{1}{2}\int d\theta^+d\bar{\theta}^+(\Lambda^{(0,2)a})^{\dagger }\Lambda^{(0,2)}_a-\frac{1}{\sqrt{2}}\int d\theta^+\Lambda^{(0,2)}_aJ^{(0,2)a}|_{\bar{\theta}^+=0}-h.c.
\end{split}
\end{equation}
There is a symmetry under exchanging $J^{(0,2)a}\leftrightarrow E^{(0,2)}_a$, which corresponds to exchanging $\Lambda_a^{(0,2)}\leftrightarrow (\Lambda^{(0,2)})^{\dagger a}$.

In order to express the above Lagrangian terms for $\mathcal{N}=(0,2)$ Fermi multiplets in $\mathcal{N}=(0,1)$ superspace, we first decompose $\Phi^{(0,2)}_m$ chiral fields into $\mathcal{N}=(0,1)$ complex scalar multiplets $\Phi_m$, as in~\eqref{(0,2) chiral in (0,1) complex scalar}. Then, we introduce $\mathcal{N}=(0,1)$ complex scalar multiplets $E_a(\Phi_m)$ and $J^a(\Phi_m)$ as functions of $\Phi_m$. The field components of $E_a(\Phi_m)$ and $J^a(\Phi_m)$ are given by
\begin{equation}
\begin{split}
	E_a(\Phi_m)=E_a(\phi_m)-\theta^+\frac{\partial E_a}{\partial\phi_m}\psi_+^m,\\
	J^a(\Phi_m)=J^a(\phi_m)-\theta^+\frac{\partial J^a}{\partial \phi_m}\psi_+^m.
\end{split}
\end{equation}
where $\phi_m$ and $\psi^m_+$ are component fields of the $\mathcal{N}=(0,1)$ complex scalar multiplet $\Phi_m$. 
The terms for an $\mathcal{N}=(0,2)$ Fermi multiplet $\Lambda_a^{(0,2)}$ in the Lagrangian can then be expressed in terms of $\mathcal{N}=(0,1)$ superspace and multiplets as 
 \begin{equation}
\begin{split}
 	\mathcal{L}_{\text{Fermi}}^{(0,2)}+\mathcal{L}^{(0,2)}_J\rightarrow&\, \mathcal{L}_{F}+\int d\theta^+W^{(0,1)}\\
 	\rightarrow&\,-\frac{1}{2}\int d \theta^+(\Lambda_a\mathcal{D}_+\Lambda_a)+\\
 	&+\int d\theta^+[\Lambda_a(J^a(\Phi_m)+E^{\dagger a}(\Phi^\dagger_m))+\Lambda^{\dagger a}(E_a(\Phi_m)+J^\dagger_a(\Phi_m^\dagger))].
 \end{split}
 \end{equation}

\paragraph{$\mathcal{N}=(0,1)$ superpotential of $\mathcal{N}=(0,2)$ gauge theories}\mbox{}

\smallskip

To conclude this section, for an $\mathcal{N}=(0,2)$ field theory with vector multiplets $V^{(0,2)}_i$, chiral multiplets $\Phi^{(0,2)}_m$ and Fermi multiplets $\Lambda^{(0,2)}_a$, the generic $\mathcal{N}=(0,2)$ Lagrangian can be expressed in terms of $\mathcal{N}=(0,1)$ multiplets and superspace as 
\begin{equation}
\mathcal{L}=\mathcal{L}_{\text{gauge}}+\mathcal{L}_s+\mathcal{L}_F+\int d \theta^+ W^{(0,1)},
\end{equation}
where $\mathcal{L}_{\text{gauge}}$, $\mathcal{L}_s$ and $\mathcal{L}_{F}$ are the usual kinetic terms for vector, scalar and Fermi superfields. The $\mathcal{N}=(0,1)$ superpotential $W^{(0,1)}$ reads
\begin{equation}\label{(0,1) superpotential of (0,2) theory}
	W^{(0,1)}=\sum_{i}\sum_{n}\Lambda_{i}^R\Phi^\dagger_n\Phi_n+\sum_a\int d\theta^+[\Lambda_a(J^a(\Phi_m)+E^{\dagger a}(\Phi^\dagger_m))+\Lambda^{\dagger a}(E_a(\Phi_m)+J^\dagger_a(\Phi_m^\dagger))],
\end{equation}
where the sum over $n$ in the first term means the sum over all complex scalar multiplets transforming under a given gauge group $i$.

\section{2d $\mathcal{N}=(0,1)$ Theories and Orientifolds}
\label{sec:2dN01orientifolds}

In this section, we discuss the construction of Spin(7) and Spin(7) orientifolds starting from 
CY 4-folds. We also explain the general structure of the $\mathcal{N}=(0,1)$ theories on D$1$-branes probing Spin(7) orientifolds, which are obtained from the 
$\mathcal{N}=(0,2)$ gauge theories associated to the parent CY$_4$ via a $\mathbb{Z}_2$ orientifold quotient. While we will focus on the case in which the CY$_4$ is toric, our construction applies in general. Concrete examples will be covered in Sections~\ref{sec:examplesenginN=01} to \ref{sec:HiggsingPartResol}.

\subsection{Spin(7) Cones and Spin(7) Orientifolds from CY$_4$}
\label{section_Spin(7)_from_CY4}

Our aim in this section is to set the stage for $Spin(7)$ orientifolds probed by D$1$-branes. The construction of the corresponding gauge theories on D$1$-branes will be introduced in~\ref{sec:orientN01gaugeth}, \ref{sec:HSreview} and \ref{sec:N01theorfromorienquot}.

We start discussing $Spin(7)$ manifolds, which are eight dimensional Riemannian manifolds of special holonomy group $Spin(7)$. Every $Spin(7)$ manifold is equipped with a globally well-defined $4$-form $\Omega^{(4)}$, called Cayley $4$-form. 

$Spin(7)$ manifolds are interesting because they lead to minimally supersymmetric theories. For instance, consider Type IIB string theory on a $M_2\times X_8$, where $M_2$ is $2$d Minkowski space and $X_8$ is a Spin(7) manifold. The number of supercharges is broken from $32$ real supercharges to $2$, since Spin(7) manifolds preserves $1/16$ of the original supersymmetry.\footnote{For more details of why Spin(7) preserves $1/16$ SUSY, we refer the reader to \cite{Becker:2000jc} and \cite{Heckman:2018mxl}.} 

Probing the singularity of such Spin(7) manifold with a stack of $N$ D$1$-branes breaks SUSY even further. We would be left with only $1$ real supercharge on the $2$d worldvolume, hence engineering $2$d $\mathcal{N}=(0,1)$ theories.

However, in this chapter we focus on an alternative, yet related, way to achieve $2$d $\mathcal{N}=(0,1)$ theories, using an orientifold construction based on the following observation. An explicit construction of Spin(7) manifolds was introduced by Joyce in \cite{Joyce:1999nk}. Start with a Calabi-Yau $4$-fold $M_8$ equipped with the holomorphic (4,0)-form $\Omega^{(4,0)}$ and K\"ahler form $J^{(1,1)}$. One can always define a $4$-form 
\begin{equation}\label{Cayley-form from CY4}
\Omega^{(4)}=\text{Re}\left(
\Omega^{(4,0)}\right)+\frac{1}{2}J^{(1,1)}\wedge J^{(1,1)},
\end{equation}
which is stabilized by a Spin(7) subgroup of the general $SO(8)$ holonomy of a $8$d Riemannian manifold.

We can now consider the parent CY$_4$ geometry, and perform an orientifold by $\Omega \sigma$, where $\Omega$ denotes worldsheet parity\footnote{We hope the context suffices for the reader not to confuse it with the holomorphic $4$-form.} and $\sigma$ is an anti-holomorphic involution keeping the real 4-form (\ref{Cayley-form from CY4}) invariant. It is easy to check, in analogy with the above arguments, that the supersymmetry preserved by D1-brane probing this orientifold singularity is $2$d $\mathcal{N}=(0,1)$. Hence, we refer to this construction as \textit{Spin(7) orientifolds}. One motivation for considering these orientifolds is that they naturally realize the ``real projection” of the ``complex” $\mathcal{N}=(0,2)$ theories mentioned in \cite{Gukov:2019lzi}. The theories on D1-branes probing Spin(7) cones, without the orientifold projection, are also interesting and we plan to investigate them in future work.

\subsection{Spin(7) Orientifolds in the Field Theory}
\label{sec:orientN01gaugeth}

We now discuss the field theory implementation of the Spin(7) orientifold construction. The field theory involution must act anti-holomorphically on the chiral fields of the parent gauge theory. Its connection to $\sigma$ will be addressed in Section~\ref{sec:HSreview}. Further details on the theory obtained via the orientifold projection will be given in Section~\ref{sec:N01theorfromorienquot}. The construction follows the standard orientifolding procedure. Anti-holomorphic orientifolds have appeared in the literature in other contexts, see, e.g., \cite{Blumenhagen:2000wh, Aldazabal:2000dg}.

Such involution must be a $\mathbb{Z}_2$ symmetry of the parent gauge theory, namely a symmetry of both its quiver and superpotential. Given the anti-holomorphicity of the transformation, it is convenient to write the superpotential in $\mathcal{N}=(0,1)$ language, as in \eqref{(0,1) superpotential of (0,2) theory}.

We will use indices $i,j=1,\ldots, g$, to label gauge groups in the parent theory. We will also use $\alpha_i,\beta_j=1,\ldots,N_i$ for Chan-Paton indices, equivalently (anti) fundamental color indices of $U(N_i)$ in the gauge theory. Every bifundamental field $\Phi_{ij}$ in the gauge theory (adjoint if $i=j$) should be regarded as an $N_i \times N_j$ matrix to be contracted with the corresponding Chan-Paton factors, namely open string states are of the form $\Phi_{ij, \alpha_i \beta_j} |\alpha_i,\beta_j\rangle$. In what follows, we will keep the color/Chan-Paton indices implicit.

Below, we present the transformation properties of each type of field under the generator of the orientifold group.

\paragraph{Vector multiplets}\mbox{}

\smallskip

Gauge fields transform as follows
\begin{equation}
A^i_{\mu} \to -\gamma_{\Omega_{i'}} A^{i'T}_{\mu} \gamma_{\Omega_{i'}}^{-1} \, ,
\label{A_involution}
\end{equation}
where the transposition acts on color indices and $\gamma_{\Omega_{i}}$ is a matrix encoding the action of worldsheet parity on the Chan-Paton degrees of freedom at the node $i$. All matrices in this expression are $N_i \times N_i$ dimensional, with $N_i=N_{i'}$.

For gauge groups that are mapped to themselves, i.e., when $i=i'$, the fact that the involution squares to the identity gives rise to the standard constraint
\begin{equation}
\gamma_{\Omega_i}^T \gamma_{\Omega_i}^{-1} = \pm \mathbb{I}_{N_i}.
\end{equation}
The two canonical solutions to this equation are the identity matrix $\gamma_{\Omega_i}=\mathbb{I}_{N_i}$, for the positive sign, and the symplectic matrix $\gamma_{\Omega_i}=J=i\epsilon_{N_i/2}$, for the negative sign. Plugging each of them back into \eqref{A_involution}, they respectively lead to gauge fields in the antisymmetric or symmetric representation, namely in the adjoint representations of the resulting $SO(N_i)$ or $USp(N_i)$ gauge groups. The corresponding gaugino is projected accordingly, completing an $\mathcal{N}=(0,1)$ vector multiplet. Our general discussion allows for independent ranks for different gauge groups. That said, in the explicit examples considered later, we will assume that the ranks in the parent theory are such that all the ranks in the orientifolded theory are equal.

\paragraph{Scalar multiplets}\mbox{}

\smallskip

Let us consider complex $\mathcal{N}=(0,1)$ scalar fields or, equivalently, the $\mathcal{N}=(0,2)$ chiral fields in the parent theory. The anti-holomorphicity of the geometric involution implies that we have to take their Hermitian conjugate and their transformation becomes
\begin{equation}
X^m_{ij} \to \eta_{mn} \gamma_{\Omega_{i'}} \bar{X}^{n}_{i'j'} \gamma_{\Omega_{j'}}^{-1} \, ,
\label{scalar_involution}
\end{equation}
where the bar indicates conjugation. We can understand the conjugation as the net result of two operations. First, we have the transposition of the matrix $X^{n}_{i'j'}$, which effectively exchanges its two endpoints. This corresponds to the usual orientation reversal between fields and their images, which is characteristic of orientifolds and is also present in holomorphic orientifolds. In addition, we take the Hermitian conjugate, which is the matrix counterpart of the conjugation involved in the anti-holomorphic involution. This leads to an additional orientation flip.

While expressions like \eqref{scalar_involution} are rather standard, this is a good point to carefully state the meaning of each of the matrices in it. Color indices are implicit. As mentioned earlier, $\gamma_{\Omega_{i'}}$ and $\gamma_{\Omega_{j'}}$ encode the action of worldsheet parity on the color indices at nodes $i'$ and $j'$, and they are $N_{i'}\times N_{i'}$ and $N_{j'}\times N_{j'}$ matrices, respectively. $X^{n}_{i'j'}$ is an $N_{i'}\times N_{j'}$ matrix, for which the transposition and Hermitian conjugation, independently, transpose the color indices. We also include the indices $m,n=1,\ldots,n^\chi_{ij}$, with $n^\chi_{ij}$ the number of $\mathcal{N}=(0,2)$ chiral fields between nodes $i$ and $j$. $\eta$ is an $n^\chi_{ij} \times n^\chi_{ij}$ matrix corresponding to the representation of the $\mathbb{Z}_2$ group generated by the field theory involution under which the $X^{m}_{ij}$ fields transform.\footnote{In principle, this representation might be reducible. The irreducible representations of $\mathbb{Z}_2$ are either 1- or 2-dimensional.} We sum over the repeated index $n$. $i'$ and $j'$ indicate the nodes connected by the field, and are clearly not summed over. Eq.~\eqref{scalar_involution} also applies to fields that are mapped to themselves.

The condition that the orientifold action is an involution implies that $\eta \cdot \eta^T = \mathbb{I}$. In the explicit examples presented later, we will mostly use $\eta=\pm 1$ (in the 1-dimensional representation case) or $\eta=\pm \left(\begin{smallmatrix} 0 & 1 \\ 1 & 0 \end{smallmatrix}\right)$, which implements a non-trivial exchange between two pairs of fields. In view of this, from now on we will reduce $\eta_{mn} \bar{X}^{n}_{i'j'}$ to  $\pm \bar{X}^{m'}_{i'j'}$, in order to simplify expressions.

The transformation \eqref{scalar_involution} and the ones for Fermi superfields that we present below, simplify considerably in the case of Abelian parents. As usual, this is sufficient for connecting the gauge theories to the probed geometries, along the lines that will be discussed in Section~\ref{sec:N01theorfromorienquot}.

\paragraph{Fermi multiplets}\mbox{}

\smallskip

Contrary to scalar fields, whose transformation always involves conjugation in order to account for the anti-holomorphicity of the geometric involution, Fermi fields may or may not be conjugated. 

Let us first consider the $\mathcal{N}=(0,1)$ complex Fermi multiplets in the parent, i.e. the $\mathcal{N}=(0,2)$ Fermi multiplets in the original theory. Their transformation is either\footnote{Here we use the simplified notation introduced earlier in the case of scalar multiplets, instead of including an $\eta$ matrix as in \eqref{scalar_involution}.}
\begin{equation}
\Lambda^m_{ij} \to \pm \gamma_{\Omega_{i'}} \bar{\Lambda}^{m'}_{i'j'} \gamma_{\Omega_{j'}}^{-1},
\label{Lambda_involution-antih}
\end{equation}
or
\begin{equation}
\Lambda^m_{ij} \to \pm \gamma_{\Omega_{i'}} \Lambda^{m' \, T}_{j'i'} \gamma_{\Omega_{j'}}^{-1}.
\label{Lambda_involution-h}
\end{equation}
Notice that the second transformation only involves transposition, without complex conjugation.

The signs and the presence or absence of complex conjugation in the transformations of each Fermi in \ref{Lambda_involution-antih} and \ref{Lambda_involution-h} are determined by imposing the transformation of the chirals and requiring the invariance of the superpotential $W^{(0,1)}$ of the parent theory. As mentioned earlier, focusing on the Abelian theory is sufficient for this. 

The decomposition of $\mathcal{N}=(0,2)$ vector multiplets gives rise to additional $\mathcal{N}=(0,1)$ adjoint Fermi fields $\Lambda_i^R$, as explained in Section~\ref{sec:N02inN01Form}. Invariance of $W^{(0,1)}$ in the parent fully determines the transformation of the $\Lambda^R_{i}$, which is given by
\begin{equation}
\Lambda_i^R \to \gamma_{\Omega_{i'}} \Lambda_{i'}^{R \, \, T} \gamma_{\Omega_{i'}}^{-1} \, .
\label{Lambda^R_projection}
\end{equation}
The relative sign between \ref{A_involution} and \ref{Lambda^R_projection} implies that for $i=i'$, an $SO$ or $USp$ projection of the gauge group is correlated with a projection of $\Lambda_i^R$ into a symmetric or antisymmetric representation, respectively.

The construction of the Spin(7) orientifolds we have just presented exclusively uses information from the gauge theory. In coming sections, we will explain how it can be connected to the geometry.\footnote{In the case of toric CY$_4$, perfect matchings of the corresponding brane brick models are powerful tools in connecting gauge theory and geometry \cite{Franco:2015tya}. It is therefore natural to ask whether and, if so, how the anti-holomorphic involution translates into perfect matchings. Preliminary investigations suggest that, at least, the involution of chiral fields maps to an anti-holomorphic involution of the perfect matchings. It would be interesting to study this question in the future.} The anti-holomorphic involution of the generators of the parent CY$_4$ geometry can be mapped to an action on scalars. This, combined with the invariance of the parent superpotential, determines the transformation of the Fermi superfields.

\subsection{Orientifold Projection of the Quiver}
\label{sec:N01theorfromorienquot}

\subsubsection*{Quiver}

In this section we explicitly discuss all possible orientifold projections of the quiver following from the rules in Section~\ref{sec:orientN01gaugeth}. The different types of $\mathcal{N}=(0,1)$ superfields, combined with their various transformations, lead to several possibilities.

\subsubsection*{Gauge groups}

\begin{figure}[!htp]
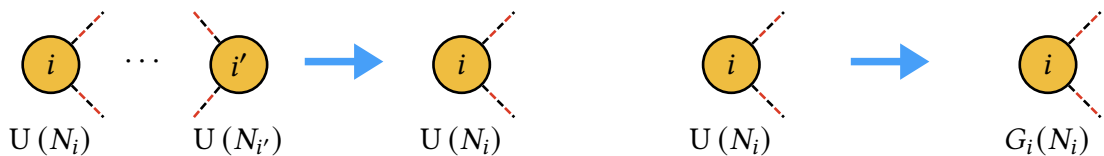

    \centering
    \begin{subfigure}[t]{0.49\textwidth}
    \centering
    
    \caption{Two nodes mapped  
    according to Rule~\ref{rule:1a}.}
    	\label{fig:rule1a}
    \end{subfigure}\hfill
    \begin{subfigure}[t]{0.49\textwidth}
    	\centering 
    	%
    	};
    	\draw[-Triangle,blueX,line width=1mm] (0.75,0.17) -- (1.25,0.17);
    	\end{tikzpicture}
    	\caption{A node mapped according to Rule~\ref{rule:1b}.}
    	\label{fig:rule1b}
    \end{subfigure}
    \caption{The two possible identifications of gauge groups. The group $G_i(N_i)$ can be either $\SO(N_i)$ or $\USp(N_i)$. Dashed black and red lines represent fields that can be either scalar or Fermi fields.}
    \label{fig:gaugegroupident}
\end{figure}

The orientifold projections for gauge groups can be one of the following two possibilities:

\begin{enumerate}[label=1\alph*.,ref=1\alph*]
\item\label{rule:1a} Every node $i\neq i'$ gives rise to a gauge factor $\U(N_i)$, as shown in Figure~\ref{fig:rule1a}. 
\item\label{rule:1b} Every node $i=i'$ gives rise to a gauge factor $\SO(N_i)$ or $\USp(N_i)$, for $\gamma_{\Omega_i}=\ID$ or $J$, respectively, as schematically shown in Figure~\ref{fig:rule1b}. 
\end{enumerate}

\subsubsection*{Matter fields}

\bigskip

\noindent\underline{$\mathcal{N}=(0,2)$ Chiral and Fermi fields}

\medskip

We start with the projection of $\mathcal{N}=(0,2)$ chiral and Fermi multiplets, equivalently $\mathcal{N}=(0,1)$ complex scalar and Fermi multiplets. Unless explicitly mentioned, the rules below apply to both scalar and Fermi fields. In figures, we will use dashed black and red lines to indicate fields that can be of the two types and we use $\mathcal{N}=(0,1)$ language. To organize the presentation, we will distinguish between the case in which a field is mapped to a different image and when it is mapped to itself.

\bigskip 

\begin{center}
{\it Fields mapped to other fields}
\end{center}

\smallskip

The two rules that follow apply to both to fields transforming anti-holomorphically, as in \eqref{scalar_involution} and \eqref{Lambda_involution-antih}, or holomorphically, as in \eqref{Lambda_involution-h}. While the resulting quiver does not depend on the presence of conjugation, such details do affect how the final fields precisely emerge from the original theory and, therefore, the projection of the superpotential. 

\begin{enumerate}[label=2\alph*.,ref=2\alph*]
\item\label{rule:2a} Consider a bifundamental or adjoint field $X_{ij}$ of the parent theory, for $j \neq i'$, which transforms into (the conjugate of) a different image field $X_{i'j'}$. The two fields, $X_{ij}$ and $X_{i'j'}$, are projected down to a single complex bifundamental (or adjoint) $X_{ij}$.\footnote{There is no distinction between $\fund$ and $\antifund$ whenever the resulting gauge group is $\SO$ or $\USp$.} Various possibilities are shown in Figure~\ref{fig:rule2a}.
\end{enumerate} 
\newpage
\begin{figure}[H]
\centering
    \begin{subfigure}[t]{\textwidth}
    \vspace*{-2.7cm} 
    \centering

    	\caption{Pairs of bifundamentals that do not share any node.}
    	\label{fig:two_bifundamentals_four_groups}
    \end{subfigure}
    \begin{subfigure}[t]{\textwidth}
    \vspace*{0.7cm} 
    \centering
    	%
    	\caption{Pairs of bifundamentals with a common node.}
    	\label{fig:two_bifundamentals_three_groups}
    \end{subfigure}
    \begin{subfigure}[t]{\textwidth}
    \vspace*{0.7cm} 
    \centering
    	%
    	\caption{Pairs of bifundamentals sharing both nodes.}
    	\label{fig:two-bifundamentals_two_groups}
    \end{subfigure}
    \end{figure}
\begin{figure}[H]\ContinuedFloat
    \begin{subfigure}[t]{\textwidth}
    \centering
    %
    };
    \draw[-Triangle,blueX,line width=1mm] (0.82,0) -- (1.17,0);
    \end{tikzpicture}
    	\caption{A pair of adjoint fields whose nodes are mapped to each other.}
    	\label{fig:two_adjoint_two_groups}
    \end{subfigure}
    \caption{Various instances of Rule~\ref{rule:2a}. 
    These pictures apply to both fields that are mapped anti-holomorphically (via \eqref{scalar_involution} or \eqref{Lambda_involution-antih}) or holomorphically (via \eqref{Lambda_involution-h}). The group $G_i(N_i)$ can be either $\SO(N_i)$ or $\USp(N_i)$.}
    \label{fig:rule2a}
\end{figure} 
 
 \begin{enumerate}[label=2\alph*.,ref=2\alph*,resume]
\item\label{rule:2b} Consider two bifundamental or adjoint fields $X_{ii'}$ and $Y_{i'i}$, which transform into (the conjugate of) each other. They give rise to two complex fields, one in the symmetric representation and the other one in the antisymmetric representation of the resulting unitary (for $i\neq i'$) or $\SO/\USp$ (for $i=i'$) node.\footnote{We thank Massimo Porrati for discussions on this point.} From now on, we indicate symmetric and antisymmetric representations with star and diamond symbols, respectively. This rule is illustrated in Figure~\ref{fig:rule2b}.
\end{enumerate} 

\begin{figure}[H]
    \centering
    \begin{subfigure}[t]{\textwidth}
    \vspace*{-0.6cm} 
    \centering

	\caption{Two bifundamental fields connecting a node and its image.}
        	\label{fig:two_bifundamentals_one_group}
    \end{subfigure}  \end{figure}%
\begin{figure}[H]
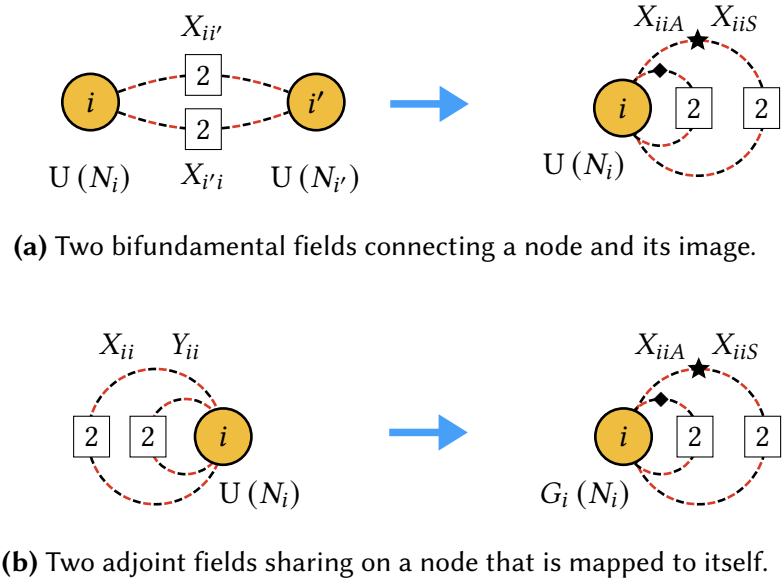
\ContinuedFloat
    \begin{subfigure}[t]{\textwidth}
    \vspace*{-0.4cm} 
    \centering
    %
    };
    \draw[-Triangle,blueX,line width=1mm] (0.82,0) -- (1.17,0);
    \end{tikzpicture}
    	\caption{Two adjoint fields sharing on a node that is mapped to itself.}
    	\label{fig:two_adjoint_one_group}
    \end{subfigure}
     \caption{The two instances of Rule~\ref{rule:2b}, depending on whether the original fields are bifundamental ($i\neq i'$) or adjoint ($i=i'$). This picture applies to both fields that are mapped anti-holomorphically (via \eqref{scalar_involution} or \eqref{Lambda_involution-antih}) or holomorphically (via \eqref{Lambda_involution-h}). The group $G_i(N_i)$ can be either $\SO(N_i)$ or $\USp(N_i)$.}
    \label{fig:rule2b}
\end{figure}

\bigskip 

\begin{center}
{\it Fields mapped to themselves}
\end{center}

\smallskip

In this case, the transformation of the quiver depends crucially on whether the map is anti-holomorphic or holomorphic. Therefore, in the figures we indicate it over the arrow connecting the parent to the orientifolded theory.

\begin{figure}[H]
\centering
    	\begin{tikzpicture}[scale=2]
    	\begin{scope}
    	\node[draw=black,line width=1pt,circle,fill=yellowX,minimum width=0.75cm,inner sep=1pt,label={[yshift=-1.75cm]:$\U\left(N_i\right)$}] (A) at (0,0) {$i$};
    	\node[draw=black,line width=1pt,circle,fill=yellowX,minimum width=0.75cm,inner sep=1pt,label={[yshift=-1.75cm]:$\U\left(N_j\right)$}] (B) at (1.5,0) {$j$};
    	\draw[line width=1pt,postaction={draw,redX,dash pattern= on 3pt off 5pt,dash phase=4pt}] [line width=1pt,black,dash pattern= on 3pt off 5pt] (A) -- node[above,midway,yshift=0.3cm] {$X_{ij}$} node[fill=white,text opacity=1,fill opacity=1,draw=black,rectangle,thin,dash pattern= on 0pt off 0pt] {$2$} (B);
    	\node[draw=black,line width=1pt,circle,fill=yellowX,minimum width=0.75cm,inner sep=1pt,label={[yshift=-1.75cm]:$G_i\left(N_i\right)$}] (C) at (3.5,0) {$i$};
    	\node[draw=black,line width=1pt,circle,fill=yellowX,minimum width=0.75cm,inner sep=1pt,label={[yshift=-1.75cm]:$G_j\left(N_j\right)$}] (D) at (5,0) {$j$};
    	\draw[line width=1pt,postaction={draw,redX,dash pattern= on 3pt off 5pt,dash phase=4pt}] [line width=1pt,black,dash pattern= on 3pt off 5pt] (C) -- node[above,midway] {$X_{ij}^R$} (D);
    	\draw[-Triangle,blueX,line width=1mm] (2.25,0) -- node[above,midway,yshift=0.1cm] {\color{black}{$X\rightarrow \pm \bar{X}$}} (2.75,0);
    	\end{scope}
    	\end{tikzpicture}
        \caption{Rule~\ref{rule:3a}, in which a complex bifundamental scalar or Fermi is mapped to itself anti-holomorphically via \eqref{scalar_involution} or \eqref{Lambda_involution-antih}. $G_i$ and $G_j$ are the same type of $\SO$ or $\USp$ gauge group.}
  	\label{fig:rule3a}
\end{figure}
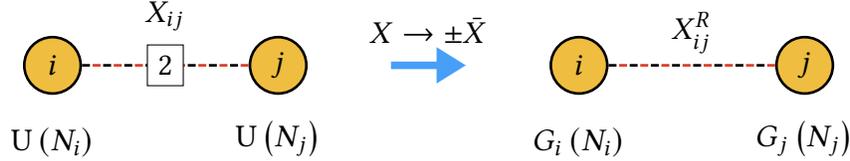

\begin{enumerate}[label=3\alph*.,ref=3\alph*]
\item\label{rule:3a} A bifundamental field $X_{ij}$ that is mapped to itself anti-holomorphically via \eqref{scalar_involution} or \eqref{Lambda_involution-antih}, with the nodes $i$ and $j$ also being their own images, gives rise to a real $\mathcal{N}=(0,1)$ field  transforming under the bifundamental of $G_i\left(N_i\right)\times G_j(N_{j})$, where $G_i$ and $G_j$ are the same type of $\SO$ or $\USp$ gauge group.\footnote{We will later elaborate on why these two gauge groups should be of the same type.} Figure~\ref{fig:rule3a} illustrates this rule.
\end{enumerate}

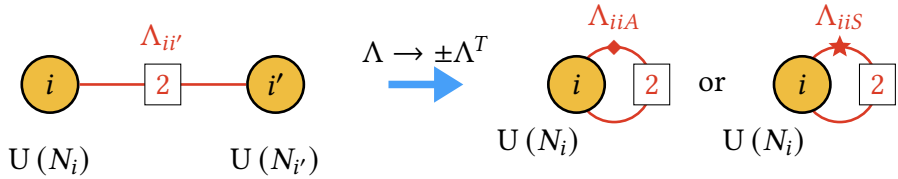
\begin{figure}[H]
    \centering
    	\begin{tikzpicture}[scale=2]
    	\node[draw=black,line width=1pt,circle,fill=yellowX,minimum width=0.75cm,inner sep=1pt,label={[yshift=-1.75cm]:$\U\left(N_i\right)$}] (A) at (0,0) {$i$};
    	\node[draw=black,line width=1pt,circle,fill=yellowX,minimum width=0.75cm,inner sep=1pt,label={[yshift=-1.75cm]:$\U\left(N_{i'}\right)$}] (B) at (1.5,0) {$i'$};
    	\draw[line width=1pt,redX] (A) -- node[above,midway,yshift=0.3cm] {$\Lambda_{ii'}$} node[fill=white,text opacity=1,fill opacity=1,draw=black,rectangle,thin] {$2$} (B);
	\draw[line width=1pt,redX] (3.75,0) circle (0.25)  node[yshift=0.5cm] {\scriptsize{$\quadro$}} node[fill=white,text opacity=1,fill opacity=1,draw=black,rectangle,thin,xshift=0.5cm] {$2$} node[yshift=0.9cm] {$\Lambda_{iiA}$};
	\node[draw=black,line width=1pt,circle,fill=yellowX,minimum width=0.75cm,inner sep=1pt,label={[xshift=-0.5cm,yshift=-1.5cm]:$\U\left(N_{i}\right)$}] (A) at (3.5,0) {$i$};
	\node at (4.4,0) {or};
	\draw[line width=1pt,redX] (5.25,0) circle (0.25)  node[yshift=0.5cm,star,star points=5, star point ratio=2.25, inner sep=1pt, fill=redX, draw] {} node[fill=white,text opacity=1,fill opacity=1,draw=black,rectangle,thin,xshift=0.5cm] {$2$} node[yshift=0.9cm] {$\Lambda_{iiS}$};
	\node[draw=black,line width=1pt,circle,fill=yellowX,minimum width=0.75cm,inner sep=1pt,label={[xshift=-0.5cm,yshift=-1.5cm]:$\U\left(N_{i}\right)$}] (A) at (5,0) {$i$};
    	\draw[-Triangle,blueX,line width=1mm] (2.25,0) -- node[above,midway,yshift=0.1cm] {\color{black}{$\Lambda\rightarrow \pm \Lambda^T$}} (2.75,0);
    	\end{tikzpicture}
        \caption{Rule~\ref{rule:3b}, in which a Fermi connecting a node to its image is mapped to itself holomorphically via \eqref{Lambda_involution-h}.}
        \label{fig:rule3b}
\end{figure}

\begin{enumerate}[label=3\alph*.,ref=3\alph*,resume]
\item\label{rule:3b} There is another possibility for a bifundamental Fermi field $\Lambda_{ii'}$ stretching between a node and its image. Such a field can only be mapped to itself in the case of a holomorphic transformation \eqref{Lambda_involution-h}.\footnote{For this reason, there is no analogue of this rule for chiral or Fermi fields transforming anti-holomorphically.} This gives rise to a complex Fermi superfield in the symmetric/antisymmetric representation of the resulting $\U\left(N_{i}\right)$ group for a $+/-$ sign, respectively, as shown in Figure~\ref{fig:rule3b}.
\end{enumerate}

\begin{figure}[H]
    \centering
    \begin{tikzpicture}[scale=2]
   \begin{scope}[yshift=-1.25cm]
   \draw[line width=1pt,redX] (1.25,0) circle (0.25)  node[yshift=0.75cm] {$\Lambda_{ii}$} node[fill=white,text opacity=1,fill opacity=1,draw=black,rectangle,thin,xshift=-0.5cm] {$2$};
	\node[draw=black,line width=1pt,circle,fill=yellowX,minimum width=0.75cm,inner sep=1pt] (A) at (1.5,0) {$i$};
	\draw[line width=1pt,redX] (3.75,0) circle (0.25)  node[yshift=0.5cm] {\color{redX}{\scriptsize{$\quadro$}}} node[fill=white,text opacity=1,fill opacity=1,draw=black,rectangle,thin,xshift=0.5cm] {$2$} node[yshift=0.9cm] {$\Lambda_{iiA}$};;
	\node[draw=black,line width=1pt,circle,fill=yellowX,minimum width=0.75cm,inner sep=1pt,label={[xshift=-0.5cm,yshift=-1.5cm]:$G_i(N_{i})$}] (A) at (3.5,0) {$i$};
	\node at (4.4,0) {or};
	\draw[line width=1pt,redX] (5.25,0) circle (0.25)  node[yshift=0.5cm,star,star points=5, star point ratio=2.25, inner sep=1pt, fill=redX, draw=redX] {} node[fill=white,text opacity=1,fill opacity=1,draw=black,rectangle,thin,xshift=0.5cm] {$2$} node[yshift=0.9cm] {$\Lambda_{iiS}$};;
	\node[draw=black,line width=1pt,circle,fill=yellowX,minimum width=0.75cm,inner sep=1pt,label={[xshift=-0.5cm,yshift=-1.5cm]:$G_i(N_{i})$}] (A) at (5,0) {$i$};
   \draw[-Triangle,blueX,line width=1mm] (2.25,0) -- node[above,midway,yshift=0.1cm] {\color{black}{$\Lambda\rightarrow \pm \Lambda^T$}} (2.75,0);
   \end{scope}
    \end{tikzpicture}
       \caption{Rule~\ref{rule:3c}, in which a complex adjoint Fermi is mapped to itself holomorphically via \eqref{Lambda_involution-h}. The group $G_i(N_i)$ can be either $\SO(N_i)$ or $\USp(N_i)$.}
    \label{fig:rule3c}
\end{figure}
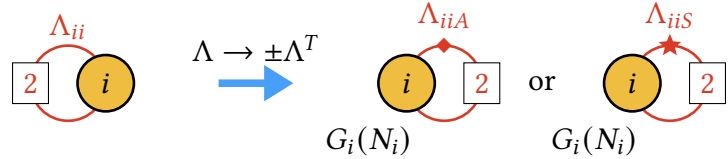

\begin{enumerate}[label=3\alph*.,ref=3\alph*,resume]
\item\label{rule:3c} Closely related to Rule~\ref{rule:3b}, consider an adjoint complex Fermi field $\Lambda_{ii}$ that is mapped to itself via the holomorphic transformation \eqref{Lambda_involution-h}. As shown in Figure~\ref{fig:rule3c}, this gives rise to a complex Fermi field in the symmetric or antisymmetric representation of the resulting gauge group for a $+/-$ sign, respectively. In this case, the $\pm$ sign in \eqref{Lambda_involution-h} correlates the projection of such Fermi with the one of the corresponding vector multiplet, which is controlled by \eqref{A_involution}. In particular, a $+$ sign implies the opposite projection, and hence we obtain symmetric/antisymmetric for $\SO/\USp$. Similarly, a $-$ sign implies the same projection, and we obtain antisymmetric/symmetric for $\SO/\USp$. 
\end{enumerate}

\begin{figure}[H]
    \centering
    \begin{tikzpicture}[scale=2]
    \begin{scope}
	\draw[line width=1pt,postaction={draw,redX,dash pattern= on 3pt off 5pt,dash phase=4pt}] [line width=1pt,black,dash pattern= on 3pt off 5pt] (1.25,0) circle (0.25)  node[yshift=0.75cm] {$X_{ii}$} node[fill=white,text opacity=1,fill opacity=1,draw=black,rectangle,thin,xshift=-0.5cm,dash pattern= on 0pt off 0pt,thin] {$2$};
	\node[draw=black,line width=1pt,circle,fill=yellowX,minimum width=0.75cm,inner sep=1pt] (A) at (1.5,0) {$i$};
	\draw[line width=1pt,postaction={draw,redX,dash pattern= on 3pt off 5pt,dash phase=4pt}] [line width=1pt,black,dash pattern= on 3pt off 5pt] (4,0) circle (0.45) node[xshift=0.90cm,xshift=0cm,star,star points=5, star point ratio=2.25, inner sep=1pt, fill=black, draw,dash pattern= on 0pt off 0pt] {}   node[yshift=1.2cm,xshift=0.5cm] {$X_{iiS}^R$};
	\draw[line width=1pt,postaction={draw,redX,dash pattern= on 3pt off 5pt,dash phase=4pt}] [line width=1pt,black,dash pattern= on 3pt off 5pt] (3.75,0) circle (0.25)  node[xshift=0.5cm] {\scriptsize{$\quadro$}} node[yshift=1.2cm,xshift=0cm] {$X_{iiA}^R$};
	\node[draw=black,line width=1pt,circle,fill=yellowX,minimum width=0.75cm,inner sep=1pt,label={[xshift=-0.5cm,yshift=-1.5cm]:$G_i(N_{i})$}] (A) at (3.5,0) {$i$};
   \draw[-Triangle,blueX,line width=1mm] (2.25,0) -- node[above,midway,yshift=0.1cm] {\color{black}{$X\rightarrow \pm \bar{X}$}} (2.75,0);
   \end{scope}
    \end{tikzpicture}
       \caption{Rule~\ref{rule:3d}, in which a complex adjoint scalar or Fermi is mapped to itself anti-holomorphically via \eqref{scalar_involution} or \eqref{Lambda_involution-antih}. The group $G_i(N_i)$ can be either $\SO(N_i)$ or $\USp(N_i)$.}
    	\label{fig:rule3d}
\end{figure}
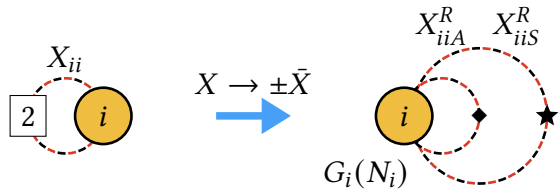

\begin{enumerate}[label=3\alph*.,ref=3\alph*,resume]
\item\label{rule:3d} Consider an adjoint complex scalar or Fermi field that is mapped to itself via the anti-holomorphic transformation in \eqref{scalar_involution} or \eqref{Lambda_involution-antih}. This gives rise to two real scalar or Fermi fields, one symmetric and one antisymmetric of node $i$. This can be understood as projecting the real and imaginary parts of the parent field with opposite signs. The sign in \eqref{A_involution} determines the projection of the real part relative to the $\SO$ or $\USp$ projection of the gauge group as in Rule~\ref{rule:3c}. This case is shown in Figure~\ref{fig:rule3d}.
\end{enumerate}

\medskip

\noindent\underline{$\mathcal{N}=(0,1)$ real Fermi fields from $\mathcal{N}=(0,2)$ vector multiplets}

\medskip

Finally, let us consider the projection of the $\mathcal{N}=(0,1)$ adjoint real Fermi fields $\Lambda_{ii}^R$ coming from the $\mathcal{N}=(0,2)$ vector multiplets. Such fields always transform according to \eqref{Lambda^R_projection}. Therefore, there are only two possibilities, depending on whether the corresponding node is mapped to a different node or to itself.

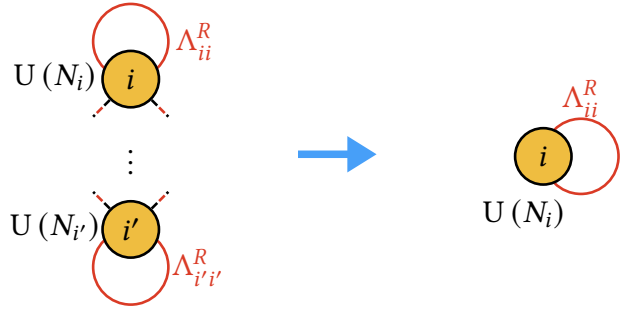
\begin{figure}[H]
\centering
    \begin{tikzpicture}[scale=3]
    \draw[help lines,white] (-1,0) grid (2,0);
    \node (L) at (-0.3,0) {
    \begin{tikzpicture}[scale=2]
    \draw[line width=1pt,redX]  (0,0.75) circle (0.25) node[xshift=0.85cm] {$\Lambda^R_{ii}$};
    \draw[line width=1pt,redX] (0,-0.75) circle (0.25) node[xshift=0.9cm] {$\Lambda^R_{i'i'}$};
    \node[draw=black,line width=1pt,circle,fill=yellowX,minimum width=0.75cm,inner sep=1pt,label={[xshift=-1cm,yshift=-0.7cm]:$\U\left(N_i\right)$}] (A) at (0,0.5) {$i$};
    	\node[draw=black,line width=1pt,circle,fill=yellowX,minimum width=0.75cm,inner sep=1pt,label={[xshift=-1cm,yshift=-0.7cm]:$\U\left(N_{i'}\right)$}] (B) at (0,-0.5) {$i'$};
    	\draw[line width=1pt,postaction={draw,redX,dash pattern= on 3pt off 5pt,dash phase=4pt}] [line width=1pt,black,dash pattern= on 3pt off 5pt] (A) -- (-0.25,0.25);
    	\draw[line width=1pt,postaction={draw,redX,dash pattern= on 3pt off 5pt,dash phase=4pt}] [line width=1pt,black,dash pattern= on 3pt off 5pt] (A) -- (0.25,0.25);
    	\draw[line width=1pt,postaction={draw,redX,dash pattern= on 3pt off 5pt,dash phase=4pt}] [line width=1pt,black,dash pattern= on 3pt off 5pt] (B) -- (-0.25,-0.25);
    	\draw[line width=1pt,postaction={draw,redX,dash pattern= on 3pt off 5pt,dash phase=4pt}] [line width=1pt,black,dash pattern= on 3pt off 5pt] (B) -- (0.25,-0.25);
    	\node at (0,0) {$\vdots$};
    \end{tikzpicture}
    };
    \node (R) at (1.6,0) {
    \begin{tikzpicture}[scale=2]
	\draw[line width=1pt,redX]  (0.25,0) circle (0.25) node[yshift=0.77cm] {$\Lambda^R_{ii}$};  
	\node[draw=black,line width=1pt,circle,fill=yellowX,minimum width=0.75cm,inner sep=1pt,label={[xshift=-0.25cm,yshift=-1.5cm]:$\U\left(N_{i}\right)$}] (A) at (0,0) {$i$};
    \end{tikzpicture}
    };
    \draw[-Triangle,blueX,line width=1mm] (0.5,0) -- (0.85,0);
    \end{tikzpicture}
    \caption{Rule~\ref{rule:4a}, in which two real Fermi fields are mapped to each other into a single real Fermi field.}
    \label{fig:rule4a}
\end{figure}

\begin{enumerate}[label=4\alph*.,ref=4\alph*]
\item\label{rule:4a} Consider a real Fermi $\Lambda^R_{ii}$ which transforms via \eqref{Lambda^R_projection} into $\Lambda^R_{i'i'}$, with $i'\neq i$. The two fields are projected down to a single real Fermi $\Lambda^R_{ii}$, as in Figure~\ref{fig:rule4a}.
\end{enumerate}

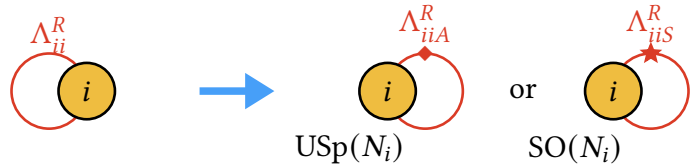
\begin{figure}[H]
\centering
    \begin{tikzpicture}[scale=2]
   \begin{scope}[yshift=-1.25cm]
   \draw[line width=1pt,redX] (1.25,0) circle (0.25)  node[yshift=0.75cm] {$\Lambda^R_{ii}$};
	\node[draw=black,line width=1pt,circle,fill=yellowX,minimum width=0.75cm,inner sep=1pt] (A) at (1.5,0) {$i$};
	\draw[line width=1pt,redX] (3.75,0) circle (0.25)  node[yshift=0.5cm] {\color{redX}{\scriptsize{$\quadro$}}}  node[yshift=0.9cm] {$\Lambda^R_{iiA}$};;
	\node[draw=black,line width=1pt,circle,fill=yellowX,minimum width=0.75cm,inner sep=1pt,label={[xshift=-0.5cm,yshift=-1.5cm]:$\USp(N_{i})$}] (A) at (3.5,0) {$i$};
	\node at (4.4,0) {or};
	\draw[line width=1pt,redX] (5.25,0) circle (0.25)  node[yshift=0.5cm,star,star points=5, star point ratio=2.25, inner sep=1pt, fill=redX, draw=redX] {} node[yshift=0.9cm] {$\Lambda^R_{iiS}$};;
	\node[draw=black,line width=1pt,circle,fill=yellowX,minimum width=0.75cm,inner sep=1pt,label={[xshift=-0.5cm,yshift=-1.5cm]:$\SO(N_{i})$}] (A) at (5,0) {$i$};
   \draw[-Triangle,blueX,line width=1mm] (2.25,0) -- 
   (2.75,0);
   \end{scope}
    \end{tikzpicture}
    \caption{Rule~\ref{rule:4b}, in which one real Fermi field is mapped to itself.}
    \label{fig:rule4b}
\end{figure}

\begin{enumerate}[label=4\alph*.,ref=4\alph*,resume]
\item\label{rule:4b} Consider a real Fermi $\Lambda^R_{ii}$ which is mapped to itself, with $i'\neq i$. Due to the relative sign between \eqref{A_involution} and \eqref{Lambda^R_projection}, this gives rise to a symmetric or antisymmetric real Fermi for an $\SO$ or $\USp$ projection of the node $i$, respectively. We show the result in Figure~\ref{fig:rule4b}.
\end{enumerate}

In general, it is possible for the theories constructed with the orientifolding procedure described above to suffer from gauge anomalies. Such anomalies can be canceled by the addition of appropriate scalar or Fermi flavors. In string theory, this corresponds to introducing flavor D$5$/D$9$-branes to cancel the local RR tadpole arising when orientifold planes are present. In Section~\ref{sec:exampC4Z2beyonduni}, we present an example in which flavor fields are needed in order to cancel the gauge anomalies. 

The gauge theories obtained from D1-branes probing Spin(7) orientifolds may provide a large class of superconformal field theories in the infrared. Determining whether this is indeed the case is an interesting direction, whose exploration we leave for future work, possibly along the lines of \cite{Gukov:2019lzi}.

\subsubsection*{Superpotential}

The superpotential of the orientifold theory is obtained from the parent superpotential by keeping the invariant terms and projecting out half of the other terms, which are identified in pairs. In the surviving terms, the parent fields must be replaced by their images under the orientifold projection.

\subsubsection*{A constraint on the relative projections of nodes connected by matter}

Requiring that the orientifold group acts on the gauge theory as an involution, leads to interesting relations between the transformation of matter fields and gauge groups. In particular, focusing on bifundamental fields, applying the transformations \eqref{scalar_involution}, \eqref{Lambda_involution-antih} or \eqref{Lambda_involution-h} twice and demanding that they amount to the identity, leads to correlations between the $\eta$ and $\gamma_{\Omega}$ matrices. For example, for a pair of nodes $i$ and $j$ connected by a single field or by a pair of fields with $\eta=\pm \left(\begin{smallmatrix} 0 & 1 \\ 1 & 0 \end{smallmatrix}\right)$ which transform anti-holomorphically, we must have $\gamma_{\Omega_i}=\gamma_{\Omega_j}$. Most of the examples we will consider later are of these two types. On the other hand, $\eta=\left(\begin{smallmatrix} 0 & 1 \\ -1 & 0 \end{smallmatrix}\right)$ implies that $\gamma_{\Omega_i}$ and $\gamma_{\Omega_j}$ are of opposite types.

\subsection{Anti-Holomorphic Involutions from the Mesonic Moduli Space}
\label{sec:HSreview}

The anti-holomorphic involution $\sigma$ of a CY$_4$ underlying Joyce's construction can be beautifully connected to the anti-holomorphic involution of the associated $\mathcal{N}=(0,2)$ gauge theory. The CY$_4$ arises as the mesonic moduli space of the parent gauge theory. Consequently, the complex coordinates parameterizing the CY$_4$ correspond to mesonic operators. Below, we present an algorithmic procedure for identifying anti-holomorphic involutions of CY$_4$ cones leading to $\Spin(7)$ manifolds. Combined with the map of generators to the gauge theory, this provides an alternative method for constructing $\Spin(7)$ orientifolds. This approach is analogous to the one introduced in \cite{Franco:2007ii} for 4d orientifolds. As usual, the construction focuses on the Abelian case of the gauge theories, but the results extend to general ranks.

In general, we can define the moduli space as the polynomial ring of the chiral fields modded by the ideal generated by the $J$- and $E$-terms, i.e.
\begin{equation}
    \mathcal{M}=\left(\CC[X_1,\ldots, X_n]/\left\langle J_{ij},E_{ij}\right\rangle\right)//\U(1)^G\coma 
\end{equation}
where $G$ is the number of $\U(1)$ gauge groups in the theory, and $n$ is the number of chiral fields. It is then possible to associate a GLSM to such a moduli space, given by a set of fields $p_a$ such that
\begin{equation}
    \mathcal{M}=\left(\CC[p_1,\ldots, p_m]//Q_{EJ}\right)//Q_{D}\coma
\end{equation}
where $Q_{EJ}$ and $Q_D$ are matrices containing $\U(1)$ charges of the $p_a$ that implement the $J$-, $E$- and $D$-terms. The mesonic moduli space is obtained by considering combinations of fields $p_a$ that are invariant under the action of these $\U(1)$'s. For details on this construction we refer to~\cite{Franco:2015tya}.

A tool that has proven to be powerful to compute such gauge invariant operators is the Hilbert series (HS)~\cite{Benvenuti:2006qr,Feng:2007ur}. The explicit expression of the HS is
\begin{equation}
    \HS(\mathbf{x},\mathbf{p})= \PE\left[\sum_{a=1}^m \mathbf{x}^{Q^a}p_a\right]\coma
\end{equation}
where $Q^a=(Q^a_{EJ},Q^a_{D})$ are the charges of the field $p_a$ represented by the collective fugacity $\mathbf{x}$. The function $\PE$ is called the Plethystic Exponential (PE) and is defined as 
\begin{equation}
    \PE\left[f(t)\right] = \PE\left[\sum_{k=0}^\infty c_k t^k\right]=\exp\left[\sum_{k=1}^\infty \frac{1}{k}\left(f\left(t^k\right)-f(0)\right)\right]=\prod_{k=1}^\infty\frac{1}{(1-t^k)^{c_k}}\fstop
\end{equation}
Performing the Molien integral over the fugacities $\mathbf{x}$, we obtain the HS of the mesonic moduli space $\mathcal{M}$:
\begin{equation}
    \HS(\mathbf{p};\mathcal{M})=\oint_{|\mathbf{x}|=1}\frac{d\mathbf{x}}{2\pi i \mathbf{x}} \HS(\mathbf{x},\mathbf{p})\fstop
\end{equation}
Such HS contains the generators of the mesonic moduli space and their relations. This information can be extracted using the Plethystic Logarithm (PL):
\begin{equation}
    \PL[\HS(\mathbf{p};\mathcal{M})]=\sum_{k=1}^\infty \frac{\mu(k)}{k}\ln\left[\HS\left(\mathbf{p}^k;\mathcal{M}\right)\right]\coma
\end{equation}
where $\mu$ is the M\"obius function. The resulting series can be finite, and in that case, the mesonic moduli space is said to be a \textit{complete intersection}, or it can be an infinite sum of positive and negative monomials in $\mathbf{p}$. The generators are identified with the positive monomials, while the relations are associated with the negative monomials. The generators for all examples in this chapter have been computed using such HS techniques.

The generators, subject to their relations, are the coordinates that parameterize the toric CY$_4$ under consideration. From the point of view of the gauge theory, these coordinates are mesons and we call them $M_a$, with $a$ running from $1$ to the number of mesons. The anti-holomorphic involution $\sigma$ acts on these coordinates by mapping each $M_a$ to a possibly different $\bar{M}_b$, with $\bar{M}_b$ being the complex conjugate of $M_b$, i.e.
\begin{equation}
    M_a \rightarrow \pm \bar{M}_b\fstop
\end{equation}
This transformation must be consistent with the relations among the generators. 
 
As explained in Section~\ref{section_Spin(7)_from_CY4}, in order to obtain a $\Spin(7)$ structure, $\sigma$ must preserve the Cayley $4$-form. A sufficient condition for this to happen is that $\Omega^{(4,0)}\rightarrow \bar{\Omega}^{(0,4)}$ \cite{Joyce:1999nk}. Consider a CY$_4$ with $n$ generators $M_a$, $a=1, \cdots, n$ and $k$ relations among them $F_\alpha(M_a)=0$, $\alpha=1,\cdots, k$. The holomorphic 4-form is computed in terms of the Poincar\'{e} residue
\begin{equation}
	\Omega^{(4,0)}=\text{Res}\frac{d M_1\wedge \cdots \wedge d M_n}{\prod_{\alpha =1}^{k}F_\alpha(M_a)} \fstop
\end{equation}
With this formula, it is straightforward to verify that all the involutions considered in this chapter satisfy $\Omega^{(4,0)}\rightarrow \bar{\Omega}^{(0,4)}$. In the following sections, we will show this explicitly in some examples.\footnote{When the HS is not a complete intersection, the number of relations is redundant. It is then possible to reduce them to their effective number, and $F_\alpha(M_a)$ represents the minimal number of relations that are necessary in order to get a $4$-form, given $n$ generators, i.e. $\alpha=1,\ldots,n-4$. Moreover, given an $\Omega^{(4,0)}$, after applying the anti-holomorphic involution, it might be necessary to use such relations to obtain the corresponding $\bar{\Omega}^{(0,4)}$. Generically, the resulting $(0,4)$-form that is obtained by the involution, is not simply the complex conjugate of $\Omega^{(4,0)}$. An explicit example of this is given in Section~\ref{sec:D3examp}.}

The procedure outlined above provides a geometric criterion for identifying an anti-holomorphic involution $\sigma$ leading to a $\Spin(7)$ orientifold. Using the definition of the generators as gauge invariant chiral operators in the field theory, we can translate $\sigma$ into the anti-holomorphic involution that acts on the chiral fields. Finally, we can complete such involution with the transformations of Fermi fields in the form of (\ref{Lambda_involution-antih}), (\ref{Lambda_involution-h}) and (\ref{Lambda^R_projection}) such that it corresponds to a $\mathbb{Z}_2$ symmetry of the $\mathcal{N}=(0,2)$ gauge theory, as discussed in Section~\ref{sec:orientN01gaugeth}.

An important observation is that the relation between the geometric anti-holomorphic involution $\sigma$ that accompanies the orientifold action, and the action on the $\mathcal{N}=(0,2)$ theory, is not one-to-one. In particular, this non-uniqueness goes beyond the obvious one due to choices of signs and $\gamma_\Omega$'s in (\ref{scalar_involution}), (\ref{Lambda_involution-antih}), (\ref{Lambda_involution-h}) and (\ref{Lambda^R_projection}). Indeed, certain orientifolded geometries defined by an involution $\sigma$ of CY$_4$ correspond to a unique action on the $\mathcal{N}=(0,2)$ quiver (up to those obvious choices), but others can admit several genuinely different possible actions from the field theory point of view. These are distinguished by the action of the orientifold on the gauge factor, in particular by the presence or absence of groups mapped to themselves. In more mathematical terms, this is related  to the presence or the absence of vector structure in type IIB singularities with orientifolds. We will discuss this in more detail and present illustrative examples in Section~\ref{sec:vectorstructure}. 

\section{$\mathbb{C}^4$ and its Orbifolds}
\label{sec:examplesenginN=01}

In this section, we construct the $2$d gauge theories on D1-branes over $\Spin(7)$ orientifolds of $\mathbb{C}^4$ and its Abelian orbifold $\mathbb{C}^4/\mathbb{Z}_2$.

\subsection{$\mathbb{C}^4$}
\label{sec:C4example}

Let first consider the simplest CY$_4$, i.e. $\CC^4$, and construct its $\Spin(7)$ orientifold. Its toric diagram is shown in Figure~\ref{fig:C4toricdiagram}.

\begin{figure}[!htp]
    \centering

	\caption{Toric diagram for $\CC^4$.}
	\label{fig:C4toricdiagram}
\end{figure}

The parent $2$d worldvolume theory on D1-branes over $\mathbb{C}^4$ is the dimensional reduction of $4$d $\mathcal{N}=4$ super Yang-Mills (SYM) and has $\mathcal{N}=(8,8)$ SUSY. In $\mathcal{N}=(0,2)$ language, this theory is given by the quiver shown in Figure~\ref{fig:C4quivN02}, and the following $J$- and $E$-terms for the Fermi fields:
\begin{equation}
\renewcommand{\arraystretch}{1.1}
%
	\caption{$\mathcal{N}=(0,2)$ language.}
	\label{fig:C4quivN02}
	\end{subfigure}
	\begin{subfigure}[t]{0.49\textwidth}
	\centering
	%
	\caption{$\mathcal{N}=(0,1)$ language.}
	\label{fig:C4quivN01}
	\end{subfigure}
	\caption{Quiver diagrams for $\CC^4$ in $\mathcal{N}=(0,2)$ and $\mathcal{N}=(0,1)$ language. $\Lambda^R$ is the real Fermi coming from the $\mathcal{N}=(0,2)$ vector multiplet.}
	\label{fig:C4quiv}
\end{figure}

Before performing the orientifold quotient, it is useful to rewrite this theory in $\mathcal{N}=(0,1)$ superspace. In $\mathcal{N}=(0,1)$ language, this theory has a vector multiplet associated with the $\U(N)$ gauge group, four complex scalar multiplets $\left(X,Y,Z\right.$ and $\left.W\right)$, three complex Fermi multiplets $\left(\Lambda^{i}\coma i=1,2,3\right)$ and one real Fermi multiplet $\left(\Lambda^R\right)$ from the $\mathcal{N}=(0,2)$ vector multiplet. The quiver is shown in Figure~\ref{fig:C4quivN01}. 
The corresponding $\mathcal{N}=(0,1)$ superpotential is given by 
\begin{equation}
\begin{split}
    W^{(0,1)}= &\,W^{(0,2)}+\Lambda^{4R}(X^\dagger X+Y^\dagger Y+Z^\dagger Z+W^\dagger W) \\
    =& \,\Lambda^1(YZ-ZY)+ \Lambda^{1\dagger}(WX-XW)+\hc\\
    +& \,\Lambda^2(ZX-XZ)+\Lambda^{2\dagger}(WY-YW)+\hc\\
    +& \,\Lambda^3(XY-YX)+\Lambda^{3\dagger}(WZ-ZW)+\hc\\
    +& \,\Lambda^{4R}(X^\dagger X+Y^\dagger Y+Z^\dagger Z+W^\dagger W) \coma
\end{split}
\label{eq:W01C4}
\end{equation}
where $W^{(0,2)}$ indicates the superpotential obtained from the $J$- and $E$-terms in \eqref{J_E_C4}.

For this theory, computing the HS for identifying the generators parameterizing the moduli space is not necessary, since these mesons are in one-to-one correspondence with the chiral superfields. The four complex coordinates $(x,y,z,w)$ of $\CC^4$ map to the four $\mathcal{N}=(0,1)$ complex scalar fields 
\begin{equation}
    (x,y,z,w)\Leftrightarrow (X,Y,Z,W)\fstop
    \label{eq:coordC4}
\end{equation}
In the Abelian case, the space is freely generated, i.e. there are no relations among the generators. This can be easily understood in $\mathcal{N}=(0,2)$ language, where the $J$- and $E$-terms in~\eqref{J_E_C4} are automatically vanishing.

Now we are ready to find an anti-holomorphic involution $\sigma$ of $\mathbb{C}^4$ and construct the gauge theory for the corresponding $\Spin(7)$ orientifold. We will choose a specific form of $\sigma$. All other possible $\sigma$'s are in fact equivalent to it via the $\SO(8)$ global symmetry of $\mathbb{C}^4$.

\subsubsection{The Orientifold Theory}

Let us consider the anti-holomorphic involution under which the $\U(N)$ gauge group is mapped to itself and the chiral fields transform as 
\begin{equation}\label{involution of c4 chiral}
    X\rightarrow \gamma_{\Omega} \bar{X} \gamma_{\Omega}^{-1}\coma  Y\rightarrow \gamma_{\Omega} \bar{Y}\gamma_{\Omega}^{-1}\coma  Z\rightarrow \gamma_{\Omega} \bar{Z}\gamma_{\Omega}^{-1}\coma W\rightarrow \gamma_{\Omega} \bar{W}\gamma_{\Omega}^{-1}\fstop
\end{equation}

Requiring the invariance of the superpotential $W^{(0,1)}$ in \eqref{eq:W01C4}, we obtain the action on the Fermi multiplets 
\begin{equation}\label{involution of c4 Fermi}
    \Lambda^1\rightarrow \gamma_{\Omega} \bar{\Lambda}^1\gamma_{\Omega}^{-1} \coma \Lambda^2\rightarrow \gamma_{\Omega} \bar{\Lambda}^2\gamma_{\Omega}^{-1} \coma \Lambda^3 \rightarrow \gamma_{\Omega} \bar{\Lambda}^3\gamma_{\Omega}^{-1}\coma \Lambda^{4R}\rightarrow\gamma_{\Omega} \Lambda^{4R\,\,T}\gamma_{\Omega}^{-1} \fstop
\end{equation}

From a geometric point of view, the anti-holomorphic involution $\sigma$ is simply given by  
\begin{equation}
     (x,y,z,w)\mapsto (\bar{x},\bar{y},\bar{z},\bar{w})\fstop
    \label{eq:univ_C4sigma}
\end{equation}
The holomorphic $4$-form $\Omega^{(4,0)}$ and K\"ahler form $J^{(1,1)}$ of $\mathbb{C}^4$ are given by
\begin{equation}
    \Omega^{(4,0)}=dx\wedge dy\wedge dz\wedge dw\coma J^{(1,1)}=\sum_{x_i\in \{ x,y,z,w \}} dx_i\wedge d\bar{x}_i
    \label{eq:OmegaJdef}
\end{equation}
They transform under $\sigma$ as 
\begin{equation}
    \Omega^{(4,0)}\rightarrow \bar{\Omega}^{(0,4)}\coma J^{(1,1)}\rightarrow -J^{(1,1)}.
    \label{eq:sigmaonOmegaJ}
\end{equation}
One can then easily check that the Cayley $4$-form defined in~\eqref{Cayley-form from CY4} is indeed invariant under this involution $\sigma$.

The orientifold theory can be derived by projecting over the involution. As discussed in Section~\ref{sec:orientN01gaugeth}, $\gamma_{\Omega}$ equal to $\ID_{N_a}$ or $J$ corresponds to the $\SO(N)$ or $\USp(N)$ gauge group after projection. We will construct the $\SO(N)$ theory in detail below. The $\USp(N)$ theory can be derived following the same procedure. 

The $\SO(N)$ gauge theory contains four real scalar superfields in the symmetric representation and four real scalar superfields in the antisymmetric representation. We will use subscripts $S$ and $A$ to keep track of representations. There are also four real Fermi superfields $\left(\Lambda^a_S\right.$ with $a=1,2, 3$ and $\Lambda^{4R}$) in the symmetric representation, and three real Fermi superfields $\left(\Lambda^a_A\right.$ with $a=1,2, 3$) in the antisymmetric representation. The origin of these matter multiplets from the parent theory is as follows
\begin{equation}
\begin{array}{ccccccc}
    X&\Rightarrow & X_S^R,X_A^R\coma & \ \ \ \ \ \ & \Lambda^1&\Rightarrow & \Lambda^{1R}_S,\Lambda^{1R}_A\coma \\[.1cm]
    Y&\Rightarrow & Y_S^R,Y_A^R\coma & & \Lambda^2&\Rightarrow &  \Lambda^{2R}_S,\Lambda^{2R}_A\coma \\[.1cm]
    Z&\Rightarrow & Z_S^R,Z_A^R\coma & & \Lambda^3&\Rightarrow &  \Lambda^{3R}_S,\Lambda^{3R}_A\coma \\[.1cm]
    W&\Rightarrow & W_S^R,W_A^R\coma & & \Lambda^{4R}&\Rightarrow & \Lambda^{4R}_S\fstop\\
\end{array}
\label{eq:C4fieldredefinuniv}
\end{equation}
The field content of the resulting $\SO(N)$ gauge theory is summarized by the quiver in Figure~\ref{fig:univ_o_theory_c4_SO}. The quiver for the $\USp(N)$ theory is shown in Figure~\ref{fig:univ_o_theory_c4_USp}. Redefining the fields according to Eq.~\eqref{eq:C4fieldredefinuniv}, it is possible to derive the $W^{(0,1)}$ after the involution from Eq.~\eqref{eq:W01C4}.

\begin{figure}[H]
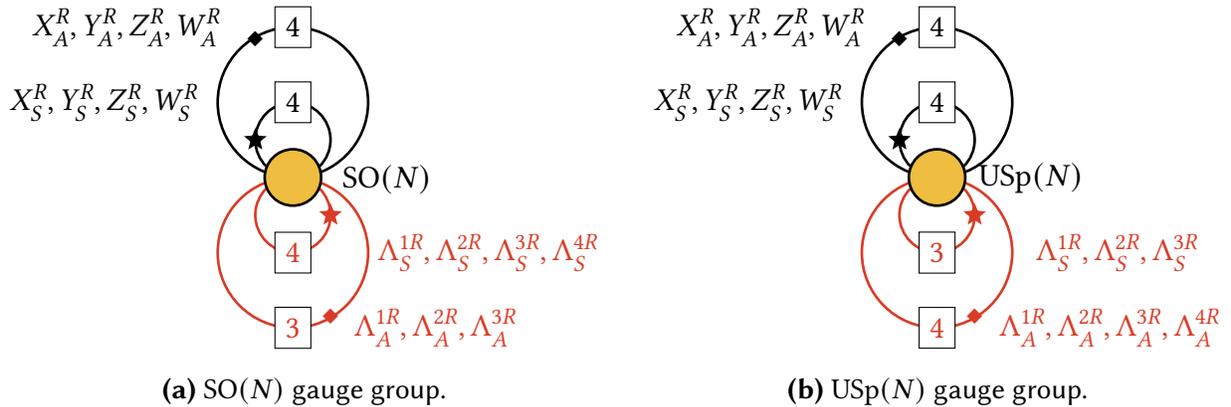

	\centering
	\begin{subfigure}[t]{0.49\textwidth}
	\centering

	\caption{$\SO(N)$ gauge group.}
	\label{fig:univ_o_theory_c4_SO}
	\end{subfigure}\hfill
	\begin{subfigure}[t]{0.49\textwidth}
	\centering
	%
	\caption{$\USp(N)$ gauge group.}
	\label{fig:univ_o_theory_c4_USp}
	\end{subfigure}
	\caption{Quiver diagrams for the orientifold theories associated with the anti-holomorphic involution of $\mathbb{C}^4$ in  \eqref{eq:univ_C4sigma}.}
	\label{fig:univ_o_theory_c4}
\end{figure}

Finally, computing the $\SO(N)^2$ anomaly contributions from different $\mathcal{N}=(0,1)$ fields using Table \ref{tab:SOUSpanomaly}, we obtain
\begin{equation}
    \underbrace{-(N-2)}_{\shortstack{Vector}}\underbrace{-4(N+2)-3(N-2)}_{\shortstack{Fermi}}\underbrace{+4(N+2)+4(N-2)}_{\shortstack{Scalar}}=0\fstop
\end{equation}
Therefore, this theory is free of gauge anomalies.

While the Spin(7) orientifold construction generically produces 2d $\mathcal{N}=(0,1)$ theories, special cases such as this one can have enhanced SUSY. This theory in fact enjoys $\mathcal{N}=(4,4)$ SUSY. To see this more explicitly, let us define the four complex coordinates of $\mathbb{C}^4$ in terms of the $8$d space transverse to the D1-branes as 
\begin{equation}
	(x,y,z,w)\equiv (x_2+ix_6, x_3+ix_7, x_4+ix_8, x_5+ix_9)\coma
\end{equation}
where $x_i$, $i=0,1,\cdots, 9$ are real spacetime coordinates. In terms of them, the geometric involution \eqref{eq:univ_C4sigma} becomes 
\begin{equation}
	(x_2,x_3,x_4,x_5,x_6,x_7,x_8,x_9)\rightarrow (x_2,x_3,x_4,x_5,-x_6,-x_7,-x_8,-x_9)\coma
\end{equation}
giving rise to a codimension-4 fixed locus, i.e., an O5-plane. The brane setup is therefore,
\begin{equation}\label{D1 on O5}
	\begin{array}{c|cccccccccc}
	 		& \, 0 \, & \, 1 \, & \, 2 \, & \, 3 \, & \, 4 \, & \, 5 \, & \, 6 \, & \, 7 \, & \, 8 \, & \, 9 \, \\
	 		\hline
		\text{D1} & \bullet & \bullet & \times & \times & \times & \times & \times & \times & \times & \times  \\
		\text{O5}& \bullet & \bullet & \bullet & \bullet & \bullet & \bullet & \times & \times & \times & \times 
	\end{array}
\end{equation}
where $\bullet$ and $\times$ indicate directions in which an object extends or does not extend, respectively. The configuration preserves $\mathcal{N}=(4,4)$ SUSY in the $2$d spacetime of the gauge theory, given by $(x_0, x_1)$. The field theory has $\SO(N)$ or $\USp(N)$ gauge symmetry, depending on the charge of the O5-plane. 

The extended SUSY can also be seen at the level of the gauge theory. The field content can be organized into $\mathcal{N}=(4,4)$ multiplets. For example, in the $\SO(N)$ case, we have
\begin{equation}
\begin{array}{rcc}
	V\oplus \Lambda^{(1,2,3)R}_A \oplus X^R_A,Y^R_A,Z^R_A,W^R_A & \rightarrow & \mathcal{N}=(4,4) ~ \text{vector multiplet} \\ & & \text{(adjoint=antisymmetric)} \\[.4cm]
	\Lambda^{4R}\oplus \Lambda_{S}^{(1,2,3)R}\oplus X^R_S,Y^R_S,Z^R_S,W^R_S &\rightarrow & \mathcal{N}=(4,4) ~ \text{hypermultiplet} \\ & & \text{(symmetric)}
\end{array}
\end{equation}
where $V$ is the $\mathcal{N}=(0,1)$ vector multiplet of the $\SO(N)$ gauge group. 

Note also that the $\SO(4)\times \SO(4)$ R-symmetry group of $\mathcal{N}=(4,4)$ supersymmetry is completely manifest in our realization. An $\SO(4)$ factor corresponds to geometric rotations in the directions transverse to the D1-branes and along the O5-plane, i.e. 2345 in \eqref{D1 on O5}.  On the other hand, the second $\SO(4)$ corresponds to rotations in the directions transverse to the O5-plane, i.e. 6789 in \eqref{D1 on O5}. The above multiplets fill out representations of $\SO(4)^2$ (noticing that the representation including the 3 Fermi multiplets must be completed by including the gauginos in the $\mathcal{N}=(0,1)$ vector multiplet, as befits an R-symmetry). It is easy to check that the interactions are also compatible with this symmetry.

Naively, one can consider seemingly different involutions $\sigma$ preserving the Cayley 4-form and construct the corresponding orientifold theories. However, the resulting theories will always be the same $2$d $\mathcal{N}=(4,4)$ $\SO(N)/\USp(N)$ gauge theory worked out above. All such anti-holomorphic involutions are equivalent, since they are connected by $\SO(8)$ rotations of the eight real coordinates of $\mathbb{C}^4$ and lead to the same brane configuration with D1-branes on top of an O5-plane. 

For example, consider the anti-holomorphic involution $(x,y,z,w)\rightarrow(\bar{y},\bar{x},\bar{z},-\bar{w})$, under which the Cayley 4-form is also invariant. Using the $\SO(8)$ global symmetry, we can redefine the eight real coordinates of $\mathbb{C}^4$ as 
\begin{equation}
\left(x_2^\prime,x_3^\prime,x_4^\prime,x_5^\prime,x_6^\prime,x_7^\prime,x_8^\prime,x_9^\prime\right)\equiv\left(\frac{x_2+x_3}{2},\frac{x_6-x_7}{2},x_4,x_9,\frac{x_2-x_3}{2},\frac{x_6+x_7}{2},x_8,x_5\right)\fstop    
\end{equation}
Then, the fixed locus of the involution corresponds to an O5-plane extended along $x_i^\prime$, $i=2,\ldots,5$. This is exactly the same orientifold configuration in \eqref{D1 on O5}. Therefore, despite the seemingly different involution, the $2$d gauge theory on D1-branes is the same up to field redefinitions.

\subsection{A Universal Involution}
\label{sec:universal involution}

Interestingly, the anti-holomorphic involution of $\mathbb{C}^4$ can be generalized to any CY$_4$. Consider the gauge theory associated to a generic toric CY$_4$. From the field theory perspective, it is always possible to define an involution as follows. First, all gauge groups are mapped to themselves. In addition, all chiral fields transform as 
\begin{equation}\label{eq:universal involution of chiral}
	X_{ij}\rightarrow \gamma_{\Omega_i}\bar{X}_{ij}\gamma_{\Omega_j}^{-1}\coma
\end{equation}
i.e. every chiral field is mapped to itself anti-holomorphically. This in turn implies that the $J$- and $E$-terms for every Fermi $\Lambda_{ij}$ transform as
\begin{equation}
	J_{ji}\rightarrow \gamma_{\Omega_j}\bar{J}_{ji}\gamma_{\Omega_i}^{-1}, E_{ij}\rightarrow \gamma_{\Omega_i}\bar{E}_{ij}\gamma_{\Omega_j}^{-1}\fstop
\end{equation}
Invariance of the superpotential $W^{(0,1)}$ implies that the action on the Fermi fields must be 
\begin{equation}\label{eq:universal involution of Fermi}
	\Lambda_{ij}\rightarrow \gamma_{\Omega_i}\bar{\Lambda}_{ij} \gamma_{\Omega_j}^{-1}\fstop
\end{equation}
Finally, as usual, the $\mathcal{N}=(0,1)$ adjoint Fermi fields coming from $\mathcal{N}=(0,2)$ vector multiplets transform as in \eqref{Lambda^R_projection}. 

This field theoretic involution translates into a simple action on the generators of CY$_4$
\begin{equation}
	\sigma_0 \,:\, M_{a}\rightarrow \bar{M}_{a}\coma
\end{equation}
namely an involution that maps every generator to its conjugate. The holomorphic 4-form $\Omega^{(4,0)}$ then transforms as $\Omega^{(4,0)}\rightarrow \bar{\Omega}^{(0,4)}$, based on the discussion in Section~\ref{section_Spin(7)_from_CY4}. This, in turn, implies the invariance of the Cayley 4-form. Therefore, $\sigma_0$ combined with worldsheet parity leads to a $\Spin(7)$ orientifold. Since $\sigma_0$ applies to any CY$_4$, we refer to it as the {\it universal involution}. The resulting gauge theory is derived using the rules in Section~\ref{sec:N01theorfromorienquot}. 

In general, depending on the geometry, other involutions can also exist. In the coming sections, we will present various examples of such involutions. $\mathbb{C}^4$ is special in that, as we have previously discussed, all its anti-holomorphic involutions are equivalent to the universal one.
 
The universal involution explicitly realizes the idea 
of 
$\mathcal{N}=(0,1)$ theories as ``real slices" of  $\mathcal{N}=(0,2)$ gauge theories \cite{Gukov:2019lzi}. Moreover, in this context, the real slicing admits a beautiful geometric interpretation as the $\Spin(7)$ orientifold of a CY$_4$. We can similarly think about other involutions as different real slices of the parent theories.

\subsection{$\mathbb{C}^4/\mathbb{Z}_2$}
\label{sec:C4Z2example}

Let us consider the $\CC^4/\ZZ_2$ orbifold with action $(x,y,z,w)\rightarrow (-x,-y,-z,-w)$ as the parent geometry. Its toric diagram is shown in Figure~\ref{fig:C4Z2toricdiagram}.

\begin{figure}[H]
    \centering

	\caption{Toric diagram for $\CC^4/\ZZ_2$.}
	\label{fig:C4Z2toricdiagram}
\end{figure}

The corresponding 2d $\mathcal{N}=(0,2)$ theory was constructed in \cite{Franco:2015tna}. Its quiver is shown in Figure~\ref{fig:C4Z2quivN02}. 

\begin{figure}[H] 
	\centering
	\begin{subfigure}[t]{\textwidth}
	\centering
	%
	\caption{$\mathcal{N}=(0,2)$ language.}
	\label{fig:C4Z2quivN02}
	\end{subfigure}\\
	\begin{subfigure}[t]{\textwidth}
	\centering
	%
	\caption{$\mathcal{N}=(0,1)$ language.}
	\label{fig:C4Z2quivN01}
	\end{subfigure}
	\caption{Quiver diagram for $\CC^4/\ZZ_2$ in $\mathcal{N}=(0,2)$ and $\mathcal{N}=(0,1)$ language.}
	\label{fig:C4Z2quiv}
\end{figure}

The $J$- and $E$-terms are:
\begin{equation}
\renewcommand{\arraystretch}{1.1}
\begin{array}{lclc}
& J                  &\text{\hspace{.5cm}} & E                    \\
\Lambda^1_{11} \,:\, & Y_{12}Z_{21}-Z_{12}Y_{21} & \quad               & W_{12}X_{21}-X_{12}W_{2 1}  \\
\Lambda^2_{11} \,:\, & Z_{12}X_{21}-X_{12}Z_{21} & \quad               & W_{12}Y_{21}-Y_{12}W_{21}   \\
\Lambda^3_{11} \,:\, & X_{12}Y_{21}-Y_{12}X_{21} & \quad               & W_{12}Z_{21}-Z_{12}W_{21}   \\
\Lambda^1_{22} \,:\, & Y_{21}Z_{12}-Z_{21}Y_{12} & \quad               & W_{21}X_{12}-X_{21}W_{12}   \\
\Lambda^2_{22} \,:\, & Z_{21}X_{12}-X_{21}Z_{12} & \quad               & W_{21}Y_{12}-Y_{21}W_{12}   \\
\Lambda^3_{22} \,:\, & X_{21}Y_{12}-Y_{21}X_{12} & \quad               & W_{21}Z_{12}-Z_{21}W_{12}   
\end{array}
\label{eq:JEterms_c4z2}
\end{equation}

Figure~\ref{fig:C4Z2quivN01} shows the quiver for this theory in $\mathcal{N}=(0,1)$ language. Denoting $W^{(0,2)}$ the superpotential obtained from \eqref{eq:JEterms_c4z2}, $W^{(0,1)}$ is given by
\begin{equation}
     W^{(0,1)} = W^{(0,2)}+
     \sum_{i,j,k=1}^2\Lambda^{4R}_{kk}( X^\dagger_{ij}X_{ij}+ Y^\dagger_{ij}Y_{ij}+ Z^\dagger_{ij}Z_{ij}+ W^\dagger_{ij}W_{ij})\fstop
\label{eq:C4Z2W01}
\end{equation}

Since the $J$- and $E$-terms in~\eqref{eq:JEterms_c4z2} are more involved, we use the HS to extract the generators of the moduli space. In Table~\ref{tab:GenerC4Z2} we present their expression in terms of chiral fields of the gauge theory.

 \begin{table}[!htp]
	\centering
	\renewcommand{\arraystretch}{1.1}
	\begin{tabular}{c|c}
		Meson    & Chiral fields  \\
		\hline
		$M_1$    & $Y_{12}Y_{21}$ \\
		$M_2$    & $Y_{12}Z_{21}=Y_{21}Z_{12}$\\
		$M_3$    & $Z_{12}Z_{21}$\\
		$M_4$    & $X_{12}Y_{21}=X_{21}Y_{12}$\\
		$M_5$    & $X_{12}Z_{21}=X_{21}Z_{12}$\\
		$M_6$    & $X_{12}X_{21}$\\
		$M_7$    & $Y_{12}W_{21}=Y_{21}W_{12}$\\
		$M_8$    & $Z_{12}W_{21}=Z_{21}W_{12}$\\
		$M_9$    & $X_{12}W_{21}=X_{21}W_{12}$\\
		$M_{10}$ & $W_{12}W_{21}$\\
	\end{tabular}
	\caption{Generators of $\CC^4/\ZZ_2$.}
	\label{tab:GenerC4Z2}
\end{table}

 The mesonic moduli space is not a complete intersection, so the PL of the HS does not terminate. We can, however, extract the relations among the generators composing the following ideal:
\begin{equation}
    \begin{split}
        \mathcal{I} = & \left\langle M_1M_3=M_2^2\coma M_1M_5=M_2M_4\coma M_3M_4=M_2M_5\coma M_1M_6=M_4^2\coma \right.\\
        & \left. M_2M_6=M_4M_5\coma M_3M_6=M_5^2\coma M_1M_8=M_2M_7\coma M_3M_7=M_2M_8\coma  \right.\\
        & \left. M_1M_9=M_4M_7\coma M_2M_9=M_4M_8\coma M_5M_7=M_2M_9\coma M_3M_9=M_5M_8\coma  \right.\\
        & \left. M_6M_7=M_4M_9\coma M_6M_8=M_5M_9\coma M_1M_{10}=M_7^2\coma M_2M_{10}=M_7M_8\coma  \right.\\
        & \left. M_3M_{10}=M_8^2\coma M_4M_{10}=M_7M_9\coma M_5M_{10}=M_8M_9\coma M_6M_{10}=M_9^2\right\rangle\fstop 
    \end{split}
\end{equation}

We now have everything necessary for identifying anti-holomorphic involutions and constructing the corresponding Spin(7) orientifolds, both from the gauge theory and from geometry.

\subsubsection{Universal Involution}

\label{sec:exampC4Z2uni} 

Let us consider the universal involution defined in Section~\ref{sec:universal involution}. It maps the two gauge groups to themselves. Chiral fields transform according to \eqref{eq:universal involution of chiral}, i.e. 
\begin{equation}
\begin{array}{cccc}
    X_{12}\rightarrow \gamma_{\Omega_1}\bar{X}_{12}\gamma_{\Omega_2}^{-1}\coma & Y_{12}\rightarrow \gamma_{\Omega_1}\bar{Y}_{12}\gamma_{\Omega_2}^{-1}\coma & Z_{12}\rightarrow \gamma_{\Omega_1}\bar{Z}_{12}\gamma_{\Omega_2}^{-1}\coma & W_{12}\rightarrow \gamma_{\Omega_1}\bar{W}_{12}\gamma_{\Omega_2}^{-1}\coma \\
    X_{21}\rightarrow \gamma_{\Omega_2}\bar{X}_{21}\gamma_{\Omega_1}^{-1}\coma & Y_{21}\rightarrow \gamma_{\Omega_2}\bar{Y}_{21}\gamma_{\Omega_1}^{-1}\coma & Z_{21}\rightarrow \gamma_{\Omega_2}\bar{Z}_{21}\gamma_{\Omega_1}^{-1}\coma & W_{21}\rightarrow \gamma_{\Omega_2}\bar{W}_{21}\gamma_{\Omega_1}^{-1}\fstop
\end{array}
\label{eq:chiralC4Z2uniinvol}
\end{equation}
The $\mathcal{N}=(0,2)$ Fermi fields transform as in \eqref{eq:universal involution of Fermi}, namely
\begin{equation}
\begin{array}{ccc}
    \Lambda^1_{11}\rightarrow \gamma_{\Omega_1}\bar{\Lambda}^1_{11}\gamma_{\Omega_1}^{-1}\coma & \Lambda^2_{11}\rightarrow \gamma_{\Omega_1}\bar{\Lambda}^2_{11}\gamma_{\Omega_1}^{-1}\coma & \Lambda^3_{11}\rightarrow \gamma_{\Omega_1}\bar{\Lambda}^3_{11}\gamma_{\Omega_1}^{-1}\coma \\
    \Lambda^1_{22}\rightarrow \gamma_{\Omega_2}\bar{\Lambda}^1_{22}\gamma_{\Omega_2}^{-1}\coma & \Lambda^2_{22}\rightarrow \gamma_{\Omega_2}\bar{\Lambda}^2_{22}\gamma_{\Omega_2}^{-1}\coma & \Lambda^3_{22}\rightarrow \gamma_{\Omega_2}\bar{\Lambda}^3_{22}\gamma_{\Omega_2}^{-1} \fstop
\end{array}
\label{eq:fermiC4Z2uniinvol}
\end{equation}
Finally, the Fermi superfields coming from the $\mathcal{N}=(0,2)$ vector multiplets transform according to \eqref{Lambda^R_projection}
\begin{equation}
\Lambda^{4R}_{11}\rightarrow \gamma_{\Omega_1}\Lambda^{4R\,\,T}_{11}\gamma_{\Omega_1}^{-1}\coma \Lambda^{4R}_{22}\rightarrow \gamma_{\Omega_2}\Lambda^{4R\,\,T}_{22}\gamma_{\Omega_2}^{-1}\fstop
\label{eq:fermiRC4Z2uniinvol}
\end{equation}
As argued in full generality in Section~\ref{sec:universal involution}, these transformations leave the superpotential $W^{(0,1)}$ in \eqref{eq:C4Z2W01} invariant.

Using Table~\ref{tab:GenerC4Z2}, we can translate this field theory involution into the geometric involution, whose action on the generators of $\mathbb{C}^4/\mathbb{Z}_2$ becomes 
\begin{equation}
	M_{a}\rightarrow \bar{M}_{a}\coma a=1,\cdots, 10 \coma
	\label{eq:C4Z2mesonuniinvol}
\end{equation}
as expected for the universal involution.

The gauge symmetry and the projections of matter fields in the orientifolded theory are controlled by $\gamma_{\Omega_1}$ and $\gamma_{\Omega_2}$. According to the discussion in Section~\ref{sec:N01theorfromorienquot}, the choices of $\gamma_{\Omega_1}$ and $\gamma_{\Omega_2}$ are not independent. In this case, they should satisfy $\gamma_{\Omega_1}=\gamma_{\Omega_2}$. To show this correlation, we consider the effect of acting with the involution twice. For example, acting on $X_{12}$ we obtain
\begin{equation}\label{eq:square of involution}
	X_{12}\rightarrow \gamma_{\Omega_1}\bar{\gamma}_{\Omega_1}X_{12}\bar{\gamma}_{\Omega_2}^{-1}\gamma_{\Omega_2}^{-1}\coma
\end{equation}
which should be equal to the identity transformation. Since $\gamma_{\Omega_i}$ is equal to $\ID_N$ or $J$, this implies that  $\gamma_{\Omega_1}=\gamma_{\Omega_2}$. Repeating this analysis for any other bifundamental field leads to the same condition. We conclude that the gauge symmetry of the orientifolded theory is 
either $\SO(N)\times \SO(N)$ or $\USp(N)\times \USp(N)$.

For concreteness, let us focus on the $\SO(N)\times \SO(N)$ case. Figure~\ref{fig:C4Z2uniinvol} shows the corresponding quiver. There are eight real bifundamental scalars, coming from the bifundamental chiral fields in the parent.\footnote{In what follows, we will use the term bifundamental in the case of matter fields that connect pairs of nodes, even when one or both of them is either SO or USp.} Every adjoint complex Fermi in the parent is projected to one symmetric and one antisymmetric real Fermi fields, while the adjoint real Fermi fields from the $\mathcal{N}=(0,2)$ vector multiplets are projected to the symmetric representation. It is rather straightforward to write the projected superpotential but, for brevity, we omit it here and in the examples that follow. Finally, it is easy to verify the vanishing of gauge anomalies.

\begin{figure}[H]
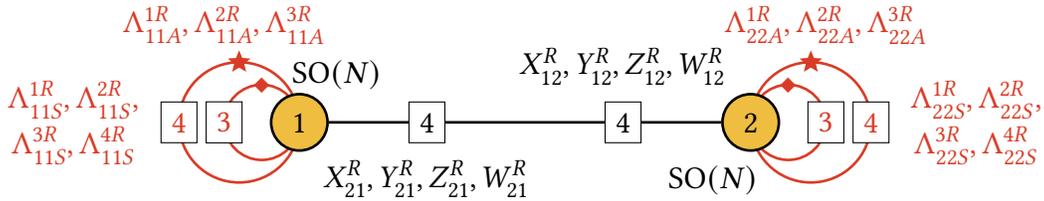

	\centering

	\caption{Quiver for the Spin(7) orientifold of $\CC^4/\ZZ_2$ using the universal involution.}
	\label{fig:C4Z2uniinvol}
	\end{figure}
	
We would like to mention that, although the above models are built as orientifolds of the $\mathbb{C}^4/\mathbb{Z}_2$ theory, they can be equivalently regarded as $\mathbb{Z}_2$ orbifolds of the orientifolds of $\mathbb{C}^4$ in Section \ref{sec:C4example}. This viewpoint is useful to display that the models inherit the $\SO(4)^2$ global symmetry of the $\mathbb{C}^4$ orientifolds, since the $\mathbb{Z}_2$ orbifold acts in the same way on the coordinates within each 4-plet. In fact, it is easy to gather the different multiplets in $\SO(4)^2$ representations (including the gauginos in the $\mathcal{N}=(0,1)$ vector multiplet, as befits an R-symmetry). We leave the check of the $\SO(4)^2$ invariance of the interactions as an exercise for the interested reader. Similar remarks apply to other orientifolds of $\mathbb{C}^4/\mathbb{Z}_2$ in coming sections.


\subsubsection{Beyond the Universal Involution: an $\SO(N)\times \USp(N)$ Theory}
\label{sec:exampC4Z2beyonduni}

Let us now consider another involution, which also maps the two gauge groups to themselves but transforms chiral fields differently, according to
\begin{equation}
\begin{array}{cccc}
    X_{12}\rightarrow \gamma_{\Omega_1}\bar{Y}_{12}\gamma_{\Omega_2}^{-1}\coma & Y_{12}\rightarrow -\gamma_{\Omega_1}\bar{X}_{12}\gamma_{\Omega_2}^{-1}\coma & Z_{12}\rightarrow \gamma_{\Omega_1}\bar{W}_{12}\gamma_{\Omega_2}^{-1}\coma & W_{12}\rightarrow -\gamma_{\Omega_1}\bar{Z}_{12}\gamma_{\Omega_2}^{-1}\coma \\
    X_{21}\rightarrow \gamma_{\Omega_2}\bar{Y}_{21}\gamma_{\Omega_1}^{-1}\coma & Y_{21}\rightarrow -\gamma_{\Omega_2}\bar{X}_{21}\gamma_{\Omega_1}^{-1}\coma & Z_{21}\rightarrow \gamma_{\Omega_2}\bar{W}_{21}\gamma_{\Omega_1}^{-1}\coma & W_{21}\rightarrow -\gamma_{\Omega_2}\bar{Z}_{21}\gamma_{\Omega_1}^{-1}\fstop
\end{array}
\label{eq:chiralC4Z2beyuniinvol}
\end{equation}

Invariance of $W^{(0,1)}$ in \eqref{eq:C4Z2W01} implies that the Fermi fields transform as 
\begin{equation}
\begin{array}{ccc}
 \Lambda^1_{11}\rightarrow \gamma_{\Omega_1}\Lambda^{1\,\,T}_{11}\gamma_{\Omega_1}^{-1}\coma & \Lambda^2_{11}\rightarrow \gamma_{\Omega_1}\Lambda^{2\,\,T}_{11}\gamma_{\Omega_1}^{-1}\coma & \Lambda^3_{11}\rightarrow \gamma_{\Omega_1}\bar{\Lambda}^3_{11}\gamma_{\Omega_1}^{-1}\coma \\
    \Lambda^1_{22}\rightarrow \gamma_{\Omega_2}\Lambda^{1\,\,T}_{22}\gamma_{\Omega_2}^{-1}\coma & \Lambda^2_{22}\rightarrow \gamma_{\Omega_2}\Lambda^{2\,\,T}_{22}\gamma_{\Omega_2}^{-1}\coma & \Lambda^3_{22}\rightarrow \gamma_{\Omega_2}\bar{\Lambda}^3_{22}\gamma_{\Omega_2}^{-1}\coma
\end{array}
\label{eq:fermiC4Z2beyuniinvol}
\end{equation}
and
\begin{equation}
\Lambda^{4R}_{11}\rightarrow \gamma_{\Omega_1}\Lambda^{4R\,\,T}_{11}\gamma_{\Omega_1}^{-1}\coma \Lambda^{4R}_{22}\rightarrow \gamma_{\Omega_2}\Lambda^{4R\,\,T}_{22}\gamma_{\Omega_2}^{-1}\fstop
\label{eq:fermiRC4Z2beyuniinvol}
\end{equation}

Using Table~\ref{tab:GenerC4Z2}, this translates into the following geometric involution
\begin{equation}
    \begin{array}{c}
    \left(M_1,M_2,M_3,M_4,M_5,M_6,M_7,M_8,M_9,M_{10},\right)\\
    \downarrow\\
 \left(\bar{M}_6,-\bar{M}_9,\bar{M}_{10},-\bar{M}_4,\bar{M}_7,\bar{M}_1,\bar{M}_5,-\bar{M}_{8},-\bar{M}_2,\bar{M}_3\right)\fstop
 \end{array}
 \label{eq:C4Z2mesonbeyonuniinvol}
 \end{equation}

As in the previous example, the choices of $\gamma_{\Omega_1}$ and $\gamma_{\Omega_2}$ are correlated because they are connected by matter fields. From \eqref{eq:chiralC4Z2beyuniinvol}, we conclude that for each pair of chiral fields that are mapped to each other, the involution corresponds to the case $\eta=\pm\left(%
\right)$ in \eqref{scalar_involution}. Following to the discussion in Section~\ref{sec:N01theorfromorienquot}, in this case the gauge groups project to $\SO(N)\times \USp(N)$. We can explicitly see this constraint by considering the square of the involution on, e.g., $X_{12}$, for which we obtain
\begin{equation}\label{eq:square of involution2}
	X_{12}\rightarrow -\gamma_{\Omega_1}\bar{\gamma}_{\Omega_1}X_{12}\bar{\gamma}_{\Omega_2}^{-1}\gamma_{\Omega_2}^{-1}\coma
\end{equation}
which should be equal to the identity. This implies that $\gamma_{\Omega_1}=\ID_N$ and  $\gamma_{\Omega_2}=J$ or $\gamma_{\Omega_1}=J$ and $\gamma_{\Omega_2}=\ID_N$. The other chiral fields lead to the same condition.

The resulting quiver is shown in Figure~\ref{fig:C4Z2beyuniinvol}. This theory suffers from gauge anomalies, which can be canceled by adding eight scalar flavors to the SO group and eight Fermi flavors to the USp group.

\begin{figure}[!htp]
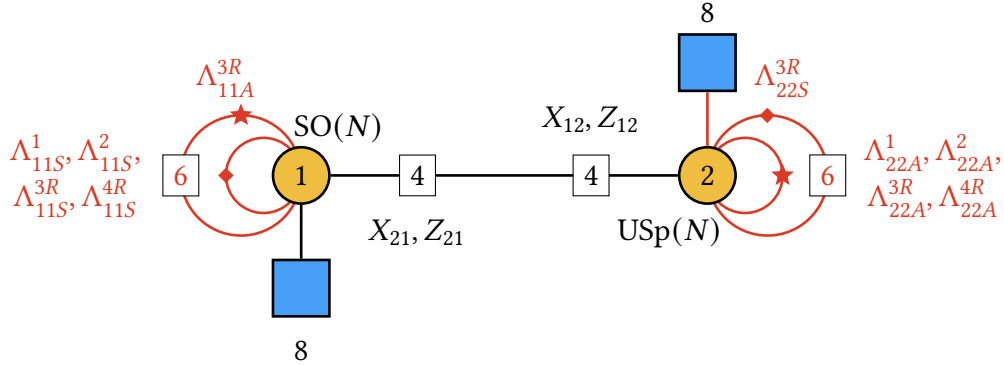

	\centering

	\caption{Quiver for the Spin(7) orientifold of $\CC^4/\ZZ_2$ using the involution in \eqref{eq:chiralC4Z2beyuniinvol}, \eqref{eq:fermiC4Z2beyuniinvol} and \eqref{eq:fermiRC4Z2beyuniinvol}. The squares indicate the number of flavors necessary to cancel gauge anomalies.}
	\label{fig:C4Z2beyuniinvol}
	\end{figure}

We would like to emphasize the fact that most of the orientifold theories in this chapter actually do not require flavor branes to cancel their anomalies. Our expectation is that this is due to the relative simplicity of the singularities considered, at the level of their structure of collapsed cycles (for instance, their toric diagrams have no collapsed cycles), and that orientifold of more general singularities are likely to require flavor branes. This is somewhat similar to the CY$_3$ case, in which ``simple" singularities (i.e. not having interior points) generically lead to theories not requiring flavor branes, and only specific cases require them \cite{Park:1999eb}. Hence, the above example is particularly remarkable, and possibly illustrates, in a relatively simple setup, a feature which may be generic in orientifolds of more involved CY$_4$ singularities.

\section{Choice of Vector Structure}
\label{sec:vectorstructure}

\subsection{Vector Structure in Type IIB Orientifold Construction}

The $\mathbb{C}^4/\ZZ_2$ example serves to address an important point, which will apply to many others of our more general examples discussed later. As already pointed out at the end of Section~\ref{sec:HSreview}, when orientifolding by a certain geometric action, there are certain discrete choices which lead to different orientifolds for the same geometric action. One such choice is the already mentioned $\SO/\USp$ projection; in this section we discuss a second (and independent) choice, corresponding to the existence or not of vector structure in certain singularities.

This possibility was first uncovered for singularities obtained as orbifolds of flat space by even order groups, e.g. $\mathbb{C}^n/\ZZ_{2N}$, triggered by the analysis in \cite{Polchinski:1996ry} of 6d orientifold models \cite{Pradisi:1988xd,Gimon:1996rq}, in particular orientifolds of $\mathbb{C}^2/\mathbb{Z}_2$. The key observation is that in such orbifolds, the orientifold acts by mapping a sector twisted by an element $\theta^k$ to the $\theta^{-k}$-twisted sector, and hence for even order $\mathbb{Z}_{2N}$, the $\theta^N$-twisted sector is mapped to itself and there are two possible choices of sign in this action. In the open string perspective, the two possibilities correspond to choices of Chan-Paton actions satisfying
\begin{equation}
\gamma_{\theta^N}=\pm \gamma_{\Omega}\gamma_{\theta^N}^T\gamma_{\Omega}^{-1}\fstop
\end{equation}
The relation with vector structure (namely, the possibility that the gauge bundle defined by the Chan-Paton matrices admits objects in the vector representation or not) was further clarified in \cite{Berkooz:1996iz} (also \cite{Witten:1997bs}). 

Although these ideas arose in the 6d orbifold context, they are far more general. For instance, the choice of vector structure has appeared in the construction of orientifolds of toroidal orbifolds in \cite{Aldazabal:1998mr}. In such compact setups, the choice of orientifolds with vector structure sometimes requires the introduction of anti-branes~\cite{Antoniadis:1999xk, Aldazabal:1999jr}; however, this is due to untwisted RR tadpoles, and hence any choice of vector structure leads to consistent orientifolds of non-compact $\mathbb{C}^2/\mathbb{Z}_{2N}$ singularities (see e.g. the constructions in 6d in \cite{Blum:1997fw,Blum:1997mm} and in 4d in \cite{Park:1998zh,Park:1999eb}). An even more important generalization is that the existence of a discrete choice of vector structure in the orientifold action generalizes beyond orbifold singularities, and applies to a far broader set of singularities. This was tacitly included in the construction of general orientifolds of general toric Calabi-Yau 3-fold singularities in \cite{Franco:2007ii}. 

In practical terms,  the appearance of the choices of vector structure in orientifolding  arises when, for a given geometry, there are different $\ZZ_2$ symmetries on the underlying quiver gauge theory, which differ in the action on the quiver nodes: an orientifold whose action on nodes is pairwise exchange, with no nodes mapped to themselves, corresponds to an action without vector structure, whereas the presence of nodes mapped to themselves corresponds to an action with vector structure.\footnote{There are cases, e.g. orbifolds by products of cyclic groups $\mathbb{Z}_N\times \mathbb{Z}_M$ etc., in which the orientifold may act with vector structure with respect to the $\mathbb{Z}_N$ and without vector structure with respect to the $\mathbb{Z}_M$. For simplicity, we ignore these more subtle possibilities and stick to the stated convention, as a practical reference device.}

We thus expect that the choice of vector structure will arise in our present setup of $\Spin(7)$ orientifolds of Calabi-Yau 4-fold singularities. In particular, for orbifolds of $\CC^4$, this should already follow from the early analysis in \cite{Polchinski:1996ry}. From this perspective, the orientifold of $\mathbb{C}^4/\ZZ_2$ constructed in Section \ref{sec:C4Z2example} corresponds to an orientifold action with vector structure, since each of the two gauge factors of the underlying $\mathcal{N}=(0,2)$ theory are mapped to themselves. Our discussion suggests that it should be possible to construct an orientifold of the same geometry, with the same orientifold geometric action, but without vector structure. This corresponds to  the symmetry of the $\mathcal{N}=(0,2)$ theory that exchanges pairwise the two gauge factors. We will indeed build this orientifold without vector structure in the following section.

This brings about an important observation. The universal involution in Section \ref{sec:universal involution} maps each gauge factor of the $\mathcal{N}=(0,2)$ theory to itself, hence it corresponds to actions with vector structure. Therefore, in geometries admitting it, the choice of orientifold action without vector structure must correspond to orientifolds actions beyond the universal involution. Thus, the possibility of choosing the vector structure is already ensuring that the set of orientifold theories is substantially larger than the class provided by the universal involution.

\subsection{$\mathbb{C}^4/\mathbb{Z}_2$ Revisited: an Orientifold Without Vector Structure} 
\label{sec:C4Z2-revisited}

Let us revisit the $\CC^4/\ZZ_2$ theory, but this time consider an anti-holomorphic involution that maps one gauge group to the other. A possible involution of the chiral fields reads
\begin{equation}
 \begin{array}{cccc}
    X_{12}\rightarrow  \gamma_{\Omega_2}\bar{X}_{21}\gamma_{\Omega_1}^{-1}\coma & Y_{12}\rightarrow  \gamma_{\Omega_2}\bar{Y}_{21}\gamma_{\Omega_1}^{-1}\coma & Z_{12}\rightarrow  \gamma_{\Omega_2}\bar{Z}_{21}\gamma_{\Omega_1}^{-1}\coma & W_{12}\rightarrow \gamma_{\Omega_2}\bar{W}_{21}\gamma_{\Omega_1}^{-1}\coma \\
    X_{21}\rightarrow  \gamma_{\Omega_1}\bar{X}_{12}\gamma_{\Omega_2}^{-1}\coma & Y_{21}\rightarrow  \gamma_{\Omega_1}\bar{Y}_{12}\gamma_{\Omega_2}^{-1}\coma & Z_{21}\rightarrow  \gamma_{\Omega_1}\bar{Z}_{12}\gamma_{\Omega_2}^{-1}\coma & W_{21}\rightarrow \gamma_{\Omega_1}\bar{W}_{12}\gamma_{\Omega_2}^{-1}\fstop
 \end{array}
    \label{eq:chirfieldsC4Z2-revisited}
\end{equation}

Invariance of $W^{(0,1)}$ in \eqref{eq:C4Z2W01} implies that Fermi fields transform as
\begin{equation}
\begin{array}{ccc}
    \Lambda^1_{11}\rightarrow \gamma_{\Omega_2}\bar{\Lambda}^1_{22}\gamma_{\Omega_2}^{-1}\coma & \Lambda^2_{11}\rightarrow \gamma_{\Omega_2}\bar{\Lambda}^2_{22}\gamma_{\Omega_2}^{-1}\coma & \Lambda^3_{11}\rightarrow \gamma_{\Omega_2}\bar{\Lambda}^3_{22}\gamma_{\Omega_2}^{-1}\coma\\
    \Lambda^1_{22}\rightarrow \gamma_{\Omega_1}\bar{\Lambda}^1_{11}\gamma_{\Omega_1}^{-1}\coma & \Lambda^2_{22}\rightarrow \gamma_{\Omega_1}\bar{\Lambda}^2_{11}\gamma_{\Omega_1}^{-1}\coma & \Lambda^3_{22}\rightarrow \gamma_{\Omega_1}\bar{\Lambda}^3_{11}\gamma_{\Omega_1}^{-1}\coma
    \end{array}
    \label{eq:fermifieldsC4Z2-revisited}
\end{equation}
and
\begin{equation}
    \Lambda^{4R}_{11}\rightarrow \gamma_{\Omega_2}\Lambda^{4R\,\,T}_{22}\gamma_{\Omega_2}^{-1}\coma
    \Lambda^{4R}_{22}\rightarrow \gamma_{\Omega_1}\Lambda^{4R\,\,T}_{11}\gamma_{\Omega_1}^{-1}\fstop
    \label{eq:realfermifieldsC4Z2-revisited}
\end{equation}

Using Table~\ref{tab:GenerC4Z2}, \eqref{eq:chirfieldsC4Z2-revisited} translates into the following geometric involution
\begin{equation}
	M_{a}\rightarrow \bar{M}_{a}\coma a=1,\cdots, 10 \coma
	\label{eq:C4Z2mesonuniinvol_without_vector_structure}
\end{equation}
which coincides with \eqref{eq:C4Z2mesonuniinvol}. This model and the one in Section~\ref{sec:exampC4Z2uni} provide concrete examples in which the same geometric action but different choices of vector structure lead to different Spin(7) orientifolds. The resulting quiver is shown in Figure~\ref{fig:C4Z2-revisited}. It is free of gauge anomalies.

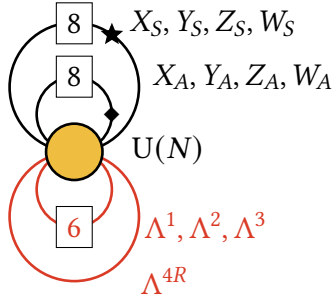
\begin{figure}[!htp]
	\centering
	\begin{tikzpicture}[scale=2]
	\draw[line width=1pt] (0,0.43) circle (0.43) node[yshift=0.7cm,xshift=0.5cm,star,star points=5, star point ratio=2.25, inner sep=1pt, fill=black, draw] {}  node[fill=white,text opacity=1,fill opacity=1,draw=black,rectangle,yshift=0.85cm,thin] {$8$} node[yshift=0.85cm,xshift=1.85cm] {$X_S,Y_S,Z_S,W_S$};
	\draw[line width=1pt] (0,0.25) circle (0.25)  node[xshift=0.5cm] {\scriptsize{$\quadro$}} node[fill=white,text opacity=1,fill opacity=1,draw=black,rectangle,yshift=0.5cm,thin] {$8$} node[yshift=0.5cm,xshift=2.25cm] {$X_A,Y_A,Z_A,W_A$};
	\draw[line width=1pt,redX] (0,-0.25) circle (0.25)  node[fill=white,text opacity=1,fill opacity=1,draw=black,rectangle,yshift=-0.5cm,thin] {\color{redX}{$6$}} node[yshift=-0.5cm,xshift=1.75cm] {$\Lambda^1,\Lambda^2,\Lambda^3$};
	\draw[line width=1pt,redX] (0,-0.43) circle (0.43) node[yshift=-0.85cm,xshift=1.2cm] {$\Lambda^{4R}$};
	\node[draw=black,line width=1pt,circle,fill=yellowX,minimum width=0.75cm,inner sep=1pt,label={[xshift=1.25cm,yshift=-0.7cm]:$\U(N)$}] (A) at (0,0) {};
	\end{tikzpicture}
	\caption{Quiver for the Spin(7) orientifold of $\CC^4/\ZZ_2$ using the involution in \eqref{eq:chirfieldsC4Z2-revisited}, \eqref{eq:fermifieldsC4Z2-revisited} and \eqref{eq:realfermifieldsC4Z2-revisited}. The underlying geometric involution coincides with the one for the model in Figure~\ref{fig:C4Z2uniinvol}, but both theories differ in the vector structure.}
	\label{fig:C4Z2-revisited}
\end{figure}

 We would like to conclude this discussion with an interesting observation: in our example, the orientifold models with/without vector structure differ also in the fact that one requires flavor branes to cancel anomalies, while the other does not. In fact, this feature has also been encountered in the $4$d case of D$3$-branes at (orientifolds of) CY$_3$. For instance, in the $4$d ${\cal N}=1$ orientifolds of even order orbifolds $\mathbb{C}^2/{\mathbb{Z}_k}$  theories in \cite{Park:1999eb}, models without/with vector structure were shown to require/not require flavor D$7$-branes.\footnote{In a T-dual type IIA picture with D4-branes suspended between $k$ NS-branes, in the presence of two O6$^\prime$-planes, the two possibilities differ in having the NS-branes splitting/not splitting the O6$^\prime$-planes in halves. In the former case, the orientifold plane charge flips sign across the NS-brane and charge conservation requires the introduction of additional half-D6-branes (i.e. flavor branes) for consistency \cite{Brunner:1998jr} (see also \cite{Brunner:1997gf,Hanany:1997gh}).}
 
 \smallskip
 
\section{Beyond Orbifold Singularities} 
\label{sec:BeyondOrbif}

In this section, we construct Spin(7) orientifolds in which the parent theory is a non-orbifold toric CY$_4$.

\subsection{$D_3$}
\label{sec:D3examp}

Let us consider the CY$_4$ with toric diagram shown in Figure~\ref{fig:D3toricdiagram}. This geometry is often referred to as $D_3$.

\begin{figure}[H]
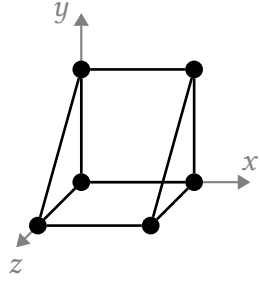

   	\centering

	\caption{Toric diagram for $D_3$.}
	\label{fig:D3toricdiagram}
\end{figure}

The $\mathcal{N}=(0,2)$ theory on D1-branes probing $D_3$ was first derived in \cite{Franco:2015tna}. Its quiver diagram is shown in Figure~\ref{fig:D3quivN02}. 

\begin{figure}[H]
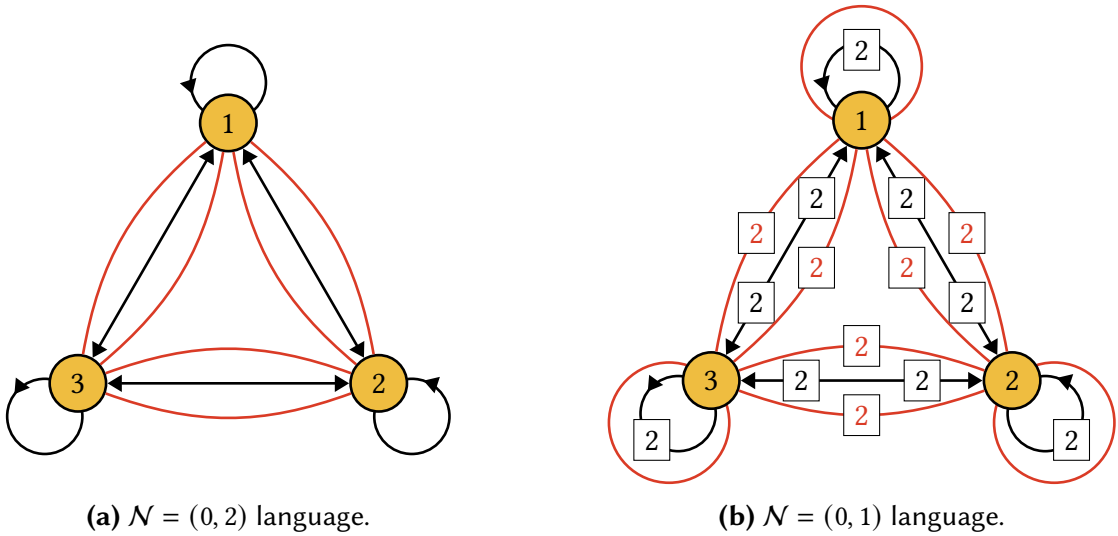

	\centering
	\begin{subfigure}[t]{0.49\textwidth}
	\centering
	%
	\caption{$\mathcal{N}=(0,2)$ language.}
	\label{fig:D3quivN02}
	\end{subfigure}\hfill
	\begin{subfigure}[t]{0.49\textwidth}
	\centering
	%
	\caption{$\mathcal{N}=(0,1)$ language.}
	\label{fig:D3quivN01}
	\end{subfigure}
	\caption{Quiver diagram for $D_3$ in $\mathcal{N}=(0,2)$ and $\mathcal{N}=(0,1)$ language.}
	\label{fig:D3quiv}
\end{figure}

The $J$- and $E$-terms read
\begin{equation}
\renewcommand{\arraystretch}{1.1}
\begin{array}{lclc}
& J                             &\text{\hspace{.5cm}} & E                             \\
\Lambda_{12}     \,:\, & X_{21} X_{13} X_{31} -X_{22} X_{21}  & \quad               & X_{11} X_{12} - X_{12} X_{23} X_{32} \\
\Lambda_{21}     \,:\, & X_{13} X_{31} X_{12}-X_{12} X_{22}   & \quad               & X_{23} X_{32} X_{21}-X_{21} X_{11}   \\
\Lambda_{23} \,:\, & X_{33} X_{32} - X_{32} X_{21} X_{12} & \quad               & X_{23} X_{31} X_{13} -X_{22} X_{23}  \\
\Lambda_{32}	    \,:\, & X_{23} X_{33} - X_{21} X_{12} X_{23} & \quad               & X_{32} X_{22} - X_{31} X_{13} X_{32} \\
\Lambda_{31}     \,:\, & X_{13} X_{33} - X_{12} X_{21} X_{13} & \quad               & X_{31} X_{11} - X_{32} X_{23} X_{31} \\
\Lambda_{13}     \,:\, & X_{31} X_{12} X_{21} -X_{33} X_{31}  & \quad               & X_{11} X_{13} - X_{13} X_{32} X_{23} 
\end{array}
\label{eq:D3-JEterms}
\end{equation}

Figure~\ref{fig:D3quivN01} shows the quiver for this theory in $\mathcal{N}=(0,1)$ language. The $W^{(0,1)}$ associated to \eqref{eq:D3-JEterms} is
\begin{equation}
    \begin{split}
       W^{(0,1)}=& W^{(0,2)}+\sum_{i=1}^3\Lambda_{ii}^R X^\dagger_{ii}X_{ii}+\sum_{\substack{i,j=1\\ j\neq i}}^3\Lambda_{ii}^R\left( X^\dagger_{ij}X_{ij}+ X^\dagger_{ji}X_{ji}\right)\fstop
        \end{split}
        \label{eq:D3superW01}
\end{equation}

Table~\ref{tab:GenerD3} shows the generators, which were obtained using the HS.
\begin{table}[H]
	\centering
	\renewcommand{\arraystretch}{1.1}
	\begin{tabular}{c|c}
		Meson    & Chiral fields  \\
		\hline
		$M_1$    & $X_{23}X_{32}=X_{11}$ \\
		$M_2$    & $X_{13}X_{31}=X_{22}$\\
		$M_3$    & $X_{12}X_{21}=X_{33}$\\
		$M_4$    & $X_{23}X_{31}X_{12}$\\
		$M_5$    & $X_{13}X_{32}X_{21}$
	\end{tabular}
	\caption{Generators of $D_3$.}
	\label{tab:GenerD3}
\end{table}
They satisfy the following relation
\begin{equation}
    \mathcal{I} = \left\langle M_{1}M_{2}M_{3}=M_{4}M_{5}\right\rangle\fstop
    \label{eq:D3ideal}
\end{equation}

Of course, as for all cases, we can consider the universal involution. However, in this section we will consider another involution, which gives rise to an $\SO(N)\times \U(N)$ (or $\USp(N)\times \U(N)$) gauge theory. 

\bigskip

\paragraph{$\SO(N)\times \U(N)$ Orientifold}\mbox{}

\smallskip

Let us consider an involution which, roughly speaking, acts as a reflection with respect to a vertical axis going through the middle of Figure~\ref{fig:D3quiv}. Node 1 maps to itself, while nodes 2 and 3 get identified.

Chiral fields transform according to 
\begin{equation}
\begin{array}{ccc}
     X_{11}\rightarrow  \gamma_{\Omega_1}\bar{X}_{11}\gamma_{\Omega_1}^{-1} \coma & X_{22}\rightarrow  \gamma_{\Omega_3}\bar{X}_{33}\gamma_{\Omega_3}^{-1} \coma & X_{33}\rightarrow  \gamma_{\Omega_2}\bar{X}_{22}\gamma_{\Omega_2}^{-1} \coma \\
     X_{12}\rightarrow  \gamma_{\Omega_1}\bar{X}_{13}\gamma_{\Omega_3}^{-1} \coma & X_{21}\rightarrow  \gamma_{\Omega_3}\bar{X}_{31}\gamma_{\Omega_1}^{-1}\coma & X_{23}\rightarrow  \gamma_{\Omega_3}\bar{X}_{32}\gamma_{\Omega_2}^{-1}\coma \\ X_{32}\rightarrow  \gamma_{\Omega_2}\bar{X}_{23}\gamma_{\Omega_3}^{-1} \coma & X_{31}\rightarrow  \gamma_{\Omega_2}\bar{X}_{21}\gamma_{\Omega_1}^{-1} \coma & X_{13}\rightarrow  \gamma_{\Omega_1}\bar{X}_{12}\gamma_{\Omega_2}^{-1}\fstop   
    \end{array}
    \label{eq:D3chiralinvolution}
\end{equation}
Invariance of $W^{(0,1)}$ in \eqref{eq:D3superW01} implies that Fermi fields transform as
\begin{equation}
    \begin{array}{ccc}
    \Lambda_{12}\rightarrow \gamma_{\Omega_1}\bar{\Lambda}_{13}\gamma_{\Omega_3}^{-1}\coma & 
    \Lambda_{21}\rightarrow -\gamma_{\Omega_3}\bar{\Lambda}_{31}\gamma_{\Omega_1}^{-1}\coma & 
    \Lambda_{23}\rightarrow-\gamma_{\Omega_2}\Lambda_{23}^{T}\gamma_{\Omega_3}^{-1}\coma\\
    \Lambda_{32}\rightarrow \gamma_{\Omega_3}\Lambda_{32}^{T}\gamma_{\Omega_2}^{-1}\coma & 
    \Lambda_{31}\rightarrow -\gamma_{\Omega_2}\bar{\Lambda}_{21}\gamma_{\Omega_1}^{-1}\coma & 
    \Lambda_{13}\rightarrow \gamma_{\Omega_1}\bar{\Lambda}_{12}\gamma_{\Omega_2}^{-1} \coma
    \end{array}
    \label{eq:D3fermiinvolution}
\end{equation}
and 
\begin{equation}
\begin{array}{ccc}
\Lambda^{R}_{11}\rightarrow \gamma_{\Omega_1}\Lambda^{R\,\,T}_{11}\gamma_{\Omega_1}^{-1} \coma & 
    \Lambda^{R}_{22}\rightarrow \gamma_{\Omega_3}\Lambda^{R\,\,T}_{33}\gamma_{\Omega_3}^{-1} \coma & 
    \Lambda^{R}_{33}\rightarrow \gamma_{\Omega_2}\Lambda^{R\,\,T}_{22}\gamma_{\Omega_2}^{-1} \fstop
\end{array}
\label{eq:D3fermiRinvolution}
\end{equation}

Using Table~\ref{tab:GenerD3}, we derive the corresponding geometric involution $\sigma$ on the generators of $D_3$
\begin{equation}
    \begin{array}{c}
    \left(M_1,M_2,M_3,M_4,M_5\right)\\
    \downarrow\\
 \ \left(\bar{M}_1,\bar{M}_3,\bar{M}_2,\bar{M}_5,\bar{M}_4\right)\fstop
\end{array}
    \label{eq:D3invol}
\end{equation}
Since $D_3$ is a complete intersection, we can easily check that $\sigma$ maps the holomorphic $4$-form $\Omega^{(4,0)}$ to $\bar{\Omega}^{(0,4)}$. We define
\begin{equation}
    \Omega^{(4,0)}=\Res \frac{dM_1\wedge dM_2\wedge dM_3 \wedge dM_4 \wedge dM_5}{M_{1}M_{2}M_{3}-M_{4}M_{5}} = \frac{dM_1\wedge dM_2\wedge dM_3 \wedge dM_4}{M_4}\coma
\end{equation}
which maps to 
\begin{equation}\label{eq:antiholo 4-form of D3}
    \frac{d\bar{M}_1\wedge d\bar{M}_3\wedge d\bar{M}_2 \wedge d\bar{M}_5}{\bar{M}_5}\fstop
\end{equation}
Either choosing the residue with respect to $M_4$ to express the holomorphic $4$-form, or by using the relation in the ideal~\eqref{eq:D3ideal}, one can show that (\ref{eq:antiholo 4-form of D3}) is exactly the anti-holomorphic $4$-form $\bar{\Omega}^{(0,4)}$ of $D_3$. Based on the discussion in Section~\ref{sec:HSreview}, we conclude that $\Omega\sigma$ with $\sigma$ in (\ref{eq:D3invol}) indeed gives rise to a $\Spin(7)$ orientifold. 

Returning to the gauge theory, we obtain an $\mathcal{N}=(0,1)$ theory with gauge symmetry $\SO(N)\times\U(N)$ or $\USp(N)\times \U(N)$, depending on whether $\gamma_{\Omega_1}=\ID_N$ or $J$. The quivers for both choices are shown in Figure~\ref{fig:o_theory_d3}. Both theories are free of gauge anomalies. 

\begin{figure}[H]
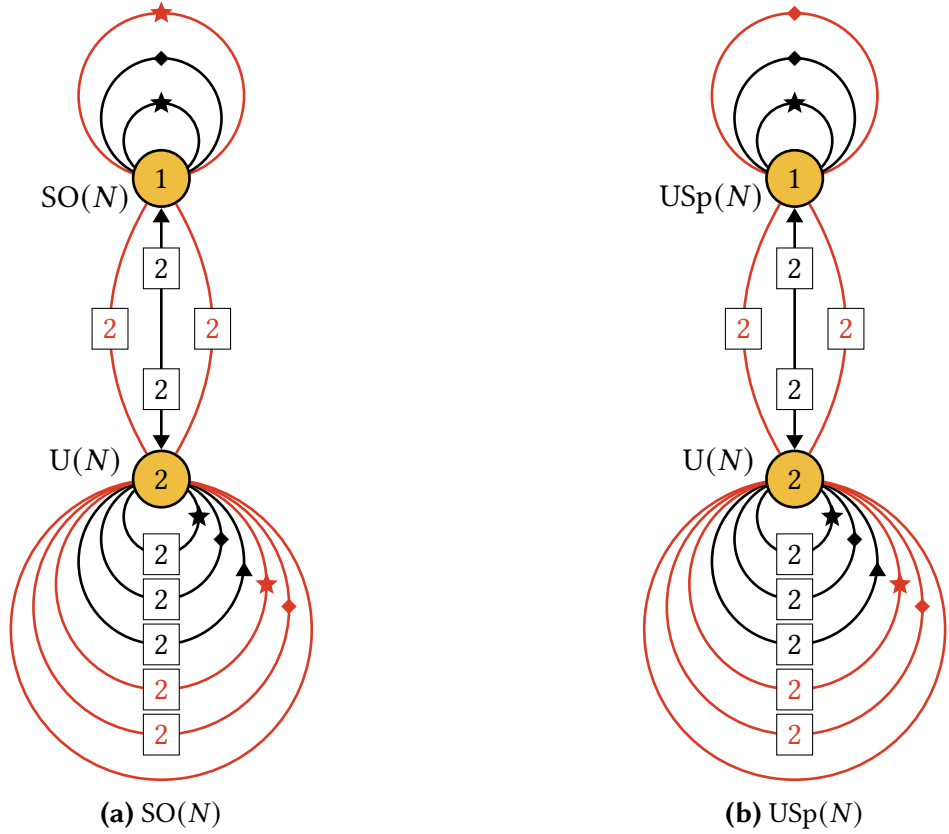

    \centering
    \begin{subfigure}[t]{0.49\textwidth}
    \centering

    \caption{$\SO(N)$}
    \label{fig:o_theory_d3_SO}
    \end{subfigure}\hfill
    \begin{subfigure}[t]{0.49\textwidth}
    \centering
    %
    \caption{$\USp(N)$}
    \label{fig:o_theory_d3_USp}
    \end{subfigure}
    \caption{Quivers for  the  Spin(7)  orientifolds  $D_3$ using the involution in \eqref{eq:D3chiralinvolution}, \eqref{eq:D3fermiinvolution} and \eqref{eq:D3fermiRinvolution}. The two different theories correspond to the choices $\gamma_{\Omega_1}=\ID_N$ or $J$.}
    \label{fig:o_theory_d3}
\end{figure}

\subsection{$H_4$}

\label{sec:H4examp}

Another example that we are going to discuss is $H_4$. We show its toric diagram in Figure~\ref{fig:H4toricdiagram}. 

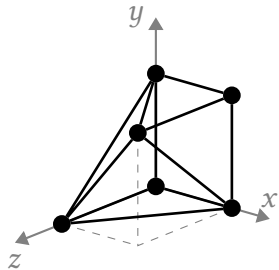
\begin{figure}[H]
    	\centering
	\begin{tikzpicture}[scale=1.5, rotate around y=-30]
	\draw[thick,gray,-Triangle] (0,0,0) -- node[above,pos=1] {$x$} (1.5,0,0);
	\draw[thick,gray,-Triangle] (0,0,0) -- node[left,pos=1] {$y$} (0,1.5,0);
	\draw[thick,gray,-Triangle] (0,0,0) -- node[below,pos=1] {$z$} (0,0,1.5);
	\draw[thin,dashed,gray] (1,0,0) -- (1,0,1) -- (1,1,1);
	\draw[thin,dashed,gray] (0,0,1) -- (1,0,1);
	\node[draw=black,line width=1pt,circle,fill=black,minimum width=0.2cm,inner sep=1pt] (O) at (0,0,0) {};
	\node[draw=black,line width=1pt,circle,fill=black,minimum width=0.2cm,inner sep=1pt] (A) at (1,0,0) {};
	\node[draw=black,line width=1pt,circle,fill=black,minimum width=0.2cm,inner sep=1pt] (B) at (0,1,0) {};
	\node[draw=black,line width=1pt,circle,fill=black,minimum width=0.2cm,inner sep=1pt] (C) at (0,0,1) {};
	\node[draw=black,line width=1pt,circle,fill=black,minimum width=0.2cm,inner sep=1pt] (D) at (1,1,1) {};
	\node[draw=black,line width=1pt,circle,fill=black,minimum width=0.2cm,inner sep=1pt] (E) at (1,1,0) {};
	\draw[line width=1pt] (O)--(A)--(E)--(B)--(O);
	\draw[line width=1pt] (O)--(C)--(B)--(D)--(A)--(C)--(D)--(E);
	\end{tikzpicture}
	\caption{Toric diagram for $H_4$.}
	\label{fig:H4toricdiagram}
\end{figure}

In particular, we will consider two $\mathcal{N}=(0,2)$ gauge theories associated with $H_4$, denoted as Phase A and Phase B. These two phases are related by $\mathcal{N}=(0,2)$ triality and were first introduced in \cite{Franco:2017cjj}. While they have different matter content and $J$- and $E$-terms, they share the same moduli space. Therefore, the generators of their moduli space and the relations among them are the same.

\subsubsection{Phase A}
\label{sec:H4PhaseA}

The quiver diagram for Phase A is shown in Figure~\ref{fig:H4_A_quiver}, both in $\mathcal{N}=(0,2)$ and $\mathcal{N}=(0,1)$ languages. 

\begin{figure}[!htp]
	\centering
	\begin{subfigure}[t]{0.49\textwidth}
	\centering

	\caption{$\mathcal{N}=(0,2)$ language.}
	\label{fig:H4AquivN02}
	\end{subfigure}\hfill
	\begin{subfigure}[t]{0.49\textwidth}
	\centering
	%
	\caption{$\mathcal{N}=(0,1)$ language.}
	\label{fig:H4AquivN01}
	\end{subfigure}
	\caption{Quiver for $H_4$ in phase A.}
	\label{fig:H4_A_quiver}
\end{figure}

The $J$- and $E$-terms are 
\begin{equation}
\renewcommand{\arraystretch}{1.1}
\begin{array}{lclc}
& J                           &\text{\hspace{.5cm}}& E                                \\
\Lambda_{11}^1 \,:\, & X_{14}X_{41} - X_{13}X_{32}Z_{21}  &                    & Y_{13}X_{34}Z_{41}-X_{12}Y_{21}         \\
\Lambda_{11}^2 \,:\, & X_{14}Y_{41} - Y_{13}X_{32}Z_{21}  &                    & X_{12}X_{21}  - X_{13}X_{34}Z_{41}      \\
\Lambda_{11}^3 \,:\, & X_{14}Z_{41} - X_{12}Z_{21}        &                    & X_{13}X_{32}Y_{21} - Y_{13}X_{34}X_{41} \\
\Lambda_{13}^1 \,:\, & X_{32}X_{21}  - X_{34}X_{41}       &                    & Y_{13}X_{33} - X_{14}Z_{41}Y_{13}       \\
\Lambda_{13}^2 \,:\, & X_{32}Y_{21}  - X_{34}Y_{41}       &                    & X_{12}Z_{21}X_{13} - X_{13}X_{33}       \\
\Lambda_{42}^1 \,:\, & X_{21}X_{14}  - Z_{21}X_{13}X_{34} &                    & Z_{41}Y_{13}X_{32} -Y_{41}X_{12}        \\
\Lambda_{42}^2 \,:\, & Y_{21}X_{14}  - Z_{21}Y_{13}X_{34} &                    & X_{41}X_{12} - Z_{41}X_{13}X_{32}       \\
\Lambda_{23} \,:\,   & X_{33}X_{32}  - X_{32}Z_{21}X_{12} &                    & Y_{21}X_{13} -X_{21}Y_{13}              \\
\Lambda_{43} \,:\,   & X_{33}X_{34}  - X_{34}Z_{41}X_{14} &                    & X_{41}Y_{13} - Y_{41}X_{13}             
\end{array}
\label{eq:H4PhaseAJEterms}
\end{equation}
The $W^{(0,1)}$ superpotential becomes
\begin{equation}
    \begin{split}
      W^{(0,1)}=&\,W^{(0,2)}
     +\Lambda_{11}^{4R}( X^\dagger_{12}X_{12}+ X^\dagger_{14}X_{14}+ X^\dagger_{21}X_{21}+ Y^\dagger_{21}Y_{21}+ Z^\dagger_{21}Z_{21}+\\
      &+ X^\dagger_{41}X_{41}+ Y^\dagger_{41}Y_{41}+ Z^\dagger_{41}Z_{41}+ X^\dagger_{13}X_{13}+ Y^\dagger_{13}Y_{13})+\\
     &+\Lambda_{22}^{R}( X^\dagger_{12}X_{12}+ X^\dagger_{32}X_{32}+ X^\dagger_{21}X_{21}+ Y^\dagger_{21}Y_{21}+ Z^\dagger_{21}Z_{21})+\\
     &+\Lambda_{33}^{R}( X^\dagger_{33}X_{33}+ X^\dagger_{32}X_{32}+ X^\dagger_{34}X_{34}+ X^\dagger_{13}X_{13}+ Y^\dagger_{13}Y_{13})+\\
     &+\Lambda_{44}^{R}( X^\dagger_{14}X_{14}+ X^\dagger_{34}X_{34}+ X^\dagger_{41}X_{41}+ Y^\dagger_{41}Y_{41}+ Z^\dagger_{41}Z_{41})\fstop
    \end{split}
    \label{eq:H4AW01sup}
\end{equation}

Table~\ref{tab:GenerH4PhaseA} shows the generators of the moduli space, which were computed using the HS, and their expression in terms of the chiral fields in phase A.
\begin{table}[H]
	\centering
	\renewcommand{\arraystretch}{1.1}
	\begin{tabular}{c|c}
		Meson & Chiral fields  \\
		\hline
		$M_1$ & $X_{33}=X_{14}Z_{41}=Z_{21}X_{12}$ \\
		$M_2$ & $Y_{21}X_{12}=Z_{41}Y_{13}X_{34}$ \\
		$M_3$ & $X_{14}Y_{41}=Z_{21}Y_{13}X_{32}$ \\
		$M_4$ & $X_{32}Y_{21}Y_{13}=X_{34}Y_{41}Y_{13}$ \\
		$M_5$ & $X_{21}X_{12}=Z_{41}X_{13}X_{34}$ \\
		$M_6$ & $X_{14}X_{41}=Z_{21}X_{13}X_{32}$ \\
		$M_7$ & $X_{32}Y_{21}X_{13}=X_{32}X_{21}Y_{13}=X_{34}Y_{41}X_{13}=X_{34}X_{41}Y_{13}$ \\
		$M_8$ & $X_{32}X_{21}X_{13}=X_{34}X_{41}X_{13}$ 
	\end{tabular}
	\caption{Generators of $H_4$ in Phase A.}
	\label{tab:GenerH4PhaseA}
\end{table}

The relations among the generators are
\begin{equation}
\begin{split}
    \mathcal{I} = &\left\langle M_1M_4=M_2M_3\coma M_1M_7=M_2M_6\coma M_1M_7=M_3M_5\coma M_2M_7=M_4M_5\coma\right.\\
    &\left.M_3M_7=M_4M_6\coma M_1M_8=M_5M_6\coma M_2M_8=M_5M_7\coma M_3M_8=M_6M_7\coma\right.\\
    &\left.M_4M_8=M_7^2\right\rangle\fstop
    \end{split}
\end{equation}

\bigskip

\paragraph{$\SO(N)\times U(N)\times \SO(N)$ Orientifold}\mbox{}

\medskip

Let us consider an anti-holomorphic involution of phase A which acts on Figure~\ref{fig:H4_A_quiver} as a reflection with respect to the diagonal connecting nodes $1$ and $3$. Then, nodes $1$ and $3$ map to themselves, while nodes $2$ and $4$ are identified.

The involution on chiral fields is
\begin{equation}
\begin{array}{cccccccccccc}
        X_{33} & \rightarrow  &  \gamma_{\Omega_3}\bar{X}_{33}\gamma_{\Omega_3}^{-1}\coma &  
        X_{14} & \rightarrow  &  \gamma_{\Omega_1}\bar{X}_{12}\gamma_{\Omega_2}^{-1} \coma &  
        Z_{41} & \rightarrow  &  \gamma_{\Omega_2}\bar{Z}_{21}\gamma_{\Omega_1}^{-1}\coma &  
        Y_{21} & \rightarrow  & \gamma_{\Omega_4}\bar{X}_{41}\gamma_{\Omega_1}^{-1}\coma \\
        Y_{13} & \rightarrow  & \gamma_{\Omega_1}\bar{X}_{13}\gamma_{\Omega_3}^{-1}\coma &  
        X_{34} & \rightarrow  &  \gamma_{\Omega_3}\bar{X}_{32}\gamma_{\Omega_2}^{-1}\coma &  
        Y_{41} & \rightarrow  &  \gamma_{\Omega_2}\bar{X}_{21}\gamma_{\Omega_1}^{-1}\coma & 
        X_{12} & \rightarrow  &  \gamma_{\Omega_1}\bar{X}_{14}\gamma_{\Omega_4}^{-1} \coma \\
        Z_{21} & \rightarrow  &  \gamma_{\Omega_4}\bar{Z}_{41}\gamma_{\Omega_1}^{-1}\coma &  
        X_{41} & \rightarrow  & \gamma_{\Omega_2}\bar{Y}_{21}\gamma_{\Omega_1}^{-1}\coma &  
        X_{13} & \rightarrow  & \gamma_{\Omega_1}\bar{Y}_{13}\gamma_{\Omega_3}^{-1}\coma &  
        X_{32} & \rightarrow  &  \gamma_{\Omega_3}\bar{X}_{34}\gamma_{\Omega_4}^{-1}\coma\\ 
        \multicolumn{12}{c}{X_{21}\rightarrow  \gamma_{\Omega_4}\bar{Y}_{41}\gamma_{\Omega_1}^{-1}\fstop}
    \end{array}
    \label{eq:chiralH4PhaseA_o_theory}
\end{equation}

It is interesting to note that since phase A is a chiral theory, it clearly illustrates a characteristic feature of anti-holomorphic involutions: they map chiral fields to images with the same orientation, as it follows from the discussion in Section~\ref{sec:orientN01gaugeth}.

From the invariance of $W^{(0,1)}$, we obtain the transformations of the Fermi fields
\begin{equation}
    \begin{array}{ccccccccc}
    \Lambda^1_{11} & \rightarrow &  -\gamma_{\Omega_1}\Lambda^{1\,\,T}_{11}\gamma_{\Omega_1}^{-1}\coma &
    \Lambda^2_{11} & \rightarrow &  \gamma_{\Omega_1}\Lambda^{2\,\,T}_{11}\gamma_{\Omega_1}^{-1}\coma &  
    \Lambda^3_{11} & \rightarrow &  -\gamma_{\Omega_1}\bar{\Lambda}^{3}_{11}\gamma_{\Omega_1}^{-1}\coma \\
    \Lambda^1_{13} & \rightarrow &  -\gamma_{\Omega_1}\bar{\Lambda}^{2}_{13}\gamma_{\Omega_3}^{-1}\coma &
    \Lambda^2_{13} & \rightarrow &  -\gamma_{\Omega_1}\bar{\Lambda}^{1}_{13}\gamma_{\Omega_3}^{-1}\coma & 
     \Lambda^1_{42} & \rightarrow &  -\gamma_{\Omega_4}\Lambda^{1\,\,T}_{42}\gamma_{\Omega_2}^{-1}\coma \\
    \Lambda^2_{42} & \rightarrow &  \gamma_{\Omega_4}\Lambda^{2\,\,T}_{42}\gamma_{\Omega_1}^{-1}\coma & 
    \Lambda_{23}   & \rightarrow &  \gamma_{\Omega_4}\bar{\Lambda}_{43}\gamma_{\Omega_3}^{-1}\coma & 
    \Lambda_{43}    &\rightarrow&    \gamma_{\Omega_2}\bar{\Lambda}_{23}\gamma_{\Omega_3}^{-1}\coma
    \end{array}
    \label{eq:fermiH4PhaseA_o_theory}
\end{equation}
and 
\begin{equation}
    \Lambda^{4R}_{11}\rightarrow \gamma_{\Omega_1}\Lambda^{4R\,\,T}_{11}\gamma_{\Omega_1}^{-1}\coma  
    \Lambda^{R}_{22}\rightarrow \gamma_{\Omega_4}\Lambda^{R\,\,T}_{44}\gamma_{\Omega_4}^{-1}\coma
    \Lambda^{R}_{33}\rightarrow \gamma_{\Omega_3}\Lambda^{R\,\,T}_{33}\gamma_{\Omega_3}^{-1}\coma
     \Lambda^{R}_{44}\rightarrow \gamma_{\Omega_2}\Lambda^{R\,\,T}_{22}\gamma_{\Omega_2}^{-1}\fstop
    \label{eq:realfermiH4PhaseA_o_theory}
\end{equation}

Using Table~\ref{tab:GenerH4PhaseA}, we find the corresponding geometric involution $\sigma$ on the generators of $H_4$  
\begin{equation}
\begin{array}{c}
     \left(M_1,M_2,M_3,M_4,M_5,M_6,M_7,M_8\right)\\
    \downarrow\\
 \left(\bar{M}_1,\bar{M}_6,\bar{M}_5,\bar{M}_8,\bar{M}_3,\bar{M}_2,\bar{M}_7,\bar{M}_4\right)\fstop
    \end{array}
    \label{eq:H4PhaseA-invol}
\end{equation}

The orientifolded theory has gauge group $G_1(N)\times \U(N) \times  G_3(N)$. The involution of the fields connecting nodes 1 and 3 implies that in this case we must have $\gamma_{\Omega_1}=\gamma_{\Omega_3}$. Then, $G_1(N)$ and $G_3(N)$ can be either both $\SO$ or both $\USp$ gauge groups, but cannot be of different types. For example, Figure~\ref{fig:o_theory_h4_A} shows the quiver for $G_2(N)=G_3(N)=\SO(N)$. The theory is free of gauge anomalies.

\begin{figure}[!htp]
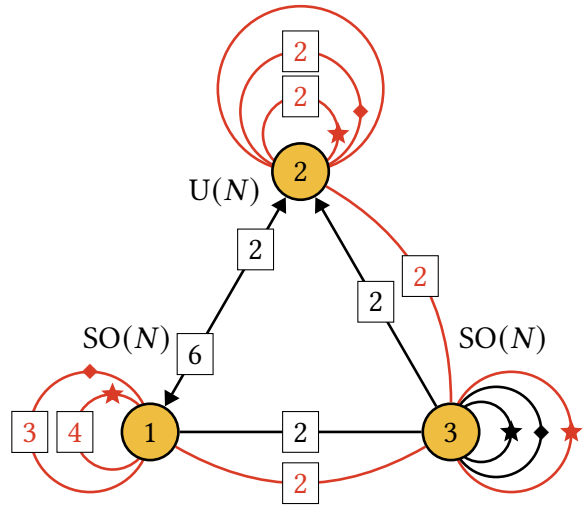

    \centering

    \caption{Quiver for a  Spin(7)  orientifold of phase A of $H_4$ using the involution in \eqref{eq:chiralH4PhaseA_o_theory}, \eqref{eq:fermiH4PhaseA_o_theory} and \eqref{eq:realfermiH4PhaseA_o_theory}.}
    \label{fig:o_theory_h4_A}
\end{figure}

\subsubsection{Phase B}

\label{sec:H4PhaseB}

Figure~\ref{fig:H4phaseBquiv} shows the quiver for phase $B$ of $H_4$.  

	\begin{figure}[H]
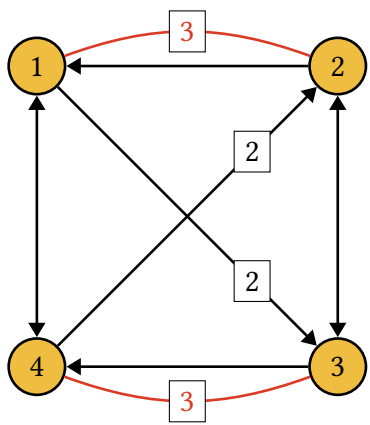
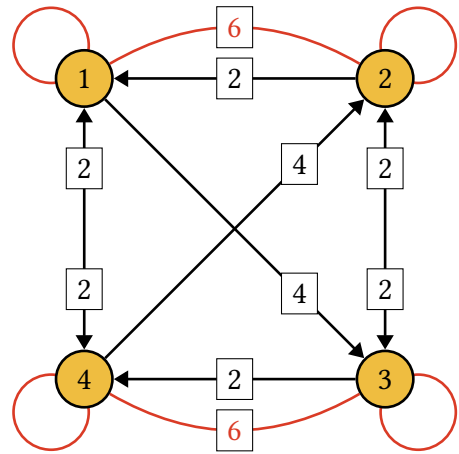

		\centering
		\begin{subfigure}[t]{0.49\textwidth}
			\centering

			\caption{$\mathcal{N}=(0,2)$ language.}
			\label{fig:H4BN02}
		\end{subfigure}\hfill
		\begin{subfigure}[t]{0.49\textwidth}
			\centering
			%
			\caption{$\mathcal{N}=(0,1)$ language.}
			\label{fig:H4BN01}
		\end{subfigure}
		\caption{Quiver for $H_4$ in phase B.}
		\label{fig:H4phaseBquiv}
	\end{figure}

The $J$- and $E$-terms are 
\begin{equation}
\renewcommand{\arraystretch}{1.1}
\begin{array}{lclc}
& J                                                                    &\text{\hspace{.5cm}}&  E                            \\
  \Lambda_{21} \,:\,     & X_{1 3}X_{34}Y_{42} - Y_{1 3}X_{34}X_{42}              &                    &  X_{21}X_{1 4}X_{41} - X_{23}X_{32}X_{21} \\
  \Lambda_{1 2}^{1} \,:\, & X_{23}X_{34}Y_{42}X_{21} - X_{21}Y_{1 3}X_{34}X_{41}  &                    &  X_{1 3}X_{32} - X_{1 4}X_{42}             \\
  \Lambda_{1 2}^{2} \,:\, & X_{21}X_{1 3}X_{34}X_{41} - X_{23}X_{34}X_{42}X_{21}  &                    &  Y_{1 3}X_{32} - X_{1 4}Y_{42}             \\
  \Lambda_{34} \,:\,     & Y_{42}X_{21}X_{1 3} - X_{42}X_{21}Y_{1 3}              &                    &  X_{34}X_{41}X_{1 4} - X_{32}X_{23}X_{34} \\
  \Lambda_{43}^{1} \,:\, & X_{34}Y_{42}X_{21}X_{1 4} - X_{32}X_{21}Y_{1 3}X_{34}  &                    &  X_{42}X_{23} - X_{41}X_{1 3}             \\                
  \Lambda_{43}^{2} \,:\, & X_{32}X_{21}X_{1 3}X_{34} - X_{34}X_{42}X_{21}X_{1 4}  &                    &  Y_{42}X_{23} - X_{41}Y_{1 3}  
\end{array}
\label{eq:H4PhaseB-JEterms}
\end{equation}

The $W^{(0,1)}$ superpotential is
\begin{equation}
    \begin{split}
        	W^{(0,1)}=&\,W^{(0,2)}+\Lambda^R_{11} (X^\dagger_{21}  X_{21}+X^\dagger_{41}  X_{41}+X^\dagger_{1 4}  X_{1 4}+X^\dagger_{1 3}  X_{1 3}+Y^\dagger_{1 3}  Y_{1 3})+\\
&+\Lambda^R_{22} (X^\dagger_{23}  X_{23}+X^\dagger_{21}  X_{21}+X^\dagger_{42}  X_{42}+X^\dagger_{32}  X_{32}+Y^\dagger_{42}  Y_{42})+\\
&+\Lambda^R_{33} (X^\dagger_{23}  X_{23}+X^\dagger_{32}  X_{32}+X^\dagger_{34}  X_{34}+X^\dagger_{1 3}  X_{1 3}+Y^\dagger_{1 3}  Y_{1 3})+\\	
&+\Lambda^R_{44} (X^\dagger_{42}  X_{42}+X^\dagger_{41}  X_{41}+X^\dagger_{34}  X_{34}+X^\dagger_{1 4}  X_{1 4}+Y^\dagger_{42}  Y_{42})\fstop
    \end{split}
    \label{eq:H4BW01sup}
\end{equation}

Table~\ref{tab:GenerH4PhaseB} shows the generators of $H_4$ in terms of the chiral fields in phase B. 
\begin{table}[H]
	\centering
	\renewcommand{\arraystretch}{1.1}
	\begin{tabular}{c|c}
		 Meson    & Chiral fields  \\
		\hline
$M_1$ & $X_{23}X_{32}=X_{41}X_{1 4}$ \\
$M_2$ & $X_{34}Y_{42}X_{23}=X_{34}X_{41}Y_{1 3}$ \\
$M_3$ & $X_{21}X_{1 4}Y_{42}=X_{21}Y_{1 3}X_{32}$ \\
$M_4$ & $X_{34}Y_{42}X_{21}Y_{1 3}$ \\
$M_5$ & $X_{34}X_{42}X_{23}=X_{34}X_{41}X_{1 3}$ \\
$M_6$ & $X_{21}X_{1 4}X_{42}=X_{21}X_{1 3}X_{32}$ \\
$M_7$ & $X_{42}X_{21}Y_{1 3}X_{34}=Y_{42}X_{21}X_{1 3}X_{34}$ \\
$M_8$ & $X_{42}X_{21}X_{1 3}X_{34}$
	\end{tabular}
	\caption{Generators of $H_4$ in Phase B.}
	\label{tab:GenerH4PhaseB}
\end{table}

They satisfy the following relations
\begin{equation}
\begin{split}
    \mathcal{I} = &\left\langle M_1M_4=M_2M_3\coma M_1M_7=M_2M_6\coma M_1M_7=M_3M_5\coma M_2M_7=M_4M_5\coma\right.\\
    &\left.M_3M_7=M_4M_6\coma M_1M_8=M_5M_6\coma M_2M_8=M_5M_7\coma M_3M_8=M_6M_7\coma\right.\\
    &\left.M_4M_8=M_7^2\right\rangle\fstop
    \end{split}
\end{equation}

This can be seen not only geometrically, but also from the gauge theory. While, as already mentioned, the generators and their relations are common to all the phases, their realizations in terms of chiral superfields in each of them are different. 

\bigskip

\paragraph{$\U(N)\times \U(N)$ Orientifold}\mbox{}

\smallskip

Let us consider an anti-holomorphic involution of phase B which acts on Figure~\ref{fig:H4phaseBquiv} as a reflection with respect to a horizontal line through the middle of the quiver. Nodes are mapped as $1 \leftrightarrow 4$ and $2 \leftrightarrow 3$.

The involution on chiral fields is
\begin{equation}
    \begin{array}{cccccccccccc}
        X_{23} & \rightarrow &   \gamma_{\Omega_3}\bar{X}_{32}\gamma_{\Omega_2}^{-1}\coma &   
        X_{41} & \rightarrow &   \gamma_{\Omega_1}\bar{X}_{14}\gamma_{\Omega_4}^{-1}\coma &   
        X_{34} & \rightarrow &   \gamma_{\Omega_2}\bar{X}_{21}\gamma_{\Omega_1}^{-1}\coma &   
        Y_{42} & \rightarrow &   \gamma_{\Omega_1}\bar{X}_{13}\gamma_{\Omega_3}^{-1}\coma \\
        Y_{13} & \rightarrow &   \gamma_{\Omega_4}\bar{X}_{42}\gamma_{\Omega_2}^{-1}\coma & 
        X_{32} & \rightarrow &   \gamma_{\Omega_2}\bar{X}_{23}\gamma_{\Omega_3}^{-1}\coma &  
        X_{14} & \rightarrow &   \gamma_{\Omega_4}\bar{X}_{41}\gamma_{\Omega_1}^{-1}\coma &  
        X_{21} & \rightarrow &   \gamma_{\Omega_3}\bar{X}_{34}\gamma_{\Omega_4}^{-1}\coma \\
        & & & 
        X_{13} & \rightarrow &   \gamma_{\Omega_4}\bar{Y}_{42}\gamma_{\Omega_2}^{-1}\coma &
        X_{42} &\rightarrow & \gamma_{\Omega_1}\bar{Y}_{13}\gamma_{\Omega_3}^{-1}\fstop
    \end{array}
    \label{eq:chiralH4PhaseBorient}
\end{equation}
Requiring the invariance of $W^{(0,1)}$, we obtain the transformations for the Fermi fields
\begin{equation}
    \begin{array}{ccccccccc}
        \Lambda_{21}     & \rightarrow &  \gamma_{\Omega_3}\bar{\Lambda}_{34} \gamma_{\Omega_4}^{-1}\coma &  
        \Lambda^1_{12}   & \rightarrow &  \gamma_{\Omega_4}\bar{\Lambda}^{2}_{43} \gamma_{\Omega_3}^{-1}\coma &  
        \Lambda^{2}_{12} & \rightarrow &  \gamma_{\Omega_4}\bar{\Lambda}^{1}_{43} \gamma_{\Omega_3}^{-1}\coma \\
        \Lambda_{34}     & \rightarrow &  \gamma_{\Omega_2}\bar{\Lambda}_{21}\gamma_{\Omega_1}^{-1}\coma &  
        \Lambda^1_{43}   & \rightarrow &  \gamma_{\Omega_1}\bar{\Lambda}^{2}_{12}\gamma_{\Omega_2}^{-1}\coma &  
        \Lambda^{2}_{43} & \rightarrow &  \gamma_{\Omega_1}\bar{\Lambda}^{1}_{12}\gamma_{\Omega_2}^{-1}\coma
    \end{array}
    \label{eq:fermiH4PhaseBorient}
\end{equation}
and 
\begin{equation}
        \Lambda^{R}_{11}\rightarrow \gamma_{\Omega_4}\Lambda^{R\,\,T}_{44}\gamma_{\Omega_4}^{-1}\coma \Lambda^{R}_{22}\rightarrow \gamma_{\Omega_3}\Lambda^{R\,\,T}_{33}\gamma_{\Omega_3}^{-1}\coma
        \Lambda^{R}_{33}\rightarrow \gamma_{\Omega_2}\Lambda^{R\,\,T}_{22}\gamma_{\Omega_2}^{-1}\coma 
        \Lambda^{R}_{44}\rightarrow \gamma_{\Omega_1}\Lambda^{R\,\,T}_{11}\gamma_{\Omega_1}^{-1}\fstop 
    \label{eq:realfermiH4PhaseBorient}
\end{equation}

Using Table~\ref{tab:GenerH4PhaseB}, we get the corresponding geometric involution $\sigma$ on the generators of $H_4$
\begin{equation}
\begin{array}{c}
     \left(M_1,M_2,M_3,M_4,M_5,M_6,M_7,M_8\right)\\
    \downarrow\\
 \left(\bar{M}_1,\bar{M}_6,\bar{M}_5,\bar{M}_8,\bar{M}_3,\bar{M}_2,\bar{M}_7,\bar{M}_4\right)\coma
    \end{array}
    \label{eq:H4PhaseB-invol}
\end{equation}
which is exactly the same anti-holomorphic involution in \eqref{eq:H4PhaseA-invol}. In Section~\ref{section_vector_structure_H4} we elaborate on the relation between both theories and the role of vector structure.

Figure~\ref{fig:o_theory_h4_B} shows the quiver for the orientifolded theory, which is anomaly free. 

\begin{figure}[!htp]
    \centering
    %
    \caption{Quiver for a  Spin(7)  orientifold of phase B of $H_4$ using the involution in \eqref{eq:chiralH4PhaseBorient}, \eqref{eq:fermiH4PhaseBorient} and \eqref{eq:realfermiH4PhaseBorient}.}
    \label{fig:o_theory_h4_B}
\end{figure}

\subsubsection{Vector Structure Explanation}

\label{section_vector_structure_H4}

On general grounds, one can expect that considering orientifolds by the same anti-holomorphic involution on geometries in different toric phases of the same geometry, should lead to equivalent $\mathcal{N}=(0,1)$ theories. Indeed, this can lead to a systematic construction of $\mathcal{N}=(0,1)$ theories related by 2d trialities, as we will discuss in Chapter \ref{cha:ch3}.

On the other hand, this is not the case for the two orientifolds constructed in the previous section. We have seen that the $H_4$ theory admits several orientifold quotients which nevertheless correspond to the same anti-holomorphic involution, see \eqref{eq:H4PhaseA-invol} and \eqref{eq:H4PhaseB-invol}.  In this section, we show that the resulting theories are different because they correspond to orientifold quotients with or without vector structure, realized in the context of a non-orbifold singularity. 

Indeed, the structure of the orientifold action on the gauge factors follows the pattern described in Section \ref{sec:vectorstructure} for orientifolds of orbifolds of $\mathbb{C}^4$. Namely, the orientifold in Section~\ref{sec:H4PhaseA} acts on the quiver of the $H_4$ theory (in the toric phase A) by swapping the two nodes 2 and 4, while mapping nodes 1 and 3 to themselves; this corresponds to an orientifold with vector structure. On the other hand, the orientifold in Section~\ref{sec:H4PhaseB} acts on the quiver of the $H_4$ theory (in the toric phase B) by swapping 1 $\leftrightarrow$ 4 and 2 $\leftrightarrow$ 3; this corresponds to an orientifold without vector structure.

Hence, even though the two models correspond to the same underlying geometry, with an orientifold action associated to the same anti-holomorphic involution, the resulting orientifold theories are associated to genuinely different actions of the orientifold on the gauge degrees of freedom, and lead to inequivalent models.

An interesting observation is that the orientifolds with and without vector structure are obtained as orientifold quotients of the theory in two different toric phases. This effect did not arise in the context of orbifolds of $\mathbb{C}^4$, since these do not admit multiple toric phases; on the other hand, it is actually an expected phenomenon in non-orbifold singularities, as it already occurs in the context of 4d ${\cal N}=1$ theories with D3-branes at orientifold singularities. We illustrate this with the following simple example.

Consider a set of D3-branes at the tip of the non-compact CY 3-fold singularity described by the equation
\begin{equation}
    xy=z^2w^2 \fstop
\end{equation}
This corresponds to a $\ZZ_2$ quotient of the conifold, of the kind introduced in \cite{Uranga:1998vf} as T-duals of 4d Hanany-Witten (HW) configurations of D4-branes suspended between NS and rotated NS-branes (aka NS$^\prime$-branes). This T-dual picture allowed to recover the same geometry from different Seiberg dual phases, as explicitly discussed in Section 3 of \cite{Feng:2001bn}. In particular, we can describe a phase A as corresponding to the type IIA configuration of D4-branes suspended in intervals separated by NS-branes ordered as NS - NS - NS$^\prime$ - NS$^\prime$ on the circle, and a phase B as corresponding to D4-branes suspended between NS-branes ordered as NS$^\prime$ - NS - NS$^\prime$ - NS on the circle. 

\begin{figure}[!htp]
\begin{center}
\begin{subfigure}[t]{0.49\textwidth}
\centering
\begin{tikzpicture}[scale=1.75]
\draw[thick] (0,0) circle (1);
\draw[dashed,greenX,thick] (0,0.5) -- node[black,pos=0.80,right] {O6} (0,1.5);
\draw[dashed,greenX,thick] (0,-0.5) -- node[black,pos=0.80,right] {O6} (0,-1.5);
\node[draw=blueX,line width=1pt,circle,fill=blueX,minimum width=0.3cm,inner sep=1pt,label={above right:NS}] (NS1) at (45:1) {};
\node[draw=blueX,line width=1pt,circle,fill=blueX,minimum width=0.3cm,inner sep=1pt,label={above left:NS}] (NS2) at (135:1) {};
\node[draw=redX,line width=1pt,circle,fill=redX,minimum width=0.3cm,inner sep=1pt,label={below left:NS$^\prime$}] (NSp1) at (225:1) {};
\node[draw=redX,line width=1pt,circle,fill=redX,minimum width=0.3cm,inner sep=1pt,label={below right:NS$^\prime$}] (NSp2) at (315:1) {};
\node at (0:1.2) {D4};
\node at (110:1.2) {D4};
\node at (180:1.2) {D4};
\node at (250:1.2) {D4};
\end{tikzpicture}
\caption{Phase A.}
\label{fig:NSNSpPhaseA}
\end{subfigure}\hfill
\begin{subfigure}[t]{0.49\textwidth}
\centering
\begin{tikzpicture}[scale=1.75]
\draw[thick] (0,0) circle (1);
\draw[dashed,greenX,thick] (0,0.5) -- node[black,pos=0.80,right] {O6} (0,1.5);
\draw[dashed,greenX,thick] (0,-0.5) -- node[black,pos=0.80,right] {O6} (0,-1.5);
\node[draw=blueX,line width=1pt,circle,fill=blueX,minimum width=0.3cm,inner sep=1pt,label={above left:NS}] (NS1) at (90:1) {};
\node[draw=blueX,line width=1pt,circle,fill=blueX,minimum width=0.3cm,inner sep=1pt,label={below left:NS}] (NS2) at (270:1) {};
\node[draw=redX,line width=1pt,circle,fill=redX,minimum width=0.3cm,inner sep=1pt,label={right:NS$^\prime$}] (NSp1) at (0:1) {};
\node[draw=redX,line width=1pt,circle,fill=redX,minimum width=0.3cm,inner sep=1pt,label={left:NS$^\prime$}] (NSp2) at (180:1) {};
\node at (45:1.2) {D4};
\node at (135:1.2) {D4};
\node at (225:1.2) {D4};
\node at (315:1.2) {D4};
\end{tikzpicture}
\caption{Phase B.}
\label{fig:NSNSpPhaseB}
\end{subfigure}
\caption{Configurations of D4-branes suspended between NS- and NS$^\prime$-branes in the presence of O6-planes leading to two different 4d ${\cal N}=1$ theories from orientifold quotients of the same CY$_3$ geometry differing only on the existence (Figure~\ref{fig:NSNSpPhaseA}) or not (Figure~\ref{fig:NSNSpPhaseB}) of vector structure.}
\label{fig:toric-phases-4d}
\end{center}
\end{figure}
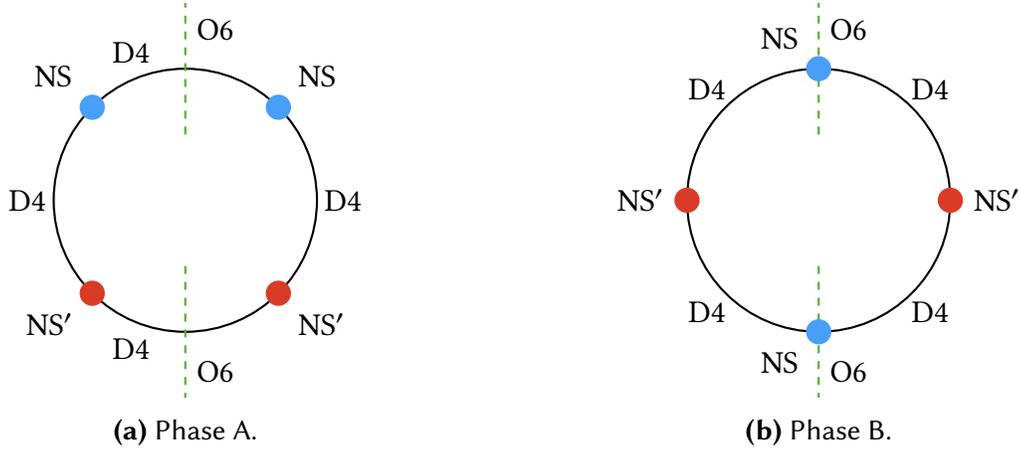

Let us now perform an orientifold quotient in the type IIB geometry, which corresponds to, e.g., introducing O6-planes in the type IIA T-dual; this can map NS-branes to NS-branes, and NS$^\prime$-branes to NS$^\prime$-branes, and cannot swap NS- and NS$^\prime$-branes. Hence, for phase A, the only $\mathbb{Z}_2$-invariant configuration must have the orientifold swapping the two NS branes, and swapping the two NS$^\prime$-branes, see Figure \ref{fig:NSNSpPhaseA}; hence, the interval between the two NS-branes and the interval between the two NS$^\prime$-branes are both mapped to themselves under the orientifold action, while the intervals between NS- and  NS$^\prime$-branes are swapped. The result corresponds to an orientifold with vector structure.

On the other hand, for phase B, a $\mathbb{Z}_2$-invariant configuration has e.g. NS-branes mapped to themselves under the orientifold action, and the two NS$^\prime$-branes swapped, see Figure \ref{fig:NSNSpPhaseB}; hence, no interval is mapped to itself, rather the four intervals are swapped pairwise. The result corresponds to an orientifold without vector structure (there is an equivalent model obtained by having NS$^\prime$-branes on top of the orientifold plane, and the two NS-branes swapped under the orientifold action).

This illustrates the fact that the construction of orientifolds with or without vector structure, for a given geometric involution, may require their realization in different toric phases.

We have thus shown that, in order for equivalent orientifold geometric involutions to produce equivalent theories, it is necessary that they also agree on the choice of vector structure they implicitly define. This is an important ingredient in the application of orientifold quotients to $\mathcal{N}=(0,2)$ trialities to generate examples of theories displaying $\mathcal{N}=(0,1)$ triality \cite{Franco:2021vxq}.

\section{Partial Resolution and Higgsing}
\label{sec:HiggsingPartResol}

In this section, we study partial resolutions connecting two different $\Spin(7)$ orientifolds, which translate into higgsings between the corresponding gauge theories.

\subsection{General Idea}
\label{sec:HiggsingOrient}

Consider two CY$_4$'s, CY$_4^{(1)}$ and CY$_4^{(2)}$, connected via partial resolution. Let us call the gauge theories on D1-branes probing them $\mathcal{T}_1^{(0,2)}$ and $\mathcal{T}_2^{(0,2)}$, respectively.\footnote{More precisely, we mean one of the various phases related via $\mathcal{N}=(0,2)$ triality for each CY$_4$.} Partial resolution translates into higgsing connecting the two gauge theories, in which the scalar component of one or more chiral fields gets a non-zero VEV (as usual, this is meant in the Born-Oppenheimer approximation in 2d). In the process, part of the gauge symmetry is higgsed and some matter fields may become massive. We refer to \cite{Franco:2015tna} for a more detailed discussion and explicit examples.

Let us now consider a $\Spin(7)$ orientifold $\mathcal{O}_1$ of CY$_4^{(1)}$ associated to a given anti-holomorphic involution $\sigma$. If the partial resolution considered above is symmetric under $\sigma$, it gives rise to a partial resolution of $\mathcal{O}_1$ into a $\Spin(7)$ orientifold $\mathcal{O}_2$ of CY$_4^{(2)}$. At the field theory level, the VEVs that higgs $\mathcal{T}_1^{(0,2)}\to \mathcal{T}_2^{(0,2)}$ are symmetric under the involution and project onto a higgsing between the orientifold gauge theories, $\mathcal{T}_1^{(0,1)}\to \mathcal{T}_2^{(0,1)}$. Figure~\ref{partial resolution, orientifolding and higgsing} illustrates the interplay between partial resolution, orientifolding and higgsing.

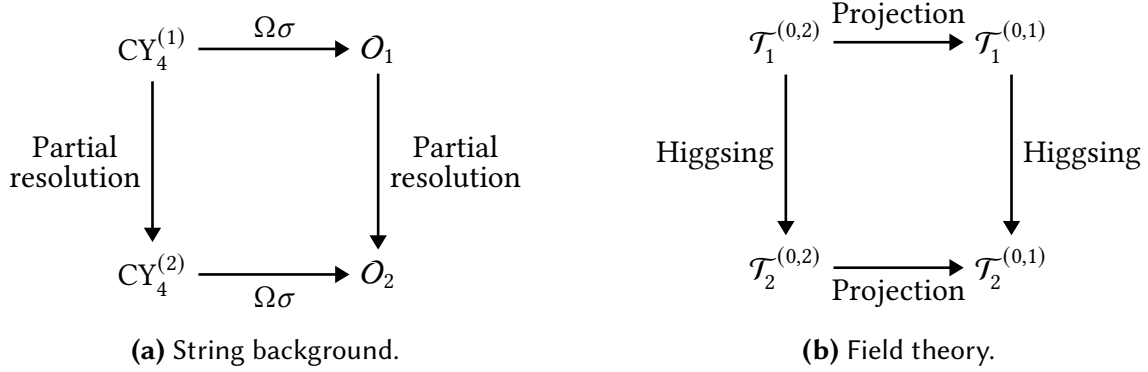
\begin{figure}[ht!]
    \centering
    \begin{subfigure}[t]{0.49\textwidth}
    \centering
       \begin{tikzpicture}[scale=2]
    \def\x{1.5};
    \node (A) at (0,0) {CY$_4^{(1)}$};
    \node (B) at (1*\x,0) {$\mathcal{O}_1$};
    \node (C) at (1*\x,-\x) {$\mathcal{O}_2$};
    \node (D) at (0,-\x) {CY$_4^{(2)}$};
    \draw[line width=1pt,-Triangle] (A)-- node[above,midway]{$\Omega\sigma$} (B);
    \draw[line width=1pt,-Triangle] (B)-- node[right,midway] {\shortstack{Partial\\resolution}} (C);
    \draw[line width=1pt,Triangle-] (C)-- node[below,midway]{$\Omega\sigma$}(D);
    \draw[line width=1pt,Triangle-] (D)-- node[left,midway] {\shortstack{Partial\\resolution}} (A);
    \end{tikzpicture}
    \caption{String background.}
    \label{fig:UniInvoParRes}
    \end{subfigure}\hfill
    \begin{subfigure}[t]{0.49\textwidth}
    \centering
    \begin{tikzpicture}[scale=2]
    \def\x{1.5};
    \node (A) at (0,0) {$\mathcal{T}_1^{(0,2)}$};
    \node (B) at (\x,0) {$\mathcal{T}_1^{(0,1)}$};
    \node (C) at (\x,-\x) {$\mathcal{T}_2^{(0,1)}$};
    \node (D) at (0,-\x) {$\mathcal{T}_2^{(0,2)}$};
    \draw[line width=1pt,-Triangle] (A)-- node[above,midway]{Projection} (B);
    \draw[line width=1pt,-Triangle] (B)-- node[right,midway] {Higgsing} (C);
    \draw[line width=1pt,Triangle-] (C)-- node[below,midway]{Projection}(D);
    \draw[line width=1pt,Triangle-] (D)-- node[left,midway] {Higgsing} (A);
    \end{tikzpicture}
    \caption{Field theory.}
    \label{fig:UniInvoHiggs}
    \end{subfigure}
    \caption{Interplay between partial resolution, orientifolding and higgsing.}
    \label{partial resolution, orientifolding and higgsing}
\end{figure}

\subsection{Partial Resolution and the Universal Involution}

Interestingly, for theories obtained via the universal involution, {\it every} partial resolution between CY$_4$'s maps to a partial resolution between $\Spin(7)$ orientifolds. In this case, every field in $\mathcal{T}_1^{(0,2)}$ and $\mathcal{T}_2^{(0,2)}$ is its own orientifold image. Therefore, the condition that chiral fields and their images get VEVs simultaneously is automatically satisfied. 

Under the universal involution, VEVs and the resulting higgsing of the gauge symmetry and mass terms for some matter fields straightforwardly map from the parent to the orientifolded theory. In other words, higgsing survives the ``real slicing" of the universal involution.

\subsection{Beyond the Universal Involution: $\mathbb{C}^4/\mathbb{Z}_2\times \mathbb{Z}_2 \rightarrow \SPP\times \mathbb{C}$}
\label{sec:C4Z2Z2toSPPC}

The interplay between partial resolutions and orientifolds that we discussed above is not limited to the universal involution. 

\subsubsection*{The Parent}

Let us consider the $\mathbb{C}^4/\mathbb{Z}_2\times \mathbb{Z}_2$ orbifold, with the two $\ZZ_2$ groups generated by the actions $(1,1,0,0)$ and $(1,0,1,0)$ on $\mathbb{C}^4$, as phase rotations (in units of $\pi$). From now on, we will omit these vectors. This orbifold can be partially resolved to SPP$\times \mathbb{C}$, where SPP denotes the complex cone over the suspended pinch point. The toric diagrams and gauge theories for both geometries can be found in Appendices~\ref{app:otheory for c4z2z2} and \ref{app:otheory for sppXC}. This partial resolution and its translation into higgsing of the gauge theory has been discussed in detail in \cite{Franco:2015tna}.

The two theories are connected by turning on a VEV for $X_{13}$.\footnote{There are other choices of the chiral field getting a VEV that lead to the same resolution. They are equivalent to this choice by symmetries.} As a result, the $\U(N)_1\times \U(N)_3$ gauge groups are broken to the diagonal $\U(N)_{1/3}$. In addition, the following Fermi-chiral pairs become massive
\begin{equation}\label{eq:massive fields in (0,2)}
    \{ \Lambda_{21}, X_{32}\},~ \left\{ \Lambda_{13}, \frac{{X}_{11}-X_{33}}{2}\right\},~\{ \Lambda_{41}, X_{34}\},~\{\Lambda_{32}, X_{21}\},~\{\Lambda_{34}, X_{41}\}\fstop
\end{equation}
Integrating out the massive fields leads to the gauge theory for SPP$\times \mathbb{C}$.

\subsubsection*{The Spin(7) Orientifold}

In Appendix~\ref{app:otheory for c4z2z2}, we present a $\Spin(7)$ orientifold of $\mathbb{C}^4/\mathbb{Z}_2\times \mathbb{Z}_2$ constructed using a non-universal involution, given in~\cref{eq:C4Z2Z2chiral,eq:C4Z2Z2fermi,eq:C4Z2Z2realfermi}. The crucial point of that involution for the discussion in this section is that it maps $X_{13}$ to itself. Following the discussion in Section~\ref{sec:HiggsingOrient}, the resolution/higgsing of the parent is therefore projected onto one for the $\Spin(7)$ orientifold.

In the $\Spin(7)$ orientifold of $\mathbb{C}^4/\mathbb{Z}_2 \times \mathbb{Z}_2$, the higgsing associated to this partial resolution proceeds by giving a VEV to $X_{13}^R$. This breaks the $\SO(N)_1\times \SO(N)_3$ gauge symmetry into the diagonal $\SO(N)_{1/3}$. In addition, the combination of real Fermi fields $\frac{\Lambda_{11S}^R+\Lambda_{33S}^R}{2}$, coming from the $\mathcal{N}=(0,2)$ vector multiplets of gauge groups 1 and 3, become massive. Finally, The following fields also become massive
\begin{equation}\label{eq:massive fields in (0,1)}
   \Lambda_{21}, ~X_{32},~\Lambda_{13}^R,~ \frac{{X}_{11S}^R-X_{33S}^R}{2},~\frac{{X}_{11A}^R-X_{33A}^R}{2},~\Lambda_{32}, X_{21}\fstop
\end{equation}
 Integrating them out, each of the surviving bifundamentals of $\SO(N)_1\times \SO(N)_3$ becomes a symmetric and an antisymmetric of $\SO(N)_{1/3}$. The resulting theory is exactly the one associated for the $\Spin(7)$ orientifold of SPP$\times \mathbb{C}$ in Figure~\ref{fig:o_theory_SPP}, generated by the anti-holomorphic involution in \cref{eq:SPPchiral,eq:SPPfermi,eq:SPPrealfermi}.

\section{Conclusions}

\label{section_conclusions}

In this chapter, we initiated the geometric engineering of 2d $\mathcal{N}=(0,1)$ gauge theories by means of D1-branes probing (orientifolds of) Spin(7) cones. In particular, we introduced Spin(7) orientifolds, which are constructed by starting from a CY$_4$ cone and quotienting it by a combination of an anti-holomorphic involution leading to a Spin(7) cone and worldsheet parity. 

We illustrated this construction with various examples, including theories coming from both orbifold and non-orbifold parent singularities, discussed the r\^ole of the choice of vector structure in the orientifold quotient, and studied partial resolutions. 

Spin(7) orientifolds explicitly realize the perspective on 2d $\mathcal{N}=(0,1)$  theories as real slices of $\mathcal{N}=(0,2)$ ones. Remarkably, this projection is mapped to Joyce’s construction of Spin(7) manifolds as quotients of CY$_4$’s by anti-holomorphic involutions.

We envision multiple directions for future research. To name a few: 
\begin{itemize}

\item In general, the map between Spin(7) orientifolds and 2d $\mathcal{N}=(0,1)$ gauge theories is not one-to-one but one-to-many. We will investigate this issue in \cite{Franco:2021vxq}, showing that this non-uniqueness provides a geometric understanding of $\mathcal{N}=(0,1)$ triality.

\item Another interesting direction is to construct the gauge theories on D1-branes probing Spin(7) manifolds obtained from CY$_4$’s via Joyce’s construction, without the additional quotient by worldsheet parity leading to Spin(7) orientifolds. A significant part of the results of this chapter would also be useful for such setups. We plan to study this problem in a future work. 

\item In Section~\ref{sec:HiggsingPartResol}, we considered resolutions of Spin(7) orientifolds. It would be interesting to investigate deformations and their gauge theory counterpart.  Understanding deformations for CY$_4$’s and their translation to the associated 2d $\mathcal{N}=(0,2)$ gauge theories would be a useful preliminary step, which is interesting in its own right. 

\end{itemize}

We hope that the novel perspective on 2d $\mathcal{N}=(0,1)$ introduced in this chapter will provide a useful tool for understanding their dynamics.



%% file: chapters/3.tex
\section{Introduction}
\label{sec:Intro}
	
2d $\mathcal{N}=(0,1)$ quantum field theories are extremely interesting, since they are barely supersymmetric and live at the borderline between non-SUSY theories and others with higher amounts of SUSY, for which powerful tools such as holomorphy become applicable. Due to the reduced SUSY, they enjoy a broad range of interesting dynamics. While there has been recent progress in their understanding, they remain relatively unexplored.

In \cite{Gadde:2013lxa}, it was discovered that $2$d $\mathcal{N}= (0, 2)$ theories exhibit IR dualities reminiscent of Seiberg duality in 4d $\mathcal{N}=1$ gauge theories \cite{Seiberg:1994pq}. This low-energy equivalence was dubbed {\it  triality} since, in its simplest incarnation, three SQCD-like theories become equivalent at low energies. Recently, an IR {\it triality} between 2d $\mathcal{N} = (0,1)$ theories with $\SO$ and $\USp$ gauge groups was proposed in \cite{Gukov:2019lzi}. Evidence supporting the proposal includes matching of anomalies and elliptic genera. This new triality can be regarded as a relative of its $\mathcal{N}= (0, 2)$ counterpart. 

The geometric engineering of $2$d $\mathcal{N} = (0, 1)$ gauge theories on D1-branes probing singularities was initiated in \cite{Franco:2021ixh}, where a new class of backgrounds denoted {\it Spin(7) orientifolds} was introduced. These orientifolds are quotients of Calabi-Yau (CY) 4-folds by a combination of an anti-holomorphic involution leading to a Spin(7) cone and worldsheet parity. They provide a beautiful correspondence between the perspective of $\mathcal{N} = (0, 1)$ theories as real slices of $\mathcal{N} = (0, 2)$  theories and Joyce’s geometric construction of Spin(7) manifolds starting from CY 4-folds. This geometric perspective provides a new approach for studying $2$d $\mathcal{N}=(0,1)$ theories.

For branes at singularities, a single geometry often corresponds to multiple gauge theories. Such non-uniqueness is the manifestation of gauge theory dualities in this context. Examples of this phenomenon abound in different dimensions. The various 4d $\mathcal{N}=1$ gauge theories on D3-branes over the same CY 3-fold are related by Seiberg duality \cite{Seiberg:1994pq,Beasley:2001zp,Feng:2001bn}. The triality of 2d $\mathcal{N}=(0,2)$ gauge theories on D1-branes over CY 4-folds and the quadrality of 0d $\mathcal{N}=1$ gauge theories on D$(-1)$-branes over CY 5-folds can be similarly understood \cite{Franco:2016nwv,Franco:2016tcm}. These ideas were further extended to the $(m +1)$-dualities of the $m$-graded quivers that describe the open string sector of the topological B-model on CY $(m + 2)$-folds for arbitrary $m\geq 0$ \cite{Franco:2017lpa,Closset:2018axq,Franco:2020ijt}. In this chapter, we will show that the engineering of $2$d $\mathcal{N}=(0,1)$ gauge theories in terms of D1-branes probing Spin(7) orientifolds leads to a similar perspective on $\mathcal{N}=(0,1)$ triality.

The chapter is organized as follows. In Section \ref{sec:N02N01trial} we review $\mathcal{N}=(0,2)$ and $\mathcal{N}=(0,1)$ trialities in their original formulations and comment on their generalizations to quivers. We discuss Spin(7) orientifolds and the corresponding $2$d $\mathcal{N}=(0,1)$ field theories arising on D$1$-branes probing them in Section \ref{sec:Spin7Orient}. In Section \ref{sec:01trial-UI} we explain how the basic $\mathcal{N}=(0,1)$ triality arises from the \textit{universal involution}. In Section \ref{sec:BeyonUI} we investigate how (generalizations of) $\mathcal{N}=(0,1)$ triality arise in the case of Spin(7) orientifolds based on more general involutions. We present our conclusions in Section \ref{sec:conclusions}. In Appendix \ref{app:Q111Z2-details} we give all the necessary details for the phases of $Q^{1,1,1}/\ZZ_2$ involved in the triality web introduced in Section \ref{sec:Q111Z2}.


\section{$\mathcal{N}=(0,2)$ and $\mathcal{N}=(0,1)$ Triality}
\label{sec:N02N01trial}

In this section, we review the trialities of 2d $\mathcal{N}=(0,2)$ \cite{Gadde:2013lxa} and $\mathcal{N}=(0,1)$ \cite{Gukov:2019lzi} gauge theories. Discussing $\mathcal{N}=(0,2)$ triality first is not only useful for setting the stage since both trialities share various features, but it is also convenient since Spin(7) orientifolds connect them.

\subsection{$\mathcal{N}=(0,2)$ Triality}
\label{sec:N02triality}

Here we present a quick review of 2d $\mathcal{N}=(0,2)$ triality. A detailed discussion can be found in \cite{Gadde:2013lxa}. Additional developments, including connections to 4d, its realization in terms of D1-branes at CY$_4$ singularities, brane brick models and mirror symmetry, appear in \cite{Gadde:2015wta,Franco:2016nwv,Franco:2016qxh,Franco:2016fxm,Franco:2018qsc}.

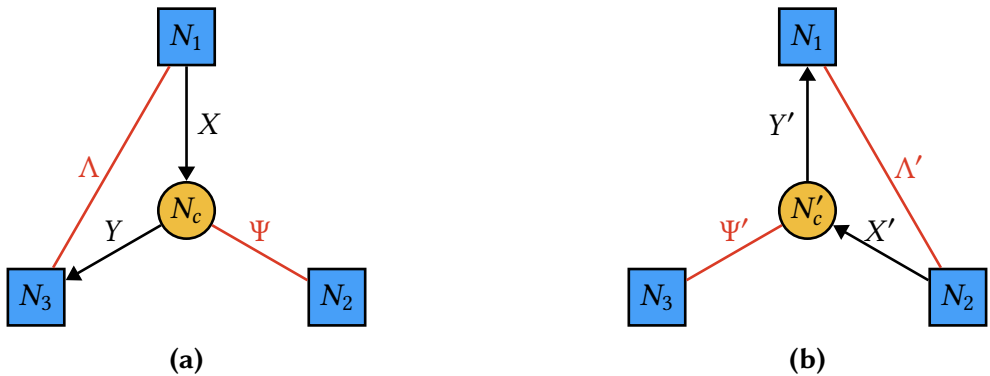
\begin{figure}[H]
	\centering
	\begin{subfigure}{0.5\textwidth}
		\centering
		\begin{tikzpicture}[scale=2]
			\def\L{2};
			\node[draw=black,line width=1pt,circle,fill=yellowX,minimum width=0.75cm,inner sep=1pt] (Nc) at (0,-0.289) {$N_c$};
			\node[draw=black,line width=1pt,fill=blueX,square,minimum width=0.75cm,inner sep=1pt] (N1) at (0,\L*0.433) {$N_1$};
			\node[draw=black,line width=1pt,fill=blueX,square,minimum width=0.75cm,inner sep=1pt] (N2) at (\L/2,-\L*0.433) {$N_2$};
			\node[draw=black,line width=1pt,fill=blueX,square,minimum width=0.75cm,inner sep=1pt] (N3) at (-\L/2,-\L*0.433) {$N_3$};
			\draw[line width=1pt,Triangle-] (Nc) -- node[right,midway] {$X$}  (N1);
			\draw[line width=1pt,redX] (Nc) -- node[above,midway] {$\Psi$} (N2);
			\draw[line width=1pt,redX] (N1) -- node[left,midway] {$\Lambda$} (N3);
			\draw[line width=1pt,-Triangle] (Nc) -- node[above,midway] {$Y$} (N3);
		\end{tikzpicture}
		\caption{}
		\label{fig:SQCDN02}
	\end{subfigure}\hfill
	\begin{subfigure}{0.5\textwidth}
		\centering
		\begin{tikzpicture}[scale=2]
			\def\L{2};
			\node[draw=black,line width=1pt,circle,fill=yellowX,minimum width=0.75cm,inner sep=1pt] (Nc) at (0,-0.289) {$N_c^\prime$};
			\node[draw=black,line width=1pt,fill=blueX,square,minimum width=0.75cm,inner sep=1pt] (N1) at (0,\L*0.433) {$N_1$};
			\node[draw=black,line width=1pt,fill=blueX,square,minimum width=0.75cm,inner sep=1pt] (N2) at (\L/2,-\L*0.433) {$N_2$};
			\node[draw=black,line width=1pt,fill=blueX,square,minimum width=0.75cm,inner sep=1pt] (N3) at (-\L/2,-\L*0.433) {$N_3$};
			\draw[line width=1pt,-Triangle] (Nc) -- node[left,midway] {$Y^\prime$}  (N1);
			\draw[line width=1pt,Triangle-] (Nc) -- node[above,midway] {$X^\prime$} (N2);
			\draw[line width=1pt,redX] (N1) -- node[right,midway] {$\Lambda^\prime$} (N2);
			\draw[line width=1pt,redX] (Nc) -- node[above,midway] {$\Psi^\prime$} (N3);
		\end{tikzpicture}
		\caption{}
		\label{fig:SQCDN02prime}
	\end{subfigure}
	\caption{$2$d $\mathcal{N}=(0,2)$ SQCD and its triality dual. The central nodes have ranks given in \eqref{eq:N02NcNcprank}.}
	\label{fig:SQCDN02transfor}
\end{figure}

Without loss of generality, we can focus on the quiver shown in Figure \ref{fig:SQCDN02}, which can be regarded as 2d $\mathcal{N}=(0,2)$ SQCD. The yellow node represents the $\SU(N_c)$ gauge group that undergoes triality, while the blue nodes are flavor $\SU(N_i)$ groups, $i=1,\ldots,3$.\footnote{More generally, as in theories arising on D1-branes probing CY$_4$ singularities, such groups can have additional matter charged under them and be gauged.} We have absorbed the multiplicities of flavor fields in the ranks of the flavor nodes. In $\mathcal{N}=(0,2)$ quivers, we adopt the convention that the head and tail of the arrow associated to a chiral field correspond to fundamental and antifundamental representations, respectively. A Fermi field connecting the flavor nodes 1 and 3 has been included to make the original and dual theories more similar.

The triality dual is shown in Figure \ref{fig:SQCDN02prime}. The rank of the central node in both theories is determined by anomaly cancellation to be
\begin{equation}
	N_c = {N_1 +N_3 - N_2 \over 2} \coma
	N_c' =  {N_2 +N_1 - N_3 \over 2} \fstop
	\label{eq:N02NcNcprank}
\end{equation}
The transformation of the rank can also be written as
\begin{equation}
	N_c'= N_1 -N_c \fstop
	\label{triality_transformation_rank}
\end{equation}
Both theories in Figure \ref{fig:SQCDN02transfor} have $J$-/$E$- terms associated to the triangular loops in the quivers.

\begin{figure}[H]
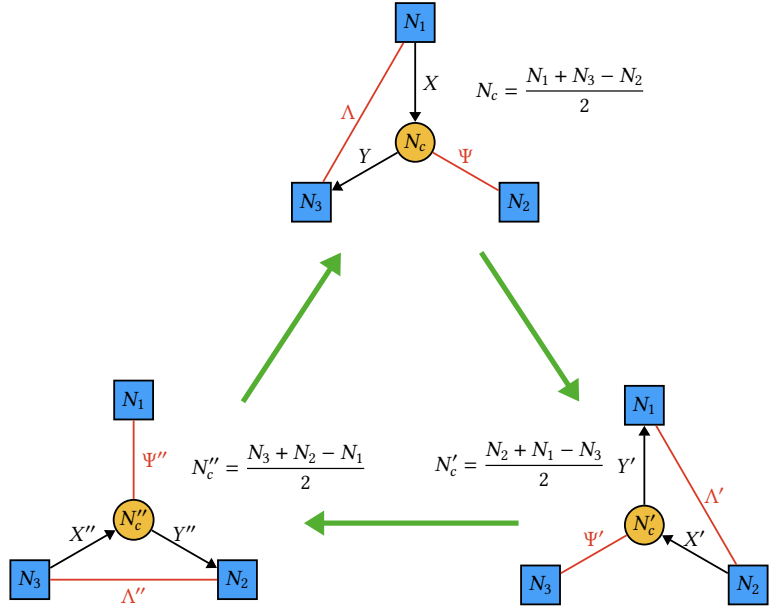

	\centering
	\scalebox{0.69}{

				}
			};
			\draw[line width=1mm, greenX,-Triangle] (-0.51,-0.22) -- (-0.3,0.1);
			\draw[line width=1mm, greenX,Triangle-] (0.21,-0.22) -- (0,0.1);
			\draw[line width=1mm, greenX,-Triangle] (0.08,-0.48) -- (-0.38,-0.48);
		\end{tikzpicture}  
	}
	\caption{Triality loop for $2$d $\mathcal{N}=(0,2)$ SQCD.}
	\label{fig:SCQDN02trialoop}
\end{figure}

Taking the dual theory as the new starting point and acting on it with triality, we obtain the theory shown on the bottom left of Figure \ref{fig:SCQDN02trialoop}. Applying triality a third time takes us back to the original theory. We can therefore think about this second dual as connected to the original theory by inverse triality.\footnote{The distinction between triality and inverse triality is just a convention.} The triality among these three theories can be viewed as a cyclic permutation of $N_1$, $N_2$ and $N_3$.

We will later use $\mathcal{N}=(0,2)$ gauge theories engineered on D1-branes probing CY 4-folds as starting points of orientifold constructions. Such theories have $\U(N)$ gauge groups. A $\U(N_c)$ version of $\mathcal{N}=(0,2)$ triality was also introduced in \cite{Gadde:2013lxa}. It only differs from the $\SU(N_c)$ triality depicted in Figure \ref{fig:SCQDN02trialoop} by the presence of additional Fermi fields in the determinant representation of the gauge group, which are necessary for the cancellation of the Abelian anomaly. It is expected that Abelian anomalies of gauge theories on D1-branes are cancelled via a generalized Green-Schwarz mechanism (see \cite{Ibanez:1998qp,Mohri:1997ef} for 4d ${\cal N}=1$ and $2$d ${\cal N}=(0,2)$ theories realized on D-branes probing orbifolds/orientifolds singularities). For this reason, the determinant Fermi fields are not present in such theories and triality reduces to the one considered in this section.

$\mathcal{N}=(0,2)$ triality can be extended to general quivers (see e.g. \cite{Gadde:2013lxa,Franco:2016nwv,Franco:2016qxh,Franco:2016fxm,Franco:2018qsc,Franco:2021elb}). It acts as a local operation on the dualized node, with the part of the quiver that is not connected to it acting as a spectator. The transformation of such a theory under triality on a gauge node $k$ can be summarized as follows. The rank of node $k$ changes according to
\begin{align}
	N'_k = \sum_{j\neq k} n_{jk}^\chi N_j - N_k \,, 
	\label{rank-rule}
\end{align}
where $n^\chi_{jk}$ is the number of chiral fields from node $j$ to node $k$. All other ranks remain the same. The field content around node $k$ changes according to the following rules:

\begin{enumerate}[label={(R.\arabic*)}]
	\item\label{rule1} {\bf Dual Flavors}. Replace each of $(\rightarrow k)$, $(\leftarrow k)$, $(\textcolor{red}{\text{ --- }}  k)$ by  $(\leftarrow k)$, $( \textcolor{red}{\text{ --- }}  k)$, $(\rightarrow k)$, respectively.
	\item\label{rule2} {\bf Chiral-Chiral Mesons}. For each subquiver $i\rightarrow k \rightarrow j$, add a new chiral field $i\rightarrow j$.
	\item\label{rule3} {\bf Chiral-Fermi Mesons}. For each subquiver $i\rightarrow k \textcolor{red}{\text{ --- }} j$, add a new Fermi field $i \textcolor{red}{\text{ --- }}  j$.
	\item\label{rule4} Remove all chiral-Fermi massive pairs generated in the previous steps.
\end{enumerate}

For a detailed discussion of the transformation of $J$- and $E$-terms, see e.g., \cite{Franco:2017lpa}.

\subsection{$\mathcal{N}=(0,1)$ Triality}
\label{sec:N01triality}

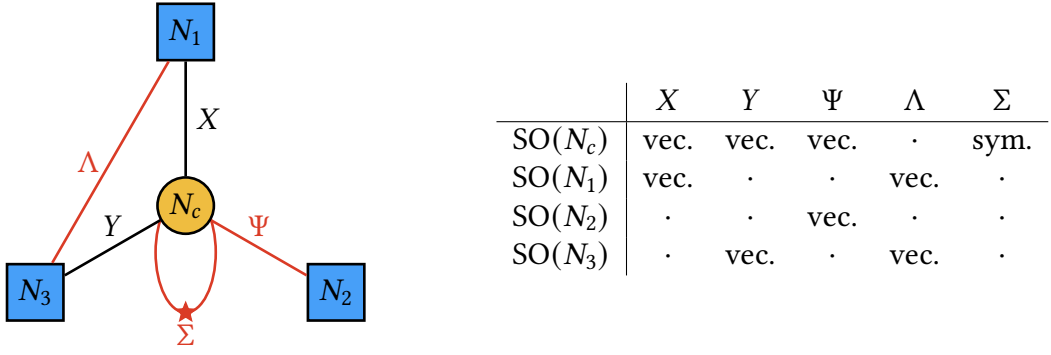
\begin{figure}[H]
	\centering
	\begin{tikzpicture}[scale=2]
		\node (Q) at (-3.2,0) {
			\begin{tikzpicture}[scale=2]
				\def\L{2};
				\draw[line width=1pt,redX] (0,-0.6) ellipse (0.2 and 0.4) node[yshift=-1.1cm] {$\Sigma$} node[xshift=0cm,yshift=-0.8cm,star,star points=5, star point ratio=2.25, inner sep=1pt, fill=redX, draw=redX] {};
				\node[draw=black,line width=1pt,circle,fill=yellowX,minimum width=0.75cm,inner sep=1pt] (Nc) at (0,-0.289) {$N_c$};
				\node[draw=black,line width=1pt,fill=blueX,square,minimum width=0.75cm,inner sep=1pt] (N1) at (0,\L*0.433) {$N_1$};
				\node[draw=black,line width=1pt,fill=blueX,square,minimum width=0.75cm,inner sep=1pt] (N2) at (\L/2,-\L*0.433) {$N_2$};
				\node[draw=black,line width=1pt,fill=blueX,square,minimum width=0.75cm,inner sep=1pt] (N3) at (-\L/2,-\L*0.433) {$N_3$};
				\draw[line width=1pt] (Nc) -- node[right,midway] {$X$}  (N1);
				\draw[line width=1pt,redX] (Nc) -- node[above,midway] {$\Psi$} (N2);
				\draw[line width=1pt,redX] (N1) -- node[left,midway] {$\Lambda$} (N3);
				\draw[line width=1pt] (Nc) -- node[above,midway] {$Y$} (N3);
			\end{tikzpicture}
		};        
		\node (T) at (0.7,0) {
			\begin{tabular}{c|ccccc}
				              & $X$     &  $Y$     & $\Psi$ & $\Lambda$ & $\Sigma$ \\
				\hline                              
				$\SO(N_c)$    & vec.    &  vec.    & vec. & $\cdot$ & sym. \\
				$\SO(N_1)$    & vec.    &  $\cdot$ & $\cdot$ & vec. & $\cdot$ \\
				$\SO(N_2)$    & $\cdot$ &  $\cdot$ & vec. & $\cdot$ & $\cdot$ \\
				$\SO(N_3)$    & $\cdot$ &  vec.    & $\cdot$ & vec. & $\cdot$ \\
			\end{tabular}
		};
	\end{tikzpicture}
	\caption{$2$d $\mathcal{N}=(0,1)$ SQCD. $N_c$ is given in Eq.~\eqref{eq:Nc01}.}
	\label{fig:SQCDN01}
\end{figure}

A similar triality for 2d $\mathcal{N}=(0,1)$ gauge theories was introduced in \cite{Gukov:2019lzi}. The primary example in which the proposal was investigated is $2$d $\mathcal{N}=(0,1)$ SQCD with $\SO(N_c)$ gauge group, whose quiver diagram is shown in Figure \ref{fig:SQCDN01}.\footnote{When drawing $\mathcal{N}=(0,1)$ quivers, black and red lines correspond to real $\mathcal{N}=(0,1)$ scalar and Fermi fields, respectively. In addition, we indicate symmetric and antisymmetric representations with star and diamond symbols, respectively.} The theory has $N_1+N_3$ scalar multiplets in the vector representation of $\SO(N_c)$. These scalar fields are further divided into two sets, $X$ and $Y$, transforming under $\SO(N_1)$ and $\SO(N_3)$ flavor groups, respectively. A bifundamental Fermi multiplet $\Lambda$ connects $\SO(N_1)$ and $\SO(N_3)$.\footnote{We will use the term bifundamental in the case of matter fields that connect pairs of nodes, even when one or both of them is either $\SO$ or $\USp$.} There are also $N_2$ Fermi multiplets $\Psi$ in the vector representation of $\SO(N_c)$ and a Fermi multiplet $\Sigma$ in the symmetric representation of $\SO(N_c)$.

Anomaly cancellation for the $\SO(N_c)$ gauge group requires that\footnote{The anomaly contributions of $\mathcal{N}=(0,1)$ multiplets in various representations are listed in Table \ref{tab:SUanomaly} and \ref{tab:SOUSpanomaly}.} 
\begin{equation}
	N_c=\frac{N_1+N_3-N_2}{2}\fstop
	\label{eq:Nc01}
\end{equation} 	

The theory also has the following superpotential consistent with its symmetries
\begin{equation}
	\label{eq:SDCD superpotential}
	W^{(0,1)}=\sum_{\alpha,\beta=1}^{N_c}\Sigma_{\alpha\beta}\left(\sum_{a=1}^{N_1}X^a_{\alpha}X^a_{\beta}+\sum_{b=1}^{N_3}Y^b_{\alpha}Y^b_{\beta}-\delta_{\alpha\beta}\right)+\sum_{a=1}^{N_1}\sum_{b=1}^{N_3}\sum_{\alpha=1}^{N_c}\Lambda_{ab}X^a_\alpha Y^b_\alpha\fstop
\end{equation}

\begin{figure}[H]
	\centering
	\begin{tikzpicture}[scale=2]
		\node (Q) at (-3.2,0) {
			\begin{tikzpicture}[scale=2]
				\def\L{2};
				\draw[line width=1pt,redX] (0,-0.6) ellipse (0.2 and 0.4) node[yshift=-1.1cm] {$\Sigma^\prime$} node[xshift=0cm,yshift=-0.8cm,star,star points=5, star point ratio=2.25, inner sep=1pt, fill=redX, draw=redX] {};
				\node[draw=black,line width=1pt,circle,fill=yellowX,minimum width=0.75cm,inner sep=1pt] (Nc) at (0,-0.289) {$N_c^\prime$};
				\node[draw=black,line width=1pt,fill=blueX,square,minimum width=0.75cm,inner sep=1pt] (N1) at (0,\L*0.433) {$N_1$};
				\node[draw=black,line width=1pt,fill=blueX,square,minimum width=0.75cm,inner sep=1pt] (N2) at (\L/2,-\L*0.433) {$N_2$};
				\node[draw=black,line width=1pt,fill=blueX,square,minimum width=0.75cm,inner sep=1pt] (N3) at (-\L/2,-\L*0.433) {$N_3$};
				\draw[line width=1pt] (Nc) -- node[left,midway] {$Y^\prime$}  (N1);
				\draw[line width=1pt] (Nc) -- node[above,midway] {$X^\prime$} (N2);
				\draw[line width=1pt,redX] (N1) -- node[right,midway] {$\Lambda^\prime$} (N2);
				\draw[line width=1pt,redX] (Nc) -- node[above,midway] {$\Psi^\prime$} (N3);
			\end{tikzpicture}
		};        
		\node (T) at (0.7,0) {
			\begin{tabular}{c|ccccc}
				              & $X^\prime$ &  $Y^\prime$ & $\Psi^\prime$ & $\Lambda^\prime$ & $\Sigma^\prime$ \\
				\hline                                     
				$\SO(N_c)$    & vec.       &  vec.       & vec. & $\cdot$ & sym. \\
				$\SO(N_1)$    & $\cdot$    &  vec.       & $\cdot$ & vec. & $\cdot$ \\
				$\SO(N_2)$    & vec.       &  $\cdot$    & $\cdot$ & vec. & $\cdot$ \\
				$\SO(N_3)$    & $\cdot$    &  $\cdot$    & vec. & $\cdot$ & $\cdot$ \\
			\end{tabular}
		};
	\end{tikzpicture}
	\caption{2d $\mathcal{N}=(0,1)$ triality dual of the theory in Figure~\ref{fig:SQCDN01}. $N^\prime_c$ is given in Eq.~\eqref{eq:Ncp01}.}
	\label{fig:SQCDN01-dual}
\end{figure}
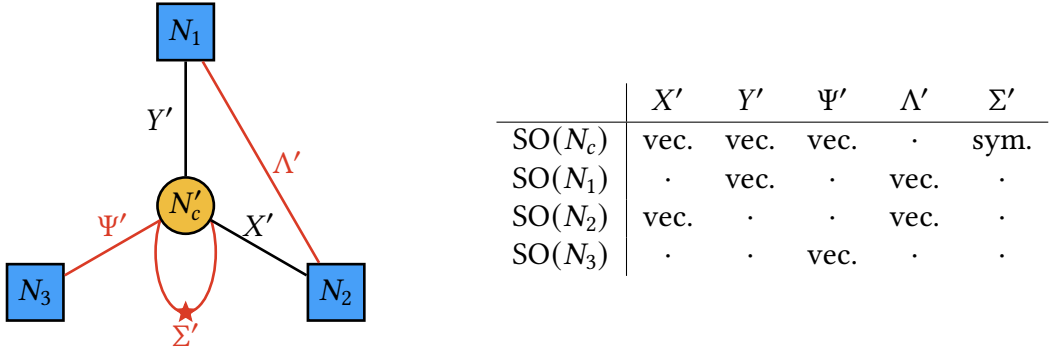

Figure \ref{fig:SQCDN01-dual} shows the dual under triality. The transformation is rather similar to the $\mathcal{N}=(0,2)$ triality discussed in the previous section. Once again, in this simple example, the structure of the dual theory is identical to the original one up to a cyclic permutation of $N_1$, $N_2$ and $N_3$. For the flavors, scalar multiplets $X$, $Y$ and Fermi multiplets $\Psi$ are replaced by scalar multiplets $Y^\prime$, Fermi multiplets $\Psi^\prime$, and scalar multiplets $X^\prime$, respectively. The new theory also contains a Fermi field $\Sigma'$ in the symmetric representation of the gauge group.

The gauge group is $\SO(N_c^\prime)$, with the rank determined by anomaly cancellation
\begin{equation}
	N_c^\prime=\frac{N_2+N_1-N_3}{2} \coma
	\label{eq:Ncp01}
\end{equation}
which can be expressed as
\begin{equation}
	N_c^\prime=N_1-N_c \fstop
\end{equation}

Very much like Rule \ref{rule3} of the previous section,
the new Fermi $\Lambda^\prime$ in the bifundamental representation of $\SO(N_1)\times \SO(N_2)$ can be regarded as a scalar-Fermi meson in terms of the fields in the initial theory, i.e. $\Lambda^\prime=X\Psi$. Similarly, we can interpret the disappearance of the original Fermi $\Lambda$ between Figures \ref{fig:SQCDN01} and \ref{fig:SQCDN01-dual} as the result of it becoming massive via its superpotential coupling to the scalar-scalar meson $X Y$, which is analogous to the chiral-chiral mesons of Rule \ref{rule2}. An interesting difference with respect to $\mathcal{N}=(0,2)$ SQCD follows from the fact that SO representations are real. Equivalently, the quivers under consideration are not oriented. It is therefore natural to ask why, in addition to $\Lambda^\prime=X\Psi$, Figure \ref{fig:SQCDN01-dual} does not {\it simultaneously} have another scalar-Fermi meson $Y \Psi$ in the bifundamental representation of $\SO(N_2)\times \SO(N_3)$. Its absence can be interpreted as descending from $\mathcal{N}=(0,2)$ triality, in which the orientation of chiral fields prevent the formation of such a gauge invariant. Additional thoughts on the connection between $\mathcal{N}=(0,2)$ and $\mathcal{N}=(0,1)$ trialities will be presented in Section \ref{sec:N01triality_quivers}. Also related to this issue, in the coming section, we will discuss scalar-Fermi mesons in more general quivers.

The superpotential is identical to \eqref{eq:SDCD superpotential} upon replacing all fields by the primed counterparts and permuting $N_1$, $N_2$ and $N_3$.

\begin{figure}[H]
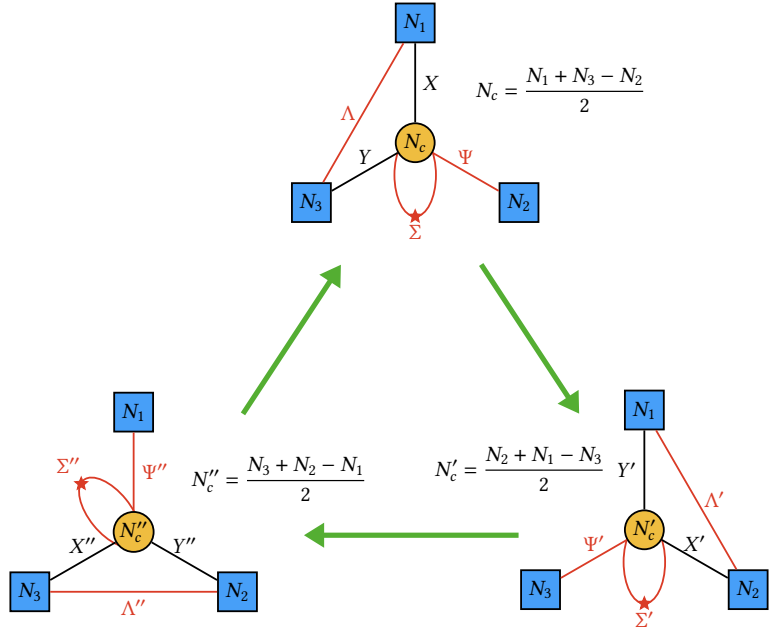

	\centering
	\scalebox{0.69}{

				}
			};
			\draw[line width=1mm, greenX,-Triangle] (-0.51,-0.22) -- (-0.3,0.1);
			\draw[line width=1mm, greenX,Triangle-] (0.21,-0.22) -- (0,0.1);
			\draw[line width=1mm, greenX,-Triangle] (0.08,-0.48) -- (-0.38,-0.48);
		\end{tikzpicture}  
	}
	\caption{Triality loop for $2$d $\mathcal{N}=(0,1)$ SQCD.} 
	\label{fig:SCQDN01trialoop}
\end{figure}

Acting with triality again gives rise to the theory shown on the bottom left of Figure \ref{fig:SCQDN01trialoop}. A third triality takes us back to the original theory. 

There is also a symplectic version of $\mathcal{N}=(0,1)$ triality \cite{Gukov:2019lzi}. The corresponding SQCD has $\USp(N_c)$ gauge group and $\USp(N_1)\times \USp(N_2)\times \USp(N_3)$ global symmetry.\footnote{Differently from~\cite{Gukov:2019lzi}, we adopt the convention $\USp(2)\simeq \SU(2)$ in order to be consistent with the notation of the orientifold theories we construct later.} The matter content is almost the same as in the $\SO(N_c)$ SQCD quiver shown in Figure \ref{fig:SQCDN01}, with the exception that the Fermi field $\Sigma$ instead transforms in the antisymmetric representation of $\USp(N_c)$. The rank of the gauge group is $N_c=\frac{N_1+N_3-N_2}{2}$ to cancel gauge anomalies. In this case, the triality loop is identical to the one shown in Figure \ref{fig:SCQDN01trialoop}.

Evidence for the $\mathcal{N}=(0,1)$ triality proposal includes matching of anomalies and elliptic genera \cite{Gukov:2019lzi}. In the coming sections, we will provide further support for this idea, by realizing 2d $\mathcal{N}=(0,1)$ theories via $\Spin(7)$ orientifolds.

\subsection{$\mathcal{N}=(0,1)$ Triality for Quiver Gauge Theories}
\label{sec:N01triality_quivers}

Let us consider the extension of $\mathcal{N}=(0,1)$ triality to general quivers. To do so, it is useful to first draw some lessons from Seiberg duality and $\mathcal{N}=(0,2)$ triality. In both cases, incoming chiral fields at the dualized gauge group play a special role.\footnote{This is a general phenomenon that applies e.g. to the order $(m+1)$ dualities of $m$-graded quivers \cite{Franco:2017lpa}. Seiberg duality and $\mathcal{N}=(0,2)$ triality correspond to the $m=1$ and $2$ cases, respectively.} They control the rank of the dual gauge group and, for triality, determine which mesons are formed. Since $\mathcal{N}=(0,1)$ quivers are unoriented, how to split the scalar fields terminating on a dualized node into two sets analogous to ``incoming" and ``outgoing" flavors is not clear. This issue was hinted in our discussion in the previous section.

In \cite{Gukov:2019lzi}, a generalization of triality to a simple class of quiver theories with $\SO(N_{c_1})\times \SO(N_{c_2})\times \ldots$ gauge group (or the symplectic counterpart) was briefly discussed. Theories in this family are obtained by combining various $\mathcal{N}=(0,1)$ SQCD building blocks, which are glued by identifying any of the three global nodes of a given theory with the gauge node of another one. Locally, the resulting theories have the same structure of basic SQCD. Namely, every gauge node is connected to three other nodes, to two of them via scalar fields and to the remaining one via Fermi fields. Due to this simple structure, the dualization of any of the gauge groups is unambiguous and proceeds as in basic triality. For every node, the two possible choices of scalar fields acting as ``incoming" or ``outgoing" corresponds to acting with triality or inverse triality.   

For general quivers, in which a given node can be connected to multiple others, how to separate the flavor scalar fields at every gauge group into two sets is an open question. All the theories that we will construct later using $\Spin(7)$ orientifolds are indeed beyond the above special class. However, this ambiguity is resolved in them by inheriting the separation of flavors from the parent $\mathcal{N}=(0,2)$ theories. 

\section{Spin(7) Orientifolds}
\label{sec:Spin7Orient}

In this section we review the construction of Spin(7) orientifolds introduced in \cite{Franco:2021ixh} and the 2d $\mathcal{N}=(0,1)$ field theories arising on D1-branes probing them. We focus the overview on a few key points relevant for subsequent sections, and refer the reader to this reference for additional details. 

Our starting point is a toric CY $4$-fold singularity $\IM_8$. When probed by a stack of D1-branes at the singular point, the worldvolume theory corresponds to an $\mathcal{N}=(0,2)$ quiver gauge field theory. When $\IM_8$ is toric, the structure of gauge groups, matter content and interactions of these theories is nicely encoded by brane brick models \cite{Franco:2015tya,Franco:2016qxh,Franco:2017cjj} (see \cite{GarciaCompean:1998kh} for an early related construction). Nevertheless, for our purposes it suffices to use the quiver description, supplemented by the explicit expression of the interaction terms ($J$- and $E$-terms).

We then perform an orientifold quotient by the action $\Omega \sigma$, where $\Omega$ is worldsheet parity and $\sigma$ is an anti-holomorphic involution of $\IM_8$ leaving a specific 4-form, that we call $\Omega^{(4)}$, invariant. Such $4$-form is constructed from the CY holomorphic $4$-form $\Omega^{(4,0)}$ and the K\"ahler form $J^{(1,1)}$ as 
\begin{equation}
    \Omega^{(4)}=\text{Re}\left(\Omega^{(4,0)}\right)+\frac{1}{2}J^{(1,1)}\wedge J^{(1,1)}\fstop
\end{equation}
If the quotient did not involve worldsheet parity, this quotient corresponds to Joyce's construction of Spin(7) geometries, with $\Omega^{(4)}$ defining the invariant Cayley 4-form of such varieties. To keep this connection in mind, the above orientifold quotients were dubbed \textit{Spin(7) orientifolds} in \cite{Franco:2021ixh}.

This orientifold quotient has a natural counterpart on the D1-brane systems, and naturally realizes a ``real projection" of the $2$d $\mathcal{N}=(0,2)$ theories in~\cite{Gukov:2019lzi}, resulting in a 2d $\mathcal{N}=(0,1)$ gauge field theory. Its structure is determined by a set of rules analogous to  those of orientifold field theories in higher dimensions (see e.g. \cite{Franco:2007ii} in the 4d context), and which were explicitly determined in \cite{Franco:2021ixh}. Morally, it corresponds to identifying the gauge factors and matter fields of the parent $\mathcal{N}=(0,2)$ theory under an involution symmetry ${\tilde \sigma}$ of the quiver, compatible with the set of interactions.

To describe the orientifold action on the field theory in more detail, we label the different nodes by an index $i$, and their orientifold images by $i'$ (with $i'=i$ corresponding to nodes mapped to themselves under the orientifold action), and denote $X_{ij}$ and $\Lambda_{ij}$ the bifundamental $\mathcal{N}=(0,2)$ chiral or Fermi multiplets charged under the gauge factors $i$ and $j$ (with $j=i$ corresponding to adjoints). The results of \cite{Franco:2021ixh} are:

\begin{enumerate}[label=1\alph*.,ref=1\alph*]
\item\label{rule:1a}  Two gauge factors $\U(N)_i$, $\U(N)_{i'}$ mapped to each other under the orientifold action (namely $i\neq i'$) are identified and give rise to a single $\U(N)$ factor in the orientifold theory. 
\item\label{rule:1b} On the other hand, a gauge factor $\U(N)_i$ mapped to itself (namely, $i'=i$) is projected down to $\SO(N)$ or $\USp(N)$. 
\end{enumerate}

\begin{enumerate}[label=2\alph*.,ref=2\alph*]
\item\label{rule:2a}   Two {\em different} chiral or Fermi bifundamental fields $X_{ij}$  and $X_{i'j'}$, mapped to each other under the orientifold action, become identified\footnote{\label{the-eta-issue} In the presence of multiple sets of fields in these representations, the mapping may include a non-trivial action on the flavor index, encoded in a matrix $\eta$. As explained in \cite{Franco:2021ixh}, the choice can impact on the orientifold projection of the relevant gauge factors. We will encounter a non-trivial use of this freedom in the example in Section \ref{sec:Q111}.} and lead to a single (chiral or Fermi) bifundamental field. This holds even in special cases for the gauge factors, such as $i'=i$, or simultaneously $i'=i$ and $j'=j$, and for the special case of fields in the adjoint, $j=i$, $j'=i'$. 

\item\label{rule:2b} Two {\em different} chiral or Fermi bifundamental fields $X_{ii'}$ and $Y_{i'i}$, related each to the (conjugate of the) other under the orientifold action, give rise to one field in the two-index symmetric and one field in the two-index antisymmetric representation of the corresponding $\SO/\USp$ $i^{th}$ gauge factor in the orientifold quotient. The rule holds also in the case of adjoint fields, namely $i'=i$.  
\end{enumerate}

\begin{enumerate}[label=3\alph*.,ref=3\alph*]
\item\label{rule:3a}  A bifundamental field $X_{ij}$ that is mapped to itself by the orientifold action gives rise to a real $\mathcal{N}=(0,1)$ field  transforming under the bifundamental of $G_i\times G_j$, where $G_i$ and $G_j$ are the same type of $\SO$ or $\USp$ gauge group.

\item\label{rule:3b} A bifundamental Fermi field $\Lambda_{ii'}$ can only be mapped to itself (resp. minus itself) in the case of a holomorphic transformation, and gives rise to a complex Fermi superfield in the symmetric (resp. antisymmetric) representation of the resulting $\U(n)_i$ group. 

\item\label{rule:3c} Closely related to Rule~\ref{rule:3b}, an adjoint complex Fermi field $\Lambda_{ii}$ that is mapped to itself (resp. minus itself) via a holomorphic transformation, gives rise to a complex Fermi field in the symmetric/antisymmetric (resp. antisymmetric/symmetric) representation of $\SO/\USp$. 

\item\label{rule:3d} An adjoint complex scalar or Fermi field that is mapped to itself gives rise to two real scalar or Fermi fields, one symmetric and one antisymmetric.
\end{enumerate}

\begin{enumerate}[label=4\alph*.,ref=4\alph*]
\item\label{rule:4a}  A real Fermi $\Lambda^R_{ii}$ which transforms into $\Lambda^R_{i'i'}$, with $i'\neq i$, are projected down to a single real Fermi $\Lambda^R_{ii}$.
\item\label{rule:4b} A real Fermi $\Lambda^R_{ii}$ mapped to itself (resp. minus itself), with $i'\neq i$, gives rise to a symmetric (resp. antisymmetric) real Fermi for an $\SO$ (resp. $\USp$) projection of the node $i$.   
\end{enumerate}

These rules suffice to construct large classes of examples of 2d $\mathcal{N}=(0,1)$ field theories, in particular the explicit examples in coming sections.

Note that the $\mathcal{N}=(0,1)$ theory obtained upon orientifolding the parent $\mathcal{N}=(0,2)$ may have non-abelian gauge anomalies. In such cases, the models require the introduction of extra flavor branes (namely, D5- or D9-branes extending in the non-compact dimensions of the CY 4-fold) for consistency. As already remarked in \cite{Franco:2021ixh}, very often the orientifolded theories happen to be non-anomalous, and hence do not require flavor branes. This will be the case in our examples later on.

\subsection*{The Universal Involution}

We would like to conclude this overview by recalling from \cite{Franco:2021ixh} that any $\mathcal{N}=(0,2)$ quiver gauge theory from D1-branes at toric CY 4-fold singularities admits a universal anti-holomorphic involution. It corresponds to mapping each gauge factor to itself (maintaining all with the same $\SO$ or $\USp$ projection), and mapping every $\mathcal{N}=(0,2)$ chiral or Fermi field  to itself anti-holomorphically.

To be more explicit, let us introduce a set of matrices $\gamma_{\Omega_i}$ implementing the action of the orientifold on the gauge degrees of freedom of the $i^{th}$ node.\footnote{Actually, the matrices $\gamma_{\Omega_i}$ are a useful ingredient in implementing the orientifold projection, even in examples beyond the universal involution, as we will exploit in explicit examples in later sections.\label{foot:gamma-omega}} Then, the orientifold projections for the universal involution read
\begin{equation}
    X_{ij}\rightarrow \gamma_{\Omega_i}\bar{X}_{ij}\gamma_{\Omega_j}^{-1} \coma \Lambda_{ij}\rightarrow \gamma_{\Omega_i}\bar{\Lambda}_{ij}\gamma_{\Omega_j}^{-1}\coma
    \label{eq:genUImap}
\end{equation}
where, by $X_{ij}$ or $\Lambda_{ij}$ we mean any chiral or Fermi field present in the gauge theory. In addition, the $\mathcal{N}=(0,1)$ adjoint Fermi fields coming from $\mathcal{N}=(0,2)$ vector multiplets transform as 
\begin{equation}
\Lambda_i^R \to \gamma_{\Omega_{i}} \Lambda_{i'}^{R \, \, T} \gamma_{\Omega_{i}}^{-1} \, .
\label{Lambda^R_projection}
\end{equation}
There is relative sign between this projection and the one for gauge fields, which implies that an $\SO$ or $\USp$ projection of the gauge group is correlated with a projection of $\Lambda_i^R$ into a symmetric or antisymmetric representation, respectively. These projections are consistent with the invariance of the $\mathcal{N}=(0,1)$ superpotential. Modding out by this orientifold action, the resulting $\mathcal{N}=(0,1)$ field theory is determined by applying the above rules.

From the geometric perspective, this universal involution corresponds to the conjugation of all generators of the toric CY 4-fold. The action on the holomorphic 4-form is $\Omega^{(4,0)}\rightarrow \bar{\Omega}^{(0,4)}$, suitable for the realization of an Spin(7) orientifold. The following section focuses on models obtained via the universal involution.

\section{$\mathcal{N}=(0,1)$ Triality and the Universal Involution}
\label{sec:01trial-UI}

Let us consider what happens when the universal involution is applied to two gauge theories associated to the same CY 4-fold, which are therefore related by $\mathcal{N}=(0,2)$ triality. Remarkably, we obtain two theories connected by precisely $\mathcal{N}=(0,1)$ triality. By construction, the two theories correspond to the same underlying Spin(7) orientifold, realizing the general idea of $\mathcal{N}=(0,1)$ triality arising from the non-uniqueness of the map between Spin(7) orientifolds and gauge theories.

\begin{figure}[H]
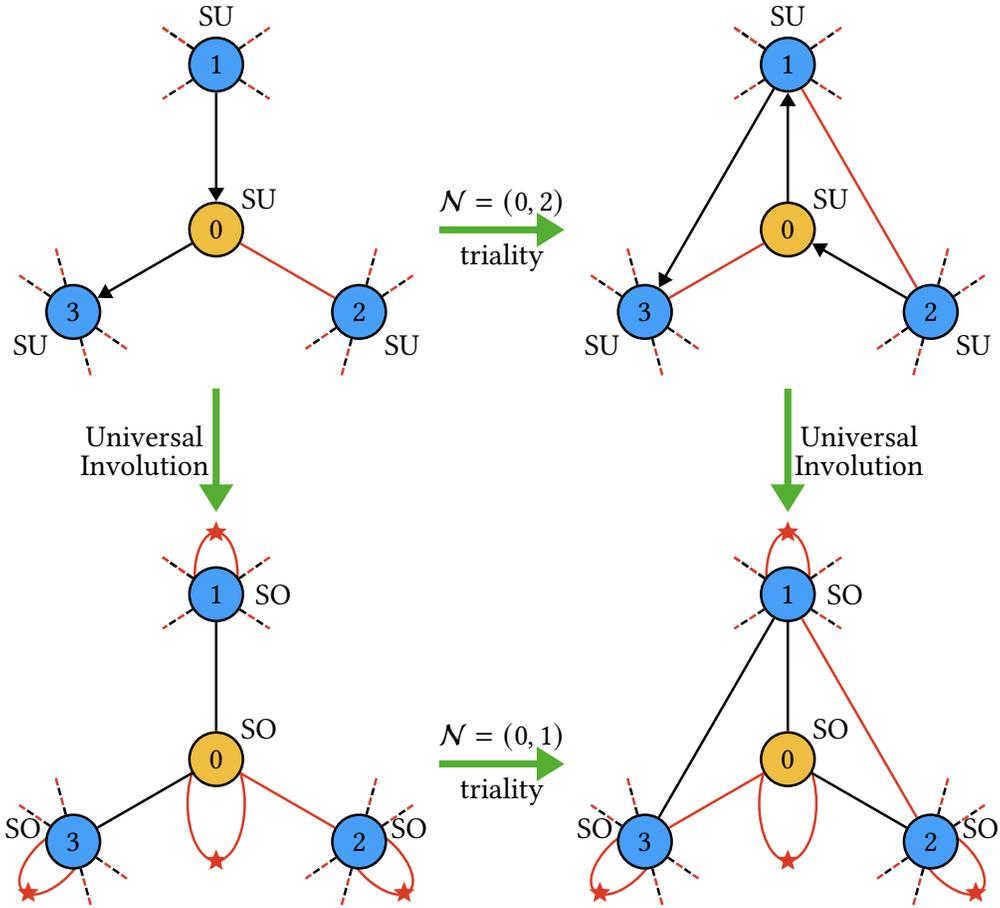

    \centering
    \scalebox{0.95}{

    };
    \draw[-Triangle,greenX,line width=1mm] (0.78,1.70) -- node[above,midway,black] {$\mathcal{N}=(0,2)$} node[below,midway,black] {triality} (1.22,1.70);
    \draw[-Triangle,greenX,line width=1mm] (0.78,-0.17) -- node[above,midway,black] {$\mathcal{N}=(0,1)$} node[below,midway,black] {triality} (1.22,-0.17);
    \draw[-Triangle,greenX,line width=1mm] (NW) -- node[xshift=-1cm,midway,black] {\shortstack{Universal\\Involution}} (SW);
    \draw[-Triangle,greenX,line width=1mm] (NE) -- node[xshift=1cm,midway,black] {\shortstack{Universal\\Involution}} (SE);
    \end{tikzpicture}
    }
 \caption{The universal involution on $\mathcal{N}=(0,2)$ triality results in $\mathcal{N}=(0,1)$ triality.}
\label{universal_involution_triality}
\end{figure}

We illustrate this projection in Figure \ref{universal_involution_triality}, which shows the neighborhood of the quiver around a dualized node $0$.\footnote{The universal involution with $\USp$ projection is analogous.} As in the previous section, nodes 1, 2 and 3 represent possibly multiple nodes which, in turn, might be connected to node $0$ by different multiplicities of fields. The red and black dashed lines represent the rest of the quiver, which might include fields stretching between nodes 1, 2 and 3. If triality generates massive fields, they can be integrated out.

An explicit example of a triality pairs associated to the universal involution will be presented in Section \ref{section_H4}. However, in Section \ref{sec:BeyonUI}, we will show how more general orientifold actions lead to interesting generalizations of the basic $\mathcal{N}=(0,1)$ triality. The general strategy will be to focus on parent CY$_4$ geometries with more than one $\mathcal{N}=(0,2)$ triality dual toric phases\footnote{We refer to a toric phase as one associated to a brane brick model \cite{Franco:2015tya}, for which the connection to the underlying CY$_4$ is considerably simplified.} (see e.g. \cite{Franco:2016nwv,Franco:2018qsc}) and to consider anti-holomorphic involutions leading to the same $\Spin(7)$ orientifold.

\subsection{The Universal Involution of $H_4$}
\label{section_H4}

\begin{figure}[H]
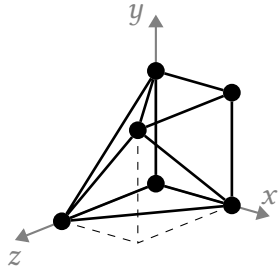

			\centering

			\caption{Toric diagram for $H_4$.}
			\label{fig:H4toricdiagram}
	\end{figure}

As explained above, the universal involution works for every CY$_4$. Therefore, it is sufficient to present one example to illustrate the main features of the construction. Let us consider the CY$_4$ with toric diagram shown in Figure \ref{fig:H4toricdiagram}, which is often referred to as $H_4$. Below we consider two toric phases for D1-branes probing $H_4$ and construct the $\mathcal{N}=(0,1)$ theories that correspond to them via the universal involution.

\subsubsection{Phase A}
	\label{sec:H4PhaseA}

\begin{figure}[H]
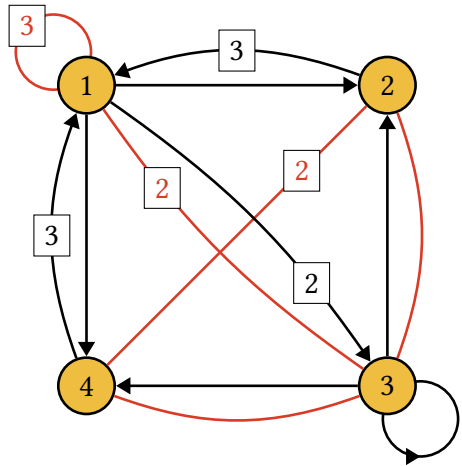

		\centering
			%
		\caption{Quiver diagram for phase A of $H_4$.}
		\label{fig:H4AquivN02}
	\end{figure}

Figure \ref{fig:H4AquivN02} shows the quiver diagram for one of the toric phases of $H_4$, which we denote phase A. This theory was first introduced in \cite{Franco:2017cjj}.
	
The corresponding $J$- and $E$-terms are 
\begin{alignat}{4}
	\renewcommand{\arraystretch}{1.1}
    & \centermathcell{J}                           &\text{\hspace{.5cm}}& \centermathcell{E                               }\nonumber \\
	\Lambda_{11}^1 \,:\, & \centermathcell{X_{14}X_{41} - X_{13}X_{32}Z_{21}  }&  & \centermathcell{Y_{13}X_{34}Z_{41}-X_{12}Y_{21}        }\nonumber \\
	\Lambda_{11}^2 \,:\, & \centermathcell{X_{14}Y_{41} - Y_{13}X_{32}Z_{21}  }&  & \centermathcell{X_{12}X_{21}  - X_{13}X_{34}Z_{41}     }\nonumber \\
	\Lambda_{11}^3 \,:\, & \centermathcell{X_{14}Z_{41} - X_{12}Z_{21}        }&  & \centermathcell{X_{13}X_{32}Y_{21} - Y_{13}X_{34}X_{41}}\nonumber \\
	\Lambda_{13}^1 \,:\, & \centermathcell{X_{32}X_{21}  - X_{34}X_{41}       }&  & \centermathcell{Y_{13}X_{33} - X_{14}Z_{41}Y_{13}      } \label{eq:H4PhaseAJEterms}\\
	\Lambda_{13}^2 \,:\, & \centermathcell{X_{32}Y_{21}  - X_{34}Y_{41}       }&  & \centermathcell{X_{12}Z_{21}X_{13} - X_{13}X_{33}      }\nonumber \\
	\Lambda_{42}^1 \,:\, & \centermathcell{X_{21}X_{14}  - Z_{21}X_{13}X_{34} }&  & \centermathcell{Z_{41}Y_{13}X_{32} -Y_{41}X_{12}       }\nonumber \\
	\Lambda_{42}^2 \,:\, & \centermathcell{Y_{21}X_{14}  - Z_{21}Y_{13}X_{34} }&  & \centermathcell{X_{41}X_{12} - Z_{41}X_{13}X_{32}      }\nonumber \\
	\Lambda_{23} \,:\,   & \centermathcell{X_{33}X_{32}  - X_{32}Z_{21}X_{12} }&  & \centermathcell{Y_{21}X_{13} -X_{21}Y_{13}             }\nonumber \\
	\Lambda_{43} \,:\,   & \centermathcell{X_{33}X_{34}  - X_{34}Z_{41}X_{14} }&  & \centermathcell{X_{41}Y_{13} - Y_{41}X_{13}            }\nonumber 
	\end{alignat}

The $\mathcal{N}=(0,1)$ superpotential is then
	\begin{equation}
	\begin{split}
	W^{(0,1)}=&\,W^{(0,2)}+ \Lambda_{11}^{4R}( X^\dagger_{12}X_{12}+ X^\dagger_{14}X_{14}+ X^\dagger_{21}X_{21}+ Y^\dagger_{21}Y_{21}+ Z^\dagger_{21}Z_{21}+\\
	&+ X^\dagger_{41}X_{41}+ Y^\dagger_{41}Y_{41}+ Z^\dagger_{41}Z_{41}+ X^\dagger_{13}X_{13}+ Y^\dagger_{13}Y_{13})+\\
	&+\Lambda_{22}^{R}( X^\dagger_{12}X_{12}+ X^\dagger_{32}X_{32}+ X^\dagger_{21}X_{21}+ Y^\dagger_{21}Y_{21}+ Z^\dagger_{21}Z_{21})+\\
	&+\Lambda_{33}^{R}( X^\dagger_{33}X_{33}+ X^\dagger_{32}X_{32}+ X^\dagger_{34}X_{34}+ X^\dagger_{13}X_{13}+ Y^\dagger_{13}Y_{13})+\\
	&+\Lambda_{44}^{R}( X^\dagger_{14}X_{14}+ X^\dagger_{34}X_{34}+ X^\dagger_{41}X_{41}+ Y^\dagger_{41}Y_{41}+ Z^\dagger_{41}Z_{41})\fstop
	\end{split}
	\end{equation}

The generators of $H_4$, which arises as the moduli space of the gauge theory, can be determined for instance using the Hilbert Series (HS) \cite{Benvenuti:2006qr,Feng:2007ur,Franco:2015tya} (see also \cite{Franco:2021ixh}). We list them in Table \ref{tab:GenerH4PhaseA}, together with their expressions as mesons in terms of chiral fields in phase A.
	\begin{table}[H]
		\centering
		\renewcommand{\arraystretch}{1.1}
		\begin{tabular}{c|c}
			Meson & Chiral superfields  \\
			\hline
			$M_1$ & $X_{33}=X_{14}Z_{41}=Z_{21}X_{12}$ \\
			$M_2$ & $Y_{21}X_{12}=Z_{41}Y_{13}X_{34}$ \\
			$M_3$ & $X_{14}Y_{41}=Z_{21}Y_{13}X_{32}$ \\
			$M_4$ & $X_{32}Y_{21}Y_{13}=X_{34}Y_{41}Y_{13}$ \\
			$M_5$ & $X_{21}X_{12}=Z_{41}X_{13}X_{34}$ \\
			$M_6$ & $X_{14}X_{41}=Z_{21}X_{13}X_{32}$ \\
			$M_7$ & $X_{32}Y_{21}X_{13}=X_{32}X_{21}Y_{13}=X_{34}Y_{41}X_{13}=X_{34}X_{41}Y_{13}$ \\
			$M_8$ & $X_{32}X_{21}X_{13}=X_{34}X_{41}X_{13}$ 
		\end{tabular}
		\caption{Generators of $H_4$ in terms of fields in phase A.}
		\label{tab:GenerH4PhaseA}
	\end{table}

The generators satisfy the following relations
	\begin{equation}
	\begin{split}
	\mathcal{I} = &\left\langle M_1M_4=M_2M_3\coma M_1M_7=M_2M_6\coma M_1M_7=M_3M_5\coma M_2M_7=M_4M_5\coma\right.\\
	&\left.M_3M_7=M_4M_6\coma M_1M_8=M_5M_6\coma M_2M_8=M_5M_7\coma M_3M_8=M_6M_7\coma\right.\\
	&\left.M_4M_8=M_7^2\right\rangle\fstop
	\label{eq:H4HSrel}
	\end{split}
	\end{equation}
	
The universal involution acts on the fields of any theory as \eqref{eq:genUImap}. This results in the expected map of the generators	
	\begin{equation}
\begin{array}{cccccccccccc}
	M_{1}  & \rightarrow & \bar{M}_{1} \coma &
	M_{2}  & \rightarrow & \bar{M}_{2} \coma &
	M_{3}  & \rightarrow & \bar{M}_{3} \coma &
	M_{4}  & \rightarrow & \bar{M}_{4} \coma \\
	M_{5}  & \rightarrow & \bar{M}_{5} \coma &
	M_{6}  & \rightarrow & \bar{M}_{6} \coma &
	M_{7}  & \rightarrow & \bar{M}_{7} \coma &
	M_{8}  & \rightarrow & \bar{M}_{8} \fstop
	\end{array} 
	\label{eq:H4PhaseA-invol}
	\end{equation}
	
	The quiver for the $2$d $\mathcal{N}=(0,1)$ orientifold theory is shown in Figure \ref{(0,1) theory of Spin(7) orientifold from h4 phase A}. It is rather straightforward to write the projected superpotential but, for brevity, we will omit it here and in the examples that follow.
	
		\begin{figure}[H]
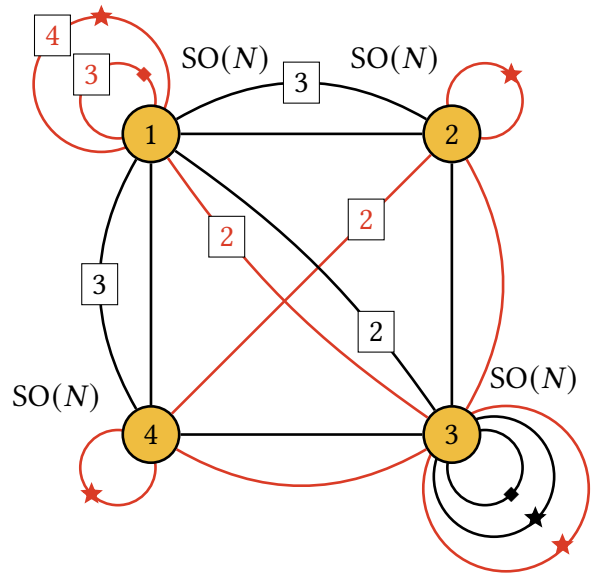

	    \centering
	   	%
	    \caption{Quiver diagram for the Spin(7) orientifold of phase A of $H_4$ using the universal involution.}
		\label{(0,1) theory of Spin(7) orientifold from h4 phase A}
	\end{figure}

\subsubsection{Phase B}
	\label{sec:H4PhaseB}

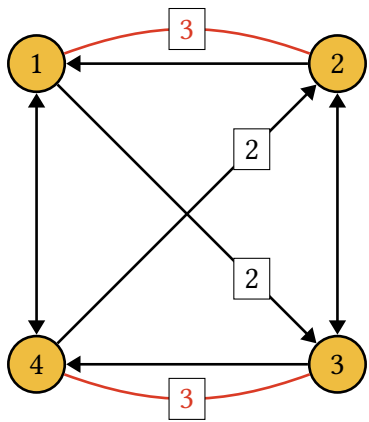
\begin{figure}[H]
		\centering
			\begin{tikzpicture}[scale=2]
			\node[draw=black,line width=1pt,circle,fill=yellowX,minimum width=0.75cm,inner sep=1pt] (A) at (0,0) {$4$};
			\node[draw=black,line width=1pt,circle,fill=yellowX,minimum width=0.75cm,inner sep=1pt] (B) at (2,0) {$3$};
			\node[draw=black,line width=1pt,circle,fill=yellowX,minimum width=0.75cm,inner sep=1pt] (C) at (2,2) {$2$};
			\node[draw=black,line width=1pt,circle,fill=yellowX,minimum width=0.75cm,inner sep=1pt] (D) at (0,2) {$1$};
			\path[-Triangle] (A) edge[line width=1pt] node[fill=white,text opacity=1,fill opacity=1,draw=black,rectangle,thin,pos=0.75] {$2$} (C);
			\path[-Triangle] (B) edge[line width=1pt] (A);
			\path[-Triangle] (D) edge[line width=1pt] node[fill=white,text opacity=1,fill opacity=1,draw=black,rectangle,thin,pos=0.75] {$2$} (B);
			\path[-Triangle] (C) edge[line width=1pt] (D);
			\draw[line width=1pt,redX] (A) to[bend right=20] node[fill=white,text opacity=1,fill opacity=1,draw=black,rectangle,thin,pos=0.5] {\color{redX}{$3$}} (B);
			\draw[line width=1pt,redX] (C) to[bend right=20] node[fill=white,text opacity=1,fill opacity=1,draw=black,rectangle,thin,pos=0.5] {\color{redX}{$3$}} (D);
			\draw[line width=1pt,Triangle-Triangle] (A) to  (D);
			\draw[line width=1pt,Triangle-Triangle] (B) to  (C);
			\node at (0,-0.42) {};
			\end{tikzpicture}
		\caption{Quiver diagram for phase B of $H_4$.}
			\label{fig:H4BN02}
	\end{figure}

Let us now consider the so-called phase $B$ of $H_4$ \cite{Franco:2017cjj}. Its quiver diagram is shown in Figure \ref{fig:H4BN02}.

\vspace{0.6cm}

The $J$- and $E$-terms are 
\begin{alignat}{4}
	\renewcommand{\arraystretch}{1.1}
    & \centermathcell{J}                           &\text{\hspace{.5cm}}& \centermathcell{E                               }\nonumber \\
\Lambda_{21} \,:\,     & \centermathcell{X_{1 3}X_{34}Y_{42} - Y_{1 3}X_{34}X_{42}       }       &    & \centermathcell{ X_{21}X_{1 4}X_{41} - X_{23}X_{32}X_{21}}\nonumber \\
\Lambda_{1 2}^{1} \,:\, & \centermathcell{X_{23}X_{34}Y_{42}X_{21} - X_{21}Y_{1 3}X_{34}X_{41} } &    & \centermathcell{ X_{1 3}X_{32} - X_{1 4}X_{42}            } \nonumber\\
\Lambda_{1 2}^{2} \,:\, & \centermathcell{X_{21}X_{1 3}X_{34}X_{41} - X_{23}X_{34}X_{42}X_{21}  }&    & \centermathcell{ Y_{1 3}X_{32} - X_{1 4}Y_{42}             } \label{eq:H4PhaseB-JEterms}\\
\Lambda_{34} \,:\,     & \centermathcell{Y_{42}X_{21}X_{1 3} - X_{42}X_{21}Y_{1 3}             } &    & \centermathcell{ X_{34}X_{41}X_{1 4} - X_{32}X_{23}X_{34} }\nonumber\\
\Lambda_{43}^{1} \,:\, & \centermathcell{X_{34}Y_{42}X_{21}X_{1 4} - X_{32}X_{21}Y_{1 3}X_{34} } &    & \centermathcell{ X_{42}X_{23} - X_{41}X_{1 3}            } \nonumber\\         
\Lambda_{43}^{2} \,:\, & \centermathcell{X_{32}X_{21}X_{1 3}X_{34} - X_{34}X_{42}X_{21}X_{1 4} } &    & \centermathcell{ Y_{42}X_{23} - X_{41}Y_{1 3}} \nonumber
\end{alignat}
	
The corresponding $W^{(0,1)}$ is
	\begin{equation}
	\begin{split}
	W^{(0,1)}=&\,W^{(0,2)}+ \Lambda^R_{1 1} (X_{21}  X^\dagger_{21}+X_{41}  X^\dagger_{41}+X_{1 4}  X^\dagger_{1 4}+X_{1 3}  X^\dagger_{1 3}+Y_{1 3}  Y^\dagger_{1 3})+\\
&+\Lambda^R_{22} (X_{23}  X^\dagger_{23}+X_{21}  X^\dagger_{21}+X_{42}  X^\dagger_{42}+X_{32}  X^\dagger_{32}+Y_{42}  Y^\dagger_{42})+\\
&+\Lambda^R_{33} (X_{23}  X^\dagger_{23}+X_{32}  X^\dagger_{32}+X_{34}  X^\dagger_{34}+X_{1 3}  X^\dagger_{1 3}+Y_{1 3}  Y^\dagger_{1 3})+\\	
&+\Lambda^R_{44} (X_{42}  X^\dagger_{42}+X_{41}  X^\dagger_{41}+X_{34}  X^\dagger_{34}+X_{1 4}  X^\dagger_{1 4}+Y_{42}  Y^\dagger_{42})\fstop
\end{split}
	\end{equation}

Table \ref{tab:GenerH4PhaseB} lists the generators of $H_4$, this time expressed in terms of chiral fields in phase B. They satisfy the same relations we presented in \eqref{eq:H4HSrel} when discussing Phase A.

\begin{table}[H]
		\centering
		\renewcommand{\arraystretch}{1.1}
		\begin{tabular}{c|c}
			Meson    & Chiral superfields  \\
			\hline
$M_1$ & $X_{23}X_{32}=X_{41}X_{1 4}$ \\
$M_2$ & $X_{34}Y_{42}X_{23}=X_{34}X_{41}Y_{1 3}$ \\
$M_3$ & $X_{21}X_{1 4}Y_{42}=X_{21}Y_{1 3}X_{32}$ \\
$M_4$ & $X_{34}Y_{42}X_{21}Y_{1 3}$ \\
$M_5$ & $X_{34}X_{42}X_{23}=X_{34}X_{41}X_{1 3}$ \\
$M_6$ & $X_{21}X_{1 4}X_{42}=X_{21}X_{1 3}X_{32}$ \\
$M_7$ & $X_{42}X_{21}Y_{1 3}X_{34}=Y_{42}X_{21}X_{1 3}X_{34}$ \\
$M_8$ & $X_{42}X_{21}X_{1 3}X_{34}$ \\
		\end{tabular}
		\caption{Generators of $H_4$ in terms of fields in phase B.}
		\label{tab:GenerH4PhaseB}
	\end{table}

Once again, we consider the universal involution, which acts on the fields of phase B as in \eqref{eq:genUImap}. This, in turn, maps the generators as in \eqref{eq:H4PhaseA-invol}. 

Figure \ref{(0,1) theory of Spin(7) orientifold from h4 phase B}, shows the resulting quiver for the orientifold theory. By construction, this gauge theory corresponds to the same Spin(7) orientifold as the one constructed from phase A in the previous section. In Section \ref{sec:H4ABtrial}, we will elaborate on the connection between both theories.

 		\begin{figure}[H]
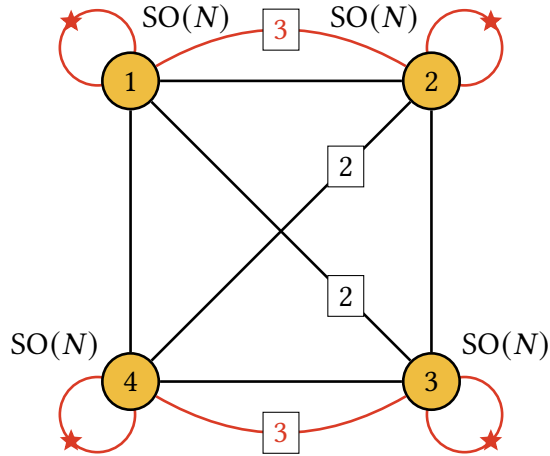

	    \centering

	    \caption{Quiver diagram for the Spin(7) orientifold of phase B of $H_4$ using the universal involution.}
		\label{(0,1) theory of Spin(7) orientifold from h4 phase B}
	\end{figure}

\subsubsection{Triality Between the Orientifolded Theories}
\label{sec:H4ABtrial}

Let us now elaborate on the connection between the two theories that we have constructed via the universal involution. Both of them correspond to the same Spin(7) orientifold of $H_4$. The parent theories, phases A and B of $H_4$, are related by $\mathcal{N}=(0,2)$ triality on either node 2 or 4 of phase A (equivalently, by inverse triality on the same nodes of phase B). This leads to a similar connection between the two orientifolded theories, this time via $\mathcal{N}=(0,1)$ triality on node 2 or 4. Figure \ref{(0,1) triality between orientifold theories from h4} summarizes the interplay between triality and orientifolding. This was expected, given our general discussion of the universal involution in Section \ref{sec:01trial-UI}.

\begin{figure}[ht!]
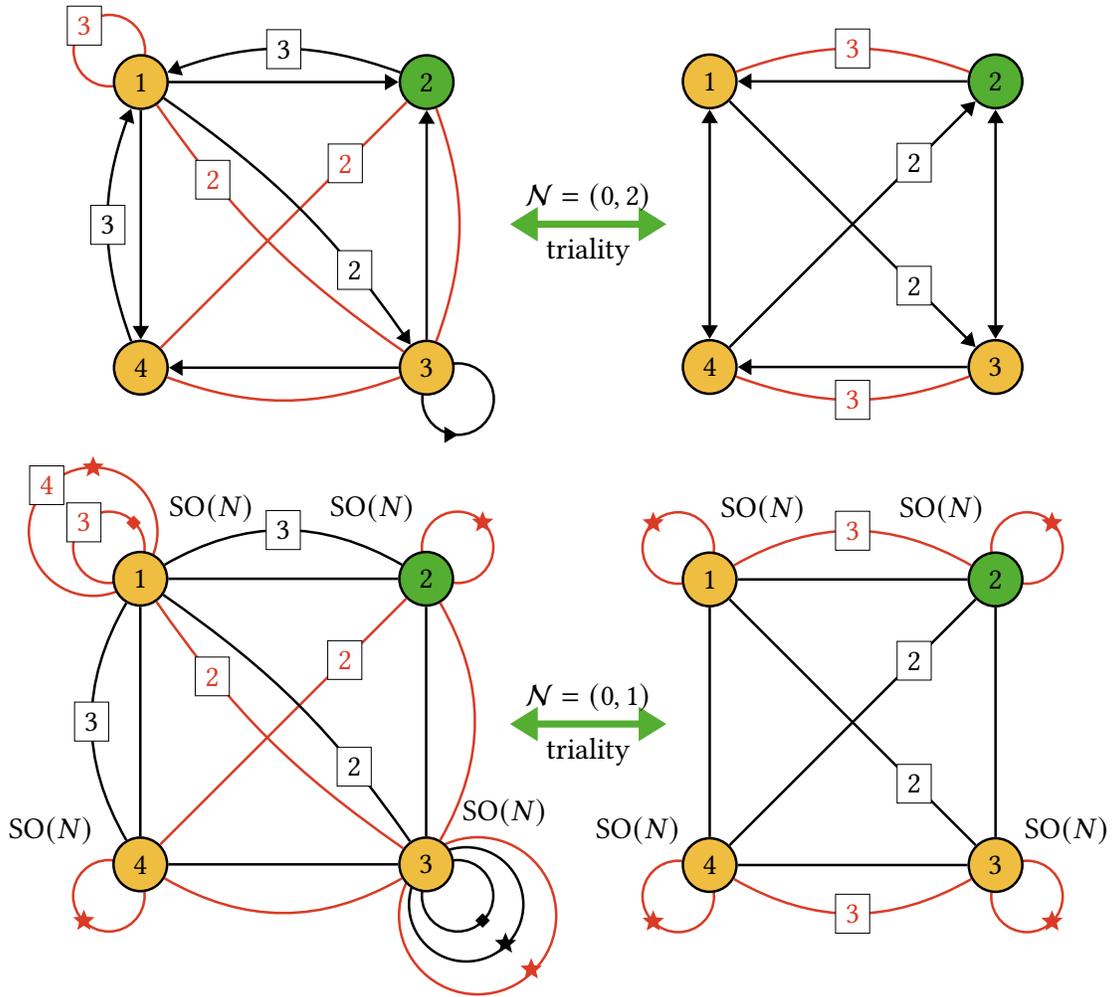

    \centering
    \scalebox{0.95}{

    };
    \draw[Triangle-Triangle,greenX,line width=1mm] (0.8,2) -- node[above,midway,black] {$\mathcal{N}=(0,2)$} node[below,midway,black] {triality} (1.35,2);
    \draw[Triangle-Triangle,greenX,line width=1mm] (0.8,0.25) -- node[above,midway,black] {$\mathcal{N}=(0,1)$} node[below,midway,black] {triality} (1.35,0.25);
    \end{tikzpicture}
    }
 \caption{Phases A and B of $H_4$ are connected by $\mathcal{N}=(0,2)$ triality on node 2 (shown in green). Upon orientifolding with the universal involution, the resulting theories are similarly connected by $\mathcal{N}=(0,1)$ triality.}
		\label{(0,1) triality between orientifold theories from h4}
\end{figure}

It is important to emphasize that it is possible for two $\Spin(7)$ orientifolds to correspond to the same geometric involution while differing in the choice of vector structure. In practical terms, the appearance of the choices of vector structure in orientifolds arises when, for a given geometry, there are different $\mathbb{Z}_2$  symmetries on the underlying quiver gauge theory, which differ in the action on the quiver nodes. Such a discrete choice generalizes beyond orbifold singularities, and it was studied in detail in \cite{Franco:2021ixh}, in anticipation of the application of $\Spin(7)$ orientifolds to triality that we carry out in this chapter. In order for equivalent orientifold geometric involutions to actually produce dual theories, it is necessary that they also agree on the choice of vector structure they implicitly define. This is the case for all the examples considered in this chapter.

Finally, it is interesting to note that, as we discussed in Section \ref{sec:N01triality_quivers}, in orientifold theories the number of ``incoming flavors” at the dualized node is inherited from the parent.

\section{Beyond the Universal Involution}
\label{sec:BeyonUI}

In this section, we present theories that are obtained from $\mathcal{N}=(0,2)$ triality dual parents by Spin(7) orientifolds that do not correspond to the universal involution. We will see that they lead to interesting generalizations of the basic $\mathcal{N}=(0,1)$ triality.\footnote{We will rightfully continue referring to the resulting equivalences between theories as trialities, due to their connections to the basic trialities of SQCD-type theories. It is reasonable to expect that we can indeed perform these transformations three times on the same quiver node. However, the three transformations, can sometimes fall outside our analysis, provided they actually exist. This is due to our restriction to the class of theories obtained as Spin(7) orientifolds of toric phases.}

\subsection{$Q^{1,1,1}$}
\label{sec:Q111}

\begin{figure}[H]
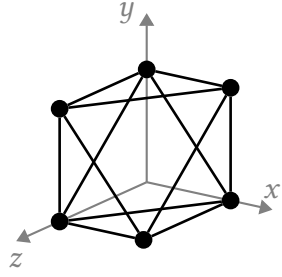

		\centering

		\caption{Toric diagram for $Q^{1,1,1}$.}
		\label{fig:Q111toricdiagram}
\end{figure}

Let us now consider the cone over $Q^{1,1,1}$, or $Q^{1,1,1}$ for short, whose toric diagram is shown in Figure \ref{fig:Q111toricdiagram}. The $\mathcal{N}=(0,2)$ gauge theories, brane brick models and the triality web relating the toric phases for this geometry have been studied at length \cite{Franco:2015tna,Franco:2015tya,Franco:2016nwv}. However, none of its Spin(7) orientifolds has been presented in the literature. Below, we construct an orientifold based on a non-universal involution.

\subsubsection{Phase A}
	\label{sec:Q111phaseA}

\begin{figure}[H]
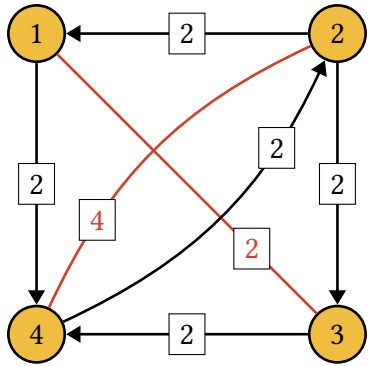

	\centering
	%
	\caption{Quiver diagram for phase A of $Q^{1,1,1}$.}
		\label{fig:Q111AN02}
\end{figure}

The toric phases for $Q^{1,1,1}$ were studied in \cite{Franco:2016nwv}. Figure \ref{fig:Q111AN02} shows the quiver for the so-called phase A.

	The $J$- and $E$-terms are
	\begin{alignat}{4}
	\renewcommand{\arraystretch}{1.1}
    & \centermathcell{J}                           &\text{\hspace{.5cm}}& \centermathcell{E                               }\nonumber \\
	\Lambda_{24}^1 \,:\, & \centermathcell{Y_{42} X_{23} Y_{34}X_{42} -Y_{42}X_{21} X_{14} X_{42} } &                     &\centermathcell{  Y_{23}X_{34}-Y_{21}Y_{14}                }\nonumber\\
	\Lambda_{24}^2 \,:\, & \centermathcell{Y_{42} Y_{21} Y_{14} X_{42}-Y_{42} Y_{23} X_{34} X_{42}} &                     &\centermathcell{ X_{23} Y_{34}-X_{21}X_{14}                }\nonumber\\
	\Lambda_{24}^3 \,:\, & \centermathcell{Y_{42} X_{14} Y_{21} X_{42}-Y_{42} X_{23} X_{34} X_{42}} &                     &\centermathcell{ X_{21} Y_{14}-Y_{23}Y_{34}                }\\
	\Lambda_{24}^4 \,:\, & \centermathcell{Y_{42} X_{21} Y_{14}X_{42} -Y_{42}Y_{23} Y_{34} X_{42} } &                     &\centermathcell{ X_{23}X_{34} -Y_{21}X_{14}                }\nonumber\\
	\Lambda_{31}^1 \,:\, & \centermathcell{Y_{14}X_{42} X_{23} -X_{14} X_{42} Y_{23}              } &                     &\centermathcell{ X_{34}Y_{42}  X_{21}-Y_{34}Y_{42}  Y_{21} }\nonumber\\
	\Lambda_{31}^2 \,:\, & \centermathcell{X_{14}Y_{42}  Y_{23}- Y_{14}Y_{42} X_{23}              } &                     &\centermathcell{  X_{34} X_{42} X_{21}-Y_{34}X_{42}  Y_{21}}\nonumber
	\end{alignat}
Finding the corresponding $W^{(0,1)}$ is a simple exercise, but we omit it here for brevity. Table~\ref{tab:GenerQ111phaseA} lists the generators for $Q^{1,1,1}$ written in terms of the gauge theory.

	\begin{table}[H]
		\centering
		\renewcommand{\arraystretch}{1.1}
		\begin{tabular}{c|c}
			Field    & Chiral superfields  \\
			\hline
			$M_1$    & $Y_{42}Y_{23}X_{34}=Y_{42}Y_{21}Y_{14}$ \\
			$M_2$    & $X_{42}Y_{23}X_{34}=X_{42}Y_{21}Y_{14}$ \\
			$M_3$    & $Y_{42}X_{23}X_{34}=Y_{42}Y_{21}X_{14}$ \\
			$M_4$    & $X_{42}X_{23}X_{34}=X_{42}Y_{21}X_{14}$ \\
			$M_5$    & $Y_{42}Y_{23}Y_{34}=Y_{42}X_{21}Y_{14}$ \\
			$M_6$    & $X_{42}Y_{23}Y_{34}=X_{42}X_{21}Y_{14}$ \\
			$M_7$    & $Y_{42}X_{23}Y_{34}=Y_{42}X_{21}X_{14}$ \\
			$M_8$    & $X_{42}X_{23}Y_{34}=X_{42}X_{21}X_{14}$ 
		\end{tabular}
		\caption{Generators of $Q^{1,1,1}$ in terms of fields in phase A.}
		\label{tab:GenerQ111phaseA}
	\end{table}

The generators satisfy the following relations
	\begin{equation}
	\begin{split}
	\mathcal{I} = &\left\langle M_1M_7=M_3M_5\coma M_3M_8=M_4M_7\coma M_1M_4=M_2M_3\coma M_5M_8=M_7M_6\coma\right.\\
	&\left.M_1M_8=M_2M_7\coma M_3M_6=M_4M_5\coma M_1M_8=M_4M_5\coma M_1M_6=M_5M_2\coma\right.\\
	&\left.M_2M_8=M_4M_6\right\rangle\fstop
	\label{eq:Q111HSrel}
	\end{split}
	\end{equation}

Let us now consider the involution that maps all the four gauge groups to themselves and has the following action on chiral fields
\begin{equation}
	\begin{array}{cccccccccccc}
	Y_{42} &\rightarrow&  -\gamma_{\Omega_4}\bar{X}_{42}\gamma_{\Omega_2}^{-1}\coma & 
	X_{42} &\rightarrow& \gamma_{\Omega_4}\bar{Y}_{42}\gamma_{\Omega_2}^{-1} \coma & 
	X_{34}&\rightarrow& \gamma_{\Omega_3}\bar{Y}_{34}\gamma_{\Omega_4}^{-1} \coma &
	Y_{34}&\rightarrow& -\gamma_{\Omega_3}\bar{X}_{34}\gamma_{\Omega_4}^{-1}\coma \\
	X_{21}&\rightarrow& -\gamma_{\Omega_2}\bar{Y}_{21}\gamma_{\Omega_1}^{-1}\coma & 
	Y_{21}&\rightarrow& \gamma_{\Omega_2}\bar{X}_{21}\gamma_{\Omega_1}^{-1}\coma & 
	Y_{23}&\rightarrow& \gamma_{\Omega_2}\bar{Y}_{23}\gamma_{\Omega_3}^{-1}\coma &  
	X_{23}&\rightarrow& \gamma_{\Omega_2}\bar{X}_{23}\gamma_{\Omega_3}^{-1}\coma \\
	& & & Y_{14}&\rightarrow& \gamma_{\Omega_1}\bar{Y}_{14}\gamma_{\Omega_4}^{-1}\coma & X_{14}&\rightarrow& \gamma_{\Omega_1}\bar{X}_{14}\gamma_{\Omega_4}^{-1}\coma
	\end{array}
	\label{eq:Q111A-chiral-invol}
	\end{equation}
	where we have used the $\gamma_{\Omega_i}$ matrices mentioned in Footnote \ref{foot:gamma-omega}.
	
	Invariance of $W^{(0,1)}$ further implies that the involution acts on Fermi fields as follows  
	\begin{equation}
	\begin{array}{ccccccccc}
	\Lambda_{24}^1&\rightarrow& -\gamma_{\Omega_2}\bar{\Lambda}^3_{24}\gamma_{\Omega_4}^{-1} \coma &
	\Lambda_{24}^2&\rightarrow& -\gamma_{\Omega_2}\bar{\Lambda}^4_{24}\gamma_{\Omega_4}^{-1} \coma &
	\Lambda_{24}^3&\rightarrow& \gamma_{\Omega_2}\bar{\Lambda}^1_{24}\gamma_{\Omega_4}^{-1} \coma \\ 
	\Lambda_{24}^4&\rightarrow& \gamma_{\Omega_2}\bar{\Lambda}^2_{24}\gamma_{\Omega_4}^{-1}\coma &
	\Lambda_{31}^1&\rightarrow& -\gamma_{\Omega_3}\bar{\Lambda}_{31}^2\gamma_{\Omega_1}^{-1}\coma & 
	\Lambda_{31}^2&\rightarrow& \gamma_{\Omega_3}\Lambda_{31}^1\gamma_{\Omega_1}^{-1}\coma
	\end{array}
	\label{eq:Q111A-Fermi-invol}
	\end{equation}
	and
	\begin{equation}
    \Lambda_{11}^R\rightarrow \gamma_{\Omega_1}\Lambda_{11}^{R\,\,T}\gamma_{\Omega_1}^{-1} \coma \Lambda_{22}^R\rightarrow \gamma_{\Omega_2}\Lambda_{22}^{R\,\,T}\gamma_{\Omega_2}^{-1} \coma \Lambda_{33}^R\rightarrow \gamma_{\Omega_3}\Lambda_{33}^{R\,\,T}\gamma_{\Omega_3}^{-1} \coma \Lambda_{44}^R\rightarrow \gamma_{\Omega_4}\Lambda_{44}^{R\,\,T}\gamma_{\Omega_4}^{-1} \fstop
    \label{eq:Q111A-RFermi-invol}
	\end{equation}
	
Interestingly, the involution in \eqref{eq:Q111A-chiral-invol} and \eqref{eq:Q111A-Fermi-invol} involves a non-trivial action on flavor indices (see e.g. the action on pairs of fields such as $(X_{21},Y_{21})$). As briefly mentioned in Section \ref{sec:Spin7Orient}, this leads to a constraint on the matrices $\gamma_{\Omega_i}$ that encode the action of the orientifold group on the gauge groups, which reads 
\begin{equation}
    \gamma_{\Omega_1}= \gamma_{\Omega_4}\neq  \gamma_{\Omega_2}= \gamma_{\Omega_3}\fstop
    \label{eq:Q111A-Omega-cond}
\end{equation}
This constraint follows for requiring that the involution squares to the identity. For a detailed discussion of this constraint and additional explicit examples, we refer the interested reader to our previous work \cite{Franco:2021ixh}. 

For concreteness, we will focus on the following solution to the constraint
\begin{equation}
\begin{array}{ccccc}
\gamma_{\Omega_1} & = & \gamma_{\Omega_4} & = & J \coma \\
\gamma_{\Omega_2} & = & \gamma_{\Omega_3} & = & \ID_N \coma
\end{array}
\label{gamma_matrices_phase_A}
\end{equation}
where $J=i\epsilon_{N/2}$ is the symplectic matrix, and $\ID_N$ is the identity matrix.
	
Using Table~\ref{tab:GenerQ111phaseA}, the involution in \eqref{eq:Q111A-chiral-invol} translates into the following action at the level of the geometry
\begin{equation}
     \begin{array}{cccccccccccc}
	M_{1}  & \rightarrow & -\bar{M}_{6} \coma &
	M_{2}  & \rightarrow & \bar{M}_{5} \coma &
	M_{3}  & \rightarrow & -\bar{M}_{8} \coma &
	M_{4}  & \rightarrow & \bar{M}_{7} \coma \\
	M_{5}  & \rightarrow & \bar{M}_{2} \coma &
	M_{6}  & \rightarrow & -\bar{M}_{1} \coma &
	M_{7}  & \rightarrow & \bar{M}_{4} \coma &
	M_{8}  & \rightarrow & -\bar{M}_{3} \coma
	\end{array} 
	\label{eq:Q111phaseA-nonuni-geom}
	\end{equation}
which is clearly not the universal involution. 

\begin{figure}[H]
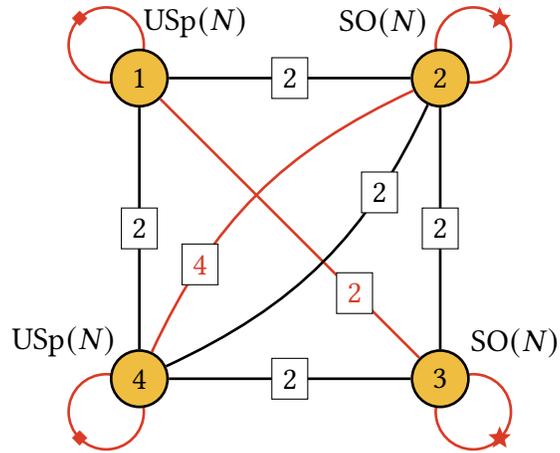

	    \centering
	  	%
	    \caption{Quiver diagram for the Spin(7) orientifold of phase A of $Q^{1,1,1}$ using the involution in \cref{eq:Q111A-chiral-invol,eq:Q111A-Fermi-invol,eq:Q111A-RFermi-invol}, together with our choice of $\gamma_{\Omega_i}$ matrices.}
		\label{o theory phase A q111}
	\end{figure}

Figure \ref{o theory phase A q111} shows the quiver for the orientifold theory, which is free of gauge anomalies. %

\newpage

\subsubsection{Phase S}
	\label{sec:Q111phaseS} 
	
\begin{figure}[H]
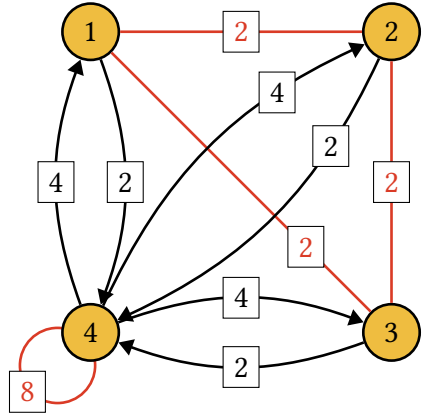

	\centering

	\caption{Quiver diagram for phase S of $Q^{1,1,1}$.}
	\label{fig:Q111SN02}
\end{figure}

Figure \ref{fig:Q111SN02} shows the quiver for phase S of $Q^{1,1,1}$ \cite{Franco:2016nwv}.

The $J$- and $E$-terms are
		\begin{alignat}{4}
	\renewcommand{\arraystretch}{1.1}
    & \centermathcell{J}                           &\text{\hspace{.5cm}}& \centermathcell{E                               }\nonumber \\
	\Lambda_{23}^1 \,:\,& \centermathcell{X_{34}Y_{42} -Y_{34}W_{42}  }& & \centermathcell{Y_{24}X_{43} -X_{24} Z_{43}}\nonumber \\
	\Lambda_{23}^2 \,:\,& \centermathcell{X_{34}X_{42}-Y_{34} Z_{42}  }& & \centermathcell{X_{24}W_{43}-Y_{24}Y_{43}  }\nonumber \\
	\Lambda_{31}^1 \,:\,& \centermathcell{X_{14} Z_{43}-Y_{14}W_{43}  }& & \centermathcell{X_{34}X_{41} -Y_{34}Z_{41} }\nonumber \\
	\Lambda_{31}^2 \,:\,& \centermathcell{X_{14}X_{43} -Y_{14} Y_{43} }& & \centermathcell{Y_{34}W_{41} -X_{34}Y_{41} }\nonumber \\
	\Lambda_{12}^1 \,:\,& \centermathcell{Y_{24}Z_{41}-X_{24}W_{41}   }& & \centermathcell{X_{14}X_{42}-Y_{14} Y_{42} }\nonumber \\
	\Lambda_{12}^2 \,:\,& \centermathcell{Y_{24}X_{41} -X_{24}Y_{41}  }& & \centermathcell{Y_{14}W_{42}-X_{14}Z_{42}  }\nonumber \\
	\Lambda_{44}^1 \,:\,& \centermathcell{Y_{43}Y_{34} -X_{41} X_{14} }& & \centermathcell{W_{41} Y_{14}-Z_{42}Y_{24} }\label{Q111PhaseSJEterms} \\
	\Lambda_{44}^2 \,:\,& \centermathcell{W_{41}Y_{14}-Z_{43}X_{34}   }& & \centermathcell{X_{41}X_{14}-Y_{42}X_{24}  }\nonumber \\
	\Lambda_{44}^3 \,:\,& \centermathcell{Y_{41}Y_{14} -X_{42}Y_{24}  }& & \centermathcell{W_{42} X_{24}-Y_{43}X_{34} }\nonumber \\
	\Lambda_{44}^4 \,:\,& \centermathcell{W_{42} X_{24}-Z_{41}X_{14}  }& & \centermathcell{X_{42} Y_{24}-Z_{43}Y_{34} }\nonumber \\
	\Lambda_{44}^5 \,:\,& \centermathcell{Z_{42}X_{24}-X_{43} X_{34}  }& & \centermathcell{Y_{24}Y_{42}-Y_{41} X_{14} }\nonumber \\
	\Lambda_{44}^6 \,:\,& \centermathcell{W_{43} Y_{34}-Y_{42}Y_{24}  }& & \centermathcell{X_{24} Z_{42}-Z_{41}Y_{14} }\nonumber \\
	\Lambda_{44}^7 \,:\,& \centermathcell{W_{43} X_{34}-W_{41} X_{14} }& & \centermathcell{X_{41}Y_{14}-X_{42} X_{24} }\nonumber \\
	\Lambda_{44}^8 \,:\,& \centermathcell{X_{41}Y_{14}-X_{43} Y_{34}  }& & \centermathcell{W_{41} X_{14}-W_{42}Y_{24} }\nonumber
	\end{alignat}
	
Table~\ref{tab:GenerQ111phaseS} shows the generators of $Q^{1,1,1}$ in terms of the gauge theory. They satisfy the same relations given in \eqref{eq:Q111HSrel}.

\begin{table}[H]
		\centering
		\renewcommand{\arraystretch}{1.1}
\begin{tabular}{c|c}
			Field    & Chiral superfields  \\
			\hline
			$M_1$    & $Z_{42}Y_{24}=Z_{43}X_{34}=W_{41}Y_{14}$ \\
			$M_2$    & $Z_{42}X_{24}=X_{43}X_{34}=Z_{41}Y_{14}$ \\
			$M_3$    & $W_{42}Y_{24}=W_{43}X_{34}=W_{41}X_{14}$ \\
			$M_4$    & $W_{42}X_{24}=Y_{43}X_{34}=Z_{41}X_{14}$ \\
			$M_5$    & $X_{42}Y_{24}=Z_{43}Y_{34}=Y_{41}Y_{14}$ \\
			$M_6$    & $X_{42}X_{24}=X_{43}Y_{34}=X_{41}Y_{14}$ \\
			$M_7$    & $Y_{42}Y_{24}=W_{43}Y_{34}=Y_{41}X_{14}$ \\
			$M_8$    & $Y_{42}X_{24}=Y_{43}Y_{34}=X_{41}X_{14}$
		\end{tabular}
		\caption{Generators of $Q^{1,1,1}$ in terms of fields in phase S.}
		\label{tab:GenerQ111phaseS}
	\end{table}	

Let us consider the involution that maps all gauge groups to themselves and acts on chiral fields as follows
\begin{equation}
	\begin{array}{cccccccccccc}
	Z_{42} & \rightarrow & \gamma_{\Omega_4}\bar{X}_{42}\gamma_{\Omega_2}^{-1}\coma &
	X_{42} & \rightarrow & -\gamma_{\Omega_4}\bar{Z}_{42}\gamma_{\Omega_2}^{-1}\coma &
	Y_{24} & \rightarrow & -\gamma_{\Omega_2}\bar{X}_{24}\gamma_{\Omega_4}^{-1}\coma &
	X_{24} & \rightarrow & \gamma_{\Omega_2}\bar{Y}_{24}\gamma_{\Omega_4}^{-1}\coma \\ 
	Z_{43} & \rightarrow & -\gamma_{\Omega_4}\bar{X}_{43}\gamma_{\Omega_3}^{-1}\coma &
	X_{43} & \rightarrow & \gamma_{\Omega_4}\bar{Z}_{43}\gamma_{\Omega_3}^{-1}\coma &
	X_{34} & \rightarrow & \gamma_{\Omega_3}\bar{Y}_{34}\gamma_{\Omega_4}^{-1}\coma &
	Y_{34} & \rightarrow & -\gamma_{\Omega_3}\bar{X}_{34}\gamma_{\Omega_4}^{-1}\coma \\
	W_{41} & \rightarrow & -\gamma_{\Omega_4}\bar{X}_{41}\gamma_{\Omega_1}^{-1}\coma &
	X_{41} & \rightarrow & -\gamma_{\Omega_4}\bar{W}_{41}\gamma_{\Omega_1}^{-1}\coma &
	Z_{41} & \rightarrow & \gamma_{\Omega_4}\bar{Y}_{41}\gamma_{\Omega_1}^{-1}\coma &
	Y_{41} & \rightarrow & \gamma_{\Omega_4}\bar{Z}_{41}\gamma_{\Omega_1}^{-1}\coma \\
	W_{42} & \rightarrow & \gamma_{\Omega_4}\bar{Y}_{42}\gamma_{\Omega_2}^{-1}\coma &
	Y_{42} & \rightarrow & -\gamma_{\Omega_4}\bar{W}_{42}\gamma_{\Omega_2}^{-1}\coma &
	W_{43} & \rightarrow & -\gamma_{\Omega_4}\bar{Y}_{43}\gamma_{\Omega_3}^{-1}\coma &
	Y_{43} & \rightarrow & \gamma_{\Omega_4}\bar{W}_{43}\gamma_{\Omega_3}^{-1}\coma \\
	& & & 
	Y_{14} & \rightarrow & \gamma_{\Omega_1}\bar{Y}_{14}\gamma_{\Omega_4}^{-1}\coma &
	X_{14} & \rightarrow & \gamma_{\Omega_1}\bar{X}_{14}\gamma_{\Omega_4}^{-1}\fstop 
	\end{array}
	\label{eq:Q111S-chiral-invol}
	\end{equation}
As we will explain shortly, we have chosen this involution in order to connect to the orientifold of phase A that we constructed in the previous section.
	
Invariance of $W^{(0,1)}$ implies the following action on Fermi fields
	\begin{equation}
	\begin{array}{cccccccccccc}
	\Lambda_{23}^1 & \rightarrow & \gamma_{\Omega_2}\bar{\Lambda}_{23}^{1}\gamma_{\Omega_3}^{-1}\coma & 
	\Lambda_{23}^2 & \rightarrow & \gamma_{\Omega_2}\bar{\Lambda}_{23}^{2}\gamma_{\Omega_3}^{-1}\coma & 
	\Lambda_{31}^1 & \rightarrow &  -\gamma_{\Omega_3}\bar{\Lambda}_{31}^{2}\gamma_{\Omega_1}^{-1}\coma & 
	\Lambda_{31}^2 & \rightarrow &  \gamma_{\Omega_3}\bar{\Lambda}_{31}^{1}\gamma_{\Omega_1}^{-1}\coma \\ 
	\Lambda_{12}^1 & \rightarrow & -\gamma_{\Omega_1}\bar{\Lambda}_{12}^{2}\gamma_{\Omega_2}^{-1}\coma & 
	\Lambda_{12}^2 & \rightarrow & \gamma_{\Omega_1}\bar{\Lambda}_{12}^{1}\gamma_{\Omega_2}^{-1}\coma & 
	\Lambda_{44}^1 & \rightarrow &  -\gamma_{\Omega_4}\bar{\Lambda}_{44}^{7}\gamma_{\Omega_4}^{-1}\coma & 
	\Lambda_{44}^7 & \rightarrow &  -\gamma_{\Omega_4}\bar{\Lambda}_{44}^{1}\gamma_{\Omega_4}^{-1}\coma \\ 
	\Lambda_{44}^2 & \rightarrow &  -\gamma_{\Omega_4}\bar{\Lambda}_{44}^{8}\gamma_{\Omega_4}^{-1}\coma & 
	\Lambda_{44}^8 & \rightarrow &  -\gamma_{\Omega_4}\bar{\Lambda}_{44}^{2}\gamma_{\Omega_4}^{-1}\coma & 
	\Lambda_{44}^3 & \rightarrow & -\gamma_{\Omega_4}\Lambda_{44}^{6\,\,T} \gamma_{\Omega_4}^{-1}\coma & 
	\Lambda_{44}^6 & \rightarrow & -\gamma_{\Omega_4}\Lambda_{44}^{3\,\,T} \gamma_{\Omega_4}^{-1}\coma \\
	& & & 
	\Lambda_{44}^4 & \rightarrow &  \gamma_{\Omega_4}\Lambda_{44}^{5\,\,T} \gamma_{\Omega_4}^{-1}\coma & 
	\Lambda_{44}^5 & \rightarrow &  \gamma_{\Omega_4}\Lambda_{44}^{4\,\,T} \gamma_{\Omega_4}^{-1}\coma 
	\end{array}
	\label{eq:Q111S-Fermi-invol}
	\end{equation}
	and
	\begin{equation}
    \Lambda_{11}^R\rightarrow \gamma_{\Omega_1}\Lambda_{11}^{R\,\,T}\gamma_{\Omega_1}^{-1} \coma \Lambda_{22}^R\rightarrow \gamma_{\Omega_2}\Lambda_{22}^{R\,\,T}\gamma_{\Omega_2}^{-1} \coma \Lambda_{33}^R\rightarrow \gamma_{\Omega_3}\Lambda_{33}^{R\,\,T}\gamma_{\Omega_3}^{-1} \coma \Lambda_{44}^R\rightarrow \gamma_{\Omega_4}\Lambda_{44}^{R\,\,T}\gamma_{\Omega_4}^{-1} \fstop
    \label{eq:Q111S-RFermi-invol}
	\end{equation}

The involution on bifundamental fields leads to the same constraints on the $\gamma_{\Omega_i}$ matrices as in~\eqref{eq:Q111A-Omega-cond}. As for phase A, we pick
\begin{equation}
\begin{array}{ccccc}
\gamma_{\Omega_1} & = & \gamma_{\Omega_4} & = & J \coma \\
\gamma_{\Omega_2} & = & \gamma_{\Omega_3} & = & \ID_N \fstop
\end{array}
\end{equation}	

\newpage

\begin{figure}[H]
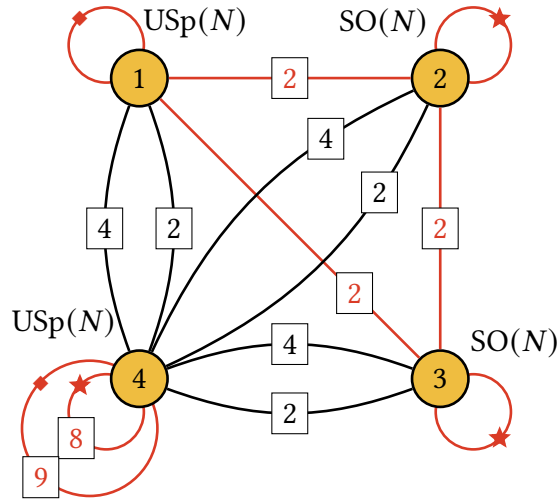

	    \centering
	    	%
		
		\vspace{0.9cm} 
		
	    \caption{Quiver  diagram  for  the  Spin(7)  orientifold  of  phase  S of $Q^{1,1,1}$ using the involution in \cref{eq:Q111S-chiral-invol,eq:Q111S-Fermi-invol,eq:Q111S-RFermi-invol}, together with our choice of $\gamma_{\Omega_i}$ matrices.}
		\label{fig:o_theory_q111_S}
	\end{figure}

Using Table~\ref{tab:GenerQ111phaseS}, \eqref{eq:Q111S-chiral-invol} translates into the following action on the generators
\begin{equation}
 \begin{array}{cccccccccccc}
	M_{1}  & \rightarrow & -\bar{M}_{6} \coma &
	M_{2}  & \rightarrow & \bar{M}_{5} \coma &
	M_{3}  & \rightarrow & -\bar{M}_{8} \coma &
	M_{4}  & \rightarrow & \bar{M}_{7} \coma \\
	M_{5}  & \rightarrow & \bar{M}_{2} \coma &
	M_{6}  & \rightarrow & -\bar{M}_{1} \coma &
	M_{7}  & \rightarrow & \bar{M}_{4} \coma &
	M_{8}  & \rightarrow & -\bar{M}_{3} \coma
	\end{array} 
\label{eq:Q111phaseS-nonuni-geom}
\end{equation} 
which is the same geometric involution that we found for phase A in \eqref{eq:Q111phaseA-nonuni-geom}. Therefore, the involutions considered on these two phases correspond to the same Spin(7) orientifold of $Q^{1,1,1}$. 
Figure \ref{fig:o_theory_q111_S} shows the quiver for the orientifold theory, which is free of gauge anomalies.

\subsubsection{Triality Between the Orientifolded Theories}
\label{sec:Q111triality}

\begin{figure}[H]
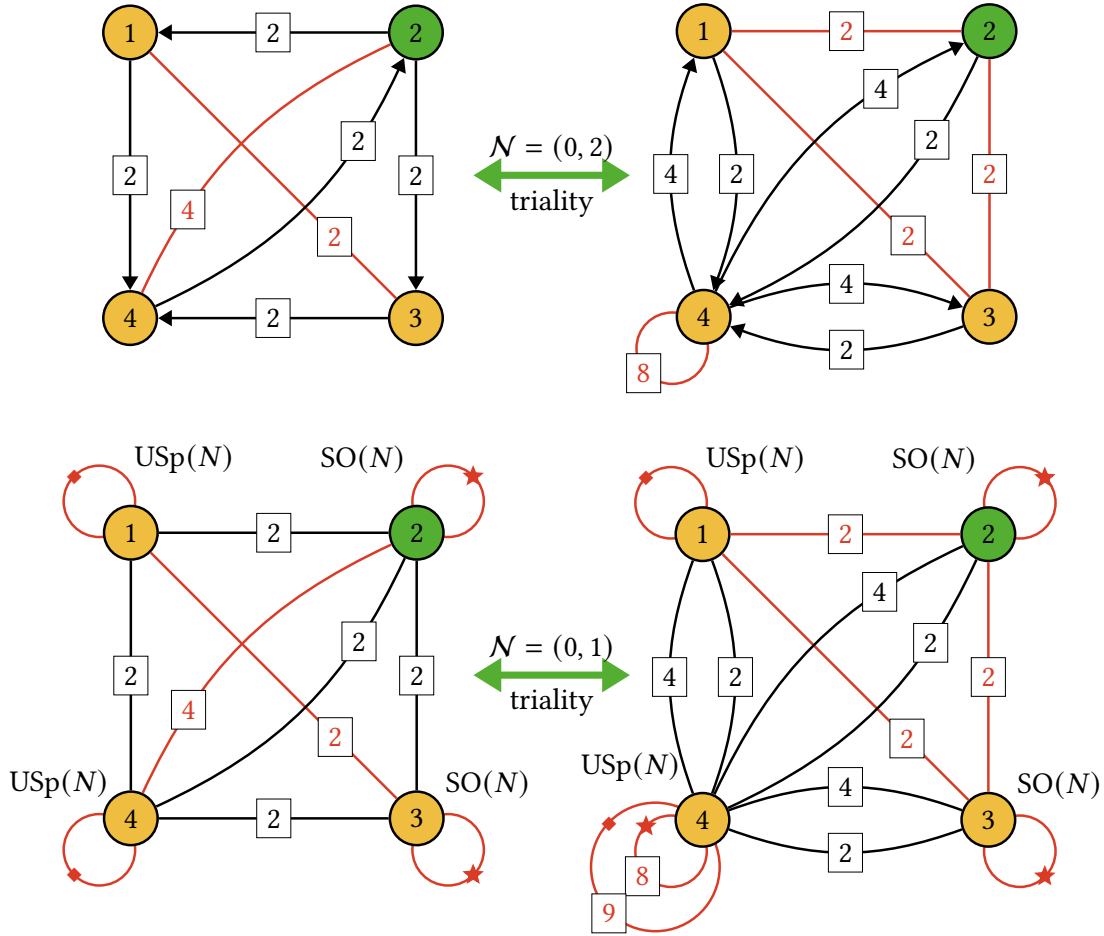

    \centering
    \scalebox{0.95}{

    };
    \draw[Triangle-Triangle,greenX,line width=1mm] (0.7,2) -- node[above,midway,black] {$\mathcal{N}=(0,2)$} node[below,midway,black] {triality} (1.25,2);
    \draw[Triangle-Triangle,greenX,line width=1mm] (0.7,0.25) -- node[above,midway,black] {$\mathcal{N}=(0,1)$} node[below,midway,black] {triality} (1.25,0.25);
    \end{tikzpicture}
    }
 \caption{Phases A and S of $Q^{1,1,1}$ are connected by $\mathcal{N}= (0,2)$ triality on node 2 (shown in green). The orientifolded theories are similarly connected by $\mathcal{N}= (0,1)$ triality.}
		\label{01_02_triality_q111}
\end{figure}

Figure \ref{01_02_triality_q111} summarizes the connections between the theories considered in this section. Again, we observe that the two theories we constructed for the same Spin(7) orientifold are related by $\mathcal{N}=(0,1)$ triality. More precisely, they are related by a simple generalization of the basic triality reviewed in Section \ref{sec:N01triality}. First, in this case, triality is applied to quivers with multiple gauge nodes. More importantly, some of the nodes that act as flavor groups are of a different type (in this example, USp) than the dualized node. As in previous examples, the orientifold construction leads to a clear prescription on how to treat scalar flavors, which is inherited from the parent theories.

\subsection{Theories with Unitary Gauge Groups: $Q^{1,1,1}/\mathbb{Z}_2$}
\label{sec:Q111Z2}

All $\mathcal{N}=(0,1)$ triality examples we constructed so far contain only $\SO(N)$ and $\USp(N)$ gauge groups. Namely, the anti-holomorphic involutions of the parent $\mathcal{N}=(0,2)$ theories, universal or not,  map all gauge groups to themselves. In this section we will construct Spin(7) orientifolds giving rise to gauge theories that include $\U(N)$ gauge groups. To do so, we focus on $Q^{1,1,1}/\mathbb{Z}_2$, whose toric diagram is shown in Figure~\ref{fig:Q111Z2toricdiagram}.\footnote{More precisely, this is the $\mathbb{Z}_2$ orbifold of the real cone over $Q^{1,1,1}$.} This CY$_4$ has a rich family of 14 toric phases. They were classified in \cite{Franco:2018qsc}, whose nomenclature we will follow. We will restrict to a subset consisting of 5 of these toric phases. In order to streamline our discussion, several details about these theories are collected in Appendix \ref{app:Q111Z2-details}.

	\begin{figure}[H]
		\centering

		\caption{Toric diagram for $Q^{1,1,1}/\ZZ_2$.}
		\label{fig:Q111Z2toricdiagram}
	\end{figure}	

Let us first consider phase D, whose quiver diagram is shown in Figure \ref{fig:Q111Z2quiverD}. We provide a 3d representation of the quiver in order to make the action of the anti-holomorphic involution that we will use to construct a Spin(7) orientifold more manifest. 

			\begin{figure}[H]
		\centering
		%
	\caption{Quiver diagram for phase D of $Q^{1,1,1}/\ZZ_2$.}
		\label{fig:Q111Z2quiverD}
	\end{figure}

The $J$- and $E$-terms for this theory are 
		\begin{alignat}{4}
	\renewcommand{\arraystretch}{1.1}
    & \centermathcell{J}                           &\text{\hspace{.5cm}}& \centermathcell{E                               }\nonumber \\
 \Lambda^1_{13}  \,:\, &  \centermathcell{W_{34} X_{41}-Y_{41} Z_{34}} & & \centermathcell{X_{18} X_{85} Y_{53}-X_{53} X_{85} Y_{18} }\nonumber\\
 \Lambda^2_{13}  \,:\, &  \centermathcell{X_{41} Y_{34}-X_{34} Y_{41} }& & \centermathcell{X_{53} Y_{18} Y_{85}-X_{18} Y_{53} Y_{85} }\nonumber\\
 \Lambda^1_{37}  \,:\, &  \centermathcell{X_{72} Y_{53} Y_{25}-X_{53} Y_{72} Y_{25} }&  & \centermathcell{X_{47} Z_{34}-X_{34} Y_{47}} \nonumber\\
 \Lambda^2_{37}  \,:\, &  \centermathcell{X_{72} X_{25} Y_{53}-X_{53} X_{25} Y_{72} }&  & \centermathcell{Y_{34} Y_{47}-W_{34} X_{47}}\nonumber \\
 \Lambda^1_{86}  \,:\, &  \centermathcell{X_{64} Y_{18} Y_{41}-X_{18} Y_{41} Y_{64} }&  & \centermathcell{X_{56} Y_{85}-X_{85} Z_{56}}\nonumber \\
 \Lambda^2_{86}  \,:\, &  \centermathcell{X_{41} X_{64} Y_{18}-X_{18} X_{41} Y_{64} }&  & \centermathcell{W_{56} X_{85}-Y_{56} Y_{85}}\nonumber \\
 \Lambda^1_{62}  \,:\, &  \centermathcell{Y_{25} Z_{56}-W_{56} X_{25} }&  & \centermathcell{X_{47} X_{64} Y_{72}-X_{72} X_{47} Y_{64}} \nonumber\\
 \Lambda^2_{62}  \,:\, &  \centermathcell{X_{25} Y_{56}-X_{56} Y_{25} }&  & \centermathcell{X_{64} Y_{72} Y_{47}-X_{72} Y_{47} Y_{64}} \label{eq:Q111Z2JEtermsD}\\
 \Lambda^1_{45}  \,:\, &  \centermathcell{W_{56} Y_{64}-W_{34} Y_{53} }&  & \centermathcell{X_{18} X_{41} X_{85}-X_{72} X_{47} X_{25}}\nonumber \\
 \Lambda^2_{45}  \,:\, &  \centermathcell{W_{56} X_{64}-W_{34} X_{53} }&  & \centermathcell{X_{47} X_{25} Y_{72}-X_{41} X_{85} Y_{18}}\nonumber \\
 \Lambda^3_{45}  \,:\, &  \centermathcell{Y_{64} Z_{56}-Y_{53} Z_{34} }&  & \centermathcell{X_{72} X_{47} Y_{25}-X_{18} X_{85} Y_{41}}\nonumber \\
 \Lambda^4_{45}  \,:\, &  \centermathcell{X_{64} Z_{56}-X_{53} Z_{34} }&  & \centermathcell{X_{85} Y_{18} Y_{41}-X_{47} Y_{72} Y_{25}}\nonumber \\
 \Lambda^5_{45}  \,:\, &  \centermathcell{Y_{56} Y_{64}-Y_{34} Y_{53} }&  & \centermathcell{X_{72} X_{25} Y_{47}-X_{18} X_{41} Y_{85}}\nonumber \\
 \Lambda^6_{45}  \,:\, &  \centermathcell{X_{64} Y_{56}-X_{53} Y_{34} }&  & \centermathcell{X_{41} Y_{18} Y_{85}-X_{25} Y_{72} Y_{47}}\nonumber \\
 \Lambda^7_{45}  \,:\, &  \centermathcell{X_{56} Y_{64}-X_{34} Y_{53} }&  & \centermathcell{X_{18} Y_{41} Y_{85}-X_{72} Y_{47} Y_{25}}\nonumber \\
 \Lambda^8_{45}  \,:\, &  \centermathcell{X_{56} X_{64}-X_{34} X_{53} }&  & \centermathcell{Y_{72} Y_{47} Y_{25}-Y_{18} Y_{41} Y_{85}}\nonumber 
	\end{alignat}

The generators of $Q^{1,1,1}/\mathbb{Z}_2$ in terms of the chiral fields in phase D are listed in Table~\ref{tab:GenQ111Z2PhaseD}.  Note that the generators and their relations are common to all the phases, but their realizations in terms of chiral superfields in each of them are different. Let us consider an anti-holomorphic involution of phase D which acts on Figure~\ref{fig:Q111Z2quiverD} as a reflection with respect to the vertical plane that contains nodes 3, 4, 5 and 6. The nodes on the plane map to themselves, while the following pairs $1\leftrightarrow 7$ and $2\leftrightarrow 8$ get identified. This leads to the anticipated mixture of $\SO/\USp$ and $\U$ gauge groups.

The involution on chiral fields is 
   \begin{equation}
\begin{array}{cccccccccccc}
	X_{18} & \rightarrow & \gamma_{\Omega_7}\bar{Y}_{72}\gamma_{\Omega_2}^{-1}\coma &
	Y_{18} & \rightarrow & \gamma_{\Omega_7}\bar{X}_{72}\gamma_{\Omega_2}^{-1}\coma &
	X_{72} & \rightarrow & \gamma_{\Omega_1}\bar{Y}_{18}\gamma_{\Omega_8}^{-1}\coma &
	Y_{72} & \rightarrow & \gamma_{\Omega_1}\bar{X}_{18}\gamma_{\Omega_8}^{-1}\coma \\ 
	X_{34} & \rightarrow & \gamma_{\Omega_3}\bar{X}_{34}\gamma_{\Omega_4}^{-1}\coma &
	Y_{34} & \rightarrow & \gamma_{\Omega_3}\bar{Z}_{34}\gamma_{\Omega_4}^{-1}\coma &
	Z_{34} & \rightarrow & \gamma_{\Omega_3}\bar{Y}_{34}\gamma_{\Omega_4}^{-1}\coma &
	W_{34} & \rightarrow & \gamma_{\Omega_3}\bar{W}_{34}\gamma_{\Omega_4}^{-1}\coma \\
	X_{41} & \rightarrow & \gamma_{\Omega_4}\bar{X}_{47}\gamma_{\Omega_7}^{-1}\coma &
	Y_{41} & \rightarrow & \gamma_{\Omega_4}\bar{Y}_{47}\gamma_{\Omega_7}^{-1}\coma &
	X_{47} & \rightarrow & \gamma_{\Omega_4}\bar{X}_{41}\gamma_{\Omega_1}^{-1}\coma &
	Y_{47} & \rightarrow & \gamma_{\Omega_4}\bar{Y}_{41}\gamma_{\Omega_1}^{-1}\coma \\
	X_{53} & \rightarrow & \gamma_{\Omega_5}\bar{Y}_{53}\gamma_{\Omega_3}^{-1}\coma &
	Y_{53} & \rightarrow & \gamma_{\Omega_5}\bar{X}_{53}\gamma_{\Omega_3}^{-1}\coma &
	X_{56} & \rightarrow & \gamma_{\Omega_5}\bar{X}_{56}\gamma_{\Omega_6}^{-1}\coma &
	Y_{56} & \rightarrow & \gamma_{\Omega_5}\bar{Z}_{56}\gamma_{\Omega_6}^{-1}\coma \\
	Z_{56} & \rightarrow & \gamma_{\Omega_5}\bar{Y}_{56}\gamma_{\Omega_6}^{-1}\coma &
	W_{56} & \rightarrow & \gamma_{\Omega_5}\bar{W}_{56}\gamma_{\Omega_6}^{-1}\coma &
	X_{64} & \rightarrow & \gamma_{\Omega_6}\bar{Y}_{64}\gamma_{\Omega_4}^{-1}\coma &
	Y_{64} & \rightarrow & \gamma_{\Omega_6}\bar{X}_{64}\gamma_{\Omega_4}^{-1}\coma \\
\end{array}   
	\label{eq:Q111Z2D-chiral-invol}
    \end{equation}

Invariance of $W^{(0,1)}$ implies the following action on Fermi fields
\begin{equation}
     \begin{array}{cccccccccccc}
	\Lambda^1_{13} & \rightarrow & -\gamma_{\Omega_3}\Lambda^2_{37}\gamma_{\Omega_7}^{-1}\coma &
	\Lambda^2_{13} & \rightarrow & \gamma_{\Omega_3}\Lambda^1_{37}\gamma_{\Omega_7}^{-1}\coma &
	\Lambda^1_{37} & \rightarrow & \gamma_{\Omega_1}\Lambda^2_{13}\gamma_{\Omega_3}^{-1}\coma &
	\Lambda^2_{37} & \rightarrow & -\gamma_{\Omega_3}\Lambda^1_{37}\gamma_{\Omega_7}^{-1}\coma \\ 
	\Lambda^1_{86} & \rightarrow & -\gamma_{\Omega_6}\Lambda^2_{62}\gamma_{\Omega_8}^{-1}\coma &
	\Lambda^2_{86} & \rightarrow & -\gamma_{\Omega_6}\Lambda^1_{62}\gamma_{\Omega_8}^{-1}\coma &
	\Lambda^1_{62} & \rightarrow & -\gamma_{\Omega_8}\Lambda^2_{86}\gamma_{\Omega_6}^{-1}\coma &
	\Lambda^2_{62} & \rightarrow & -\gamma_{\Omega_8}\Lambda^1_{86}\gamma_{\Omega_6}^{-1}\coma \\
	\Lambda^1_{45} & \rightarrow & \gamma_{\Omega_4}\bar{\Lambda}^2_{45}\gamma_{\Omega_5}^{-1}\coma &
	\Lambda^2_{45} & \rightarrow & \gamma_{\Omega_4}\bar{\Lambda}^1_{45}\gamma_{\Omega_5}^{-1}\coma &
	\Lambda^3_{45} & \rightarrow & \gamma_{\Omega_4}\bar{\Lambda}^6_{45}\gamma_{\Omega_5}^{-1}\coma &
	\Lambda^4_{45} & \rightarrow & \gamma_{\Omega_4}\bar{\Lambda}^5_{45}\gamma_{\Omega_5}^{-1}\coma \\
	\Lambda^5_{45} & \rightarrow & \gamma_{\Omega_5}\bar{\Lambda}^4_{45}\gamma_{\Omega_5}^{-1}\coma &
	\Lambda^6_{45} & \rightarrow & \gamma_{\Omega_5}\bar{\Lambda}^3_{45}\gamma_{\Omega_5}^{-1}\coma &
	\Lambda^7_{45} & \rightarrow & \gamma_{\Omega_5}\bar{\Lambda}^8_{45}\gamma_{\Omega_5}^{-1}\coma &
	\Lambda^8_{45} & \rightarrow & \gamma_{\Omega_5}\bar{\Lambda}^7_{45}\gamma_{\Omega_5}^{-1}\coma
	\end{array}   
	\label{eq:Q111Z2D-Fermi-invol}
    \end{equation}
and
	\begin{equation}
	 \begin{array}{cccccccccccc}
	\Lambda_{11}^R & \rightarrow & \gamma_{\Omega_7}\Lambda_{77}^{R\,\,T}\gamma_{\Omega_7}^{-1} \coma &
	\Lambda_{22}^R & \rightarrow & \gamma_{\Omega_8}\Lambda_{88}^{R\,\,T}\gamma_{\Omega_8}^{-1} \coma &
	\Lambda_{33}^R & \rightarrow & \gamma_{\Omega_3}\Lambda_{33}^{R\,\,T}\gamma_{\Omega_3}^{-1} \coma &
	\Lambda_{44}^R & \rightarrow & \gamma_{\Omega_4}\Lambda_{44}^{R\,\,T}\gamma_{\Omega_4}^{-1} \coma \\
	\Lambda_{55}^R & \rightarrow & \gamma_{\Omega_5}\Lambda_{55}^{R\,\,T}\gamma_{\Omega_5}^{-1} \coma &
	\Lambda_{66}^R & \rightarrow & \gamma_{\Omega_6}\Lambda_{66}^{R\,\,T}\gamma_{\Omega_6}^{-1} \coma &
	\Lambda_{77}^R & \rightarrow & \gamma_{\Omega_1}\Lambda_{11}^{R\,\,T}\gamma_{\Omega_1}^{-1} \coma &
	\Lambda_{88}^R & \rightarrow & \gamma_{\Omega_2}\Lambda_{22}^{R\,\,T}\gamma_{\Omega_2}^{-1} \fstop
	\end{array}
	\label{eq:Q111Z2D-RFermi-invol}
	\end{equation}
	
Using Table \ref{tab:GenQ111Z2PhaseD}, we find the corresponding geometric involution on the generators of $Q^{1,1,1}/\mathbb{Z}_2$
\begin{equation}
     \begin{array}{ccccccccccccccc}
	M_{1}  & \rightarrow & \bar{M}_{27} \coma &
	M_{2}  & \rightarrow & \bar{M}_{24} \coma &
	M_{3}  & \rightarrow & \bar{M}_{21} \coma &
	M_{4}  & \rightarrow & \bar{M}_{18} \coma &
	M_{5}  & \rightarrow & \bar{M}_{12} \coma \\
	M_{6}  & \rightarrow & \bar{M}_{6} \coma &
	M_{7}  & \rightarrow & \bar{M}_{26} \coma &
	M_{8}  & \rightarrow & \bar{M}_{23} \coma &
	M_{9}  & \rightarrow & \bar{M}_{20} \coma &
	M_{10} & \rightarrow & \bar{M}_{17} \coma \\
	M_{11} & \rightarrow & \bar{M}_{11} \coma &
	M_{12} & \rightarrow & \bar{M}_{5} \coma &
	M_{13} & \rightarrow & \bar{M}_{25} \coma &
	M_{14} & \rightarrow & \bar{M}_{22} \coma &
	M_{15} & \rightarrow & \bar{M}_{19} \coma \\
	M_{16} & \rightarrow & \bar{M}_{16} \coma &
	M_{17} & \rightarrow & \bar{M}_{10} \coma &
	M_{18} & \rightarrow & \bar{M}_{4} \coma &
	M_{19} & \rightarrow & \bar{M}_{15} \coma &
	M_{20} & \rightarrow & \bar{M}_{9} \coma \\ 
	M_{21} & \rightarrow & \bar{M}_{3}\coma &
	M_{22} & \rightarrow & \bar{M}_{14}\coma &
	M_{23} & \rightarrow & \bar{M}_{8}\coma &
	M_{24} & \rightarrow & \bar{M}_{2}\coma &
	M_{25} & \rightarrow & \bar{M}_{13}\coma \\
	\multicolumn{15}{c}{M_{26}   \rightarrow  \bar{M}_{7}\coma 
	M_{27}  \rightarrow  \bar{M}_{1}\fstop}
	\end{array}   
	\label{eq:Q111Z2D-Meson-invol}
    \end{equation}
    
The orientifolded theory has gauge group $\U(N)_1\times \U(N)_2\times \prod_{i=3}^6G_i(N)$. The involution of fields connecting nodes 3, 4, 5 and 6 gives rise to the constraint
\begin{equation}
\gamma_{\Omega_3}=\gamma_{\Omega_4}=\gamma_{\Omega_5}=\gamma_{\Omega_6} \fstop 
\label{gamma_matrices_Q111/Z2}
\end{equation}
Let us set the four matrices equal to $\ID_N$, i.e. project the corresponding gauge groups to $\SO(N)$. Figure~\ref{fig:orien_Q111Z2quiverD} shows the quiver for the resulting theory, which is free of gauge anomalies.

	\begin{figure}[H]
		\centering

	\caption{Quiver  diagram  for  the  Spin(7)  orientifold  of  phase  D  of $Q^{1,1,1}/\ZZ_2$ using  the involution in \cref{eq:Q111Z2D-chiral-invol,eq:Q111Z2D-Fermi-invol,eq:Q111Z2D-RFermi-invol}, together with our choice of $\gamma_{\Omega_i}$ matrices.}
		\label{fig:orien_Q111Z2quiverD}
	\end{figure}

Let us pause for a moment to think about a possible interpretation on this theory. We note that it has two distinct types of nodes. First, we have $\U(N)$ nodes with adjoint Fermi fields, which can be combined into $\mathcal{N}=(0,2)$ vector multiplets. Second, there are $\SO(N)$ nodes with symmetric Fermi fields which, contrary to the previous case, are inherently $\mathcal{N}=(0,1)$. This is because the adjoint of $\SO(N)$ is instead the antisymmetric representation. 
We can similarly consider whether it is possible to combine the bifundamental fields into $\mathcal{N}=(0,2)$ multiplets, which may or may not be broken by the superpotential. In this example, all bifundamental fields come in pairs so, leaving the superpotential aside, they can form $\mathcal{N}=(0,2)$ multiplets. Broadly speaking, we can therefore regard this theory as consisting of coupled $\mathcal{N}=(0,1)$ and $\mathcal{N}=(0,2)$ sectors.\footnote{A similar interpretation in terms of coupled $\mathcal{N}=(0,1)$ and $\mathcal{N}=(0,2)$ sectors was proposed in the analysis of non-compact models in \cite{Gukov:2019lzi}.} This discussion extends to the other orientifolds of $Q^{1,1,1}/\mathbb{Z}_2$ considered in this section and is a generic phenomenon. Interestingly, we will see below that Spin(7) orientifolds produce theories in which triality acts on either of these two sectors.

\begin{figure}[H]
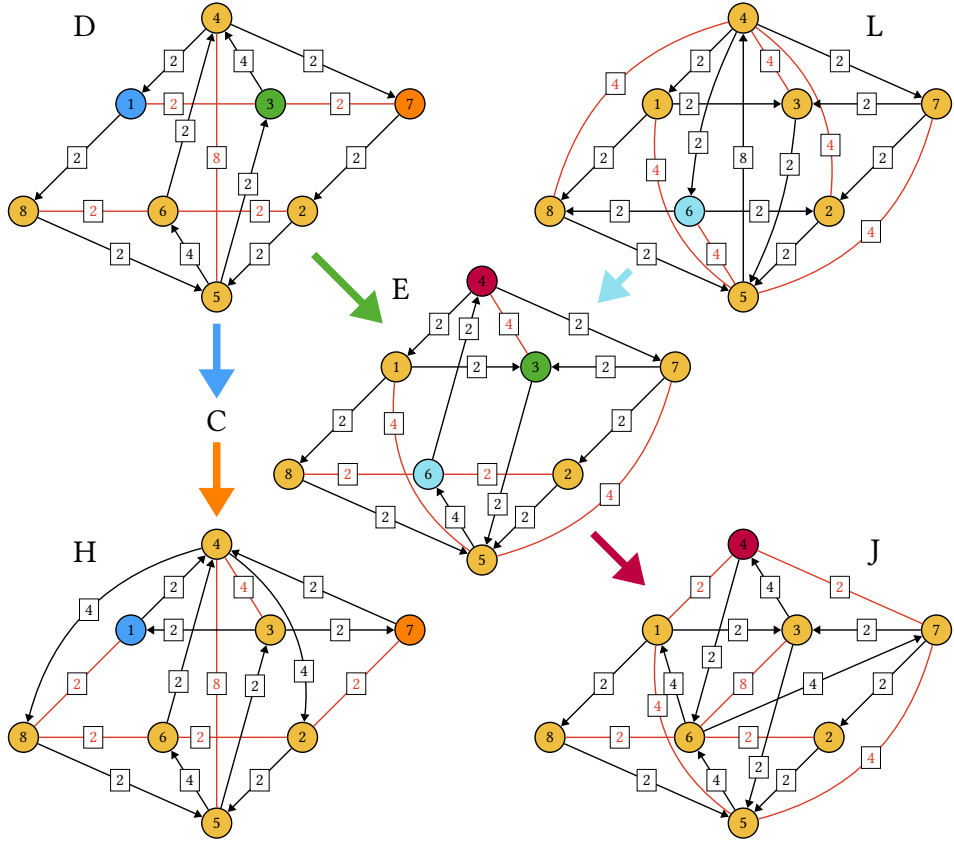

    \centering

	}
	   };
	   \node (Q111Z2C) at (-\L,0) {C};
	   \node at (-\L*1.5,\L*1.5) {D};
	   \node at (-\L*0.3,\L*0.5) {E};
	   \node at (-\L*1.5,-\L*0.5) {H};
	   \node at (\L*1.5,-\L*0.5) {J};
	   \node at (\L*1.5,\L*1.5) {L};
	   \draw[line width=1mm,blueX,-Triangle] (Q111Z2D) -- 
	   (Q111Z2C);
	   \draw[line width=1mm,orange,-Triangle] (Q111Z2C) -- 
	   (Q111Z2H);
	   \draw[line width=1mm,greenX,Triangle-] (Q111Z2D) -- 
	   (Q111Z2E);
	   \draw[line width=1mm,cyanX,-Triangle] (1,1) -- 
	   (0.78,0.78);
	   \draw[line width=1mm,purple,-Triangle] (0.75,-0.75) -- 
	   (1.1,-1.1);
	    \end{tikzpicture}
	    \caption{$\mathcal{N}=(0,2)$ triality web for phases D, E, H, J and L of $Q^{1,1,1}/\ZZ_2$.}
	    \label{fig:02trialityq111z2}
\end{figure}

In order to find other $\mathcal{N}=(0,1)$ theories associated with the same $\Spin(7)$ orientifold, one needs to find the field-theoretic involutions of other toric phases of $Q^{1,1,1}/\mathbb{Z}_2$ leading to $\U(N)^2\times \SO(N)^4$ gauge theories, whose geometric involution is the same as \eqref{eq:Q111Z2D-Meson-invol}. Scanning the 14 toric phases of $Q^{1,1,1}/\mathbb{Z}_2$, we found that only 5 of them (including phase D) admit $\mathcal{N}=(0,1)$ orientifolds with $\U(N)^2\times \SO(N)^4$ gauge symmetry. Let us first present the $\mathcal{N}=(0,2)$ triality web for these 5 phases in Figure~\ref{fig:02trialityq111z2}, which can be regarded as a portion of the whole triality web for $Q^{1,1,1}/\mathbb{Z}_2$ in \cite{Franco:2018qsc}.

Colored arrows connecting different phases indicate $\mathcal{N}=(0,2)$ triality transformations between them. Furthermore, the quiver node on which triality acts is shown in the same color as the corresponding arrow. Note that from phase D to phase H there are two triality steps, where the intermediate stage is the so-called phase C in \cite{Franco:2018qsc}. However, since phase C does not give rise to a $\U(N)^2\times \SO(N)^4$ orientifold, we do not show its quiver here.

Similarly to phase D, we consider the anti-holomorphic involutions of phases E, H, J and L which act on their quivers shown in Figure~\ref{fig:02trialityq111z2} as reflections with respect to the vertical plane that contains nodes 3, 4, 5 and 6. Then, the nodes on the plane map to themselves, while the pairs $1\leftrightarrow 7$ and $2\leftrightarrow 8$ get identified. In all these cases, we choose the $\gamma_{\Omega_i}$ matrices as for phase D, so they have $\U(N)^2\times \SO(N)^4$ gauge group. The construction of the $\mathcal{N}=(0,1)$ theories associated with the $\Spin(7)$ orientifold for these phases is detailed in Appendix~\ref{app:Q111Z2-details}. The crucial point is that they all correspond to the same $\Spin(7)$ orientifold of $Q^{1,1,1}/\mathbb{Z}_2$, since they are all associated to the same geometric involution as that of phase D, given in \eqref{eq:Q111Z2D-Meson-invol}.

\begin{figure}[ht]
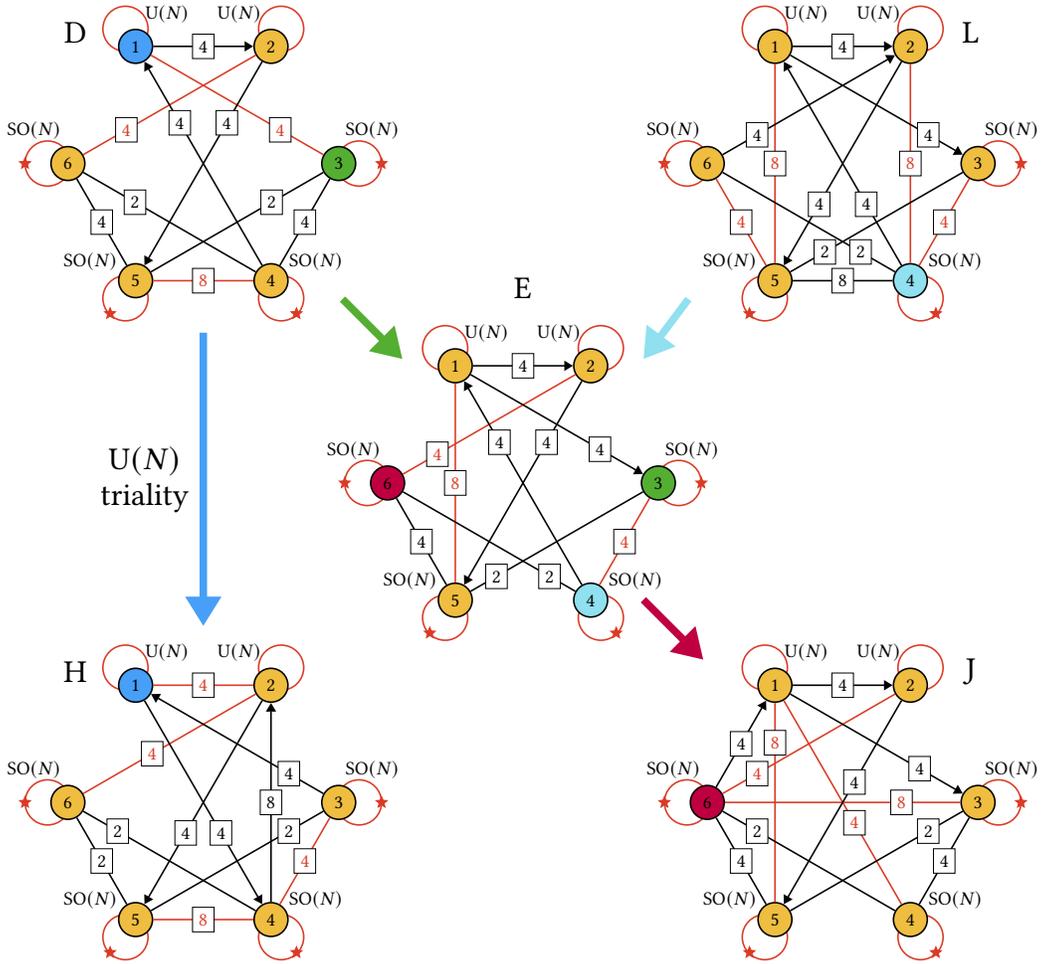

    \centering

	}
	   };
	   \node at (-\L*1.4,\L*1.4) {D};
	   \node at (-\L*0,\L*0.6) {E};
	   \node at (-\L*1.4,-\L*0.6) {H};
	   \node at (\L*1.4,-\L*0.6) {J};
	   \node at (\L*1.4,\L*1.4) {L};
	   \draw[line width=1mm,blueX,-Triangle] (Q111Z2D) -- node[midway,left,black] {\shortstack{$\U(N)$\\triality}}
	   (Q111Z2H);
	   (Q111Z2H);
	   \draw[line width=1mm,greenX,-Triangle] (-1.2,1.2) -- 
	   (-0.8,0.8);
	   \draw[line width=1mm,cyanX,-Triangle] (1.1,1.2) -- 
	   (0.8,0.8);
	   \draw[line width=1mm,purple,-Triangle] (0.8,-0.8) -- 
	   (1.2,-1.2);
	    \end{tikzpicture}
	    \caption{Triality web for the $\mathcal{N}=(0,1)$ theories associated with the Spin(7) orientifolds under consideration for phases D, E, H, J and L of $Q^{1,1,1}/\ZZ_2$.}
	    \label{fig:01trialityq111z2}
\end{figure}

From a field theory perspective, we find that the orientifolds of phases D, E, J and L are connected by $\mathcal{N}=(0,1)$ triality transformation on various $\SO(N)$ gauge groups (with the obvious generalization to more general flavor groups). These are the first examples of $\mathcal{N}=(0,1)$ triality in the presence of $\U(N)$ gauge groups. Interestingly, the orientifolds of phases D and H are not connected by the usual $\mathcal{N}=(0,1)$ triality on an $\SO(N)$ node, but by triality on node 1, which is of $\U(N)$ type. This transformation locally follows the rules of $\mathcal{N}=(0,2)$ triality. Such $\U(N)$ triality in $\mathcal{N}=(0,1)$ gauge theories is a new phenomenon which, to the best of our knowledge, has not appeared in the literature before. Following our earlier discussion, it can be nicely interpreted as $\mathcal{N}=(0,2)$ triality in the presence of an $\mathcal{N}=(0,1)$ sector. In our $\Spin(7)$ orientifold construction, the $\U(N)$ triality has a clear origin: the two $\mathcal{N}=(0,2)$ trialities that connect phases D and H passing through phase C are projected onto a single $\U(N)$ triality connecting the orientifolds of phases D and H. In the case of nodes that are not mapped to themselves, an even number of trialities in the parent is necessary in order to get a new phase that is also symmetric under the involution. Figure~\ref{fig:01trialityq111z2} summarizes the web of trialities for the Spin(7) orientifolds under consideration.

\section{Conclusions}
\label{sec:conclusions}

D1-branes probing singularities provide a powerful framework for engineering 2d gauge theories. In Chapter \ref{cha:ch2} adapted from \cite{Franco:2021ixh}, these constructions were extended to $\mathcal{N}=(0,1)$ theories with the introduction of Spin(7) orientifolds.

In this chapter we introduced a new, geometric, perspective on the triality of 2d $\mathcal{N}=(0,1)$ gauge theories, by showing that it arises from the non-uniqueness of the correspondence between Spin(7) orientifolds and the gauge theories on D1-brane probes. 

Let us reflect on how 2d trialities with different SUSY amounts manifest in D1-branes at singularities. $\mathcal{N}=(0,2)$ triality similarly arises from the fact that multiple gauge theories can be associated to the same underlying CY$_4$ \cite{Franco:2016nwv}. We explained that Spin(7) orientifolds based on the universal involution give rise to exactly the $\mathcal{N}=(0,1)$ triality of \cite{Gukov:2019lzi}. But our work shows that the space of possibilities is far richer. Indeed, general Spin(7) orientifolds extend triality to theories that can be regarded as consisting of coupled $\mathcal{N}=(0,2)$ and $(0,1)$ sectors. The geometric construction of these theories therefore leads to extensions of triality that interpolate between the pure $\mathcal{N}=(0,2)$ and $(0,1)$ cases.

On the practical side, Spin(7) orientifolds also give a precise prescription for how scalar flavors transform under triality in general quivers, which is inherited from the transformation of the corresponding chiral flavors in the parent.

%% file: chapters/4.tex
\section{Introduction}

Dualities sit at the heart of some of the deepest insights into the non-perturbative dynamics of
quantum fields and strings. In the case of quantum field theories (QFTs) engineered via string theory,
these dualities can often be recast in terms of specific geometric transformations of the extra-dimensional
geometry. A particularly notable example of this sort is the famous $SL(2,\mathbb{Z})$ duality symmetry of
type IIB string theory which descends to a duality action on the QFTs realized on the worldvolume of
probe D3-branes.

Recently it has been appreciated that symmetries themselves can be generalized in a number of different
ways. In particular, in \cite{Gaiotto:2014kfa} it was argued that symmetries can be understood in
terms of corresponding topological operators (see also \cite{Gaiotto:2010be,Kapustin:2013qsa,Kapustin:2013uxa,Aharony:2013hda}).\footnote{For a partial list 
of recent work in this direction see e.g.,
\cite{Gaiotto:2014kfa,Gaiotto:2010be,Kapustin:2013qsa,Kapustin:2013uxa,Aharony:2013hda,
DelZotto:2015isa,Sharpe:2015mja, Heckman:2017uxe, Tachikawa:2017gyf,
Cordova:2018cvg,Benini:2018reh,Hsin:2018vcg,Wan:2018bns,
Thorngren:2019iar,GarciaEtxebarria:2019caf,Eckhard:2019jgg,Wan:2019soo,Bergman:2020ifi,Morrison:2020ool,
Albertini:2020mdx,Hsin:2020nts,Bah:2020uev,DelZotto:2020esg,Hason:2020yqf,Bhardwaj:2020phs,
Apruzzi:2020zot,Cordova:2020tij,Thorngren:2020aph,DelZotto:2020sop,BenettiGenolini:2020doj,
Yu:2020twi,Bhardwaj:2020ymp,DeWolfe:2020uzb,Gukov:2020btk,Iqbal:2020lrt,Hidaka:2020izy,
Brennan:2020ehu,Komargodski:2020mxz,Closset:2020afy,Thorngren:2020yht,Closset:2020scj,
Bhardwaj:2021pfz,Nguyen:2021naa,Heidenreich:2021xpr,Apruzzi:2021phx,Apruzzi:2021vcu,
Hosseini:2021ged,Cvetic:2021sxm,Buican:2021xhs,Bhardwaj:2021zrt,Iqbal:2021rkn,Braun:2021sex,
Cvetic:2021maf,Closset:2021lhd,Thorngren:2021yso,Sharpe:2021srf,Bhardwaj:2021wif,Hidaka:2021mml,
Lee:2021obi,Lee:2021crt,Hidaka:2021kkf,Koide:2021zxj,Apruzzi:2021mlh,Kaidi:2021xfk,Choi:2021kmx,
Bah:2021brs,Gukov:2021swm,Closset:2021lwy,Yu:2021zmu,Apruzzi:2021nmk,Beratto:2021xmn,Bhardwaj:2021mzl,
Debray:2021vob, Wang:2021vki,
Cvetic:2022uuu,DelZotto:2022fnw,Cvetic:2022imb,DelZotto:2022joo,DelZotto:2022ras,Bhardwaj:2022yxj,Hayashi:2022fkw,
Kaidi:2022uux,Roumpedakis:2022aik,Choi:2022jqy,
Choi:2022zal,Arias-Tamargo:2022nlf,Cordova:2022ieu, Bhardwaj:2022dyt,
Benedetti:2022zbb, Bhardwaj:2022scy,Antinucci:2022eat,Carta:2022spy,
Apruzzi:2022dlm, Heckman:2022suy, Baume:2022cot, Choi:2022rfe,
Bhardwaj:2022lsg, Lin:2022xod, Bartsch:2022mpm, Apruzzi:2022rei,
GarciaEtxebarria:2022vzq, Cherman:2022eml, Heckman:2022muc, Lu:2022ver, Niro:2022ctq, Kaidi:2022cpf,
Mekareeya:2022spm, vanBeest:2022fss, Antinucci:2022vyk, Giaccari:2022xgs, Bashmakov:2022uek,Cordova:2022fhg,
GarciaEtxebarria:2022jky, Choi:2022fgx, Robbins:2022wlr, Bhardwaj:2022kot, Bhardwaj:2022maz, Bartsch:2022ytj, Gaiotto:2020iye,Agrawal:2015dbf, Robbins:2021ibx, Robbins:2021xce,Huang:2021zvu,
Inamura:2021szw, Cherman:2021nox,Sharpe:2022ene,Bashmakov:2022jtl, Inamura:2022lun, Damia:2022bcd, Lin:2022dhv,Burbano:2021loy, Damia:2022rxw} and \cite{Cordova:2022ruw} for a recent review.}

The fact that generalized symmetry operators are topological also suggests that the ``worldvolume'' itself may support a 
non-trivial topological field theory. One striking consequence of this fact is that the product of 
two generalized symmetry operators may produce a sum of symmetry operators, i.e., there can be a non-trivial 
fusion category (i.e., multiple summands in the product), and this is closely tied to the appearance of ``non-invertible'' symmetry operators (see e.g., \cite{Thorngren:2019iar,Komargodski:2020mxz, Gaiotto:2020iye, Nguyen:2021naa, Heidenreich:2021xpr, Thorngren:2021yso, Agrawal:2015dbf, Robbins:2021ibx, Robbins:2021xce, Sharpe:2021srf, Koide:2021zxj, Huang:2021zvu,
Inamura:2021szw, Cherman:2021nox, Kaidi:2021xfk, Choi:2021kmx, Wang:2021vki, Bhardwaj:2022yxj,
Hayashi:2022fkw, Sharpe:2022ene, Choi:2022zal, Kaidi:2022uux, Choi:2022jqy, Cordova:2022ieu,
Bashmakov:2022jtl, Inamura:2022lun, Damia:2022bcd, Choi:2022rfe, Lin:2022dhv, Bartsch:2022mpm,
Lin:2022xod, Cherman:2022eml, Burbano:2021loy, Damia:2022rxw, Apruzzi:2022rei, GarciaEtxebarria:2022vzq, Heckman:2022muc, 
Niro:2022ctq, Kaidi:2022cpf, Mekareeya:2022spm, Antinucci:2022vyk, Giaccari:2022xgs, Bashmakov:2022uek,Cordova:2022fhg,
GarciaEtxebarria:2022jky, Choi:2022fgx, Bhardwaj:2022kot, Bhardwaj:2022maz, Bartsch:2022ytj}).

Now, one of the notable places where non-invertible symmetries have been observed is in the context of certain ``duality / triality defects''.\footnote{A word on terminology: in this work we use the stringy notion of a non-abelian $SL(2,\mathbb{Z})$ duality. In particular, we will be interested in operations which generate subgroups of $SL(2,\mathbb{Z})$ with order different than two. In what follows we shall sometimes 
refer to all of these as dualities even if the order is different than two. That being said, in our field theory examples we will explain when we are dealing with a specific duality / triality defect.}
At generic points of parameter space, a duality interchanges one description of a field theory with another. However, at special points in the parameter space (such as the critical point of the 2D Ising model), this duality operation simply sends one back to the same theory. In that case, one has a 0-form symmetry, and therefore one expects a codimension one generalized symmetry operator \cite{Kaidi:2021xfk, Choi:2021kmx, Choi:2022zal, Kaidi:2022uux, Kaidi:2022cpf}. 
An intriguing feature of these symmetry operators is that they can sometimes have a non-trivial fusion rule, indicating the presence of a non-invertible symmetry. In many cases, one can argue for the existence of such a non-invertible symmetry, even without knowing the full structure of the TFT localized on a duality defect.

Given the fact that many field theoretic dualities have elegant geometric characterizations, it is natural to ask whether these topological duality defects can be directly realized in terms of objects in string theory. In particular, one might hope that performing this analysis could provide additional insight into the associated worldvolume TFTs, and provide a systematic method for extracting the corresponding fusion rules for these generalized symmetry operators. A related point is that in many QFTs of interest, a weakly coupled Lagrangian description may be unavailable and so one must seek out alternative (often geometric) characterizations of these systems.

Along these lines, it was recently shown in \cite{Apruzzi:2022rei, GarciaEtxebarria:2022vzq,Heckman:2022muc} that for QFTs engineered via localized singularities / branes, generalized symmetry operators are obtained from ``branes at infinity''. The resulting defects are topological in the sense that they do not contribute to the stress energy tensor of the localized QFT. Starting from the topological terms of a brane ``at infinity'', one can then extract the resulting TFT on its worldvolume, and consequently, extract the resulting fusion rules for the associated generalized symmetry operators.

Our aim in this chapter will be to use this perspective to propose a general prescription for duality defects where the duality of the QFT is inherited from the $SL(2,\mathbb{Z})$ duality of type IIB strings. In particular, we focus on the case of 4D QFTs realized from D3-branes probing a localized singularity of a non-compact Calabi-Yau threefold $X$ which we assume has a conical topology, namely it can be written as a cone over $\partial X$: $\mathrm{Cone}(\partial X) = X$. Such QFTs have a marginal parameter $\tau$ descending from the axio-dilaton of type IIB string theory, and 2-form potentials of a bulk 5D TFT descending from the $SL(2,\mathbb{Z})$ doublet of 2-form potentials (RR and NS-NS) which governs the 1-form electric and magnetic symmetries of the 4D probe theory.

In this setting, the ``branes at infinity'' which implement a duality transformation are simply given by specific bound states of $(p,q)$ 7-branes. 
In a general IIB / F-theory background, a bound state of $(p,q)$ 7-branes acts on the axio-dilaton and $SL(2,\mathbb{Z})$ doublet of 2-form 
potentials $\mathcal{B}^j = (C_2 , B_2)$ as:
\begin{equation}
\tau \mapsto \frac{a \tau + b}{c \tau + d} \,\,\, \text{and} 
\left[
\begin{array}
[c]{c}%
C_{2}\\
B_{2}%
\end{array}
\right]  \mapsto\left[
\begin{array}
[c]{cc}%
a & b\\
c & d
\end{array}
\right]  \left[
\begin{array}
[c]{c}%
C_{2}\\
B_{2}%
\end{array}
\right] 
\end{equation}
in the obvious notation. At the level of topology this monodromy can be localized to a branch cut whose endpoints are physical, namely the locus of a bound state of 7-branes. 

Of particular significance are the specific monodromy transformations which leave fixed particular values of $\tau$. Geometrically, these are specified by constant axio-dilaton profiles for 7-branes, which are in turn given by specific Kodaira fibers which specify how the elliptic fiber of F-theory degenerates on the locus of the 7-brane. The full list is $II, III, IV, I_{0}^{\ast}, IV^{\ast}, III^{\ast}, II^{\ast}$, which respectively support the gauge algebras $\mathfrak{su}_1, \mathfrak{su}_{2}, \mathfrak{su_3}, \mathfrak{so}_8, \allowbreak \mathfrak{e}_6, \mathfrak{e}_7, \mathfrak{e}_8$. Putting all of this together, it is natural to expect that 
the duality defects of the QFT simply lift to appropriate 7-branes 
wrapped on all of $\partial X$.

Our main claim is that wrapping 7-branes on a ``cycle at infinity'' leads to topological duality / triality defects in the 
4D worldvolume theory of the probe D3-brane. One way to see this is to consider the dimensional reduction on the boundary five-manifold $\partial X$. This results in the 5D symmetry TFT of the 4D field theory (see \cite{Apruzzi:2021nmk} as well as \cite{Aharony:1998qu, Heckman:2017uxe}). In this 5D theory, 7-branes wrapped on $\partial X$ specify codimension two defects which fill out a three-manifold in the 4D spacetime. In this 5D TFT limit where all metric data has been decoupled, the reduction of the 7-brane ``at infinity'' can be pushed into the interior, and can equivalently be viewed as specifying a codimension two defect in the bulk. In particular, as codimension two objects, they come with a branch cut structure, and this in turn impacts the structure of anomalies both in the 5D bulk as well as the 3D TFT localized on the topological defect.

In particular, we find that the choice of where to terminate the other end of the branch cut emanating from the 7-branes has a non-trivial impact on the resulting structure of the TFT. For each choice of branch cut, we get a corresponding anomaly inflow to the 7-brane defect. Doing so, we show that one choice of a branch cut gives the constructions of \cite{Kaidi:2021xfk, Kaidi:2022cpf, Antinucci:2022vyk} for Kramers-Wannier-like duality defects, while another choice produces the half-space gauging construction of \cite{Choi:2021kmx, Choi:2022zal}.
One can also entertain ``hybrid'' configurations of branch cuts, and these also produce duality / triality defects. In an Appendix we also show how these considerations are compatible with dimensional reduction of topological terms present in the 8D worldvolume of the 7-branes. We emphasize that while these analyses also make use of the 5D symmetry TFT, our analysis singles out the role of codimension two objects (and their associated branch cuts) which descend from wrapped 7-branes. Indeed, this top down perspective allows us to unify different construction techniques.

In the field theory literature, the main examples of duality / triality defects have centered on $\mathcal{N} = 4$ SYM theory and closely related examples. In the present context where this QFT arises from D3-branes probing $\mathbb{C}^3$, we see that the main ingredients for duality / triality defects readily generalize to $\mathcal{N} = 1$ SCFTs as obtained from D3-branes probing $X$ a Calabi-Yau cone with a singularity. In that setting, the IIB duality group corresponds to a duality action which is present at a specific (tuned) subspace of the conformal manifold of the SCFT. In particular, dimensional reduction of 7-branes on $\partial X$ leads to precisely the same topological defects, and thus provides us with a generalization to QFTs with less supersymmetry. On the other hand, the full 5D symmetry TFT will in this case be more involved simply because the topology $\partial X$ can in general support more kinds of objects. For example, other 0-form symmetries are present in such systems, and crossing the associated local defects charged under these discrete 0-form symmetries  through a duality / triality wall leads to non-trivial transformation rules.

The rest of this chapter is organized as follows. In section \ref{sec:SETUP} we present our general setup involving probe
D3-branes in a Calabi-Yau threefold. In particular, we show how boundary conditions ``at infinity'' specify the global form of the theory, and how duality / triality defects arise from 7-branes wrapped on the boundary geometry. In section \ref{sec:DefectTFT} we consider the 5D symmetry TFT obtained from dimensional reduction on the boundary $\partial X$. The 7-branes descend to codimension two objects with branch cuts, and the choice of how to terminate these branch cuts leads to different implementations of duality / triality defects. After this, in section \ref{sec:N4} we show that our top down considerations are compatible with the bottom up analyses in the field theory literature. Section \ref{sec:N=1} shows how these considerations generalize to systems with minimal supersymmetry. We present our conclusions and some directions for future work in section \ref{sec:CONC}. In Appendix \ref{app:other} we show how the various defects considered in the main body are implemented in other top down constructions. In Appendix \ref{app:minimalTFT7branes} we give a proposal for the relevant topological terms of a non-perturbative 7-brane which reduce to a suitable 3D TFT (after reduction on $\partial X$). Finally, in Appendix \ref{app:orbo} we give some further details on the special case of D3-branes probing $\mathbb{C}^{3} / \mathbb{Z}_3$.

\section{General Setup}\label{sec:SETUP}

We now present the general setup for implementing duality / triality interfaces and defects in the context of brane probes of singularities. The construction we present produces supersymmetric 4D quantum field theories $\mathfrak{T}^{(N)}_X$ realized as the world-volume theory of a stack of $N$ D3-branes probing a non-compact Calabi-Yau threefold $X$. 


The Calabi-Yau threefolds $X$ we are considering are of conical topology
\be 
X=\mathrm{Cone } \left( \partial X \right)
\ee 
with link $\partial X$, the asymptotic boundary of $X$. The topology of $\partial X$ therefore determines the topology of $X$ fully. The apex of the cone supports a real codimension six singularity. We introduce the radial coordinate $r\in \mathbb{R}_{\geq 0}$ so that the singularity sits at $r=0$ and the asymptotic boundary sits at $r=\infty$.

For example, $X=\mathbb{C}^3$ determines $\mathfrak{T}^{(N)}_X$ to be 4D $\mathcal{N}=4$ supersymmetric Yang-Mills theory. In cases of reduced holonomy, for example $X=\mathbb{C}^3/\Gamma$ with $\Gamma\subset SU(3)$, we preserve $\mathcal{N}=1$ supersymmetry. In all cases, $\mathfrak{T}^{(N)}_X$ is some quiver gauge theory, with quiver nodes specified by a basis of ``fractional branes'' which can be visualized at large volume (i.e., away from the orbifold point of moduli space) as a collection of D3-, D5- and D7-branes and their anti-brane counterparts wrapped on cycles in a resolution of $X$ \cite{Klebanov:1998hh, Uranga:1998vf,Aharony:1997ju}. Nodes are connected by oriented arrows which should be viewed as open strings stretching between the fractional branes. The gauge theory characterization is especially helpful at weak coupling, and serves to define the QFT in the first place. The quiver gauge theory comes with a collection of marginal couplings and we can consider tuning these parameters to ``strong coupling''. At such points in the conformal manifold, the gauge theory description is less useful, but we can still speak of the SCFT defined by the probe D3-branes.

In the quiver gauge theory, the IIB axio-dilaton descends to a particular choice of marginal couplings. Moreover, the celebrated $SL(2,\mathbb{Z})$ duality of IIB strings\footnote{The precise form of the duality group and its actions on fermions leads to some additional subtleties. For example, taking into account fermions, there is the metaplectic cover of $SL(2,\mathbb{Z})$ \cite{Pantev:2009de}, and taking into account reflections on the F-theory torus (associated with worldsheet orientation reversal and $(-1)^{F_L}$ parity, this enhances to the $\mathsf{Pin}^{+}$ cover of $GL(2,\mathbb{Z})$ \cite{Tachikawa:2018njr} (see also \cite{Debray:2021vob, Dierigl:2022reg}). These subtleties can appear if one carefully tracks the boson / fermion number of extended operators but in what follows we neglect this issue.} descends to a duality transformation at a specific point in the conformal manifold of the 4D SCFT \cite{Lawrence:1998ja, Kachru:1998ys} (for a recent discussion see, e.g., \cite{Garcia-Etxebarria:2016bpb}).

The other bulk supergravity fields of type IIB also play an important role in specifying the global structures of the field theory. Boundary conditions $P$ for such bulk fields at $\partial X$ determine an absolute theory $\mathfrak{T}^{(N)}_{X,P}$ from the relative theory $\mathfrak{T}^{(N)}_{X}$. In particular, such boundary conditions also determine the spectrum of extended objects ending at or contained within $\partial X$ which specify the defects and generalized symmetry operators of $\mathfrak{T}^{(N)}_{X,P}$ \cite{Gaiotto:2014kfa,GarciaEtxebarria:2019caf, Bhardwaj:2021mzl}.

For example, there is an $SL(2,\mathbb{Z})$ doublet $(C_2, B_2) = \mathcal{B}^{j}$ (RR and NS) of 2-form potentials which couple to D1- and F1-strings of the IIB theory respectively. Wrapping bound states of these objects compatible with $P$ along the radial direction in $X$ leads to heavy line defects of the 4D quiver gauge theory $\mathfrak{T}^{(N)}_{X,P}$. The spectrum of line defects then fixes the global form of the quiver gauge group.

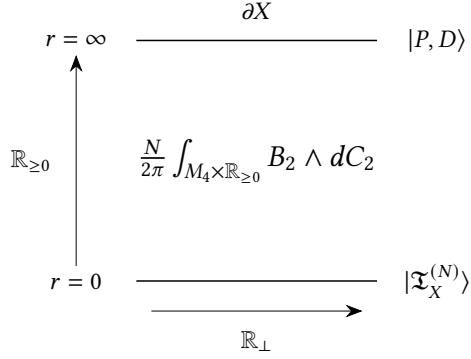
\begin{figure}[t]
    \centering
    \scalebox{0.8}{
    \begin{tikzpicture}
	\begin{pgfonlayer}{nodelayer}
		\node [style=none] (0) at (-2, 2) {};
		\node [style=none] (1) at (-2, -2) {};
		\node [style=none] (2) at (-3, 2) {$r=\infty$};
		\node [style=none] (3) at (-3, -2) {$r=0$};
		\node [style=none] (10) at (0, 2.5) {$\partial X$};
		\node [style=none] (11) at (-3, -1.75) {};
		\node [style=none] (12) at (-3, 1.75) {};
		\node [style=none] (13) at (-3.75, 0) {$\mathbb{R}_{\geq0}$};
		\node [style=none] (15) at (2, 2) {};
		\node [style=none] (16) at (2, -2) {};
		\node [style=none] (17) at (-1.75, -2.5) {};
		\node [style=none] (18) at (1.75, -2.5) {};
		\node [style=none] (19) at (0, -3) {$\mathbb{R}_\perp$};
		\node [style=none] (20) at (3, -2) {$\ket{\mathfrak{T}^{(N)}_X}$};
		\node [style=none] (21) at (3, 2) {$\ket{P,D}$};
        \node [style=none] (22) at (0, 0) {\large $\frac{N}{2\pi} \int_{M_4\times \mathbb{R}_{\geq 0}}B_2\wedge dC_2$};
	\end{pgfonlayer}
	\begin{pgfonlayer}{edgelayer}
		\draw [style=ArrowLineRight] (11.center) to (12.center);
		\draw [style=ThickLine] (0.center) to (15.center);
		\draw [style=ThickLine] (1.center) to (16.center);
		\draw [style=ArrowLineRight] (17.center) to (18.center);
	\end{pgfonlayer}
\end{tikzpicture}}
    \caption{Sketch of the symmetry TFT \eqref{eq:5d TFT terms1}. We depict the half-plane $\mathbb{R}_{\geq 0}\times\mathbb{R}_\perp$ with coordinates $(r,x_\perp)$ where $\mathbb{R}_\perp$ is some direction parallel to the D3-brane worldvolume. The boundary conditions for the symmetry TFT are denoted $\ket{\mathfrak{T}^{(N)}_X},\ket{P,D}$ respectively. }
    \label{fig:SetUp0}
\end{figure}

The possible boundary conditions $P$ are determined by the symmetry TFT \cite{Freed:2012bs,Freed:2022qnc} which follows by reduction of the 10D Chern-Simons term in IIB supergravity,\footnote{Strictly speaking, one also needs to utilize the self-dual condition for $F_5=dC_4$ in order to get the correct coefficient in \eqref{eq:5d TFT terms1}. 
For further discussion on this point, see e.g., \cite{Belov:2006jd, Belov:2006xj}.} much as in references \cite{Aharony:1998qu, Apruzzi:2021nmk} (see also \cite{Heckman:2017uxe}):
\begin{equation}
    S_{\mathrm{(CS)}}=-\frac{1}{4\kappa^2}\int_{M_4\times X}C_4\wedge dB_2\wedge dC_2.
\end{equation}
The stack of D3-branes source $N$ units of 5-form flux threading $\partial X$ and therefore the symmetry TFTs of all quiver gauge theories under consideration contain the universal term
\begin{equation}\label{eq:5d TFT terms1}
    S_{(\mathrm{SymTFT}),0}= \frac{N}{4\pi}\int_{M_4\times \mathbb{R}_{\geq 0}} \epsilon_{ij} \mathcal{B}^{i} \cup d \mathcal{B}^{j} \,,
\end{equation}
where we have integrated over the link $\partial X$. Here we have introduce a manifestly $SL(2,\mathbb{Z})$ invariant presentation of the action using the 
two-index tensor $\epsilon_{ij}$ to raise and lower doublet indices. In our conventions, $\epsilon_{21}=-\epsilon_{12}=1$. In terms of the individual components of this $SL(2,\mathbb{Z})$ doublet, 
the equations of motion for the action \eqref{eq:5d TFT terms1} are
\begin{equation}\label{eq:ZN}
    N dB_2 = N dC_2 = 0,
\end{equation}
which constrains $B_2$ and $C_2$ to be $\mathbb{Z}_N$-valued 1-form symmetry background fields.\footnote{Note that these steps are identical to the derivation of a bulk topological term in $AdS_5$ \cite{Witten:1998wy}, while here the term lives along $M_4\times \mathbb{R}_{\geq 0}$.} In general, we will denote $\mathbb{Z}_N$-valued fields using the same notation as their $U(1)$ counterparts but are related by a rescaling. For example, in conventions where the NS-NS flux $\frac{1}{2\pi}H_3$ is integrally quantized we have\footnote{Another natural choice would be to take $\int_{Q_3}H_3\in \mathbb
{Z}$ for all 3-manfolds $Q_3$ in which case we would drop the factor of $2\pi$ on the RHS of \eqref{eq:rescaling}.}
\begin{equation}\label{eq:rescaling}
    B^{U(1)}_2=\frac{2\pi}{N}B^{\mathbb{Z}_N}_2
\end{equation}
where the holonomies $\int_{\Sigma_2}B^{\mathbb{Z}_N}_2=k \; \mathrm{mod}\; N$ for some Riemann surface $\Sigma$ and integer $k$. Notice that since \eqref{eq:rescaling} is only valid when the holonomies of the $U(1)$ field are $N^{\mathrm{th}}$ roots of unity, for a $\mathbb{Z}_N$-valued field one is free to take either the LHS or RHS as normalizations. We will drop the superscripts in the future making clear which convention we are using for discrete fields when it arises. For additional details on the structure of the defect group in this theory (via related top down constructions) see Appendix \ref{app:other}. 

The relative theory $\mathfrak{T}^{(N)}_X$ sets enriched Neumann boundary conditions at $r=0$ while at $r=\infty$ we have mixed Neumann-Dirichlet boundary conditions for the fields of the symmetry TFT.  These are respectively denoted as
\begin{equation}
      \ket{\mathfrak{T}^{(N)}_X}\,, \quad \ket{P,D}\,, \qquad \quad \braket{ P,D\,|\,\mathfrak{T}^{(N)}_X}=Z_{\mathfrak{T}^{(N)}_{X,P}}(D)
\end{equation}
and contract to give the partition function of the absolute theory $\mathfrak{T}^{(N)}_{X,P}$ with background fields determined by $P$ set to the values $D$. Here $D$ is a form profile and in particular does not carry $SL(2,\mathbb{Z})$ indices (see figure \ref{fig:SetUp0}).

\begin{figure}[t]
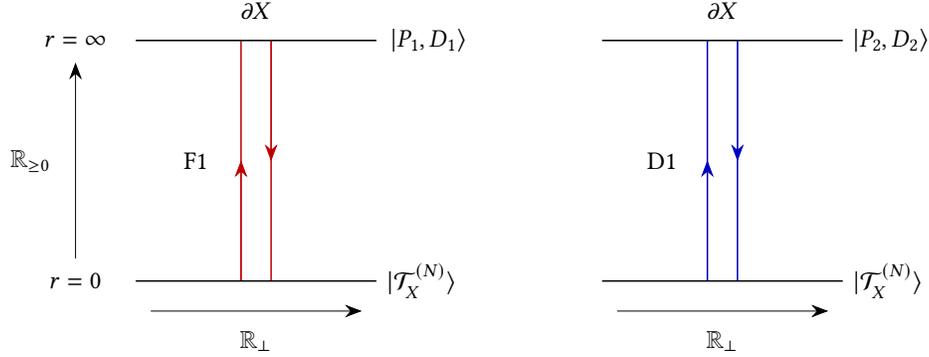

    \centering
    \scalebox{0.8}{

    }
    \caption{Boundary conditions and defects for $\mathfrak{T}_X^{(N)}$. We sketch the half-plane $\mathbb{R}_{\geq 0}\times \mathbb{R}_\perp$ parametrized by $(r,x_\perp)$. The polarization $P_1,P_2$ determine that Dirichlet boundary conditions are set for $B_2,C_2$ respectively $B_2|_{\partial X}=D_1$ and $C_2|_{\partial X}=D_2$. Line defects are realized by F1/D1-strings and correspond to Wilson and 't Hooft lines respectively. Our conventions are such that the left, radially outgoing strings are of charge $[0,-1]$ and $[-1,0]$ and the right, incoming strings are of charge $[0,1]$ and $[1,0]$ respectively.}
    \label{fig:F1D1}
\end{figure}

Consider for example $X=\mathbb{C}^3$ in which case \eqref{eq:5d TFT terms1} describes the full symmetry TFT. First note that \eqref{eq:ZN} makes it clear that we are discussing a theory with gauge algebra $\mathfrak{su}(N)$ rather than $\mathfrak{u}(N)$. This $U(1)$ factor is lifted via a Stueckelberg mechanism.\footnote{Intuitively, when we pick an origin for $\mathbb{C}^3$ we put the whole system in a ``box'' with a conformal boundary. This removes the center of mass degree of freedom for the system.} A standard set of boundary conditions include a purely electric or purely magnetic polarization via the boundary conditions:
\begin{align}\label{eq:electricbc}
 \textnormal{$B_2|_{\partial X}$ Dirichlet, $C_2|_{\partial X}$ Neumann} \quad  &\longleftrightarrow \quad \textnormal{Global electric 1-form symmetry}\\ \label{eq:magneticbc} 
 \textnormal{$B_2|_{\partial X}$ Neumann, $C_2|_{\partial X}$ Dirichlet} \quad  &\longleftrightarrow \quad \textnormal{Global magnetic 1-form symmetry}.
\end{align}
Concretely, we are considering $\mathcal{N} = 4$ SYM theory with gauge algebra $\mathfrak{su}(N)$. The electric polarization produces gauge group $SU(N)$ while the magnetic polarization produces gauge group $PSU(N) = SU(N) / \mathbb{Z}_N$. Given electric/magnetic boundary conditions we can stretch F1/D1 strings between the D3-branes and the asymptotic boundary to construct line defects in the 4D worldvolume theory (see figure \ref{fig:F1D1}).

One can also consider more general mixed boundary conditions. In general, $SL(2,\mathbb{Z}_N)$ duality transformations
\begin{equation}\label{eq:Mono}
~\begin{bmatrix} C_2 \\B_2 \end{bmatrix}\rightarrow \begin{bmatrix} a&b\\c&d \end{bmatrix}\begin{bmatrix} C_2 \\B_2 \end{bmatrix}, \qquad \mathbb{S}=\begin{bmatrix} 0&1\\-1&0 \end{bmatrix}\,, \quad \mathbb{T}=\begin{bmatrix} 1&1\\0&1 \end{bmatrix}
\end{equation}
where $ad-bc=1$, map between boundary conditions and group these into orbits. In the case $\mathcal{N}=4$ SYM with gauge algebra $\mathfrak{su}(N)$ and when $N$ has no square divisors,\footnote{See \cite{Bergman:2022otk} for details when dropping this assumption.} one can generate all possible mixed boundary conditions for $B_2$ and $C_2$, and thus all forms of the gauge group.

More generally, given a quiver gauge theory, the individual gauge group factors are all correlated due to bifundamental states which are charged under different centers of the gauge group. Indeed, note also that we can always move the D3-brane away from the singularity (i.e., to finite $r > 0$), and the center of this gauge group in the infrared will need to be compatible with the boundary conditions specified at $r = \infty$.

Let us also record our $SL(2,\mathbb{Z})$ conventions for strings and 5-branes here. The charge vector $Q$ of a $(p,q)$-string or $(p,q)$-5-brane has components, with $\epsilon_{12}=-\epsilon_{21}=-1$, following conventions laid out in \cite{Weigand:2018rez},
\begin{equation}
    Q_i=\epsilon_{ij}Q^j=[q,p]\,, \qquad Q^i=\begin{bmatrix}
        p \\ -q
    \end{bmatrix}\,,
\end{equation}
in particular S-duality maps Wilson lines $W$ and 't Hooft lines $H$ on the D3-brane worldvolume as $(W,H)\rightarrow (H,-W)$.

We further lay out our conventions for symmetry TFTs following \cite{Kaidi:2022uux,Kaidi:2022cpf}. The enriched Neumann boundary condition at $r=0$ is expanded as
\begin{equation}
    \ket{\mathfrak{T}_X^{(N)}}=\sum_{\mathbf{a}\in P} Z_{\mathfrak{T}_{X,P}^{(N)}}(a)\ket{\mathbf{a}}
\end{equation}
where $Z_{\mathfrak{T}_{P,X}^{(N)}}(a)$ is the partition function of the absolute theory derived from $\mathfrak{T}_X^{(N)}$ by choice of polarization $P$ and $a$ is a background field profile for the corresponding higher symmetry. In this chapter we are mainly concerned with 1-form symmetries of gauge theories and here $P$ fixes the global form of the gauge group and $a$ is a 1-form symmetry background field. Further we denote a background field configuration by a vector $\mathbf{a}$ which is oriented in the corresponding defect group and has a form profile of $a$. Topological Dirichlet and Neumann boundary conditions at $r=\infty$ in 4D are respectively
\begin{equation}
\begin{aligned}    \ket{P,D}_{\mathrm{Dirichlet}}&=\sum_{\mathbf{a}\in P}\delta(D-a)\ket{\mathbf{a}} \\  \ket{P,E}_{\mathrm{Neumann}}&=\sum_{\mathbf{a}\in P}\exp\left(\frac{2\pi i}{N} \int E\cup a\right)\ket{\mathbf{a}}
    \end{aligned}
\end{equation}
where we have normalized fields to take values in $\mathbb{Z}_N$. We will mainly work with Dirichlet boundary conditions throughout and omit the index `Dirichlet' when it causes no confusion. Whenever two polarizations $P,P'$ are related by a discrete Fourier transform or equivalently by gauging we have the pairing
\begin{equation}
    \braket{\mathbf{a}|\mathbf{b}}=\exp\left(\frac{2\pi i}{N}\int a \cup b \right)\qquad \forall\, \mathbf{a}\in P,\mathbf{b}\in P'\,.
\end{equation}
More generally, boundary conditions in 4D can be stacked with counterterms, so we define 
\begin{equation}
\begin{aligned}    \ket{P_{G_r},D}_{\mathrm{Dirichlet}}&=\sum_{\mathbf{a}\in P_{G_k}}\delta(D-a)\exp\left(\frac{2\pi ir}{N}\int \frac{\mathcal{P}(D)}{2}\right)\ket{\mathbf{a}} 
    \end{aligned}
\end{equation}
where $\mathcal{P}$ is the Pontryagin square. Here we have labelled a polarization $P$ by the global form of the gauge group $G$ it realizes, and the subscript $r$ counts the number of stacked counterterms. For example $SU(2)_r$ denotes $SU(2)$ theory stacked with $r$ couterterms.

\subsection{Proposal for Topological Duality Interfaces/Operators}
\label{sec:Proposal}

{\renewcommand{\arraystretch}{1.35}
\begin{table}[t]
    \centering
    \begin{center}
\begin{tabular}{||c | c | c | c | c | c ||}
 \hline
 Fiber Type $\mathfrak{F}$ & Lines & Monodromy $ \rho $ & Refined Linking $p/2k$ & $(k,m)$ & $\tau$ \\ [0.5ex] 
 \hline\hline
 $II,\,\mathfrak{su}(1)$ & $-$ & {\footnotesize $\left(\begin{array}{cc}
     0 &  1 \\
     -1  & 1
 \end{array}\right)$ } & $-$ & $-$ & $e^{i\pi/3}$\\ 
 \hline
 $III,\,\mathfrak{su}(2)$ & $\mathbb{Z}_2$ & {\footnotesize $\left(\begin{array}{cc}
     0 &  1 \\
     -1  & 0
 \end{array}\right)$ } &  $\frac{3}{4}$ & $(2,3)$ & $e^{i\pi/2}$ \\
 \hline
 $IV,\,\mathfrak{su}(3)$ & $\mathbb{Z}_3$ & {\footnotesize $\left(\begin{array}{cc}
     0 &  1 \\
     -1  & -1
 \end{array}\right)$ } &    $\frac{4}{6}$ & $(3,4)$ &$e^{i\pi/3}$ \\
 \hline
 $I_{0}^{\ast},\,\mathfrak{so}(8)$ & $\mathbb{Z}_2\oplus \mathbb{Z}_2$ &  {\footnotesize $\left(\begin{array}{cc}
     -1 &  0 \\
     0  & -1
 \end{array}\right)$ } & {\footnotesize $\left(\begin{array}{cc}
     2/4 &  3/4 \\
     3/4  & 2/4
 \end{array}\right)$ } & $-$  & $\tau$ \\
 \hline
 $IV^{\ast},\,\mathfrak{e}_6$ & $\mathbb{Z}_3$ & {\footnotesize $\left(\begin{array}{cc}
     -1 &  -1 \\
     1  & 0
 \end{array}\right)$ } &  $\frac{2}{6}$ & $(3,2)$ &$e^{i\pi/3}$ \\
 \hline
 $III^{\ast},\,\mathfrak{e}_7$ & $\mathbb{Z}_2$ & {\footnotesize $\left(\begin{array}{cc}
     0 &  -1 \\
     1  & 0
 \end{array}\right)$ } &   $\frac{1}{4}$ & $(2,1)$ &$e^{i\pi/2}$ \\
 \hline
 $II^{\ast},\,\mathfrak{e}_8$ & $-$ &  {\footnotesize $\left(\begin{array}{cc}
     1 &  -1 \\
     1  & 0
 \end{array}\right)$ } & $-$ & $-$ & $e^{i\pi/3}$ \\
  \hline
\end{tabular}
\end{center}
    \caption{Elliptic data of 7-brane profiles with constant axio-dilaton $\tau$. Their group of lines is isomorphic to $\textnormal{coker}(\rho-1)$ which is isomorphic to $\mathbb{Z}_k$ except for fiber type $I_0^*,II,II^*$. The label $m$ of these lines is determined from the refined self-linking numbers $m/2k$ which gives the spin of non-trivial lines. The refined self-linking numbers compute via the Gordon-Litherland approach laid out in \cite{Apruzzi:2021nmk} employing the divisors (Kodaira thimbles) computed in \cite{Hubner:2022kxr} or alternatively via the quadratic refinement laid out in \cite{Gukov:2020btk} and the linking number computations in \cite{Cvetic:2021sxm}. Note in particular that in all cases $\textnormal{gcd}(k,m)=1$ and $mk\in 2\mathbb{Z}$.}
    \label{tab:Fibs}
\end{table}}

Consider the spacetime $M_4=M_3\times \mathbb{R}_\perp$ with $\mathbb{R}_\perp$ parametrized by the coordinate $x_\perp$. We now argue that 7-branes wrapped on $M_3\times\partial X$ at some point $\bar x_{\perp}\in \mathbb{R}_\perp$ realize topological duality/triality interfaces and operators. A subset of our constructions work only for 7-branes with constant axio-dilaton profile and we list these in table \ref{tab:Fibs} together with their topological data. 

\begin{figure}[t]
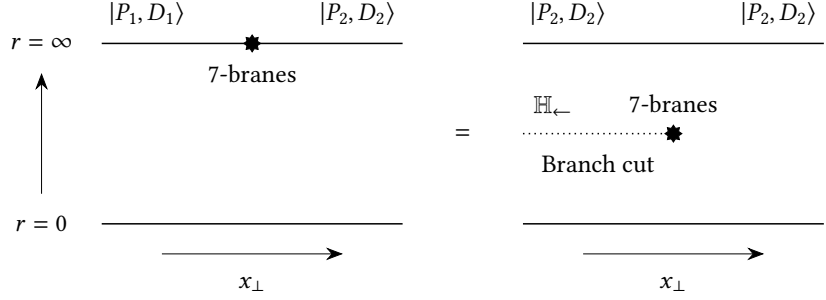

    \centering
    \scalebox{0.8}{

}
    \caption{Case\,(1), 7-branes wrapped on $M_3\times \partial X$, we sketch the plane $\mathbb{R}_{\geq0}\times \mathbb{R}_\perp$. The topological boundary conditions $\ket{P_1,D_1}$ are the monodromy transform of the boundary conditions $\ket{P_2,D_2}$ and result from stacking the branch cut with the asymptotic boundary. The branch cut is supported on $\mathbb{H}_{\leftarrow}\times \partial X$ and runs parallel to the D3-branes. Conventions are such that the monodromy matrix $\rho$ acts crossing the branch cut top to bottom. }
    \label{fig:7BraneInfinity2}
\end{figure}

\begin{figure}[t]
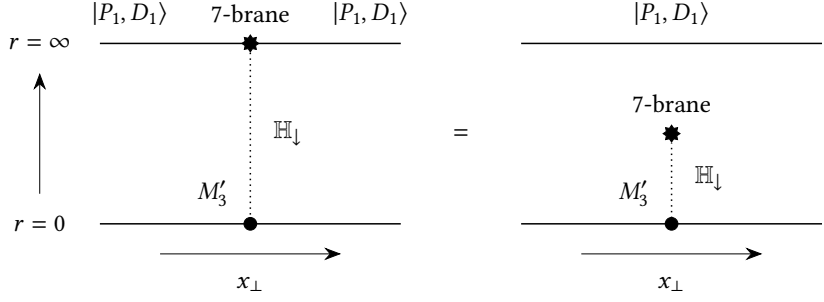

    \centering
    \scalebox{0.8}{

    }
    \caption{Case\,(2), 7-branes wrapped on $M_3\times \partial X$, we sketch the plane $\mathbb{R}_{\geq0}\times \mathbb{R}_\perp$. There is a single set of boundary conditions $\ket{P_1,D_1}$. The branch cut is supported on $\mathbb{H}_{\downarrow}\times \partial X$ and runs perpendicular to the D3-branes. Conventions are such that the monodromy matrix $\rho$ acts crossing the branch cut left to right.}
    \label{fig:Case2}
\end{figure}

\begin{figure}
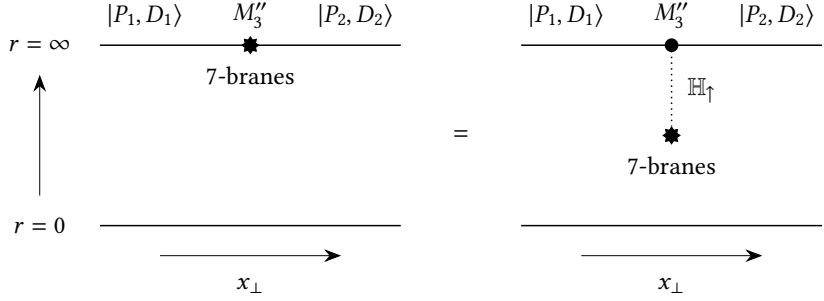

    \centering
    \scalebox{0.8}{
    %
    }
    \caption{Case\,(3), 7-branes wrapped on $M_3\times \partial X$, we sketch the plane $\mathbb{R}_{\geq0}\times \mathbb{R}_\perp$. The 7-brane insertion gives rise to two boundary conditions $\ket{P_1,D_1},\ket{P_2,D_2}$. The branch cut is supported on $\mathbb{H}_{\uparrow} \times \partial X$ and runs perpendicular to the D3-branes. Conventions are such that the monodromy matrix $\rho$ acts crossing the branch cut right to left.}
    \label{fig:Case3}
\end{figure}

\begin{figure}
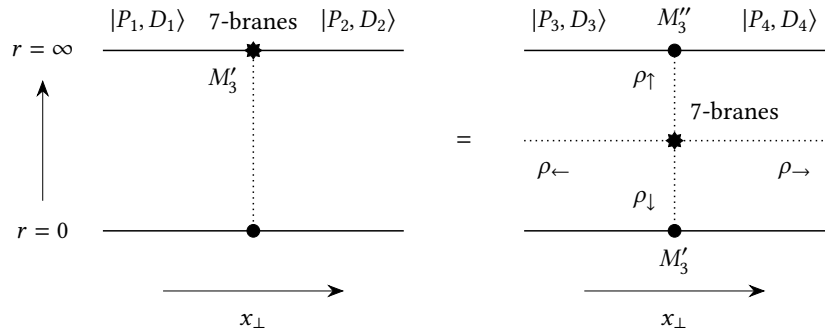

    \centering
    \scalebox{0.8}{%
}
    \caption{Case\,(4) hybrid case of cases 1,2,3, the 7-branes wrapped on $M_3\times \partial X$, we sketch the plane $\mathbb{R}_{\geq0}\times \mathbb{R}_\perp$. The overall monodromy is $\rho=\rho_{\leftarrow}\rho_{\uparrow}\rho_{\rightarrow}\rho_{\downarrow}$. Each monodromy factor $\rho_{\bullet}$ has its own branch cut separately. The branch cuts labelled $\rho_{\downarrow},\rho_{\uparrow}$ intersect the D3-brane worldvolume and asymptotic boundary in $M_3',M_3''$, respectively.}
    \label{fig:Case4}
\end{figure}

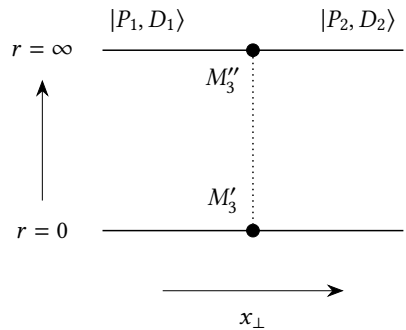
\begin{figure}
    \centering
    \scalebox{0.8}{\begin{tikzpicture}
	\begin{pgfonlayer}{nodelayer}
		\node [style=none] (0) at (-2.5, 1.5) {};
		\node [style=none] (1) at (2.5, 1.5) {};
		\node [style=none] (2) at (-2.5, -1.5) {};
		\node [style=none] (3) at (2.5, -1.5) {};
		\node [style=none] (6) at (-1.75, 2) {$\ket{P_1,D_1}$};
		\node [style=none] (7) at (1.75, 2) {$\ket{P_2,D_2}$};
		\node [style=none] (8) at (-3.5, -1.5) {$r=0$};
		\node [style=none] (9) at (-3.5, 1.5) {$r=\infty$};
		\node [style=none] (10) at (-3.5, 1) {};
		\node [style=none] (11) at (-3.5, -1) {};
		\node [style=none] (12) at (-1.5, -2.5) {};
		\node [style=none] (13) at (1.5, -2.5) {};
		\node [style=none] (14) at (0, -3) {$x_\perp$};
		\node [style=none] (15) at (-0.5, 1) {$M_3''$};
		\node [style=none] (17) at (-0.5, -1) {$M_3'$};
		\node [style=Circle] (18) at (0, 1.5) {};
		\node [style=Circle] (19) at (0, -1.5) {};
	\end{pgfonlayer}
	\begin{pgfonlayer}{edgelayer}
		\draw [style=ThickLine] (0.center) to (1.center);
		\draw [style=ThickLine] (3.center) to (2.center);
		\draw [style=ArrowLineRight] (11.center) to (10.center);
		\draw [style=ArrowLineRight] (12.center) to (13.center);
		\draw [style=DottedLine] (18) to (19);
	\end{pgfonlayer}
\end{tikzpicture}}
    \caption{S-duality is realized by cutting the symmetry TFT along the dotted line and gluing the pieces according to the desired S-duality transformation. No 7-branes are inserted. }
    \label{fig:Sduality}
\end{figure}

First consider a 7-brane wrapped at $r=\infty$ following the prescription in \cite{Heckman:2022muc}. This gives rise to a topological interface / symmetry operator in the 4D theory as the 7-brane is formally at infinite distance wrapped on a cycle of infinite volume. This decouples the non-topological interactions between the D3 and 7-brane and the non-topological degrees of freedom on the 7-brane worldvolume respectively \cite{Heckman:2022muc}. In this topological limit, for the 5D bulk theory, we can consider adding codimension two defects, i.e., the remnants of these 7-branes at infinity. Since everything is now treated as topological, we are free to insert these 7-branes anywhere in the interior, and as such we have different choices for where to extend the branch cut of this defect. We consider four distinct choices for the $SL(2,\mathbb{Z})$ monodromy branch cut:
\begin{enumerate}
    \item The branch cut is supported on $\mathbb{H}_{\leftarrow}\times \partial X$ with $\partial \mathbb{H}_{\leftarrow} =M_3$ and is oriented along $x_\perp$ parallel to the D3-branes with $x_\perp < \bar x_\perp$ along the asymptotic boundary. See the left subfigure in figure \ref{fig:7BraneInfinity2}. The branch cut ends at infinity. 
    \item The branch cut is supported on $\mathbb{H}_{\downarrow}\times \partial X$ with $\partial \mathbb{H}_{\downarrow}=M_3-M_3'$ (treated as a 3-chain) and is oriented radially inwards along $x_\perp=\bar x_\perp$ perpendicular to the D3-branes. See the left subfigure in figure \ref{fig:Case2}. The branch cut ends on $M_3'$, which is a subset of the D3-brane worldvolume, and supports an operator $\mathcal{D}(M_3')$ possibly coupled to background fields. While cuts can normally only begin / end on 7-branes, this can be made precise via a method of images procedure. The 7-brane has constant axio-dilaton profile.
    \item The branch cut is supported on $\mathbb{H}_{\uparrow}\times \partial X$ with $\partial \mathbb{H}_{\uparrow}= M_3- M_3''$ (treated as a 3-chain) and is oriented radially outwards along $x_\perp=\bar x_\perp$ perpendicular to the D3-branes (see figure \ref{fig:Case3}). The branch cut ends on $M_3''$ which is contained in the asymptotic boundary. The 7-brane has constant axio-dilaton profile.
    \item Whenever the 7-brane monodromy matrix is not prime over $\mathbb{Z}_N$, i.e., it can be factored into more than one non-trivial factor in $SL(2,\mathbb{Z}_N)$, we can consider separate branch cuts for each factor. We can then
    consider hybrid configurations of cases 1,2,3 with each branch cut realizing one of the previous setups (see figure \ref{fig:Case4}). The 7-brane has constant axio-dilaton profile.
\end{enumerate}
The cases differ in the structure of the boundary conditions and cases 2,3,4 include additional operators supported on $M_3',M_3''$ absorbing the branch cut.\footnote{Branch cuts are not physical, but their endpoints are. The branch cut arises from a choice of gauge for the background connection $A_1$ of the $SL(2,\mathbb{Z})$ duality bundle as we explain momentarily. This localizes an anomaly flow which we can either absorbed by an end point (case 2,3) or flow along a half-line (case 1).} In this picture, S-duality is realized by a vertically running branch cut without 7-brane insertions. (see figure \ref{fig:Sduality}). We discuss each case in turn. Of course we can also consider multiple 7-brane insertions.

First consider the setup for case 1. The branch cut is stacked with part of the asymptotic boundary, let us therefore move the topological 7-brane into the bulk. See the right subfigure in figure \ref{fig:7BraneInfinity2}. In the symmetry TFT formalism these two configurations are topologically equivalent. There is now a single 5D boundary condition $\ket{P_2,D_2}$ along the asymptotic boundary at $r=\infty$. Here $P_2$ denotes the choice of polarization, i.e., which combination of $B_2,C_2$ have Dirichlet boundary conditions imposed and $D_2$ denotes the boundary value for the condition, i.e., $(p_iB_2+q_iC_2)|_{r=\infty}=D_2$ for some collection of integer pairs $(p_i,q_i)$ specified by $P_2$.

\begin{figure}[t]
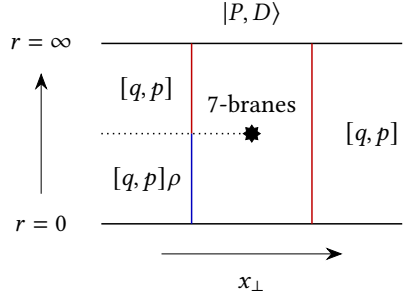

    \centering
    \scalebox{0.8}{%
}
    \caption{Half-space gauging via 7-brane insertion. Given a boundary condition $\ket{P,D}$ such that strings of type $[q,p]$ can terminate on $\partial X$ we find the admissible line defects on the D3-branes (located at $r=0$) to differ between the left and right hand side by a monodromy transformation. The D3-brane stack therefore experiences an effective change of polarization which amounts to a half-space gauging.}
    \label{fig:halfspacegauging2}
\end{figure}

Let us denote the monodromy matrix of the 7-brane by $\rho$. Crossing the branch cut, a string labelled by charges $[q,p]$ is transformed to a string with charges $[q,p]\rho$. Given a single topological boundary condition $\ket{P_2,D_2}$ which permits strings of charge $[q,p]$ to end on the boundary, it now follows that in the half-spaces $\bar x_\perp< x_\perp$ and $\bar x_\perp> x_\perp$ of the D3-brane worldvolume we have line defects whose charges are multiples of $[q,p]$ and $[q,p]\rho$, respectively (see figure \ref{fig:halfspacegauging2}).  Stacking the branch cut as shown in figure \ref{fig:7BraneInfinity2} (right to left) now clearly alters the boundary conditions by a monodromy transformation, more precisely its mod $N$ reduction $\rho\in SL(2,\mathbb{Z}_N)$. This change in boundary condition is a topological manipulation at the boundary, such as half-space gauging or the stacking of a counterterm
along $x_\perp<\bar x_\perp$ in $M_4$ and we propose:
\begin{equation}
    \textnormal{Topological Duality Interface $\mathcal{I}(M_3,\mathfrak{F})$}~~ \longleftrightarrow  
\begin{array}{c}
    \textnormal{7-brane of Type $\mathfrak{F}$ on} \\ \textnormal{ $M_3\times \partial X$ with cut $\mathbb{H}_{\leftarrow}\times \partial X$} 
    \end{array}
\end{equation}
The interface $\mathcal{I}(M_3,\mathfrak{F})$ promotes to a topological defect operator $\mathcal{U}(M_3,\mathfrak{F})$ when the theories separated by the defect are dual. Due to the branch cut not intersecting the D3-brane stack we can employ arbitrary 7-branes in this construction.

Let us now consider the setup for case 2. As above, we can move the topological 7-brane into the bulk giving a topologically equivalent configuration, see figure \ref{fig:Case2} left to right. There is a single 5D boundary condition $\ket{P_1,D_1}$. Defects running between the D3 stack and the topological boundary do not cross the branch cut and the effective polarization on the brane is identical along $\bar x_\perp < x_\perp$ and $\bar x_\perp > x_\perp$. The branch cut intersecting the D3 worldvolume gives rise to a 0-form operator realizing an S-duality transformation on local\footnote{Standard S-duality acts on both local and non-local operators and is realized as in figure \ref{fig:Sduality} by a completely vertical branch cut. Such a branch cut cannot be deformed into the 5D bulk and connects two distinct boundary conditions $\ket{P_i,D_i}$ with $i=1,2$.} operators in 4D. Therefore, upon collapsing the 5D slab to 4D we obtain for all choices of 7-branes with constant axio-dilaton profile a topological defect operator in a given theory. We propose: 
\begin{equation}
    \textnormal{Topological Defect Operator $\mathcal{V}(M_3,\mathfrak{F})$}~~ \longleftrightarrow  
    \begin{array}{c}
    \textnormal{7-brane of Type $\mathfrak{F}$ on} \\ \textnormal{ $M_3\times \partial X$ with cut $\mathbb{H}_{\downarrow}\times \partial X$} 
    \end{array}
\end{equation}


Let us now consider the setup for case 3. Again we can move the topological 7-brane into the bulk giving a topologically equivalent configuration, see figure \ref{fig:Case3} left to right. There are now two boundary conditions $\ket{P_1,D_1},\ket{P_2,D_2}$ giving rise to distinct polarizations in $\bar{x}_\perp < x_\perp$ and $\bar{x}_\perp > x_\perp$. This change in polarization is realized by a codimension one operator in the asymptotic boundary. This operator acts on symmetry operators according to the monodromy matrix of the 7-brane. It does not act on local operators and we will not have much to say about this operator.

\begin{figure}[t]
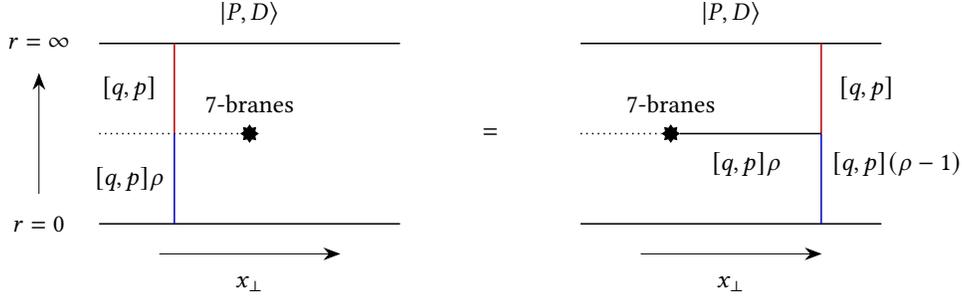

    \centering
    \scalebox{0.8}{

}
    \caption{Boundary condition $\ket{P,D}$ admitting $[q,p]$ strings to terminate at infinity. Passing a 4D line defect labelled by $[q,p]\rho$ through the symmetry defect $\mathcal{U}(M_3,\mathfrak{F})$ from left to right creates a topological surface operator of charge $[q,p](\rho-1)$ via Hanany-Witten brane creation. }
    \label{fig:HW}
\end{figure}

Next we discuss common properties of $\mathcal{I},\mathcal{V}$, independent of the choice of branch cuts, depending exclusively on the type $\mathfrak{F}$ of the 7-brane inserted. Hanany-Witten transition \cite{Hanany:1996ie} predicts the creation of a topological symmetry operator when line defects are dragged through $\mathcal{I}$, see figure \ref{fig:HW}.  Genuine defects constructed from asymptotic $[q,p]$ strings transform upon crossing $M_3$ to non-genuine defects with identical $[q,p]$ charge but with a topological operator of charge $[q,p](\rho-1)$ attached. The latter results from the corresponding string which attaches to the 7-branes stack and is otherwise embedded in the asymptotic boundary making it topological \cite{Heckman:2022muc}. Identical considerations hold for the operator $\mathcal{V}$. Such brane creation processes were already observed in D3/D5 systems in \cite{Apruzzi:2022rei}.

Let us now quantify the qualitative discussion above. The insertion of 7-branes turns on a non-trivial flat background for the $SL(2,\mathbb{Z})$ duality bundle. The topological nature of the symmetry TFT allows us to localize this background onto a choice of branch cut and in terms of the $SL(2,\mathbb{Z})$ doublet $(\mathcal{B}^i) = (C_2,B_2)$ then
\begin{equation}\label{eq:symtftcoupling}\begin{aligned}
    \mathcal{S}_{\mathrm{(SymTFT),0}}&=-\frac{N}{4\pi}\int_{M_5} \epsilon_{ij}\mathcal{B}^i \wedge  d \mathcal{B}^j \\[1em] 
     ~\xrightarrow[~\text{Insertion}~]{7-\text{brane}}~\qquad \mathcal{S}_{\mathrm{(SymTFT),1}}&=\mathcal{S}_{\mathrm{(SymTFT),0}}+\mathcal{S}_{(\mathrm{cut})} + \mathcal{S}_{\textnormal{(Defects)}}\,.
    \end{aligned}
\end{equation}
Here the term $\mathcal{S}_{\textnormal{(Defects)}}$ denotes the 3D TFT supported on $M_3\subset M_4$ resulting from wrapping 7-branes on $M_3\times \partial X$ in case 1 and in cases 2 and 3 it includes possibly an additional TFT supported on $M_3',M_3''$. We postpone the analysis of $\mathcal{S}_{\textnormal{(Defects)}}$ together with the discussion of boundary conditions for the bulk fields of the 5D symmetry TFT along such defects. The term $\mathcal{S}_{(\mathrm{cut})}$ denotes a counterterm localized to the branch cut $\mathbb{H}$ with $M_3 \subset \partial \mathbb{H}$.

\begin{figure}[t]
    \centering
    \scalebox{0.8}{
    \begin{tikzpicture}
	\begin{pgfonlayer}{nodelayer}
		\node [style=none] (0) at (-2, 0) {};
		\node [style=none] (1) at (2, 0) {};
		\node [style=none] (4) at (-3.25, 0) {Branch cut};
		\node [style=none] (5) at (0, 2) {};
		\node [style=none] (6) at (0, 0) {};
		\node [style=none] (7) at (-1, -2) {};
		\node [style=none] (8) at (1, -2) {};
		\node [style=none] (13) at (0.75, 2) {$S_{[q,p]}$};
		\node [style=none] (19) at (-1.75, -2) {$S_{[q,p]}$};
		\node [style=none] (20) at (2.25, -2) {$S_{[q,p](\rho-1)}$};
        \node [style=none] (21) at (0, -2.5) {};
	\end{pgfonlayer}
	\begin{pgfonlayer}{edgelayer}
		\draw [style=DottedLine] (0.center) to (1.center);
		\draw [style=RedLine] (5.center) to (6.center);
		\draw [style=RedLine] (6.center) to (7.center);
		\draw [style=ThickLine] (6.center) to (8.center);
	\end{pgfonlayer}
\end{tikzpicture}
    }
    \caption{5D surface symmetry operator stretching across the branch cut. $S_{[q,p]}$ transforms into $S_{[q,p]\rho}$ which can be decomposed into two surface symmetry operators as shown.}
    \label{fig:Surface}
\end{figure}
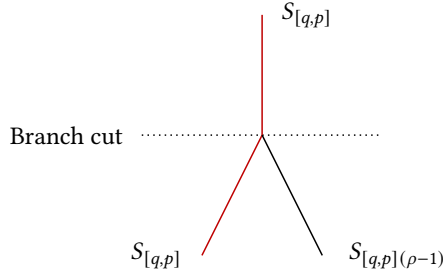

We now derive the 4D action $\mathcal{S}_{(\mathrm{cut})}$. Let us begin by determining the action on the branch cut attaching to a single D7-brane. The monodromy matrix is denoted $\mathbb{T}$ in \eqref{eq:Mono}. Clearly, a $B_2$ profile remains unchanged when crossing the branch cut and we now claim
\begin{equation}\label{eq:Cut3}
    \mathcal{S}_{(\mathrm{cut})}^{\mathrm{D7}}=\frac{2\pi i}{N}\int_\mathbb{H} \frac{\mathcal{P}(B_2)}{2} 
\end{equation}
where $\mathbb{H}$ denotes the branch cut. For a $(p,q)$ 7-brane we replace $B_2$ by $B_2^{[q,p]}=pB_2+qC_2$. Stacks of $k$ D7-branes then have action $k \mathcal{S}_{(\mathrm{cut})}^{\mathrm{D7}}$ and similarly for stacks of $(p,q)$ 7-branes.

We argue for \eqref{eq:Cut3} by considering its action on topological boundary conditions and via anomaly considerations. The former amounts to checking that the branch cut action indeed realizes the 0-form symmetry with monodromy matrix $\rho= \mathbb{T}$ on symmetry operators and defect operators in 5D. The latter shows that the branch cut carries off an anomaly sourced by the 7-brane insertion. In \cite{Antinucci:2022vyk} the action \eqref{eq:Cut3} is derived via condensation.

As a check, consider Dirichlet boundary conditions $\ket{P_{G_k},D}$ realizing $B_2|_{r=\infty}=D$ with global form $G$ and $k$ counterterms proportional to $\mathcal{P}(D)$ stacked. Then colliding the branch cut with the topological boundary amounts to acting as
\begin{equation}
    \ket{P_{G_k},D} \rightarrow \exp(\mathcal{S}_{(\mathrm{cut})})\ket{P_{G_{k}},D}=\ket{P_{G_{k+1}},D}
\end{equation}
which here stacks the boundary condition with the counterterm $(2\pi i/N)\mathcal{P}(D)/2$ without changing the polarization. This is consistent with the fact that defects for $P$ are Wilson lines constructed from fundamental strings which are not acted on upon crossing the branch cut. Therefore the polarization must remain unchanged and stacking the branch cut with the boundary can at most add counterterms. More generally this follows from T-duality \cite{Aharony:2013hda} and other cases are argued for identically, we give explicit computations in section \ref{sec:N4}.

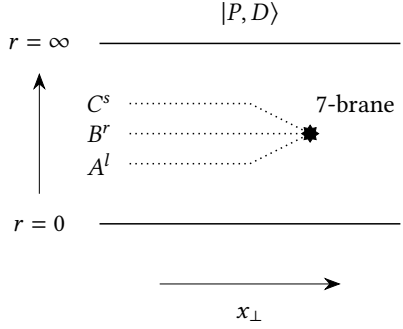
\begin{figure}
    \centering
    \scalebox{0.8}{
    \begin{tikzpicture}
	\begin{pgfonlayer}{nodelayer}
		\node [style=none] (0) at (-2.5, 1.5) {};
		\node [style=none] (1) at (2.5, 1.5) {};
		\node [style=none] (2) at (-2.5, -1.5) {};
		\node [style=none] (3) at (2.5, -1.5) {};
		\node [style=none] (4) at (0, 2) {$\ket{P,D}$};
		\node [style=none] (6) at (-3.5, -1.5) {$r=0$};
		\node [style=none] (7) at (-3.5, 1.5) {$r=\infty$};
		\node [style=none] (8) at (-3.5, 1) {};
		\node [style=none] (9) at (-3.5, -1) {};
		\node [style=none] (10) at (-1.5, -2.5) {};
		\node [style=none] (11) at (1.5, -2.5) {};
		\node [style=none] (12) at (0, -3) {$x_\perp$};
		\node [style=Star] (13) at (1, 0) {};
		\node [style=none] (14) at (0, 0.5) {};
		\node [style=none] (16) at (0, -0.5) {};
		\node [style=none] (17) at (-2, -0.5) {};
		\node [style=none] (18) at (-2, 0) {};
		\node [style=none] (19) at (-2, 0.5) {};
		\node [style=none] (20) at (-2.5, -0.5) {$A^l$};
		\node [style=none] (21) at (-2.5, 0) {$B^r$};
		\node [style=none] (22) at (-2.5, 0.5) {$C^s$};
		\node [style=none] (23) at (1.75, 0.5) {7-brane};
	\end{pgfonlayer}
	\begin{pgfonlayer}{edgelayer}
		\draw [style=ThickLine] (0.center) to (1.center);
		\draw [style=ThickLine] (3.center) to (2.center);
		\draw [style=ArrowLineRight] (9.center) to (8.center);
		\draw [style=ArrowLineRight] (10.center) to (11.center);
		\draw [style=DottedLine] (19.center) to (14.center);
		\draw [style=DottedLine] (17.center) to (16.center);
		\draw [style=DottedLine] (14.center) to (13);
		\draw [style=DottedLine] (16.center) to (13);
		\draw [style=DottedLine] (18.center) to (13);
	\end{pgfonlayer}
\end{tikzpicture}
    }
    \caption{The branch cut of attaching to any 7-brane insertion can be decomposed into branch cuts individually associated with $(p,q)$ 7-branes. }
    \label{fig:BCsplit}
\end{figure}

With this we can give the branch cut term for any 7-brane insertion (see figure \ref{fig:BCsplit}). Any 7-brane can be represented as a supersymmetric bound state of certain $(p,q)$ 7-branes denoted $A,B,C$, their $(p,q)$ charges are $(1,0),(3,1),(1,1)$ respectively. All 7-branes are then of the form $A^lB^rC^s$ and their branch cut therefore supports the operator
\begin{equation}\label{eq:BranchCutOperator}
\mathcal{O}_{(l,r,s)}=\mathcal{O}_A^l\mathcal{O}_B^r\mathcal{O}_C^s
\end{equation}
which is purely determined by the monodromy of the 7-brane and where
\begin{equation}\begin{aligned}
    \mathcal{O}_A&=\exp\left( \frac{2\pi i}{N}\int_{\mathbb{H}_A} \frac{\mathcal{P}(B_2)}{2}\right)\\
    \mathcal{O}_B&=\exp\left( \frac{2\pi i}{N}\int_{\mathbb{H}_B} \frac{\mathcal{P}(3B_2+C_2)}{2}\right)\\
    \mathcal{O}_C&=\exp\left( \frac{2\pi i}{N}\int_{\mathbb{H}_C} \frac{\mathcal{P}(B_2+C_2)}{2}\right)\,.
\end{aligned}\end{equation}
Here $B_2,C_2$ are normalized to $\mathbb{Z}_N$-valued forms and $\mathbb{H}_A,\mathbb{H}_B,\mathbb{H}_C$ denote the branch cuts localizing the monodromy of the respective brane stack. Note that it is not possible in general to present $\mathcal{O}_{(l,r,s)}$ as a single exponential, the fields $B_2,C_2$ are conjugate and do not commute. However, it is possible to present it as a TFT \cite{Antinucci:2022vyk}.

Let us discuss possible anomalies. Note first that we have the boundary condition $B_2|_{\mathrm{D7}}=0$ at the D7 locus. This follows from noting that $B_2|_{\mathrm{D7}}\neq 0$ implies that along infinitesimal loops linking the D7-brane we have $C_2\rightarrow C_2'\approx C_2+B_2|_{\mathrm{D7}}$ leading to a discontinuous, ill-defined profile for $C_2$ contradicting the D7-brane solution. In the M-theory dual picture this is equivalent to the torus cycle corresponding to $B_2$ shrinking at the D7-brane locus. Backgrounds with only $C_2$ turned on are invariant under monodromy and therefore we have no constraint on the $C_2$ profile along the D7-brane.

When considering a stack of $k$ D7-branes the boundary condition is $kB_2|_{\mathrm{D7}}=0$ and together with $B_2$ taking values in $\mathbb{Z}_N$ we have $\textnormal{gcd}(k,N)B_2|_{\mathrm{D7}}=0$. Now $B_2|_{\mathrm{D7}}$ no longer vanishes generically but takes values which are multiples of $N/\textnormal{gcd}(k,N)$. The $C_2$ profile remains unconstrained. 

Next, consider the background transformation $\mathcal{B}\rightarrow \mathcal{B}+d\lambda$ where $\lambda$ is twisted by the monodromy of the 7-brane, i.e. $d\lambda$ is subject to the same boundary conditions along the D7-brane as $\mathcal{B}_2$. The Pontryagin square term $\mathcal{P}(B_2)$ gives rise to a boundary term iff $\textnormal{gcd}(k,N)\neq 1$. This anomaly must be absorbed by the 3D TFT associated with the wrapped D7-branes. 

The 3D TFT has non-trivial lines precisely when $\textnormal{gcd}(k,N)\neq 1$. For this consider the dual M-theory geometry of a D7-brane whose normal geometry is given by an elliptic fibration $Z$ over $\mathbb{C}$ with one Kodaira $I_k$ singularity. The boundary $\partial Z$ of this normal geometry is elliptically fibered over a circle and has first homology groups $H_1(\partial Z)=\mathbb{Z}\oplus\mathbb{Z}\oplus \mathbb{Z}_k$. One free factor corresponds to the base circle and can be neglected in our discussion. This leaves us with $\mathbb{Z}\oplus \mathbb{Z}_k$. Of these, only the torsional generator of $\mathbb{Z}_k$ collapses at the $I_k$ singularity, sweeping out a non-compact two-cycle in the process. Wrapping an M2 brane on this cycle constructs a Wilson line. In 5D we work modulo $N$ and overall this gives a defect group of lines isomorphic to $\mathbb{Z}_{\textnormal{gcd}(k,N)}$ for the 3D TFT. In section \ref{sec:DefectTFT} we show that these lines organize into a 3D TFT with an anomaly sourcing the one carried by the branch cut. 

More generally the Pontryagin square terms on individual elementary branch cuts are anomalous under background transformations and we have a local anomaly flowing along each cut. Summing over branch cuts, the net anomaly is then necessarily absorbed by the 7-brane which can be viewed as an edge mode to the theories localized on the branch cuts. Alternatively, the overall anomaly is non-vanishing whenever the 7-brane sources an anomaly.

Let us therefore discuss when 7-brane insertions source anomalies in greater generality by studying the branch cut terms. For this, let us first consider in the 5D symmetry TFT the surface symmetry operators acting on the surface defects constructed from $(p,q)$-strings. They are:\footnote{Our definition differs from the definition $S_{(p,q)}'(\Sigma_2)\equiv S_{(p,0)}(\Sigma_2)S_{(0,q)}(\Sigma_2)$ as given in \cite{Kaidi:2022cpf} by a phase with argument proportional to the self-linking number of the surface $\Sigma_2$. This relative phase follows from the Baker-Campbell-Hausdorff formula and noting that $B_2,C_2$ are conjugate variables in the 5D TFT.}
\begin{equation}
    S_{[q,p]}(\Sigma_2)=\exp\left(2\pi i\oint_{\Sigma_2} (pB_2+qC_2)\right)\,, 
\end{equation}
with integers $p,q$ modulo $N$. Similarly to strings they are transformed when stretching across the $SL(2,\mathbb{Z})$ branch cut. Consider a surface $\Sigma_2$ separated by the branch cut $\mathbb{H}$ into two components $\Sigma_2 =\Sigma_2^+\cup \Sigma_2^-$, then
\begin{equation}
    S_{[q,p]}(\Sigma_2^{+})S_{[q,p]\rho}(\Sigma_2^{-})=S_{[q,p]}(\Sigma_2)S_{[q,p](\rho-1)}(\Sigma_2^{-})
\end{equation}
whenever the self-linking number of $\Sigma_2$ in the 5D bulk vanishes, see figure \ref{fig:Surface}. This is interpreted as the intersection of $S_{[q,p]}(\Sigma_2)$ with the branch cut sourcing $S_{[q,p](\rho-1)}(\Sigma_2^{-})$. Correspondingly, in terms of the $\mathbb{Z}_N$-valued 1-form symmetry background fields this relation can be expressed as
\begin{equation}\label{eq:dualityznaction}
    \delta \left( (\rho-1) \begin{bmatrix}
        C_2 \\ B_2
    \end{bmatrix}\right) = A_1\cup \begin{bmatrix}
        C_2 \\ B_2
    \end{bmatrix}
\end{equation}
where $A_1$ is the background field of the $SL(2,\mathbb{Z})$ bundle proportional to the Poincar\'e dual of the branch cut. A priori, $A_1$ takes values in $\mathbb{Z}_n$ where $n$ is the order of $\rho\in SL(2,\mathbb{Z})$, or $U(1)$ when $n$ is infinite, but in \eqref{eq:dualityznaction} this $A_1$ takes values in in $\mathbb{Z}_{\mathrm{gcd}(n,N)}$. In other words, $A_1$ takes values in $\mathbb{Z}_{\mathrm{gcd}(4,N)}$ or $\mathbb{Z}_{\mathrm{gcd}(3,N)}$ for duality\footnote{One might ask whether one should call this order four operation a ``quadrality'' operator. On local operators it is indeed order two, which accounts for the terminology. This subtlety between $2$ versus $4$ will show up later in our analysis of mixed anomalies.} and triality defects respectively\footnote{There are also hexality defects which are furnished by 7-branes of Type $II$ or $II^*$ whose background field in the 4D relative theory is $\mathbb{Z}_{\mathrm{gcd}(6,N)}$-valued. We come back to these theories at the end of the section to show that their mixed anomalies with the 1-form symmetry is always trivial.}, and similar to \eqref{eq:rescaling} we have a relation between normalizations of discrete-valued fields as (assuming $n$ is finite)
\begin{equation}\label{eq:rescaling2}
    A^{\mathbb{Z}_n}_1=\frac{1}{\mathrm{gcd}(n,N)}A^{\mathbb{Z}_{\mathrm{gcd}(k,N)}}_1
\end{equation}
where $A^{\mathbb{Z}_\ell}_1$ has holonomies that take values $p \; \mathrm{mod}\; \ell$.

Two related points to take into account are that one must first choose a polarization to have an invertible anomaly TFT, and in general $\rho-1$ does not have an inverse in $\mathbb{Z}_N$ coefficients.\footnote{In the cases when $(\rho-1)$ does have an inverse we have that $\mathrm{gcd}(2,N)=0$ or $\mathrm{gcd}(3,N)=0$ making the anomaly trivial in the first place.} We are motivated then to consider polarization choices that are duality/triality invariant, which is consistent with the fact that otherwise, the topological defect implementing the duality or triality defect is an interface between separate theories rather than a symmetry operator which may have a mixed 't Hooft anomaly. Recall that the background 1-form field for a given polarization spans a 1-dimensional subspace of $\mathrm{Span}(C_2,B_2)^\mathrm{T}$ and we say that a polarization is duality or triality invariant if there are non-trivial solutions to the following equation, posed over $\mathbb{Z}_N$,
\begin{equation}\label{eq:Eigenvalue}
    (\rho-1)\mathcal{B}_2=0\,
\end{equation}
which only has non-trivial solutions in $\mathbb{Z}_K$ subgroups of $\mathbb{Z}_N$ where $K=\mathrm{gcd}(k,N)$ and $k$ is the coefficient appearing in Table \ref{tab:Fibs}.\footnote{There are no solutions if $K$ and $N$ are coprime.} We denote such eigenvector solutions as $\mathcal{B}_2^\rho$ which is the product of an $SL(2,\mathbb{Z})$ vector and the form profile $B_2^\rho$. 

Consider for example the case of dualities that $N$ is even. We denote the lift to the full $\mathbb{Z}^{(1)}_N$-background field by $B^\rho_2$ as well where the two are related as
\begin{equation}\label{eq:rescale3}
    (B^\rho_2)^{\mathbb{Z}_{N}}=\frac{1}{2}(B^\rho_2)^{\mathbb{Z}_{2}}
\end{equation}
when the LHS only takes values in the $\mathbb{Z}_2$ subgroup of $\mathbb{Z}_N$ generated by $N/2 \; \mathrm{mod} \; N$. It will be clear that the anomaly will be invariant under this choice of lift.

We can decompose the vector space as $\mathrm{ker}(\rho-1)\oplus \mathrm{ker}(\rho-1)^\perp$ where the latter is generated by vectors $v$ such that $(\rho-1)v=v$. This allows us to define $(\mathcal{B}^\rho_2)^\perp$ which due to the $SL(2,\mathbb{Z})$ invariant pairing $\epsilon_{ij}$ in \eqref{eq:symtftcoupling}, allows us to rewrite that coupling schematically as $\int \mathcal{B}_2^\rho \cup \delta (\mathcal{B}^\rho_2)^\perp$. Substituting \eqref{eq:dualityznaction} then allows us to derive the mixed 't Hooft anomalies between 0-form duality/triality symmetries and $\mathbb{Z}^{(1)}_N$ 1-form symmetries:\footnote{In $\mathcal{S}_{\mathrm{duality}}$, observe that we take a gcd with respect to $2$ rather then $N$ because the operation is order two on local operators.}
\begin{equation}\label{eq:mixed anomaly term in the bulk}
    \begin{aligned}
       \mathcal{S}_{\mathrm{duality}}&= \frac{2\pi}{\mathrm{gcd}(2,N)}\int A_1\cup \frac{\mathcal{P}(B^\rho_2)}{2}  \\
    \mathcal{S}_{\mathrm{triality}} &= \frac{2\pi}{\mathrm{gcd}(3,N)}\int A_1\cup \frac{\mathcal{P}(B^\rho_2)}{2}
    \end{aligned}
\end{equation}
where we implemented the refinement $B_2^\rho \cup B_2^\rho \rightarrow \mathcal{P}(B_2^\rho)/2$ with $\mathcal{P}$ the Pontryagin square operation following \cite{Gaiotto:2014kfa,Kapustin:2014gua,Benini:2018reh}. The normalization of discrete-valued fields is motivated to match with \cite{Kaidi:2021xfk} which, in our presentation, means that we are using the LHS of \eqref{eq:rescaling2} for the normalization of $A_1$ and the RHS of \eqref{eq:rescaling} for the normalization of the $\mathbb{Z}_2$ or $\mathbb{Z}_3$-valued field $B^\rho_2$.\footnote{For the duality case, only the image of $A_1$ in the quotient $\mathbb{Z}_4/\mathbb{Z}_2\simeq \mathbb{Z}_2$ couples to $B^\rho_2$ which explains the factor $\mathrm{gcd}(2,N)$ rather than $\mathrm{gcd}(4,N)$.} 

We have that $A_1$ is Poincar\'e dual to the branch cut $\mathbb{H}$ and therefore
\begin{equation}\label{eq:CT}
    \begin{aligned}
    \textnormal{Duality Interface:}&\quad   S_{\mathrm{cut}}= \frac{2\pi}{\mathrm{gcd}(2,N)}\int_\mathbb{H} \frac{\mathcal{P}(B^\rho_2)}{2}  \\
    \textnormal{Triality Interface:}&\quad S_{\mathrm{cut}}= \frac{2\pi}{\mathrm{gcd}(3,N)}\int_\mathbb{H} \frac{\mathcal{P}(B^\rho_2)}{2}\,.
    \end{aligned}
\end{equation}
We conclude that upon inserting a 7-brane giving rise to the anomaly \eqref{eq:mixed anomaly term in the bulk}, we can localize this anomaly to a single branch cut with action \eqref{eq:CT}. Conversely, in cases with no net anomaly the anomaly localized to individual branch cuts cancels and therefore the branch cut term does not admit a presentation as a simple exponential.

\begin{figure}
    \centering
    \scalebox{0.8}{\begin{tikzpicture}
	\begin{pgfonlayer}{nodelayer}
		\node [style=none] (0) at (-2.5, 1.5) {};
		\node [style=none] (1) at (2.5, 1.5) {};
		\node [style=none] (2) at (-2.5, -1.5) {};
		\node [style=none] (3) at (2.5, -1.5) {};
		\node [style=none] (7) at (0, 2) {$\ket{P,D}$};
		\node [style=none] (13) at (-3.5, -1.5) {$r=0$};
		\node [style=none] (14) at (-3.5, 1.5) {$r=\infty$};
		\node [style=none] (15) at (-3.5, 1) {};
		\node [style=none] (16) at (-3.5, -1) {};
		\node [style=none] (17) at (-2.5, 0) {};
		\node [style=none] (18) at (2.5, 0) {};
		\node [style=none] (19) at (3.25, 0) {$\mathcal{O}^{[q,p]}$};
		\node [style=none] (20) at (0, -0.5) {$M_4'$};
		\node [style=none] (21) at (0, 0.25) {};
		\node [style=none] (22) at (0, 1.25) {};
		\node [style=none] (23) at (0, -2) {$\ket{\mathfrak{T}^{(N)}_X}$};
	\end{pgfonlayer}
	\begin{pgfonlayer}{edgelayer}
		\draw [style=ThickLine] (0.center) to (1.center);
		\draw [style=ThickLine] (3.center) to (2.center);
		\draw [style=ArrowLineRight] (16.center) to (15.center);
		\draw [style=DottedLine] (17.center) to (18.center);
		\draw [style=ArrowLineRight] (21.center) to (22.center);
	\end{pgfonlayer}
\end{tikzpicture}
}
    \caption{The operator $\mathcal{O}^{[q,p]}$ is defined on $M_4'$ in the 5D bulk which is homotopic to either boundary. The operator acts on the topological boundary condition $\ket{P,D}$ by colliding $M_4'$ with the corresponding boundary component. }
    \label{fig:BulkOps}
\end{figure}
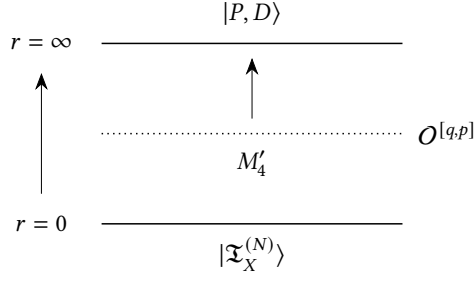

Let us now discuss another class of codimension one bulk operators of the 5D symmetry TFT. We use this to generalize the discussion of counterterm actions considered above. Along these lines, fix $B_2^{[q,p]}$ as the combination of doublet fields which pulls back to a $(p,q)$ 7-brane (i.e., is compatible with this choice of monodromy). 
Then, introduce the operator:
\begin{equation}
 \mathcal{O}^{[q,p]}=\exp\left( \frac{2\pi i}{N}\int_{M_4'}\frac{\mathcal{P}(B_2^{[q,p]})}{2}\right)\,.
\end{equation}
where $M_4'$ a copy of $M_4$ deformed into the bulk of the 5D TFT. It runs horizontally, parallel to the D3-brane worldvolume. More generally we could consider four-manifolds with boundary in conjunction with various edge modes.  Here $B_2^{[q,p]}=pB_2+qC_2$ is the form supported on the branch cut of a $(p,q)$ 7-brane. When contracting the 5D slab to 4D the branch cut is layered with the topological boundary condition and so the unitary operator $\mathcal{O}^{[q,p]}$ realizes an isomorphism on the vector space of boundary conditions generated by $\{\ket{P_i,D_i}\}$ (see figure \ref{fig:BulkOps}). Let us write
\begin{equation}\label{eq:Opq}
    \bra{P_i,D_i}\mathcal{O}^{[q,p]}= \bra{P_i^{[q,p]},D_i}\,.
\end{equation}
Note that whenever the $SL(2,\mathbb{Z})$ vector $[q,p]$ is contained in the polarization $P$ then the corresponding topological boundary condition is an eigenvector and $\mathcal{O}^{[q,p]}$ acts by stacking the boundary condition with a counterterm whose profile is determined by the Dirichlet boundary condition. However, more generally $\mathcal{O}^{[q,p]}$ is not diagonal, in particular there is no operator $\mathcal{O}^{[q,p]}$ which acts via counterterm stacking on all topological boundary conditions. 

We close this section by emphasizing that nothing about this discussion requires a large $N$ limit, the existence of a holographic dual, or having $\mathcal{N} = 4$ supersymmetry.

\section{Defect TFT}
\label{sec:DefectTFT}

We now discuss the 3D TFT supported on the defect constructed by wrapping some 7-brane on $ \partial X$ and its coupling to the bulk 5D symmetry TFT. The main idea will be to treat the wrapped 7-branes as codimension two defects in the 5D theory. See Appendix \ref{app:minimalTFT7branes} for a discussion of how dimensional reduction of the 7-brane on $\partial X$ can result in such topological terms.

\subsection{Minimal Abelian TFT from Defects}\label{ssec:linkingminimaltheory}

Consider the setup of the previous section consisting of a stack of 7-branes of type $\mathfrak{F}$ with monodromy matrix $\rho$ compactified on $\partial X$ to 3D in the presence of $N$ D3-branes located at the apex of $X$, the cone over $\partial X$. This produces a 3D TFT $\mathcal{T}$ supported on $M_3$ coupled to the worldvolume of the D3-brane stack. 

Let us begin by considering case 1 with a horizontal branch cut (along $\mathbb{H}_{\leftarrow}$) for which there is a single topological boundary condition along the asymptotic boundary. The 7-brane is located at $\bar x_\perp$ and separates the D3-worldvolume into two half-spaces $\bar x_\perp>  x_\perp$ and $\bar x_\perp<  x_\perp$.

\begin{figure}[t]
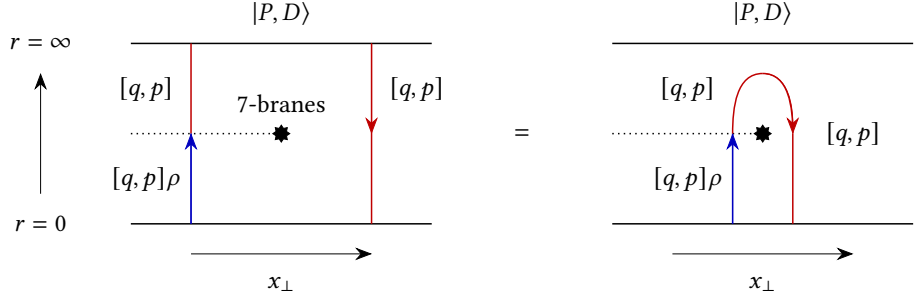

    \centering
    \scalebox{0.8}{

    }
    \caption{Collision of line defect of charge $[q,p]$ with line defect $-[q,p]\rho$. Conversely, line defects of charge $[q,p](1-\rho)$ are lines not inherent to the 3D TFT $\mathcal{T}$.}
    \label{fig:Screening}
\end{figure}

Previously we studied, via the Hanany-Witten transition, the consequence of dragging lines in $P_i$ through $M_3$ for $i=1,2$. Now consider two lines of equal but opposite asymptotic charge in each of the half-spaces and collide these. This produces a line operator and is interpreted as an open string running between the 7-branes and the D3-branes. If the lines have charges $\pm[q,p]$ then the latter is a string of charge $[q,p](1-\rho)$ (see figure \ref{fig:Screening}). These are precisely the lines not inherent to the 3D TFT $\mathcal{T}$, they can be moved off $M_3$. The inherent lines of the 3D TFT are associated with open strings running between the 7-branes and D3-branes modulo this screening effect.

With this the line defects of $\mathcal{T}$ are simply the line defects of the 7-brane stack modulo $N$. Let us study the lines of $\mathcal{T}$ in absence of the D3-brane flux, i.e., we do not impose screening modulo $N$.

In order to study the spin (in the sense of \cite{Hsin:2018vcg, Gukov:2020btk}) of such lines, consider the purely geometric M-/F-theory dual setup of an isolated stack of such 7-branes consisting of a local K3 surface $Z\rightarrow \mathbb{C}$. The boundary $\partial Z\rightarrow S^1$ has first homology $H_1(\partial Z)\cong \mathbb{Z}\oplus \textnormal{coker}(\rho-1)$ and line defects charged under the center symmetry of the system are constructed by wrapping M2-branes on cones over torsional one-cycles in $H_1(\partial Z)$. Let us assume that this homology group  is isomorphic to $\mathbb{Z}_k$ with generator $\gamma$. This torsional one-cycle is contained in the elliptic fiber, we write $\gamma=r \sigma_{a}+ s \sigma_{b}$ (in the obvious notation) for one of its representatives. This is dual to a string of charge $Q=[s,r]$. We have $k\gamma=0$ and $k$ copies of the corresponding string are screened in the F-theory setup. 

The spin of the lines associated with $\gamma$ is now determined by the refined self-linking number of $\gamma$ in $\partial Z$ as given by
\begin{equation}
    \ell(\gamma,\gamma)=\frac{1}{2k}\;\!\gamma \cdot \Sigma_\gamma\quad \mathrm{mod~1}
    \end{equation}
where $\partial \Sigma_\gamma=k\gamma$, determines the spin $h[L_\gamma]=\ell(L_\gamma,L_\gamma)=m/2k$ of the line $L_\gamma$ (see table \ref{tab:Fibs}). For a stack of $k$ $(p,q)$ 7-branes we have $m=k-1$.

Let us now take the screening effects due to the D3-brane flux into account. In this case the lines of 3D TFT $\mathcal{T}$ trivialize modulo $N$ and $k$, and therefore give rise to a 1-form symmetry $\mathbb{Z}_K$ with charged lines $L_\gamma$ and $K=\textnormal{gcd}(k,N)$. Whenever $mK\in 2\mathbb{Z}$ and $\textnormal{gcd}(m,K)=1$ it follows from the general discussion in \cite{Hsin:2018vcg} that the lines $L_\gamma$ form a consistent minimal abelian TFT denoted $\mathcal{A}^{K,m}$ \footnote{$\mathcal{A}^{K,m}$ is defined as a minimal 3D TFT with $\mathbb{Z}_K^{(1)}$ symmetry, whose symmetry lines have spins given by $h[a^s] \equiv \frac{ms^2}{2K} \mod 1$, where $a$ is a generating symmetry line such that $a^K = 1$.}. For branes of constant axio-dilaton and stacks of $(p,q)$ 7-branes we have $\textnormal{gcd}(k,m)=1$ and therefore $\textnormal{gcd}(K,m)=1$ follows. When these two conditions are met we have
\begin{equation}\label{eq:TFT}
    \mathcal{T}[B]=\mathcal{A}^{K,m}[B] \otimes \mathcal{T}'
\end{equation}
where $\mathcal{T}'$ is a decoupled TFT with lines neutral under the $\mathbb{Z}_K$ 1-form symmetry and $B$ is a 2-form background field for the 1-form symmetry which follows from the coupling of the open strings running between the D3-branes and 7-branes to $(B_2,C_2)$ and is $B=Q_i \mathcal{B}^i|_{\textnormal{7-brane}}$ which is $SL(2,\mathbb{Z})$ invariant where $Q$ is the charge vector of the strings.

\begin{figure}[t]
    \centering
    \scalebox{0.8}{\begin{tikzpicture}
	\begin{pgfonlayer}{nodelayer}
		\node [style=none] (0) at (-2.5, 1.5) {};
		\node [style=none] (1) at (2.5, 1.5) {};
		\node [style=none] (2) at (-2.5, -1.5) {};
		\node [style=none] (3) at (2.5, -1.5) {};
		\node [style=Star] (4) at (0, 0) {};
		\node [style=none] (5) at (-2.5, 0) {};
		\node [style=none] (6) at (0, 2) {$\ket{P,D}$};
		\node [style=none] (7) at (-1.5, -2) {};
		\node [style=none] (8) at (1.5, -2) {};
		\node [style=none] (9) at (0, -2.5) {$x_\perp$};
		\node [style=none] (10) at (0, 0.5) {7-branes};
		\node [style=none] (17) at (0, -1.5) {};
		\node [style=none] (18) at (-3.5, -1.5) {$r=0$};
		\node [style=none] (19) at (-3.5, 1.5) {$r=\infty$};
		\node [style=none] (20) at (-3.5, 1) {};
		\node [style=none] (21) at (-3.5, -1) {};
		\node [style=none] (22) at (0.5, -0.75) {$L_\gamma$};
	\end{pgfonlayer}
	\begin{pgfonlayer}{edgelayer}
		\draw [style=ThickLine] (0.center) to (1.center);
		\draw [style=ThickLine] (3.center) to (2.center);
		\draw [style=DottedLine] (5.center) to (4);
		\draw [style=ArrowLineRight] (7.center) to (8.center);
		\draw [style=RedLine] (4) to (17.center);
		\draw [style=ArrowLineRight] (21.center) to (20.center);
	\end{pgfonlayer}
\end{tikzpicture}
}
    \caption{Line operators of $\mathcal{T}$ coupled to the 4D theory are open strings running between the 7-branes and the D3-brane stack. The lines $L_\gamma^n$ organize into the minimal abelian TFT $\mathcal{A}^{K,m}$ where $K,m$ follow from the elliptic data of the 7-brane.}
    \label{fig:MinimalLines}
\end{figure}
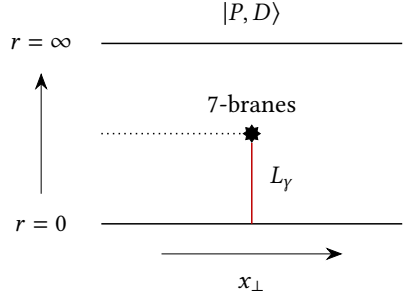

The charge vector $Q_i$ associated with $\gamma$ is defined modulo vectors in the image of $\rho-1$. Consider the charge vector $Q'_i=Q_i+(q (\rho-1))_i$ in the same coset. Then, $B$ changes as
\begin{equation}
    B=Q_i\mathcal{B}^i~\rightarrow ~ Q_i'\mathcal{B}^i=Q_i\mathcal{B}^i+(q(\rho-1))_i\mathcal{B}^i=Q_i\mathcal{B}^i+q_i((\rho-1)\mathcal{B})^i
\end{equation}
and for this coupling to be well-defined we require the profile of $\mathcal{B}$ to lie in the kernel of $\rho-1$ modulo $K$. 

For example, consider $N=2$ with insertion of a 7-brane of type $\mathfrak{F}=III^{\ast}$. We compute $k=2$ and therefore $K=2$. The eigenvector is $\mathcal{B}_2^\rho=B_2^\rho[1,1]^t$ and the background in \eqref{eq:TFT} is $B=B_2^\rho$. We have $\mathcal{T}[B_2^\rho]=\mathcal{A}^{2,1}[B_2^\rho]\otimes \mathcal{T}'$.

When $K=1$ no eigenvector exists, $B_2^\rho$ vanishes and $\mathcal{T}$ has no lines coupling to the bulk. In this case $\mathcal{T}$ does not absorb an anomaly. 

We note that this discussion of the TFT $\mathcal{T}$ is independent of branch cut choice and therefore extends to cases 2,3,4. 

Next, let us discuss case 2 with vertically running branch cut terminating on the 0-form operator $\mathcal{D}(M_3
',B_2^\rho)$ contained in the D3-worldvolume. This is another defect possible supporting a 3D TFT. We claim that for all parameter values $\mathcal{D}(M_3
',B_2^\rho)$ does not support a TFT interacting with the bulk. As discussed previously, the operator $\mathcal{D}(M_3
',B_2^\rho)$ realizes a gluing condition on the enriched Neumann boundary condition set by the D3-brane stack and, is not realized by a 7-brane. Therefore, no strings end on $M_3'$ and it does not support a defect group of its own, in contrast to the 7-brane insertion. With this we conjecture that the TFT living at the intersection is trivial or at least does not interact with the bulk.

\section{Example: $\mathcal{N} = 4$ SCFTs}
\label{sec:N4}
In the previous section we presented a general discussion of how 7-branes implement duality defects in systems obtained from D3-branes probing an isolated Calabi-Yau singularity. In particular, we saw that the structure of the branch cuts leads to distinct implementations for various sorts of duality defects (as well as triality defects). In this section we show that this matches to the available results in the literature for $\mathcal{N} = 4$ SYM theory, in particular the case where the gauge algebra is $\mathfrak{su}(2)$. Our construction readily generalizes to higher rank Lie algebras, and (deferring the classification of possible global realizations of the gauge group) this provides a uniform perspective for duality defects in other $\mathcal{N} = 4$ SCFTs realized by probe D3-branes. Combining this with the discussion in Appendix \ref{app:orbo}, we anticipate that the same considerations will also apply to the full set of $\mathcal{N} = 4$ SCFTs.

Before proceeding, let us briefly spell out our notational conventions, which essentially follow those of references 
\cite{Aharony:2013hda,Kaidi:2022uux}. All possible global forms are given by $SU(2)_i$ and $SO(3)_{\pm, i}$, $i = 0, 1$. Here $SU(2)$ is the electric polarization where only the Wilson lines with $(z_e, z_m) = (1, 0)$ are present; $SO(3)_+$ is the global form in which only the 't Hooft lines with $(z_e, z_m) = (0, 1)$ are present, and for $SO(3)_-$ only the dyonic lines with $(z_e, z_m) = (1, 1)$ are present. On top of that, we use $i = 0, 1$ to specify the absence or presence of a counterterm $\delta S = -\int \frac{\mathcal{P}(B)}{2}$, where $B$ is the background gauge field for the $\mathbb{Z}_2^{(1)}$ 1-form symmetry.

\subsection{Duality Defects in \texorpdfstring{$\mathfrak{su}(2)$}{} SYM\texorpdfstring{$_{\mathcal{N} = 4}$}{} Theory}

Duality defects for 4D $\mathcal{N}=4$ $\mathfrak{su}(2)$ supersymmetric Yang-Mills theory can be constructed field theoretically via the constructions in \cite{Kaidi:2021xfk} for Kramers-Wannier-like defects and half-space gauging \cite{Choi:2022zal}. We present the string theory construction for both emphasizing differences and similarities between the two approaches.

\subsubsection*{KOZ construction of Kramers-Wannier-like Duality Defects}
Let us first discuss the construction of \cite{Kaidi:2021xfk} for Kramers-Wannier-like duality defects and its string theory realization. We start with a lighting review of the field-theoretic construction and refer the reader to  \cite{Kaidi:2021xfk} for more details.  Consider the $SO(3)_-$ theory, with 1-form symmetry background field $B$ and mixed anomaly
\begin{equation}
    \pi \int_{M_5} A^{(1)}\cup \frac{\mathcal{P}(B)}{2}
\end{equation}
where $A^{(1)}$ is the background field for a $\mathbb{Z}_2$ 0-form symmetry, here S-duality on local operators at $\tau=i$ . Denote by $\mathcal{D}(M_3,B)$ the codimension one topological operator realizing this $\mathbb{Z}_2$ symmetry operator in the presence of a 1-form background $B$. The mixed anomaly implies that
\begin{equation}
    \mathcal{D}(M_3',B)\exp\left( i\pi \int_{\mathbb{\mathbb{H}}'}\frac{\mathcal{P}(B)}{2} \right)
\end{equation}
is invariant under background transformations of $B$. Here, $\mathbb{H}'\subset M_4$ is a half-space of the spacetime $M_4$ with $\partial \mathbb{H}'=M_3'$. Similarly, the minimal abelian TFT $\mathcal{A}^{2,1}$ is an edge mode, and the combination:
\begin{equation}
    \mathcal{A}^{2,1}(M_3,B)\exp\left( i\pi \int_{\mathbb{H}}\frac{\mathcal{P}(B)}{2} \right)
\end{equation}
is also invariant under background transformations of $B$, here $\partial \mathbb{H}=M_3$. We can therefore consider
\begin{equation}\label{eq:Fun}
    \mathcal{A}^{2,1}(M_3,B)\exp\left( i\pi \int_{\mathbb{H}''}\frac{\mathcal{P}(B)}{2} \right) \mathcal{D}(M_3',B)
\end{equation}
with $\partial \mathbb{H}''=M_3-M_3'$ and an invertible codimension one defect in $SO(3)_-$ theory is constructed contracting $\mathbb{H}''$ and setting $M_3=M_3'$. Gauging $B$ to the global form $SU(2)$, this defect becomes non-invertible.

\begin{figure}
    \centering
    \scalebox{0.8}{
    \begin{tikzpicture}
	\begin{pgfonlayer}{nodelayer}
		\node [style=none] (0) at (-2.5, 1.5) {};
		\node [style=none] (1) at (2.5, 1.5) {};
		\node [style=none] (2) at (-2.5, -1.5) {};
		\node [style=none] (3) at (2.5, -1.5) {};
		\node [style=Star] (4) at (0, 0) {};
		\node [style=none] (5) at (0, -1.5) {};
		\node [style=none] (7) at (0.75, -0.25) {$III^*$};
		\node [style=none] (8) at (0, 2) {$\ket{P_{SU(2)_0},D}$};
		\node [style=none] (9) at (-1.5, -2.5) {};
		\node [style=none] (10) at (1.5, -2.5) {};
		\node [style=none] (11) at (0, -3) {$x_\perp$};
		\node [style=Circle] (12) at (0, -1.5) {};
		\node [style=none] (13) at (0, -2) {$\mathcal{D}(M_3',B_2^\rho)$};
		\node [style=none] (14) at (-3.5, -1.5) {$r=0$};
		\node [style=none] (15) at (-3.5, 1.5) {$r=\infty$};
		\node [style=none] (16) at (-3.5, 1) {};
		\node [style=none] (17) at (-3.5, -1) {};
        \node [style=none] (29) at (0, 0.5) {$\mathcal{A}^{2,1}(B_2^\rho)\otimes \mathcal{T}'$};
        \node [style=none] (18) at (3.5, 0) {};
		\node [style=none] (19) at (5.5, 0) {};
		\node [style=none] (20) at (6.5, 0) {};
		\node [style=none] (21) at (11.5, 0) {};
		\node [style=none] (22) at (9, 0) {};
		\node [style=none] (23) at (4.5, -0.5) {5D $\rightarrow$ 4D};
		\node [style=none] (24) at (9, -0.5) {$\mathcal{N}(B_2^\rho)\otimes \mathcal{T}'$};
		\node [style=CircleRed] (25) at (9, 0) {};
		\node [style=none] (26) at (7.75, 0.5) {$Z_{SU(2)_0}(D)$};
		\node [style=none] (28) at (10.25, 0.5) {$Z_{SU(2)_0}(D)$};
	\end{pgfonlayer}
	\begin{pgfonlayer}{edgelayer}
		\draw [style=ThickLine] (0.center) to (1.center);
		\draw [style=ThickLine] (3.center) to (2.center);
		\draw [style=DottedLine] (5.center) to (4);
		\draw [style=ArrowLineRight] (9.center) to (10.center);
		\draw [style=ArrowLineRight] (17.center) to (16.center);
        \draw [style=ThickLine] (20.center) to (21.center);
		\draw [style=ArrowLineRight] (18.center) to (19.center);
	\end{pgfonlayer}
\end{tikzpicture}
    }
    \caption{Insertion of a 7-brane of type $III^*$ into the 5D symmetry TFT of 4D $\mathcal{N}=4$ SYM with vertically running branch cut and global structure $SU(2)_0$ constructs the symmetry operator $\mathcal{N}(M_3,B_2^\rho)=\mathcal{A}^{2,1}(M_3,B_2^\rho)\otimes \mathcal{D}(M_3,B_2^\rho)$ introduced in \cite{Kaidi:2021xfk}. Here tensor product denotes stacking as a consequence of contracting the branch cut. The background $B_2^\rho$ is the background field for the global form $SO(3)_-$. In particular it is not the background field for $SU(2)$ and therefore $\mathcal{N}$ couples dynamically to the theories on both half-spaces.}
    \label{fig:KOZ}
\end{figure}
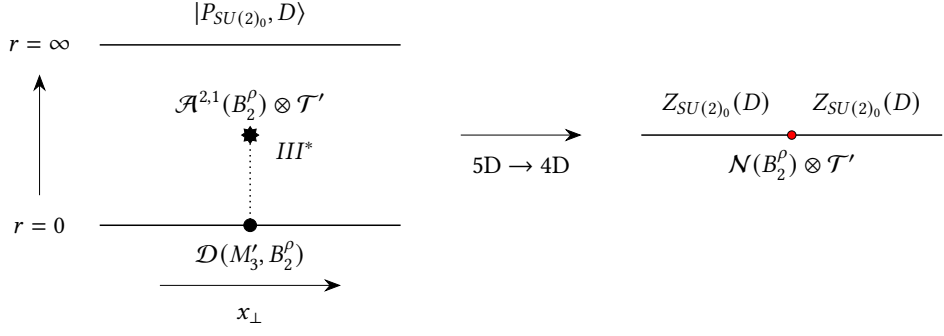

Now we turn to the string theory realization of the duality defects. We introduce a 7-brane of type $III^*$ wrapped on $M_3\times S^5$ with the branch cut intersecting with D3-branes at $M_3'$, separating the 4D spacetime into two parts. From table \ref{tab:Fibs}, $\tau=i$ is the fixed value of $III^*$ monodromy so we are able to have $\tau=i$ on the full worldvolume of the D3-branes and realize the operator $\mathcal{D}(M_3',B_2^\rho)$ on $M_3'$. The 3D TFT on $M_3$ is determined by the type of 7-brane, which can be read from table \ref{tab:Fibs} for type $III^*$ is $\mathcal{A}^{2,1}[B_2^\rho] \otimes \mathcal{T}'$. The left picture in figure \ref{fig:KOZ} illustrates this construction in terms of the 5D symmetry TFT slab.

Contracting the 5D slab to 4D then stacks $\mathcal{A}^{2,1}(M_{3}, B_{2}^{\rho})$ and $\mathcal{D}(M_{3}^{\prime}, B_{2}^{\rho})$, which gives rise to the 3D duality defect $\mathcal{N}(M_{3},B_2^\rho)$ within the 4D spacetime. Note that the whole construction so far is independent of the choice of the global structure of the theory. In other words, one can choose any boundary condition at $r=\infty$ for the 5D symmetry TFT and then contract the slab. In the case of $SU(2)$ and $SO(3)_+$, the $B_2^\rho$ is a dynamical field in the 4D bulk so $\mathcal{N}(B_2^\rho)$ is a non-invertible defect since it non-trivially couples to the 4D theory. In the case of $SO(3)_-$, $B_2^\rho$ is a background and non-dynamical, therefore $\mathcal{N}(B_2^\rho)$ is invertible. This perfectly matches the result from the field theory perspective and is identified as a non-intrinsic defect in \cite{Kaidi:2022cpf}.

\subsubsection*{Half-space gauging construction}\label{subsec:half-space gauging su(2) duality}
\begin{figure}
    \centering
    \scalebox{0.8}{\begin{tikzpicture}
	\begin{pgfonlayer}{nodelayer}
		\node [style=none] (5) at (-6, 1.5) {};
		\node [style=none] (6) at (-1, 1.5) {};
		\node [style=none] (7) at (-6, -1.5) {};
		\node [style=none] (8) at (-1, -1.5) {};
		\node [style=Star] (10) at (-2, 0) {};
		\node [style=none] (11) at (-6, 0) {};
		\node [style=none] (13) at (-1.5, 0.5) {$III^*$};
		\node [style=none] (18) at (-3.5, 2) {$\bra{P_{SU(2)_{+,0}},D}$};
		\node [style=none] (20) at (-7, 1.5) {$r=\infty$};
		\node [style=none] (21) at (-7, 1) {};
		\node [style=none] (22) at (-7, -1) {};
		\node [style=none] (26) at (-5, -2) {};
		\node [style=none] (27) at (-2, -2) {};
		\node [style=none] (28) at (-3.5, -2.5) {$x_\perp$};
		\node [style=none] (29) at (-4, -0.5) {$H$};
		\node [style=none] (31) at (0, 0) {};
		\node [style=none] (32) at (2, 0) {};
		\node [style=none] (33) at (3, 0) {};
		\node [style=none] (34) at (8, 0) {};
		\node [style=none] (35) at (5.5, 0) {};
		\node [style=none] (36) at (1, -0.5) {5D $\rightarrow$ 4D};
		\node [style=none] (37) at (5.5, -0.5) {$\mathcal{A}^{2,1}\otimes \mathcal{T}'$};
		\node [style=CircleRed] (38) at (5.5, 0) {};
		\node [style=none] (39) at (4.25, 0.5) {$Z_{P_{SO(3)_{+,0}}}(D)$};
		\node [style=none] (40) at (6.75, 0.5) {$Z_{SU(2)_{0}}(D)$};
		\node [style=none] (41) at (-4, 0.5) {$e^{i\pi \int P(B_2^\rho)/2}$};
	\end{pgfonlayer}
	\begin{pgfonlayer}{edgelayer}
		\draw [style=ThickLine] (5.center) to (6.center);
		\draw [style=ThickLine] (8.center) to (7.center);
		\draw [style=DottedLine] (11.center) to (10);
		\draw [style=ArrowLineRight] (22.center) to (21.center);
		\draw [style=ArrowLineRight] (26.center) to (27.center);
		\draw [style=ThickLine] (33.center) to (34.center);
		\draw [style=ArrowLineRight] (31.center) to (32.center);
	\end{pgfonlayer}
\end{tikzpicture}
    }
    \caption{Insertion of a 7-brane of type $III^*$ into the 5d symmetry TFT of 4D $\mathcal{N}=4$ SYM with horizontal running branch cut and topological boundary condition $SU(2)_0$ gives rise to half-space gauging as in \cite{Choi:2022zal}.}
    \label{fig:ShaoEtAl}
\end{figure}
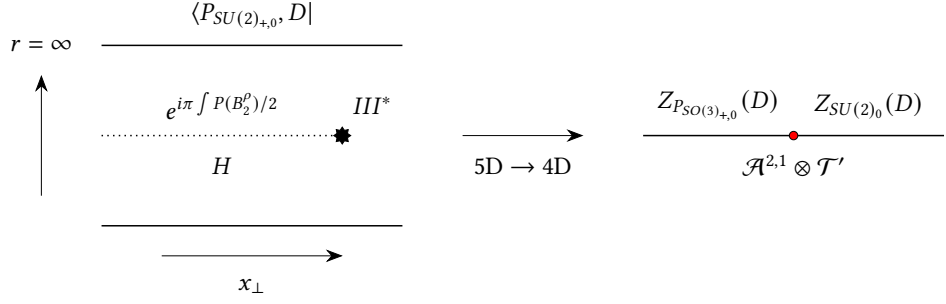

Let us now give the string theoretic setup for the half-space gauging construction in \cite{Choi:2022zal}. For example, we insert a 7-brane of type $\mathfrak{F}=III^*$ into the bulk and with a horizontally running branch cut which funnels the anomaly sourced by this insertion to infinity (see figure \ref{fig:ShaoEtAl}).

\noindent We claim that this realizes half-space gauging. This follows since $B_2^\rho$ is oriented along the polarization of the global form $SO(3)_-$, we therefore have:
\begin{equation}\label{eq:LongComp}\begin{aligned}
&~~~\, \bra{P_{SU(2)_0},D}\exp\left( i\pi \int \mathcal{P}(B_2^\rho)/2 \right)\\  &=  \sum_d \braket{P_{SU(2)_0},D\,|\,P_{SO(3)_{-,0}},d}\bra{P_{SO(3)_{-,0}},d}\exp\left( i\pi \int \mathcal{P}(d)/2 \right)\\
 &=  \sum_d \braket{P_{SU(2)_1},D\,|\,P_{SO(3)_{-,0}},d}\bra{P_{SO(3)_{-,0}},d}\exp\left( i\pi \int \mathcal{P}(D)/2+\mathcal{P}(d)/2 \right)\\
  &=  \sum_d \bra{P_{SO(3)_{-,0}},d}\exp\left( i\pi \int \mathcal{P}(D)/2 +D\cup d+\mathcal{P}(d)/2 \right)\\
&=\sum_d \bra{P_{SO(3)_{-,1}},d}\exp\left( i\pi \int D\cup d\right)\exp\left( i\pi \int \mathcal{P}(D)/2 \right)
  \\
   &= \bra{P_{SO(3)_{+,1}},D}\exp\left( i\pi \int \mathcal{P}(D)/2 \right)\\
    &=\bra{P_{SO(3)_{+,0}},D} \\
\end{aligned}
\end{equation}
as anticipated by noting that the S-duality branch cut interchanges electric and magnetic lines. Here the sums run over $\mathbb{Z}_2$-valued 2-forms $d$ (see figure \ref{fig:ShaoEtAl}). 

\begin{figure}
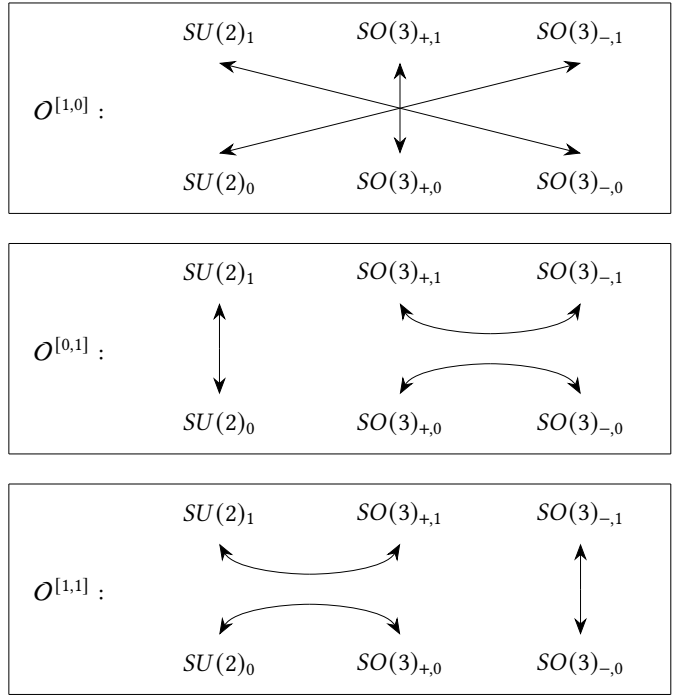

    \centering
    \scalebox{0.8}{
    }
    \caption{Operators $\mathcal{O}^{[q,p]}$ for 4D $\mathcal{N}=4$ $\mathfrak{su}(2)$ gauge theory.}
    \label{fig:Opq}
\end{figure}

Note that the operation of half-space gauging is not realized universally by one type of 7-brane. For example, in the above $\bra{P_{SO(3)_{-,r}}}$ is mapped to $\bra{P_{SO(3)_{-,r+1}}}$, with index $r$ mod 2, by branch cut stacking and the global structure is preserved. 

For this reason, it is instructive to study the other possible operators of type $\mathcal{O}^{[q,p]}$ as introduced in \eqref{eq:Opq}. The full generating set is: 
\begin{equation}\label{eq:definition of su(2) branch cut operators}
    \begin{aligned}
        \mathcal{O}^{[1,0]}&=\exp\left( i\pi \int \mathcal{P}(C_2)/2 \right)\\
        \mathcal{O}^{[0,1]}&=\exp\left( i\pi \int \mathcal{P}(B_2)/2 \right)\\
        \mathcal{O}^{[1,1]}&=\exp\left( i\pi \int \mathcal{P}(B_2+C_2)/2 \right)\\
    \end{aligned}
\end{equation}
and for the case of 7-branes of type $\mathfrak{F}=III^*$ we have $\mathcal{O}^{[1,1]}$ realized on the branch cut. Repeating computations similar to \eqref{eq:LongComp}, we find the results displayed in figure \ref{fig:Opq}. The global forms related by gauging are \cite{Kaidi:2022uux}:
\begin{equation}\label{eq:GlobalFormPair}\begin{aligned}
    SU(2)_0~&\leftrightarrow~ SO(3)_{+,0}\\
    SU(2)_1~&\leftrightarrow~ SO(3)_{-,0}\\
    SO(3)_{+,1}~&\leftrightarrow~ SO(3)_{-,1}
    \end{aligned}
\end{equation}

We can now give the completely general procedure. Pick a pair of global forms in \eqref{eq:GlobalFormPair} to be realized on two half-spaces. Then, determine the operator $\mathcal{O}^{[q,p]}$ connecting these
\begin{equation}\label{eq:GlobalFormPair2}\begin{aligned}
    \mathcal{O}^{[1,1]}\,:&\qquad SU(2)_0~\!\!\!\!\!\!&&\leftrightarrow~ SO(3)_{+,0}\\
    \mathcal{O}^{[1,0]}\,:&\qquad SU(2)_1~\!\!\!\!\!\!&&\leftrightarrow~ SO(3)_{-,0}\\
    \mathcal{O}^{[0,1]}\,:&\qquad SO(3)_{+,1}~\!\!\!\!\!\!&&\leftrightarrow~ SO(3)_{-,1}
    \end{aligned}
    \end{equation}
Next determine the 7-brane which supports $\mathcal{O}^{[q,p]}$ on their branch cut. These are for example $\mathfrak{F}=III^*,  I_1^{[0,1]},I_1^{[1,0]}$ respectively where the latter two are D7- and $[0,1]$-7-branes. Note that in all cases $B_2^\rho$ is neither the background field for the left nor for the right global form on each half-space. Consequently, the minimal abelian TFT supported on the 7-brane interacts with the degrees of freedom in both half-spaces. This leads to the defect realizing a non-invertible symmetry.

\subsection{Triality Defects in \texorpdfstring{$\mathfrak{su}(2)$}{} SYM\texorpdfstring{$_{\mathcal{N} = 4}$}{} Theory}
Let us move to triality defects in the $\mathfrak{su}(2)$ theory. We will see that in this case, only the half-space gauging construction works, which aligns with the fact that triality defects for $\mathfrak{su}(2)$ are intrinsic. 

\subsubsection*{KOZ construction of Kramers-Wannier-like Triality Defects}

We first consider the construction following \cite{Kaidi:2021xfk} and show how it does not work. In this case the branch cut of the 7-brane is vertical and ends on the worldvolume of the D3-branes. 
Therefore, we are restricted in our construction to 7-branes whose monodromy has fixed points at $\tau=\frac{i\pi}{3}$ or $\tau=\frac{2\pi i}{3}$, which are values necessary for triality defects \cite{Choi:2022zal, Kaidi:2022uux}. 

Let us take the type $IV^*$ 7-brane as an example. Consider a similar setup as in Figure \ref{fig:KOZ}, but with the type $IV^*$ instead of $III^*$ 7-brane. Naively, one may expect 3D TFTs $\mathcal{A}^{3,2}(B_2^\rho)$ and $\mathcal{D}(M_3',B_2^\rho)$ living on the two ends of the branch cut, respectively. However, in the case of $\mathfrak{su}(2)$ theory, where $B_2$ and $C_2$ are both $\mathbb{Z}_2$ fields in the bulk, there is in fact no non-trivial $B_2^\rho$ preserved by the type $IV^*$ 7-brane monodromy and hence the 7-brane does not source an anomaly. So the setup with inserted 7-branes reduces to studying the branch cut. This is nicely aligned with our discussion on the mixed anomalies shown in \eqref{eq:mixed anomaly term in the bulk}. For triality defects of $\mathfrak{su}(2)$ theory, the mixed anomaly between the triality symmetry and the 1-form symmetry is trivial.

\subsubsection*{Half-space gauging construction}

\begin{figure}
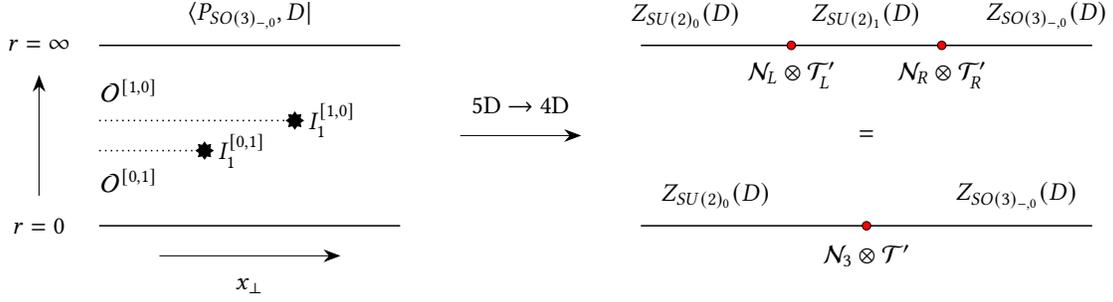

    \centering
    \scalebox{0.8}{
}
    \caption{Left: sketch of the 5D symmetry TFT for 4D $\mathcal{N}=4$ $\mathfrak{su}(2)$ theory. With two 7-brane insertions of type $\mathfrak{F}=I_1^{[1,0]},I_1^{[0,1]}$ and branch cut operators $\mathcal{O}^{[1,0]},\mathcal{O}^{[0,1]}$ respectively. Right: sketch of the 5D slab contracted to 4D, top and bottom are equivalent. The top, bottom figure shows the 4D theory when the 7-brane insertions are displaced, aligned along $x_\perp$ respectively. The latter results in the triality defect $\mathcal{N}_3$.   }
    \label{fig:Triality}
\end{figure}

All ingredients we need to build triality defects are already introduced in Section \ref{subsec:half-space gauging su(2) duality}. Let us take $SO(3)_{-,0}$ theory as an example. From (\ref{eq:GlobalFormPair2}) we know that $SO(3)_{-,0}$ turns into $SU(2)_1$ under half-space gauging, which can be realized by acting with the operator $\mathcal{O}^{[1,0]}$ living on the horizontal branch cut off a $(0,1)$-7-brane. We can then insert a D7-brane into the bulk, so that the operator $\mathcal{O}^{[0,1]}$ on the horizontal branch cut is introduced. This leads to adding a counterterm for the $SU(2)_1$ theory, so that it becomes the $SU(2)_0$ theory. According to \cite{Aharony:2013hda}, the $SU(2)_0$ theory is indeed dual to $SO(3)_{-,0}$ via the modular transformation $\mathbb{T}\cdot \mathbb{S}$. Therefore by contracting the 5D TFT slab, we indeed get a triality defect $\mathcal{N}_3$.

Furthermore, in this case we are able to determine the lines of the TFT $\mathcal{N}_3$. First, note that the insertion of a single $(p,q)$ 7-brane gives a 3D defect with no lines of its own. See the discussion in subsection \ref{sec:Proposal}. However, when inserting multiple 7-branes, as shown in figure \ref{fig:Triality}, the combined system can contain line defects which are constructed by $(p,q)$ string junctions terminating at the 7-brane insertion and the D3-brane locus. This generalizes the setup displayed in figure \ref{fig:MinimalLines}. These lines constitute the lines of the 3D TFT obtain from the fusion of the TFT supported at the individual 7-brane insertions. In the case of the triality defect we may have Y-shaped string junctions ending on $\mathcal{N}_{L},\mathcal{N}_{R}$ and at $r=0$ and these descend to the lines of the triality defect $\mathcal{N}_3$, the fusion of $\mathcal{N}_{L}$ with $\mathcal{N}_{R}$. 

The line defects of $\mathcal{N}_3$ are therefore determined by the total monodromy $\rho=\rho_L\rho_R$. In the case where $\mathcal{N}_{L},\mathcal{N}_{R}$ are respectively engineered by $I_1^{[0,1]},I_1^{[1,0]}$ type fibers the overall monodromy has trivial cokernel $\textnormal{coker}(\rho-1)$ and therefore there are no additional lines coupling to the fields $B_2,C_2$.

\begin{figure}
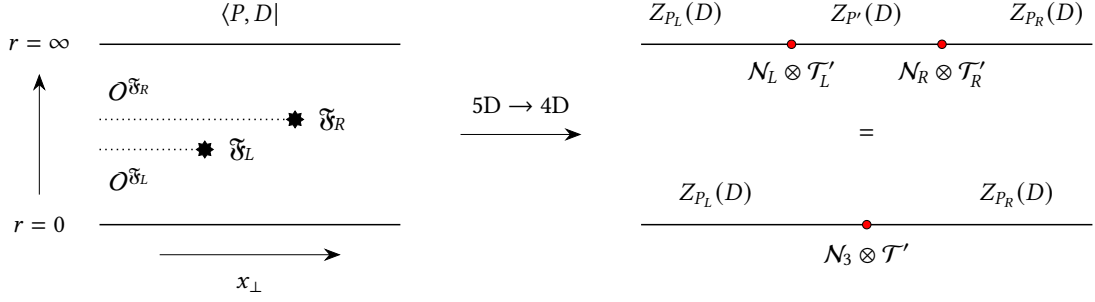

    \centering
    \scalebox{0.8}{%
}
    \caption{Left: Triality defects construction for 4D $\mathcal{N}=4$ $\mathfrak{su}(2)$ theory from 
    the 5D symmetry TFT. With two 7-brane insertions of type $\mathfrak{F}_L,\mathfrak{F}_R$ and branch cut operators $\mathcal{O}^{\mathfrak{F}_L},\mathcal{O}^{\mathfrak{F}_R}$ respectively. }
    \label{fig:Trialityforallsu2}
\end{figure}

Triality defects for other global structures of $\mathfrak{su}(2)$ can be realized following the same steps. Figure \ref{fig:Trialityforallsu2} illustrates the generic construction, for which we specify ingredients for all cases in the following table.
{\renewcommand{\arraystretch}{1.35}
\begin{table}[H]
\begin{center}
\begin{tabular}{|c|c|c|}
\hline
   $P_{G_m}$  & $\mathfrak{F}_R, \mathfrak{F}_L$ & $Z_{P_L}(D), Z_{P'}(D)$, $Z_{P_R}(D)$  \\
   \hline 
   $SU(2)_0$  & $III^*, I_1^{[1,0]}$ & $Z_{SO(3)_{+,1}}(D), Z_{SO(3)_{+,0}}(D)$, $Z_{SU(2)_0}(D)$  \\
   \hline 
   $SU(2)_1$  & $I_1^{[1,0]},III^*$ & $Z_{SO(3)_{-,1}}(D), Z_{SO(3)_{-,0}}(D)$, $Z_{SU(2)_1}(D)$  \\
   \hline
   $SO(3)_{+,0}$  & $III^*,I_1^{[0,1]}$ & $Z_{SU(2)_1}(D), Z_{SU(2)_0}(D)$, $Z_{SO(3)_{+,0}}(D)$ \\
   \hline
   $SO(3)_{+,1}$  & $I_1^{[0,1]},III^*$ & $Z_{SO(3)_{-,0}}(D), Z_{SO(3)_{-,1}}(D)$, $Z_{SO(3)_{+,1}}(D)$\\
   \hline 
   $SO(3)_{-,0}$  & $I_1^{[1,0]},I_1^{[0,1]}$ & $Z_{SU(2)_{0}}(D), Z_{SU(2)_{1}}(D)$, $Z_{SO(3)_{-,0}}(D)$ \\
   \hline
   $SO(3)_{-,1}$  & $I_1^{[0,1]},I_1^{[1,0]}$ & $Z_{SO(3)_{+,0}}(D), Z_{SO(3)_{+,1}}(D)$, $Z_{SO(3)_{-,1}}(D)$ \\
   \hline
\end{tabular}
\end{center}
\caption{Triality defect constructions for all global structures of $\mathfrak{su}(2)$, with all ingredients illustrated in Figure \ref{fig:Trialityforallsu2}.}
    \label{tab:all triality defects for su(2)}
\end{table}}

\subsection{Defects in the \texorpdfstring{$\mathfrak{su}(N)$}{} Case}
With these examples in hand, we now generalize our discussion to 
construct duality and triality defects for $\mathfrak{su}(N)$ theories with $\mathcal{N} = 4$ supersymmetry. 
In this case, a complete treatment would first require the classification of possible global forms of the theory. 
Rather than proceed in this way, we mainly focus on how things work in the purely electric case, where the gauge group is $SU(N)$, as well as the purely magnetic case, where the gauge group is $SU(N) / \mathbb{Z}_N$.

First, consider the construction of Kramers-Wannier-like duality defects from 7-brane insertions with vertical branch cuts attaching to the D3-brane world volume. For duality/triality defects of $\mathfrak{su}(N)$ theories, if there is a non-trivial $B_2^\rho$ living on the vertical branch cut, one can always choose a topological boundary condition for the 5D TFT, i.e., a global structure of the 4D gauge theory, so that $B_2^\rho$ is the non-dynamical background gauge field. One then ends up with an invertible duality/triality defect $\mathcal{N}(M_3, B_2^\rho)$ in this duality frame, so it is a non-intrinsic duality/triality defect \cite{Kaidi:2022cpf}. On the other hand, if there is no non-trivial $B_2^\rho$ that can be defined on the branch cut, the mixed anomaly is trivial, hence the construction of \cite{Kaidi:2021xfk} does not work. This leads to an intrinsic duality/triality defect \cite{Kaidi:2022cpf} which can only be realized by half-space gauging.

From now on, we focus on providing a universal realization for the half-space gauging construction with $N$ generic.

\subsubsection*{Duality defects}
\begin{figure}
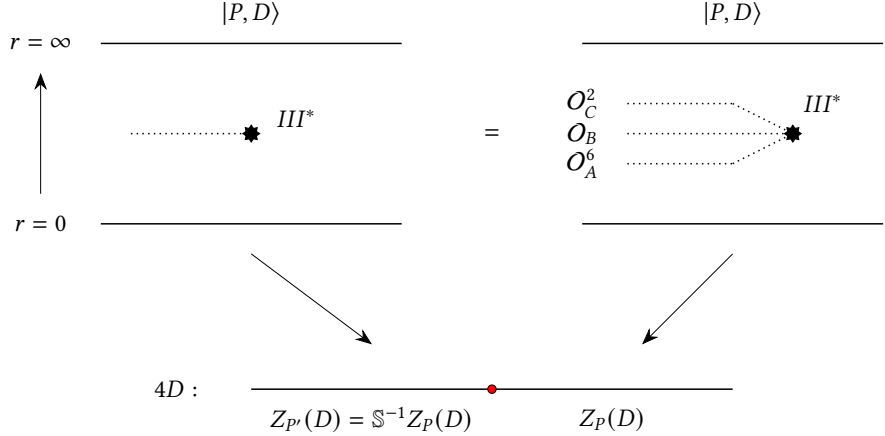

    \centering
    \scalebox{0.8}{
}
    \caption{Type $III^*$ 7-brane with monodromy inverse to S-duality. Left: the monodromy is localized onto a single branch cut. Right: the monodromy is localized onto three distinct branch cuts associated with stacks of $(p,q)$ 7-branes. Bottom: contraction of the 5D slab to 4D. }
    \label{fig:III*resolution}
\end{figure}

As we already discussed, 7-branes with the horizontal branch cut in the 5D TFT give rise to topological manipulations possibly changing the global structure, and / or adding counter-terms for the 4D theory. In order to build duality defects, we need 7-branes with topological manipulations which can be compensated exactly by that of the modular $\mathbb{S}$ or $\mathbb{S}^{-1}$ transformation. Recall the $\mathbb{S}$ transformation is defined by 
\begin{equation}
    \mathbb{S}: [q,p]\rightarrow [q,p]\begin{pmatrix} 0&1\\-1&0 \end{pmatrix}, \quad \tau \rightarrow -\frac{1}{\tau}
\end{equation}
where $p$ and $q$ are electric and magnetic charges respectively. In our string theory construction, $p$ and $q$ are charges for F1- and D1-strings respectively. Therefore, the candidate 7-branes for duality defects are type $III$ and $III^*$, whose monodromy matrices $\rho$ have same actions on the $[q,p]$ charges as $\mathbb{S}$ and $\mathbb{S}^{-1}$. Therefore, by inserting $III$ or $III^*$ 7-branes and contracting the 5D TFT slab, the 4D theory in the half-space acted by the branch cut is dual  to the original theory in the other half-space via a modular $\mathbb{S}$ or $\mathbb{S}^{-1}$ transformation. This realizes duality defects at $\tau=i$. See the left picture of Figure \ref{fig:III*resolution}.

Let us now discuss topological operators living on the branch cut. From now on, we will focus on the case of the type $III^\ast$ 7-brane, but the following discussion also works for the type $III$ 7-brane similarly. For generic values of $N$, i.e., $B_2$ and $C_2$ both $\mathbb{Z}_N$-valued, it is not always possible to define a $B_2^\rho$ which is invariant under the $III^\ast$ monodromy matrix. So it seems unclear how to build topological operators living on the branch cut. However, it is always possible to factorize the monodromy matrix into elements in $SL(2,\mathbb{Z}_N)$, such that the branch cut is correspondingly separated for each element where a non-trivial $B_2^\rho$ can be defined. In fact, the type $III^\ast$ 7-brane can be constructed by three simple types of $(p,q)$ 7-branes:
\begin{equation}\label{eq:definition ABC brane}
    A:(1,0), ~B:(3,1), ~C:(1,1),
\end{equation}
as 
\begin{equation}
    III^*: A^6BC^2.
\end{equation}
The monodromy matrix for a $(p,q)$ 7-brane is given by 
\begin{equation}
    \rho_{(p,q)}=\begin{pmatrix}1+pq&p^2\\-q^2&1-pq\end{pmatrix},
\end{equation}
under which $pB_2+qC_2$ is obviously always invariant. Therefore, based on our discussion in Section \ref{sec:SETUP}, on the branch cut with monodromy matrix $\rho_{(p,q)}$, we can define a 4D topological operator: 
\begin{equation}
    \mathcal{O}^{(p,q)}=\exp \left( {\frac{2\pi i}{N}\int \frac{\mathcal{P}(pB_2+qC_2)}{2}}\right)\,.
\end{equation}

Now we can separate the branch cut of the $III^*$ 7-brane into multiple ones corresponding to the monodromy matrices for $C^2,B$ and $A^6$, respectively. See the right picture of figure \ref{fig:III*resolution}. Stacking topological operators living on all branch cuts together, we end up with the topological operator for the type $III^\ast$ 7-brane:
\begin{equation}\label{eq:branchcut operator for e7}
    \mathcal{O}^{III^*}=\exp \left( \frac{2\pi i\times 6}{N}\int \frac{\mathcal{P}(B_2)}{2}  \right) \exp \left( \frac{2\pi i}{N}\int \frac{\mathcal{P}(3B_2+C_2)}{2}  \right)\exp \left( \frac{2\pi i\times 2}{N}\int \frac{\mathcal{P}(B_2+C_2)}{2}  \right)
\end{equation}
This is well-defined for $\mathbb{Z}_N$-valued $B_2$ and $C_2$ with $N$ generic. 

For the $\mathfrak{su}(2)$ theory, the first and the third factors in (\ref{eq:branchcut operator for e7}) are both trivial. The second term becomes 
$\exp \left( i\pi \int \mathcal{P}(B_2+C_2)/2  \right)$ because $B_2$ is $\mathbb{Z}_2$-valued. Now $\mathcal{O}^{III^*}$ exactly reduces to the operator $\mathcal{O}^{[1,1]}$ we discussed around (\ref{eq:definition of su(2) branch cut operators}), realizing non-invertible duality defects for $SU(2)$ and $SO(3)_+$. For the $\mathfrak{su}(3)$ theory, the operator reduces to 
\begin{equation}
    \mathcal{O}^{III^*}_{N=3}=\exp \left( \frac{2\pi i}{3}\int \frac{\mathcal{P}(C_2)}{2}  \right)\exp \left( \frac{4\pi i}{3}\int \frac{\mathcal{P}(B_2+C_2)}{2}  \right). 
\end{equation}
Acting with this operator on, for example, the boundary conditions of the 5D TFT which result in the $SU(3)_0$ theory, one can do a  similar computation as in (\ref{eq:LongComp}) and reach the $\overline{PSU(3)}_{0,0}$ theory\footnote{Here we denote global structures with the notation in \cite{Kaidi:2022cpf}, with the two sub-indices for different background gauge fields and counterterms respectively. The overline means an opposite sign for the background field compared to that without an overline. Explicitly, $\overline{PSU(3)}_{0,0}$ means background field purely magnetic $-C_2$ without counterterm.}, which is dual to $SU(3)_0$ theory via the modular $\mathbb{S}$ transformation \cite{Kaidi:2022cpf}. Hence, this realizes non-invertible duality defects. Note the fact that duality defects for $\mathfrak{su}(2)$ and $\mathfrak{su}(3)$ are respectively non-intrinsic and intrinsic, but our construction provides a simple unified realization for them. In fact, our construction works for all values of $N$ regardless of its divisors.

\subsubsection*{Triality defects}
Triality defects from 7-branes can be built following similar steps as in duality defects. In order to find an order 3 topological manipulation, we need to consider 7-branes with monodromy matrices acting on $[q,p]$-string charges which can be compensated by modular transformations  $\mathbb{S}^{m}\cdot \mathbb{T}^{n}$ or $\mathbb{T}^{m}\cdot \mathbb{S}^{n}$, with $m=\pm1, n=\pm1$. The candidate 7-branes with this desired property include the Kodaiar types $II, II^\ast, IV$ and $IV^\ast$.

As in the case of duality defects, we will focus on one certain type of 7-brane and all other candidates work in a similar way. Let us take the $IV^\ast$ 7-brane as an example. In order to  define the topological operator living on its branch cut consistently for all $N$, we separate its branch cut into multiple ones. Each of these branch cuts provides a monodromy as an element in $SL(2,\mathbb{Z})$. In IIB string/F-theory, the $IV^\ast$ 7-brane admits a standard decomposition  
\begin{equation}
    \mathrm{Type} \, IV^\ast : A^5BC^2,
\end{equation}
where $A,B$ and $C$ are elementary $(p,q)$ 7-branes defined in (\ref{eq:definition ABC brane}). Therefore, we can now separate the branch cut of the $IV^*$ 7-brane into those corresponding to the monodromy matrices for $C^2, B$ and $A^5$ 7-branes, respectively. Stacking topological operators living on all branch cuts together, we end up with the operator for the type $IV^\ast$ 7-brane as: 
\begin{equation}\label{eq:branchcut operator for e6}
    \mathcal{O}^{IV^*}=\exp \left( \frac{2\pi i \times 5}{N}\int \frac{\mathcal{P}(B_2)}{2}  \right) \exp \left( \frac{2\pi i }{N}\int \frac{\mathcal{P}(3B_2+C_2)}{2}  \right)\exp \left( \frac{2\pi i \times 2}{N}\int \frac{\mathcal{P}(B_2+C_2)}{2}  \right).
\end{equation}

For the $\mathfrak{su}(2)$ theory, the third factor in (\ref{eq:branchcut operator for e6}) degenerates, so the operator reduces to 
\begin{equation}
    \mathcal{O}^{IV^\ast}_{N=2}=\exp \left( \pi i\int \frac{\mathcal{P}(B_2)}{2}  \right) \exp \left( \pi i \int \frac{\mathcal{P}(B_2+C_2)}{2}  \right),
\end{equation}
which reads $\mathcal{O}^{[0,1]}\cdot \mathcal{O}^{[1,1]}$ under the definition (\ref{eq:definition of su(2) branch cut operators}) for $\mathfrak{su}(2)$. One can easily see from Figure \ref{fig:Opq} that this corresponds to triality defects. For example, starting from $SO(3)_{-,0}$, this topological operator corresponds to first adding a counterterm and then gauging. The resulting theory is $SO(3)_{+,1}$, which is dual to $SO(3)_{-,0}$ via modular $\mathbb{S}\cdot \mathbb{T}$ transformation, thus realizing triality defects. For the $\mathfrak{su}(3)$ theory, the operator for $IV^\ast$ 7-brane reduces to
\begin{equation}
    \mathcal{O}^{IV^\ast}_{N=3}=\exp \left( \frac{4\pi i}{3}\int \mathcal{P}(B_2)/2  \right) \exp \left( \frac{2\pi i}{3}\int \mathcal{P}(C_2)/2  \right)\exp \left( \frac{4\pi i}{3}\int \mathcal{P}(B_2+C_2)/2  \right).
\end{equation}
Acting with this operator on, for example, the $SU(3)_0$ theory, one can do a similar computation as in (\ref{eq:LongComp}) and reach the $\overline{PSU(3)}_{1,0}$ theory, which is dual to $SU(3)_0$ theory via a modular $\mathbb{S}\cdot \mathbb{T}^{-1}$ transformation \cite{Kaidi:2022cpf}, thus realizing non-invertible duality defects. As in the case of duality defects, triality defects for $\mathfrak{su}(2)$ and $\mathfrak{su}(3)$ are of different types, namely intrinsic for $\mathfrak{su}(2)$ and non-intrinsic for $\mathfrak{su}(3)$. However, our construction does not depend on this difference and provides a simple unified realization for them. In fact, our construction works for all values of $N$ regardless of its divisors.

\section{Example: $\mathcal{N} = 1$ SCFTs}\label{sec:N=1}

In this section we show that the considerations of the previous sections readily extend to a broad class of $\mathcal{N} = 1$ SCFTs. We focus on the case of D3-brane probes of an isolated Calabi-Yau threefold singularity. It is well-known that this can be characterized in terms of a quiver gauge theory. In this characterization, the IIB axio-dilaton descends to a specific marginal coupling of the gauge theory. In particular, the IIB $SL(2,\mathbb{Z})$ duality group descends to a specific non-abelian duality group action at this special point in the moduli space. In particular, at the special points in moduli space given by $\tau_{IIB} = i$ and $\tau_{IIB} = \exp(2 \pi i / 6)$ we expect there to be duality / triality defects which act on the resulting $\mathcal{N} = 1$ SCFTs. Of course, at these points of strong coupling the sense in which the weakly coupled Lagrangian description in terms of a quiver gauge theory is actually available is less clear, but the definition of the field theory is still clear. An important feature of this more general class of SCFTs is that we typically have more 0-form symmetries, and we can also consider passing the associated defects charged under these symmetries through the  duality / triality defects.

To begin, let us recall some basic properties of these D3-brane probes of Calabi-Yau threefold singularities.\footnote{There is a vast literature on the subject of D-brane probes of singularities, see e.g., \cite{Douglas:1996sw, Franco:2005rj, Yamazaki:2008bt} for reviews.} In general terms, we get quiver gauge theories with gauge algebra $\mathfrak{su}(N_1)\times \mathfrak{su}(N_2)\times \cdots \times \mathfrak{su}(N_M)$, with $M$ the number of quiver gauge factors. Each gauge group factor has a complexified gauge coupling $\tau_i$ which specifies a marginal coupling of the SCFT. In an SCFT realized by a probe D3-brane, there exists a linear combination of these $\tau_{i}$ which corresponds to the IIB axio-dilaton:
\begin{equation}
\tau_{IIB} = \underset{i}{\sum} n_i\tau_{i},
\end{equation}
where the $n_i$ are positive integers which depend on the details of the geometry. There is a special subspace in the conformal manifold at which the $SL(2,\mathbb{Z})$ duality group action of type IIB strings descends to the SCFT. For example, in the special case where $n_i = 1$ for all $i$, this is given by taking $\tau_{i} = \tau$.

In any case, the IIB $SL(2,\mathbb{Z})$ duality descends to a field-theoretic duality, much as in the $\mathcal{N}=4$ case, and just as there, this also changes the global structure of the theory. The 1-form symmetry depends on the global structure of the theory. For example, with gauge group $SU(N)^M$, the $\mathcal{N}=1$ theory has a purely electric $\mathbb{Z}_N$ 1-form symmetry, which is the diagonal center of all $SU(N)$ groups. The charged Wilson lines can be built, as in the $\mathcal{N}=4$ case, by F1-strings stretched between D3-branes and the asymptotic boundary $\partial X$ of Calabi-Yau threefold $X$.

The construction of duality / triality defects in $\mathcal{N}=4$ theories readily generalizes to $\mathcal{N}=1$ theories. Indeed, our analysis carries over essentially unchanged, with 7-branes wrapped on the boundary $\partial X$ of the Calabi-Yau cone $X$ probed by the stack of D3-branes. In this sense, we automatically implement duality / triality defects in these $\mathcal{N} = 1$ SCFTs just from the top down origin as D3-brane probe theories. 

However, this is not the end of the story. Compared to the $\mathcal{N} = 4$ case where $X = \mathbb{C}^3$, in the $\mathcal{N} = 1$ case, the boundary topology $\partial X$ will in general be more intricate. In particular, dimensional reduction of the 10D supergravity on $\partial X$ will result in a Symmetry TFT with additional fields and interactions terms. These additional fields indicate the presence of more symmetries for the 4D field theory, and these can also be stacked with the duality defect. Here we explain some of the main elements of this construction, focusing on how 0-form symmetry topological operators behave under crossing a duality defect.

\subsection{Symmetry TFTs via IIB String Theory}

We will now present a top down approach to extracting symmetry TFTs for $\mathcal{N}=1$ SCFTs via IIB on a Calabi-Yau threefold $X$. We will reduce the topological term of IIB action on the asymptotic boundary $\partial X$ of the internal manifold $X$, which is a Sasaki-Einstein 5-manifold. This procedure involves treating the various IIB supergravity fluxes as elements in differential cohomology (see e.g., \cite{Freed:2006yc, Freed:2006ya}). Reduction of the linking 5-manifold leads to a symmetry TFT, as explained in  \cite{Apruzzi:2021nmk} (see also \cite{Aharony:1998qu, Heckman:2017uxe}).
Compared with the $\mathcal{N}=4$ case, this reduction leads to additional fields and interaction terms.

More explicitly, consider D3-branes probing a non-compact Calabi-Yau threefold $X$ with an isolated singularity at the tip of the cone. For ease of exposition, we assume that the asymptotic boundary $\partial X$ has the cohomology classes:
\begin{equation}\label{eq:cohomology class of E-S}
    H^*(\partial X)=\{ \mathbb{Z},0,\mathbb{Z}^{b_2}\oplus \text{Tor}H^2(\partial X),\mathbb{Z}^{b_2},\text{Tor}H^4(\partial X),\mathbb{Z} \},
\end{equation}
which covers many cases of interest. In the above,
$b_2$ is the second Betti number. Although not exhaustive, the cases covered by line (\ref{eq:cohomology class of E-S}) includes an infinite family of Calabi-Yau singularities, including $\mathbb{C}^3/\Gamma$ with isolated singularities, as well as complex cones over del Pezzo surfaces. The relevant term in the IIB supergravity action descends from the IIB topological term:\footnote{Here we neglect a subtlety with the 5-form field strength being self-dual. This amounts to a choice of quadratic refinement in an auxiliary 11D spacetime. See \cite{Belov:2006jd, Belov:2006xj, Monnier:2012xd, Heckman:2017uxe}.}
\begin{equation}\label{eq:IIB TFT differential cohomology}
     -\int_{M_4\times X}C_4\wedge dB_2 \wedge dC_2 \rightarrow -\int_{M_4\times  X} \breve{F}_5\star \breve{H}_3\star \breve{G}_3.
\end{equation}
Based on (\ref{eq:cohomology class of E-S}), we can expand $\breve{F}_5, \breve{H}_3$ and $\breve{G}_3$ as 
\begin{equation}
\begin{split}
    \breve{F}_5=&\breve{f}_5\star \breve{1}+\sum_{\alpha=1}^{b_2}\breve{f}_3^{(\alpha)}\star \breve{u}_{2(\alpha)}+\sum_{\alpha=1}^{b_2}\breve{f}_{2(\alpha)}\star \breve{u}_3^{(\alpha)}+N\mathrm{\breve{v}ol}
    +\sum_{i}\breve{E}_3^{(i)}\star \breve{t}_{2(i)}+\sum_{i}\breve{E}_{1(i)}\star \breve{t}_4^{(i)},\\
    \breve{H}_3=&\breve{h}_3\star \breve{1}+\sum_{\alpha=1}^{b_2}\breve{h}_1^{(\alpha)}\star \breve{u}_{2(\alpha)}+\sum_{\alpha=1}^{b_2}\breve{h}_{0(\alpha)}\star \breve{u}_3^{(\alpha)}
    +\sum_{i}\breve{B}_1^{(i)}\star \breve{t}_{2(i)},\\
    \breve{G}_3=&\breve{g}_3\star \breve{1}+\sum_{\alpha=1}^{b_2}\breve{g}_1^{(\alpha)}\star \breve{u}_{2(\alpha)}+\sum_{\alpha=1}^{b_2}\breve{g}_{0(\alpha)}\star \breve{u}_3^{(\alpha)}
    +\sum_{i}\breve{C}_1^{(i)}\star \breve{t}_{2(i)}.
\end{split}
\end{equation}
In the above equations, $\breve{1}, \mathrm{\breve{v}ol}$ and $\breve{u}_{2(\alpha)},\breve{u}_{3}^{(\alpha)}$ correspond to the free part $\mathbb{Z}$ and $\mathbb{Z}^{b_2}$ of cohomology classes, respectively, whereas $\breve{t}_{2(i)}$ and $\breve{t}_4^{(i)}$ correspond to the torsional part $\text{Tor}H^2(\partial X)=\text{Tor}H^4(\partial X)$ respectively and $i$ runs over its generators. The index placement of $\breve{u}_{2(\alpha)}$ compared to $  \breve{u}_3^{(\alpha)}$ (and $\breve{t}_{2(i)}$ compared to $\breve{t}_{4}^{(i)}$) indicate that their star product have non-trivial integral over $\partial X$. 

In particular, since $\partial X$ has infinite volume, any fields arising as coefficients of free cocycles that are dual to free cycles of positive degree should be interpreted as infinitely massive dynamical fields, and thus should be set to zero:
\begin{equation}
  \breve{f}_3^{(\alpha)} = \breve{f}_{2(\alpha)} = \breve{h}_1^{(\alpha)} = \breve{h}_{0(\alpha)} =
    \breve{g}_1^{(\alpha)} = \breve{g}_{0(\alpha)} = 0.
\end{equation}
The IIB topological term (\ref{eq:IIB TFT differential cohomology}) can then be expanded as
\begin{equation}\label{eq:DC expansion of IIB TFT}
\begin{split}
    &-\int_{M_4\times X} \breve{F}_5\star \breve{H}_3\star \breve{G}_3 \\
    &=-\int_{\partial X} \mathrm{\breve{v}ol}\star \breve{1} \star \breve{1} \int_{M_4\times \mathbb{R}_{\ge 0}}N\breve{h}_3\star \breve{g}_3\\
    &-\sum_{i,j,k}\int_{\partial X}\breve{t}_{2(i)}\star \breve{t}_{2(j)}\star \breve{t}_{2(k)}\int_{M_4\times \mathbb{R}_{\ge 0}}\breve{E}_3^{(i)}\star \breve{B}_1^{(j)}\star \breve{C}_1^{(k)}\\
    &-\sum_{i,j}\int_{\partial X}\breve{t}_{2(i)}\star \breve{t}_4^{(j)}\int_{M_4\times \mathbb{R}_{\ge 0}} \breve{E}_{1(j)}\star \left( \breve{B}_1^{(i)}\star \breve{g}_3+\breve{h}_3\star \breve{C}_1^{(i)} \right).
\end{split}
\end{equation}

Carrying out the reduction on $\partial X$ now involves integrating over this space in differential cohomology. 
For non-torsional classes, we find:
\begin{equation}\label{eq:integral over vol and dual forms}
\int_{\partial X}\mathrm{\breve{v}ol}\star \breve{1}\star \breve{1}=1,
\end{equation}
since $\mathrm{\breve{v}ol}$ is the volume form of $\partial X$. Integrals of torsional generators over $\partial X$ are determined by linking numbers which can be derived from intersection numbers between divisors of $X$:
\begin{equation}\label{eq:integral from linking pairing}
\begin{split}
    &\mathcal{C}_{ijk}\equiv \int_{\partial X}\breve{t}_{2(i)}\star \breve{t}_{2(j)}\star \breve{t}_{2(k)},\\
    &\mathcal{C}_i^j\equiv \int_{\partial X}\breve{t}_{2(i)}\star \breve{t}_4^{(j)}.
\end{split}
\end{equation}

The IIB topological term then reduces to the 5D Symmetry TFT:
\begin{equation}\label{eq:TFT for generic N=1}
    \begin{split}
       \mathcal{S}_{5D}&=-\underset{M_4\times \mathbb{R}_{\ge 0}}{\int} \bigg\{ N\breve{h}_3\star \breve{g}_3 - \sum_{i,j,k}\mathcal{C}_{ijk}\breve{E}_3^{(i)}\star \breve{B}_1^{(j)}\star \breve{C}_1^{(k)}-\sum_{i,j}\mathcal{C}_{i}^j\breve{E}_{1(j)}\star \left( \breve{B}_1^{(i)}\star \breve{g}_3+\breve{h}_3\star \breve{C}_1^{(i)} \right) \bigg\}.
    \end{split}
\end{equation}
Here fields denoted by capital letters are from torsional classes and are background fields for discrete symmetries:
\begin{equation}
    \breve{E}_3^{(i)} \leftrightarrow G^{(2)},\quad  \breve{B}_1^{(i)} \leftrightarrow G^{(0)},\quad  \breve{C}_1^{(i)} \leftrightarrow \tilde{G}^{(0)}
\end{equation}
where $G^{(2)}  \cong G^{(0)} \cong \text{Tor} H^2(\partial X) \cong \text{Tor} H^4(\partial X)$. On the other hand, fields denoted by lowercase letters correspond to field strengths of background fields for $U(1)$ symmetries. However, due to the presence of the coefficient $N$ in the first term, they are effectively $\mathbb{Z}_N$ symmetries
\begin{equation}
    \breve{g}_3 \leftrightarrow \mathbb{Z}_{N(m)}^{(1)},\quad  \breve{h}_3 \leftrightarrow \mathbb{Z}_{N(e)}^{(1)},
\end{equation}

The correspondence between fields, global symmetries and charged operators from wrapped branes are presented in table \ref{tab:charged operators of N=1 scfts}.

\begin{table}[t!]
    \centering
    \begin{tabular}{|c|c|c|}
\hline Fields&Global symmetries&Charged operators\\
\hline $\breve{h}_3$&$\mathbb{Z}_N^{(1)}$&F1-strings along $\mathbb{R}_{\geq 0}$, D3-branes wrapping $\mathbb{R}_{\geq 0}\times \sigma_2$\\
\hline $\breve{g}_3$&$\mathbb{Z}_N^{(1)}$&D1-strings along $\mathbb{R}_{\geq 0}$, D3-branes wrapping $\mathbb{R}_{\geq 0}\times \sigma_2$\\
\hline $\breve{E}_3^{(i)}$&$[\text{Tor}H^4(\partial X)]^{(2)}$&D3-branes wrapping $\mathbb{R}_{\geq 0}\times \gamma_1^{(i)}$\\
\hline $\breve{E}_{1(i)}$&$[\text{Tor}H^2(\partial X)]^{(0)}$&D3-branes wrapping $\mathbb{R}_{\geq 0}\times \gamma_{3(i)}$\\
\hline $\breve{B}_{1}^{(i)}$&$[\text{Tor}H^4(\partial X)]^{(0)}$&F1-string wrapping $\mathbb{R}_{\geq 0}\times \gamma_1^{(i)}$\\
\hline $\breve{C}_{1}^{(i)}$&$[\text{Tor}H^4(\partial X)]^{(0)}$&D1-string wrapping $\mathbb{R}_{\geq 0}\times \gamma_1^{(i)}$\\
\hline
\end{tabular}
    \caption{Fields in the 5D TFT and their corresponding global symmetries in 4D SCFTs. The charged operators composed of various types of branes are also presented. $\sigma$ and $\gamma$ denote non-torsional and torsional cycles, respectively. We use an upper index for the 1-cycles and a lower index for the 3-cycles.}
    \label{tab:charged operators of N=1 scfts}
\end{table}

The first term in (\ref{eq:TFT for generic N=1}) is the same as (\ref{eq:5d TFT terms1}), and so in this sense, all of our analysis of the $\mathcal{N} = 4$ SYM case carries over directly to this more general setting. In the $\mathcal{N}=1$ quiver SCFTs with gauge algebra $\mathfrak{su}(N)^K$, $\breve{h}_3$ and $\breve{g}_3$ are field strengths of gauge fields for the diagonal $\mathbb{Z}_N$ electric and magnetic 1-form symmetries, respectively.  The second term encodes the mixed anomaly between a 2-form symmetry and two 0-form symmetries. The last two terms encode mixed anomalies between 1-form symmetries and two 0-form symmetries.

These additional contributions beyond $\breve{h}_3$ and $\breve{g}_3$ are the main distinction from the $\mathcal{N} = 4$ case. We now discuss in further detail their stringy origins as well some of their properties.

\subsection{Discrete 0-form symmetries}
Let us now investigate the discrete 0-form symmetries more carefully. Denote the three classes of 0-form symmetries as $\mathbf{E}_{(i)}, \mathbf{B}^{(i)}$ and $\mathbf{C}^{(i)}$, with respective background fields $\breve{E}_{1(i)},\breve{B}^{1(i)}$ and $\breve{C}^{1(i)}$. As we list in table \ref{tab:charged operators of N=1 scfts}, $\mathbf{B}$ and $\mathbf{C}$ act on charged local operators in the 4D SCFT constructed respectively by F1- and D1-strings wrapping the cone over a torsional one-cycle $\gamma_1^{(i)}$. According to \cite{Heckman:2022muc}, we can build their symmetry operators with the magnetic dual branes, i.e., NS5- and D5-branes wrapping $M_3\times \gamma_{3{(i)}}$ where $M_3\subset M_4$ are 3D symmetry defects inside the 4D spacetime. The symmetry operators are then given by:
\begin{equation}\label{eq:symmetry operators for BC}
\begin{split}
     \mathcal{U}_{\mathbf{B}^{(i)}}=\exp \left(  i \int_{M_3\times \gamma_{3(i)}} B_6+\cdots \right),\\
    \mathcal{U}_{\mathbf{C}^{(i)}}=\exp \left(  i \int_{M_3\times \gamma_{3(i)}} C_6 +\cdots \right),
\end{split}
\end{equation}
where the $\cdots$ indicate additional terms in the respective Wess-Zumino actions. We defer a treatment of these other terms to future work. These two 0-form symmetry operators are related to each other under the IIB $SL(2,\mathbb{Z})$ duality. For example, the leading order terms $B_6$ and $C_6$ transform as:
\begin{equation}\label{eq:sl2z for b6c6}
\left[
    \begin{array}
[c]{c}%
B_{6}\\
C_{6}%
\end{array}
\right]  \mapsto\left[
\begin{array}
[c]{cc}%
a & b\\
c & d
\end{array}
\right]  \left[
\begin{array}
[c]{c}%
B_{6}\\
C_{6}%
\end{array}
\right] ,
\end{equation}
which gives rise to
\begin{equation}\label{eq:s-duality between 0-form symmetries}
    \mathcal{U}_{\mathbf{B}^{(i)}}\rightarrow \mathcal{U}_{\mathbf{B}^{(i)}}^a\mathcal{U}_{\mathbf{C}^{(i)}}^b, ~\mathcal{U}_{\mathbf{C}^{(i)}}\rightarrow \mathcal{U}_{\mathbf{B}^{(i)}}^c\mathcal{U}_{\mathbf{C}^{(i)}}^d.
\end{equation}
The symmetry operator of $\mathbf{E}_{(i)}$ is built by D3-branes wrapping torsional 1-cycles:
\begin{equation}\label{eq:symmetry operator for E}
    \mathcal{U}_{\mathbf{E}_{(i)}}=\exp \left( i \int_{M_3\times \gamma_1^{(i)}}C_4 +\cdots \right),
\end{equation}
which is obviously self-dual under S-duality due to the self-dual property of D3-branes.

We comment that similar transformation rules were worked out in \cite{Gukov:1998kn} in the special case $X = \mathbb{C}^3 / \mathbb{Z}_3$ with boundary topology $S^5 / \mathbb{Z}_3$. See Appendix \ref{app:orbo} for further discussion on this point, and the relation between the present work and this analysis. As noted in \cite{Gukov:1998kn}, the symmetry generators for $\mathbf{B}^{(i)}$ and $\mathbf{C}^{(i)}$ do not quite commute in the presence of a D3-brane wrapped on a torsional 1-cycle. Essentially the same arguments used in this special case carry over to this more general case as well, and we refer the interested reader to \cite{Gukov:1998kn} for further details.

Let us now turn to the interplay of these 0-form symmetries with our duality defects. We have already seen that the $\mathbf{B}^{(i)}$ and $\mathbf{C}^{(i)}$ symmetry generators transform non-trivially under IIB dualities. As such, we expect that when passing the corresponding operators charged under these symmetries through a duality wall that the transformation will be non-trivial.

\begin{figure}[t!]
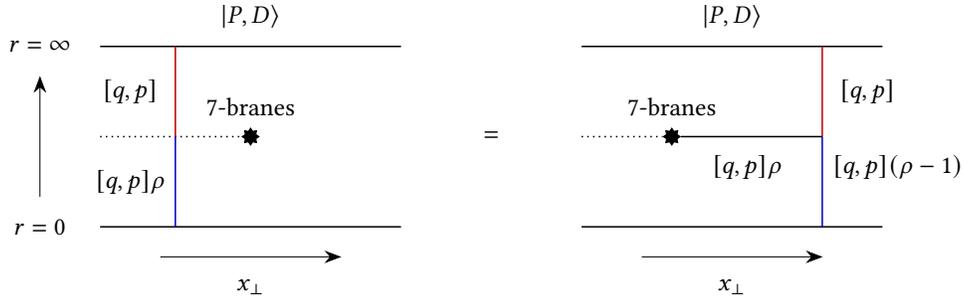

    \centering
    \scalebox{0.8}{

}
    \caption{A $[q,p]$ string wrapping the cone of a torsional 1-cycle at $r=\infty$ corresponds to a local operator of the 4D SCFT at $r=0$. Passing the  duality defect $\mathcal{U}(M_3,\mathfrak{F})$ through the local operator from right to left creates a symmetry line operator of charge $[q,p](\rho-1)$ via a Hanany-Witten transition.}
    \label{fig:HW1}
\end{figure}

Recall that charged operators for $\mathbf{B}^{(i)}$ and $\mathbf{C}^{(i)}$ are respectively non-compact D3-branes, F1- and D1-strings which stretch along the radial direction of the Calabi-Yau cone. To answer what happens when we pull such a brane through the duality wall, it is enough to track the transformation of the respective branes. The action of duality defects on these charged local operators in the 4D SCFT can be read from the Hanany-Witten transition. For example, insert a 7-brane as the duality defect in 4D theory, and consider a string wrapping Cone$(\gamma_1)$ as a local operator with charged  respectively under $\mathbf{B}^{(i)}$ and $\mathbf{C}^{(i)}$. The Hanany-Witten transition between the 7-brane and the string creates a new string attaching them, which is a 1D symmetry operator in the 4D SCFT. See figure \ref{fig:HW1} for an illustration in the case of half-space gauging construction.

This is very similar to our discussion around Figure \ref{fig:HW} how charged line operators transformed through duality defects, which from the string theory point of view obviously have the same origin as the Hanany-Witten transition. As it is for local operators, this is also analogous to the case in 2D CFTs, e.g. the critical Ising model, where the non-invertible symmetry line maps the local spin operator to the disorder operator which lives at the edge of another symmetry line \cite{Verlinde:1988sn}.

One can in principle also consider passing the topological operators for these 0-form symmetries through a duality / triality defect as well. This case is more subtle since it involves an order of limits issue concerning how fast we send the 7-branes to infinity, and how the 5-branes (by)pass the corresponding branch cuts. It would be interesting to study this issue, but it is beyond the scope of the present work.

\section{Conclusions}
\label{sec:CONC}
In this chapter, we have presented a top down construction of duality defects for 4D QFTs engineered via D3-brane probes of isolated Calabi-Yau singularities. In this description, the duality group of type IIB strings descends to a duality of the localized QFT at a special point of the conformal manifold, and topological duality defects are implemented by suitable 7-branes wrapped ``at infinity'' which implement field theoretic duality topological symmetry operators. The branch cuts of the 7-branes directly descend to branch cuts in the 5D symmetry TFT which governs the anomalies of the 4D theory. This provides a uniform perspective on various ``bottom up'' approaches to realizing duality / triality defects from the gauging of 1-form symmetries in presence of a mixed anomaly \cite{Kaidi:2021xfk}
and half-space gauging constructions \cite{Choi:2021kmx, Choi:2022zal}. This uniform perspective applies both to $\mathcal{N} = 4$ SYM theory as well as a number of $\mathcal{N} = 1$ quiver gauge theories tuned to a point of strong coupling. In the remainder of this section we discuss a few avenues which would be natural to consider further.

One of the main items in our analysis is the important role of the branch cuts present in 7-branes, and how these dictate the structure of anomalies in the 5D symmetry TFT, as well as the resulting 3D TFTs localized on the duality defects. In Appendix \ref{app:minimalTFT7branes} we take some preliminary steps in reading this data off directly from dimensional reduction of a 7-brane. It would be interesting to perform further checks on this proposal, perhaps by considering dimensional reduction on other ``internal'' manifolds.

In our extension to $\mathcal{N} = 1$ quiver gauge theories, we focused on the special case where the singularity of the Calabi-Yau cone is isolated. This was more to obtain some technical simplifications rather than there being any fundamental obstacle to performing the same computations. It would be interesting to study the structure of the resulting 5D symmetry TFT, as well as the interplay between 7-branes (and their branch cuts) with non-isolated singularities.

It would also be quite interesting to consider other special points in the conformal manifold of 4D SCFTs realized via D3-brane probes of singularities. If these special points admit a non-trivial automorphism, one could hope to similarly lift this automorphism to a geometric object in a string construction, and thus realize a broader class of symmetry operators.

Our primary focus has been on QFTs realized via D3-branes at singularities, but we have also indicated in Appendix \ref{app:other} how these considerations can be generalized to other top down constructions. In particular, it would be interesting to study the structure of duality defects in theories which cannot be obtained from D3-brane probes of singularities.

%% file: App/A.tex

This appendix is adapted from \cite{Franco:2021vxq}. 

In this appendix, we present two additional examples of Spin(7) manifolds, which are considered in Section~\ref{sec:C4Z2Z2toSPPC} to discuss partial resolutions.

\section{$\CC^4/\ZZ_2\times \ZZ_2$}
\label{app:otheory for c4z2z2}

\begin{figure}[H]
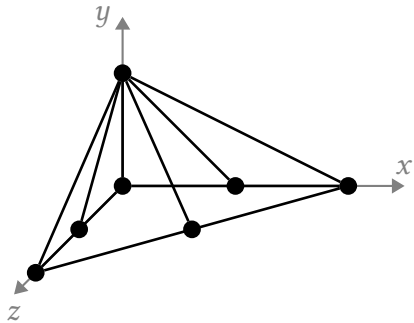

    \centering

	\caption{Toric diagram for $\mathbb{C}^4/\mathbb{Z}_2\times \mathbb{Z}_2$.}
	\label{fig:C4Z2Z2toricdiagram}
\end{figure}

Figure~\ref{fig:C4Z2Z2toricdiagram} shows the toric diagram for the $\CC^4/\ZZ_2\times \ZZ_2$ orbifold.

\begin{figure}[H]
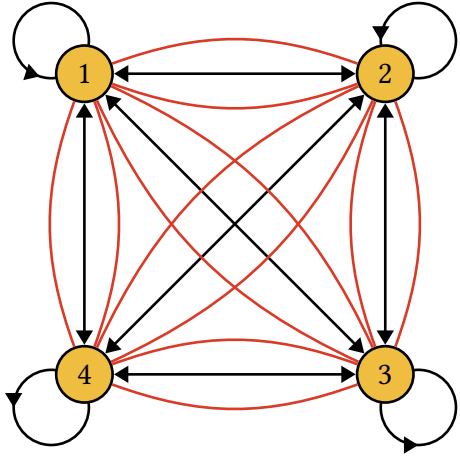
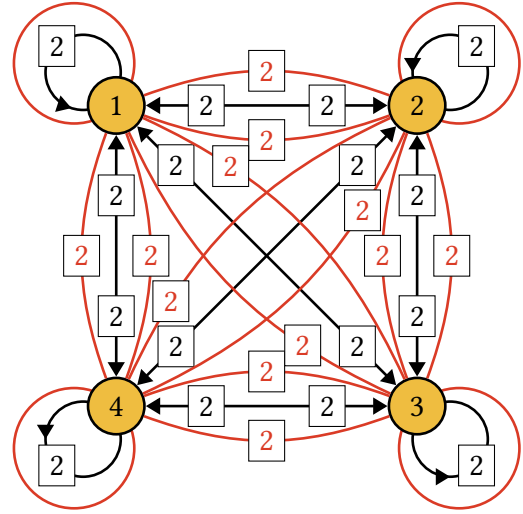

		\centering
 		\begin{subfigure}[t]{0.49\textwidth}
 			\centering
			%
			\caption{$\mathcal{N}=(0,2)$ language.}
			\label{fig:C4Z2Z2N02}
		\end{subfigure}\hfill
		\begin{subfigure}[t]{0.49\textwidth}
			\centering
			%
			\caption{$\mathcal{N}=(0,1)$ language.}
			\label{fig:C4Z2Z2N01}
		\end{subfigure}
		\caption{Quiver diagram for $\mathbb{C}^4/\mathbb{Z}_2\times \mathbb{Z}_2$ in $\mathcal{N}=(0,2)$ and $\mathcal{N}=(0,1)$ language.}
		\label{fig:C4Z2Z2quiv}
	\end{figure}

The gauge theory for D1-branes probing this orbifold was first constructed in \cite{Franco:2015tna}. Its quiver diagram is shown in Figure~\ref{fig:C4Z2Z2N02}.

The $J$- and $E$-terms are
\begin{equation}
    \begin{array}{lclc}
& J                                              &\text{\hspace{.5cm}}& E                         \\
\Lambda_{12} \,:\, & X_{23}X_{31} - X_{24}X_{41} &                    &X_{11}X_{12} - X_{12}X_{22}\\ 
\Lambda_{21} \,:\, & X_{14}X_{42} - X_{13}X_{32} &                    &X_{22}X_{21} - X_{21}X_{11}\\ 
\Lambda_{13} \,:\, & X_{34}X_{41} - X_{32}X_{21} &                    &X_{11}X_{13} - X_{13}X_{33}\\ 
\Lambda_{31} \,:\, & X_{12}X_{23} - X_{14}X_{43} &                    &X_{33}X_{31} - X_{31}X_{11}\\ 
\Lambda_{14} \,:\, & X_{42}X_{21} - X_{43}X_{31} &                    &X_{11}X_{14} - X_{14}X_{44}\\ 
\Lambda_{41} \,:\, & X_{13}X_{34} - X_{12}X_{24} &                    &X_{44}X_{41} - X_{41}X_{11}\\ 
\Lambda_{23} \,:\, & X_{31}X_{12} - X_{34}X_{42} &                    &X_{22}X_{23} - X_{23}X_{33}\\ 
\Lambda_{32} \,:\, & X_{24}X_{43} - X_{21}X_{13} &                    &X_{33}X_{32} - X_{32}X_{22}\\ 
\Lambda_{24} \,:\, & X_{43}X_{32} - X_{41}X_{12} &                    &X_{22}X_{24} - X_{24}X_{44}\\ 
\Lambda_{42} \,:\, & X_{21}X_{14} - X_{23}X_{34} &                    &X_{44}X_{42} - X_{42}X_{22}\\ 
\Lambda_{34} \,:\, & X_{41}X_{13} - X_{42}X_{23} &                    &X_{33}X_{34} - X_{34}X_{44}\\
\Lambda_{43} \,:\, & X_{32}X_{24} - X_{31}X_{14} &                    &X_{44}X_{43} - X_{43}X_{33}
\end{array}
\label{J_E_C4/Z2xZ2}
\end{equation}

Figure~\ref{fig:C4Z2Z2N01} shows the quiver for this theory in $\mathcal{N}= (0,1)$ language. The $W^{(0,1)}$ associated to \eqref{J_E_C4/Z2xZ2} is
\begin{equation}
    \begin{split}
        W^{(0,1)} = & \, W^{(0,2)}+ \sum_{\substack{i,j=1\\i\neq j}}^4\Lambda^{R}_{ii}\left(X^\dagger_{ij}X_{ij}+X^\dagger_{ji}X_{ji}\right)+\sum_{i=1}^4\Lambda^{R}_{ii}X^\dagger_{ii}X_{ii}\fstop
    \end{split}
\end{equation}

Table~\ref{tab:GenerC4Z2Z2} shows the generators of the moduli space and their expression in terms of chiral fields.

 \begin{table}[!htp]
	\centering
	\renewcommand{\arraystretch}{1.1}
	\begin{tabular}{c|c}
		Meson    & Chiral fields  \\
		\hline
		$M_1$    & $X_{12}X_{21}=X_{43}X_{34}$ \\
		$M_2$    & $X_{13}X_{31}=X_{24}Z_{42}$\\
		$M_3$    & $X_{14}X_{41}=X_{23}X_{32}$\\
		$M_4$    & $X_{14}X_{42}X_{21}=X_{23}X_{31}X_{12}=X_{32}X_{24}X_{43}=X_{41}X_{13}X_{34}=$\\ 
		         & $=X_{14}X_{43}X_{31}=X_{23}X_{34}X_{42}=X_{32}X_{21}X_{13}=X_{41}X_{12}X_{24}$\\
		$M_5$    & $X_{11}=X_{22}=X_{33}=X_{44}$\\
	\end{tabular}
	\caption{Generators of $\CC^4/\ZZ_2\times \ZZ_2$.}
	\label{tab:GenerC4Z2Z2}
\end{table}

They are subject to the following relation
\begin{equation}
    \mathcal{I}=\left\langle M_1M_2M_3=M_4^2\right\rangle\fstop
\end{equation}

\bigskip

\paragraph{$\SO(N)\times \U(N)\times \SO(N)$ Orientifold}\mbox{}

\medskip

Let us consider an anti-holomorphic involution which acts on Figure~\ref{fig:C4Z2Z2quiv} as a reflection with respect to the diagonal connecting nodes 1 and 3. Then, nodes 1 and 3 map to themselves, while nodes 2 and 4 are identified.

The involution on chiral fields is
 \begin{equation}\label{eq:C4Z2Z2chiral}
\begin{array}{cccccccccccc}
 	 X_{12} & \rightarrow &   \gamma_{\Omega_1}\bar{X}_{14}\gamma_{\Omega_4}^{-1}\coma & 
 	 X_{21} & \rightarrow &  \gamma_{\Omega_4}\bar{X}_{41}\gamma_{\Omega_1}^{-1} \coma  & 
 	 X_{13} & \rightarrow &   \gamma_{\Omega_1}\bar{X}_{13}\gamma_{\Omega_3}^{-1} \coma & 
 	 X_{31} & \rightarrow &   \gamma_{\Omega_3}\bar{X}_{31}\gamma_{\Omega_1}^{-1} \coma \\
 	 X_{14} & \rightarrow &   \gamma_{\Omega_1}\bar{X}_{12}\gamma_{\Omega_2}^{-1}\coma & 
 	 X_{41} & \rightarrow &  \gamma_{\Omega_2}\bar{X}_{21}\gamma_{\Omega_1}^{-1} \coma  & 
 	 X_{43} & \rightarrow &   \gamma_{\Omega_2}\bar{X}_{23}\gamma_{\Omega_3}^{-1} \coma & 
 	 X_{34} & \rightarrow &   \gamma_{\Omega_3}\bar{X}_{32}\gamma_{\Omega_2}^{-1} \coma \\
 	 X_{24} & \rightarrow &   \gamma_{\Omega_4}\bar{X}_{42}\gamma_{\Omega_2}^{-1}\coma & 
 	 X_{42} & \rightarrow &  \gamma_{\Omega_2}\bar{X}_{24}\gamma_{\Omega_4}^{-1} \coma  & 
 	 X_{23} & \rightarrow &   \gamma_{\Omega_4}\bar{X}_{43}\gamma_{\Omega_3}^{-1} \coma & 
 	 X_{32} & \rightarrow &   \gamma_{\Omega_3}\bar{X}_{34}\gamma_{\Omega_4}^{-1} \coma \\
 	 X_{11} & \rightarrow &   -\gamma_{\Omega_1}\bar{X}_{11}\gamma_{\Omega_1}^{-1}\coma & 
 	 X_{22} & \rightarrow &  -\gamma_{\Omega_4}\bar{X}_{44}\gamma_{\Omega_4}^{-1} \coma  & 
 	 X_{33} & \rightarrow &  -\gamma_{\Omega_3}\bar{X}_{33}\gamma_{\Omega_3}^{-1} \coma & 
 	 X_{44} & \rightarrow &   -\gamma_{\Omega_2}\bar{X}_{22}\gamma_{\Omega_2}^{-1} \fstop
 \end{array}
 \end{equation}

From the invariance of $W^{(0,1)}$, we obtain the transformations of the Fermi fields
 \begin{equation}\label{eq:C4Z2Z2fermi}
     \begin{array}{cccccccccccc}
 	 \Lambda_{12} & \rightarrow &   -\gamma_{\Omega_1}\bar{\Lambda}_{14}\gamma_{\Omega_4}^{-1}\coma & 
 	 \Lambda_{21} & \rightarrow &  -\gamma_{\Omega_4}\bar{\Lambda}_{41}\gamma_{\Omega_1}^{-1} \coma  & 
 	 \Lambda_{13} & \rightarrow &   -\gamma_{\Omega_1}\bar{\Lambda}_{13}\gamma_{\Omega_3}^{-1} \coma & 
 	 \Lambda_{31} & \rightarrow &   -\gamma_{\Omega_3}\bar{\Lambda}_{31}\gamma_{\Omega_1}^{-1} \coma \\
 	 \Lambda_{14} & \rightarrow &   -\gamma_{\Omega_1}\bar{\Lambda}_{12}\gamma_{\Omega_2}^{-1}\coma & 
 	 \Lambda_{41} & \rightarrow &  -\gamma_{\Omega_2}\bar{\Lambda}_{21}\gamma_{\Omega_1}^{-1} \coma  & 
 	 \Lambda_{43} & \rightarrow &   -\gamma_{\Omega_2}\bar{\Lambda}_{23}\gamma_{\Omega_3}^{-1} \coma & 
 	 \Lambda_{34} & \rightarrow &   -\gamma_{\Omega_3}\bar{\Lambda}_{32}\gamma_{\Omega_2}^{-1} \coma \\
 	 \Lambda_{24} & \rightarrow &   -\gamma_{\Omega_4}\bar{\Lambda}_{42}\gamma_{\Omega_2}^{-1}\coma & 
 	 \Lambda_{42} & \rightarrow &  -\gamma_{\Omega_2}\bar{\Lambda}_{24}\gamma_{\Omega_4}^{-1} \coma  & 
 	 \Lambda_{23} & \rightarrow &   -\gamma_{\Omega_4}\bar{\Lambda}_{43}\gamma_{\Omega_3}^{-1} \coma & 
 	 \Lambda_{32} & \rightarrow &   -\gamma_{\Omega_3}\bar{\Lambda}_{34}\gamma_{\Omega_4}^{-1} \coma 
 \end{array}
 \end{equation}
 and
\begin{equation}\label{eq:C4Z2Z2realfermi}
 \begin{array}{cccccccccccc}
 	 \Lambda^{R}_{11} & \rightarrow &   \gamma_{\Omega_3}\Lambda^{R\,\,T}_{11}\gamma_{\Omega_3}^{-1}\coma & 
 	 \Lambda^{R}_{22} & \rightarrow &  \gamma_{\Omega_2}\Lambda^{R\,\,T}_{44}\gamma_{\Omega_2}^{-1} \coma  & 
 	 \Lambda^{R}_{33} & \rightarrow &   \gamma_{\Omega_1}\Lambda^{R\,\,T}_{33}\gamma_{\Omega_1}^{-1} \coma & 
 	 \Lambda^{R}_{44} & \rightarrow &   \gamma_{\Omega_4}\Lambda^{R\,\,T}_{22}\gamma_{\Omega_4}^{-1} \fstop
 \end{array}
 \end{equation}

\begin{figure}[H]
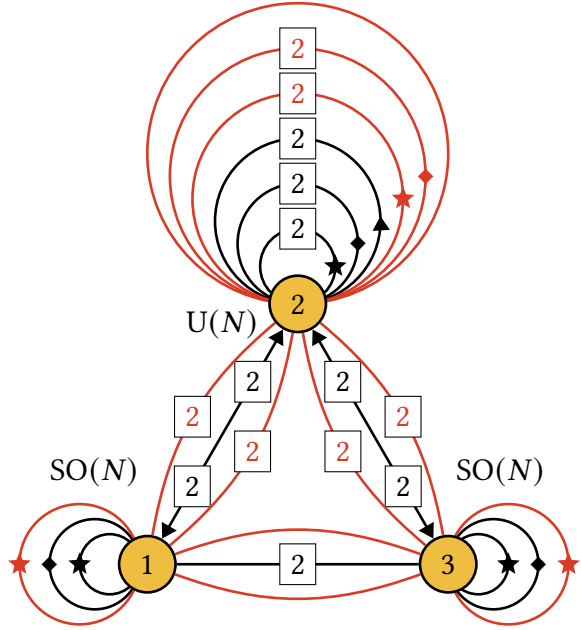

    \centering
    %
    \caption{Quiver for the  Spin(7) orientifold of $\mathbb{C}^4/\mathbb{Z}_2\times \mathbb{Z}_2$ using the involution in \eqref{eq:C4Z2Z2chiral}, \eqref{eq:C4Z2Z2fermi} and \eqref{eq:C4Z2Z2realfermi}.}
    \label{fig:o_theory_C4Z2Z2}
\end{figure}

 Using Table~\ref{tab:GenerC4Z2Z2}, the transformation in the field theory translates into the geometric involution $\sigma$ acting on the generators as
 \begin{equation}
 \begin{array}{c}
     \left(M_1,M_2,M_3,M_4,M_5\right)\\
   \downarrow\\
   \left(\bar{M}_3,\bar{M}_2,\bar{M}_1,\bar{M}_4,-\bar{M}_5\right)\coma
 \end{array}
 \end{equation}
 which, as expected, is different from and inequivalent to the universal involution $M_a\rightarrow \bar{M}_a$. The quiver for the resulting theory is shown in Figure~\ref{fig:o_theory_C4Z2Z2}, which is also free of gauge anomalies. 
 
\section{SPP$\times\CC$}

\label{app:otheory for sppXC}

Figure~\ref{fig:SPPtoricdiagram} shows the toric diagram for SPP$\times\CC$.

\begin{figure}[H]
    \centering

	\caption{Toric diagram for SPP$\times\CC$.}
	\label{fig:SPPtoricdiagram}
\end{figure}

The gauge theory for D1-branes probing this CY$_4$ was introduced in \cite{Franco:2015tna}. Its quiver diagram is shown in Figure~\ref{fig:SPPquivN02}.

\begin{figure}[H]
	\centering
	\begin{subfigure}[t]{0.49\textwidth}
	\centering
	%
	\caption{$\mathcal{N}=(0,2)$ language.}
	\label{fig:SPPquivN02}
	\end{subfigure}\hfill
	\begin{subfigure}[t]{0.49\textwidth}
	\centering
	%
	\caption{$\mathcal{N}=(0,1)$ language.}
	\label{fig:SPPquivN01}
	\end{subfigure}
	\caption{Quiver diagram for SPP$\times\CC$ in $\mathcal{N}=(0,2)$ and $\mathcal{N}=(0,1)$ language.}
	\label{fig:SPPquiv}
\end{figure}

The $J$- and $E$-terms are
\begin{equation}
    \begin{array}{lclc}
                   & J                                       &   \text{\hspace{.5cm}}&   E                     \\
\Lambda_{11} \,:\, & X_{13}X_{31} - X_{12}X_{21}             &                    &   \Phi_{11}X_{11} - X_{11}\Phi_{11} \\
\Lambda_{21} \,:\, & X_{12}X_{23}X_{32} - X_{11}X_{12}       &                    &   \Phi_{22}X_{21} - X_{21}\Phi_{11} \\
\Lambda_{12} \,:\, & X_{21}X_{11} - X_{23}X_{32}X_{21}       &                    &    X_{12}\Phi_{22}-\Phi_{11}X_{12} \\
\Lambda_{31} \,:\, & X_{13}X_{32}X_{23}- X_{11}X_{13}        &                    &    X_{31}\Phi_{11}-\Phi_{33}X_{31} \\
\Lambda_{13} \,:\, & X_{31}X_{11} - X_{32}X_{23}X_{31}       &                    &   \Phi_{11}X_{13} - X_{13}\Phi_{33} \\
\Lambda_{32} \,:\, & X_{21}X_{12}X_{23} -X_{23}X_{31}X_{13}  &                    &   \Phi_{33}X_{32} - X_{32}\Phi_{22} \\
\Lambda_{23} \,:\, & X_{32}X_{21}X_{12} - X_{31}X_{13}X_{32} &                    &   \Phi_{22}X_{23} - X_{23}\Phi_{33} 
\end{array}
\label{eq:SPPxC-JETerms}
\end{equation}

Figure~\ref{fig:SPPquivN01} shows the quiver for this theory in $\mathcal{N}= (0,1)$ language. The $W^{(0,1)}$ associated to \eqref{eq:SPPxC-JETerms} is
\begin{equation}
    W^{(0,1)} = W^{(0,2)}+ \Lambda^R_{11}X^\dagger_{11}X_{11}+ \sum_{\substack{i,j=1\\i\neq j}}^3\Lambda^R_{ii}\left(X^\dagger_{ij}X_{ij}+X^\dagger_{ji}X_{ji}\right)+\sum_{i=1}^3\Lambda^R_{ii}\Phi^\dagger_{ii}\Phi_{ii}\fstop
\end{equation}

\begin{table}[!htp]
   \centering
	\renewcommand{\arraystretch}{1.1}
	\begin{tabular}{c|c}
		Meson    & Chiral fields  \\
		\hline
	$M_1$ & $X_{13}X_{32}X_{21}$ \\	
    $M_2$ & $X_{11}=X_{23}X_{32}$ \\
    $M_3$ & $X_{31}X_{12}X_{23}$ \\
    $M_4$ & $X_{12}X_{21}=X_{13}X_{31}$ \\
    $M_5$ & $\Phi_{11}=\Phi_{22}=\Phi_{33}$ 
    \end{tabular}
    \caption{Generators of SPP$\times\CC$.}
    \label{tab:GenerSPP}
\end{table}

Table~\ref{tab:GenerSPP} shows the generators of SPP$\times\CC$ and their expression in terms of chiral fields. They satisfy the following relation
\begin{equation}
    \mathcal{I}=\left\langle M_1M_3=M_2M_4^2\right\rangle\fstop
\end{equation}

\bigskip

\paragraph{$\SO(N)\times \U(N)$ Orientifold}\mbox{}

\medskip

Let us consider an anti-holomorphic involution which acts on Figure~\ref{fig:SPPquiv} as a reflection with respect to a vertical line going through node 1. Then, node 1 maps to itself, while nodes 2 and 3 are identified.

The involution on chiral fields is
 \begin{equation}\label{eq:SPPchiral}
\begin{array}{cccccccccccc}
 	 X_{12} & \rightarrow &   \gamma_{\Omega_1}\bar{X}_{13}\gamma_{\Omega_3}^{-1}\coma & 
 	 X_{21} & \rightarrow &  \gamma_{\Omega_3}\bar{X}_{31}\gamma_{\Omega_1}^{-1} \coma  & 
 	 X_{13} & \rightarrow &   \gamma_{\Omega_1}\bar{X}_{12}\gamma_{\Omega_2}^{-1} \coma & 
 	 X_{31} & \rightarrow &   \gamma_{\Omega_2}\bar{X}_{21}\gamma_{\Omega_1}^{-1} \coma \\
 	 X_{23} & \rightarrow &   \gamma_{\Omega_3}\bar{X}_{32}\gamma_{\Omega_2}^{-1} \coma & 
 	 X_{32} & \rightarrow &   \gamma_{\Omega_2}\bar{X}_{23}\gamma_{\Omega_3}^{-1} \coma &
 	 X_{11} & \rightarrow &   \gamma_{\Omega_1}\bar{X}_{11}\gamma_{\Omega_1}^{-1}\coma & 
 	 \Phi_{11} & \rightarrow &  -\gamma_{\Omega_1}\bar{\Phi}_{11}\gamma_{\Omega_1}^{-1} \coma \\ 
 & & &
 \Phi_{22} & \rightarrow &  -\gamma_{\Omega_3}\bar{\Phi}_{33}\gamma_{\Omega_3}^{-1} \coma & 
 	 \Phi_{33} & \rightarrow &   -\gamma_{\Omega_2}\bar{\Phi}_{22}\gamma_{\Omega_2}^{-1} \fstop
& & &
 \end{array}
 \end{equation}

From the invariance of $W^{(0,1)}$, we obtain the transformations of the Fermi fields
 \begin{equation}\label{eq:SPPfermi}
     %
 \end{equation}
 
 Using Table~\ref{tab:GenerSPP}, the transformation in the field theory translates into the geometric involution $\sigma$ acting on the generators as
 \begin{equation}
 \begin{array}{c}
     \left(M_1,M_2,M_3,M_4,M_5\right)\\
   \downarrow\\
   \left(\bar{M}_3,\bar{M}_2,\bar{M}_1,\bar{M}_4,-\bar{M}_5\right)\coma
 \end{array}
 \end{equation}
which is different from and inequivalent to the universal involution $M_a\rightarrow \bar{M}_a$. The quiver for the resulting theory is shown in Figure~\ref{fig:o_theory_C4Z2Z2}, which is also free of gauge anomalies.

\begin{figure}[H]
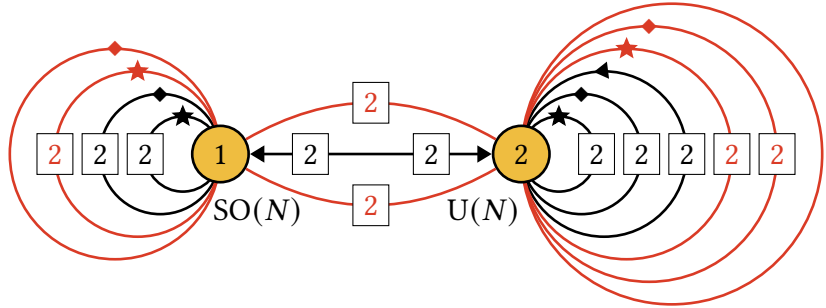

    \centering
    %
    \caption{Quiver for a  Spin(7)  orientifold of $\SPP\times \CC$ using the involution in \eqref{eq:SPPchiral}, \eqref{eq:SPPfermi} and \eqref{eq:SPPrealfermi}.}
    \label{fig:o_theory_SPP}
\end{figure}


%% file: App/B.tex
	
This Appendix is adapted from \cite{Franco:2021ixh}.

In Section \ref{sec:Q111Z2}, we introduced a web of trialities that contains a Spin(7) orientifold of phase D of $Q^{1,1,1}/\ZZ_2$ and summarized it in Figure \ref{fig:01trialityq111z2}. In this appendix, we collect all the relevant information for the other theories in this web.

\section{Phase E}
\label{app:Q111Z2PhaseE}

The quiver for phase E is shown in Figure \ref{fig:Q111Z2quiverE}. 

	\begin{figure}[H]
		\centering

	\caption{Quiver diagram for phase E of $Q^{1,1,1}/\ZZ_2$.}
	\label{fig:Q111Z2quiverE}
	\end{figure}

The corresponding $J$- and $E$-terms are given by
		\begin{alignat}{4}
	\renewcommand{\arraystretch}{1.1}
    & \centermathcell{J}                           &\text{\hspace{.5cm}}& \centermathcell{E                               }\nonumber \\
\Lambda_{86}^1 \,:\,& \centermathcell{X_{64} Y_{41} Y_{18}-Y_{64} Y_{41} X_{18} }& & \centermathcell{Y_{85} X_{56}-X_{85} Z_{56} }\nonumber \\
\Lambda_{86}^2 \,:\,& \centermathcell{Y_{64} X_{41} X_{18}-X_{64} X_{41} Y_{18} }& & \centermathcell{Y_{85} W_{56}-X_{85} Y_{56} }\nonumber \\
\Lambda_{26}^1 \,:\,& \centermathcell{Y_{64} Y_{47} X_{72}-X_{64} Y_{47} Y_{72} }& & \centermathcell{Y_{25} X_{56}-X_{25} W_{56} }\nonumber \\
\Lambda_{26}^2 \,:\,& \centermathcell{X_{64} X_{47} Y_{72}-Y_{64} X_{47} X_{72} }& & \centermathcell{Y_{25} Z_{56}-X_{25} Y_{56} }\nonumber \\
\Lambda_{75}^1 \,:\,& \centermathcell{Y_{56} Y_{64} X_{47}-W_{56} Y_{64} Y_{47} }& & \centermathcell{X_{73} X_{35}-X_{72} X_{25} }\nonumber \\
\Lambda_{75}^2 \,:\,& \centermathcell{Z_{56} X_{64} X_{47}-X_{56} X_{64} Y_{47} }& & \centermathcell{Y_{73} Y_{35}-Y_{72} Y_{25} }\nonumber \\
\Lambda_{75}^3 \,:\,& \centermathcell{X_{56} Y_{64} Y_{47}-Z_{56} Y_{64} X_{47} }& & \centermathcell{Y_{73} X_{35}-X_{72} Y_{25} }\nonumber \\
\Lambda_{75}^4 \,:\,& \centermathcell{W_{56} X_{64} Y_{47}-X_{64} Y_{56} X_{47} }& & \centermathcell{X_{73} Y_{35}-Y_{72} X_{25} }\label{eq:Q111Z2JEtermsE} \\
\Lambda_{43}^1 \,:\,& \centermathcell{Y_{35} Y_{56} X_{64}-X_{35} Y_{56} Y_{64} }& & \centermathcell{X_{47} X_{73}-X_{41} X_{13} }\nonumber \\
\Lambda_{43}^2 \,:\,& \centermathcell{Y_{35} X_{56} X_{64}-X_{35} X_{56} Y_{64} }& & \centermathcell{Y_{47} Y_{73}-Y_{41} Y_{13} }\nonumber  \\
\Lambda_{43}^3 \,:\,& \centermathcell{X_{35} W_{56} Y_{64}-Y_{35} W_{56} X_{64} }& & \centermathcell{Y_{47} X_{73}-X_{41} Y_{13} }\nonumber \\
\Lambda_{43}^4 \,:\,& \centermathcell{X_{35} Z_{56} Y_{64}-Y_{35} Z_{56} X_{64} }& & \centermathcell{X_{47} Y_{73}-Y_{41} X_{13} }\nonumber \\
\Lambda_{15}^1 \,:\,& \centermathcell{Z_{56} Y_{64} Y_{41}-Y_{56} Y_{64} X_{41} }& & \centermathcell{X_{13} X_{35}-X_{18} X_{85} }\nonumber \\
\Lambda_{15}^2 \,:\,& \centermathcell{X_{56} X_{64} Y_{41}-W_{56} X_{64} X_{41} }& & \centermathcell{Y_{13} Y_{35}-Y_{18} Y_{85} }\nonumber \\
\Lambda_{15}^3 \,:\,& \centermathcell{W_{56} Y_{64} X_{41}-X_{56} Y_{64} Y_{41} }& & \centermathcell{Y_{13} X_{35}-X_{18} Y_{85} }\nonumber \\
\Lambda_{15}^4 \,:\,& \centermathcell{X_{64} Y_{56} X_{41}-X_{64} Z_{56} Y_{41} }& & \centermathcell{X_{13} Y_{35}-Y_{18} X_{85} }\nonumber 	
\end{alignat}
	
	Finally, the generators of the moduli space expressed in terms of the chiral fields are listed in Table~\ref{tab:GenQ111Z2PhaseE}. 

\medskip
	
\paragraph{$\U(N)^2\times \SO(N)^4$ orientifold} \mbox{}

\medskip

Let us consider an anti-holomorphic involution of phase E which acts on the nodes in Figure \ref{fig:Q111Z2quiverE} as $1\leftrightarrow 7$, $2\leftrightarrow 8$ and maps all other nodes mapped to themselves. Chiral fields transform according to
  \begin{equation}
\begin{array}{cccccccccccc}
		X_{41} & \rightarrow & \gamma_{\Omega_4}\bar{X}_{47}\gamma_{\Omega_7}^{-1}\coma &
		Y_{41} & \rightarrow & \gamma_{\Omega_4}\bar{Y}_{47}\gamma_{\Omega_7}^{-1}\coma &
		X_{72} & \rightarrow & \gamma_{\Omega_1}\bar{Y}_{18}\gamma_{\Omega_8}^{-1}\coma &
		Y_{72} & \rightarrow & \gamma_{\Omega_1}\bar{X}_{18}\gamma_{\Omega_8}^{-1}\coma \\ 
		X_{18} & \rightarrow & \gamma_{\Omega_7}\bar{Y}_{72}\gamma_{\Omega_2}^{-1}\coma &
		Y_{18} & \rightarrow & \gamma_{\Omega_7}\bar{X}_{72}\gamma_{\Omega_2}^{-1}\coma &
		X_{64} & \rightarrow & \gamma_{\Omega_6}\bar{Y}_{64}\gamma_{\Omega_4}^{-1}\coma &
		Y_{64} & \rightarrow & \gamma_{\Omega_6}\bar{X}_{64}\gamma_{\Omega_4}^{-1}\coma \\
		X_{25} & \rightarrow & \gamma_{\Omega_8}\bar{X}_{85}\gamma_{\Omega_5}^{-1}\coma &
		Y_{25} & \rightarrow & \gamma_{\Omega_8}\bar{Y}_{85}\gamma_{\Omega_5}^{-1}\coma &
		X_{85} & \rightarrow & \gamma_{\Omega_2}\bar{X}_{25}\gamma_{\Omega_5}^{-1}\coma &
		Y_{85} & \rightarrow & \gamma_{\Omega_2}\bar{Y}_{25}\gamma_{\Omega_5}^{-1}\coma \\
		X_{35} & \rightarrow & \gamma_{\Omega_3}\bar{Y}_{35}\gamma_{\Omega_5}^{-1}\coma &
		Y_{35} & \rightarrow & \gamma_{\Omega_3}\bar{X}_{35}\gamma_{\Omega_5}^{-1}\coma &
		X_{47} & \rightarrow & \gamma_{\Omega_4}\bar{X}_{41}\gamma_{\Omega_1}^{-1}\coma &
		Y_{47} & \rightarrow & \gamma_{\Omega_4}\bar{Y}_{41}\gamma_{\Omega_1}^{-1}\coma \\
		X_{56} & \rightarrow & \gamma_{\Omega_5}\bar{X}_{56}\gamma_{\Omega_6}^{-1}\coma &
		Y_{56} & \rightarrow & \gamma_{\Omega_5}\bar{Y}_{56}\gamma_{\Omega_6}^{-1}\coma &
		Z_{56} & \rightarrow & \gamma_{\Omega_5}\bar{W}_{56}\gamma_{\Omega_6}^{-1}\coma &
		W_{56} & \rightarrow & \gamma_{\Omega_5}\bar{Z}_{56}\gamma_{\Omega_6}^{-1}\coma \\
\end{array}  
	\label{eq:Q111Z2E-chiral-invol}
\end{equation}

Requiring the invariance of $W^{0,1}$, the Fermi fields transform as
\begin{equation}
	\begin{array}{cccccccccccc}
		\Lambda_{86}^1 & \rightarrow &  \gamma_{\Omega_2}\bar{\Lambda}^1_{26}\gamma_{\Omega_6}^{-1}\coma &
		\Lambda_{86}^2 & \rightarrow &  \gamma_{\Omega_2}\bar{\Lambda}^2_{26}\gamma_{\Omega_6}^{-1}\coma &
		\Lambda_{26}^1 & \rightarrow &  \gamma_{\Omega_8}\bar{\Lambda}^1_{86}\gamma_{\Omega_6}^{-1}\coma &
		\Lambda_{26}^2 & \rightarrow &  \gamma_{\Omega_8}\bar{\Lambda}^2_{86}\gamma_{\Omega_6}^{-1}\coma \\ 
		\Lambda_{75}^1 & \rightarrow &  \gamma_{\Omega_1}\bar{\Lambda}^4_{15}\gamma_{\Omega_5}^{-1}\coma &
		\Lambda_{75}^2 & \rightarrow &  \gamma_{\Omega_1}\bar{\Lambda}^3_{15}\gamma_{\Omega_5}^{-1}\coma &
		\Lambda_{75}^3 & \rightarrow &  \gamma_{\Omega_1}\bar{\Lambda}^2_{15}\gamma_{\Omega_5}^{-1}\coma &
		\Lambda_{75}^4 & \rightarrow &  \gamma_{\Omega_1}\bar{\Lambda}^1_{15}\gamma_{\Omega_5}^{-1}\coma \\
		\Lambda_{43}^1 & \rightarrow & -\gamma_{\Omega_4}\bar{\Lambda}^2_{43}\gamma_{\Omega_3}^{-1}\coma &
		\Lambda_{43}^2 & \rightarrow & -\gamma_{\Omega_4}\bar{\Lambda}^1_{43}\gamma_{\Omega_3}^{-1}\coma &
		\Lambda_{43}^3 & \rightarrow & -\gamma_{\Omega_4}\bar{\Lambda}^4_{43}\gamma_{\Omega_3}^{-1}\coma &
		\Lambda_{43}^4 & \rightarrow & -\gamma_{\Omega_4}\bar{\Lambda}^3_{43}\gamma_{\Omega_3}^{-1}\coma \\
		\Lambda_{15}^1 & \rightarrow &  \gamma_{\Omega_7}\bar{\Lambda}^4_{75}\gamma_{\Omega_5}^{-1}\coma &
		\Lambda_{15}^2 & \rightarrow &  \gamma_{\Omega_7}\bar{\Lambda}^3_{75}\gamma_{\Omega_5}^{-1}\coma &
		\Lambda_{15}^3 & \rightarrow &  \gamma_{\Omega_7}\bar{\Lambda}^2_{75}\gamma_{\Omega_5}^{-1}\coma &
		\Lambda_{15}^4 & \rightarrow &  \gamma_{\Omega_7}\bar{\Lambda}^1_{75}\gamma_{\Omega_5}^{-1}\coma
	\end{array}   
	\label{eq:Q111Z2E-Fermi-invol}
\end{equation}
and
\begin{equation}
	\begin{array}{cccccccccccc}
		\Lambda_{11}^R & \rightarrow & \gamma_{\Omega_7}\Lambda_{77}^{R\,\,T}\gamma_{\Omega_7}^{-1} \coma &
		\Lambda_{22}^R & \rightarrow & \gamma_{\Omega_8}\Lambda_{88}^{R\,\,T}\gamma_{\Omega_8}^{-1} \coma &
		\Lambda_{33}^R & \rightarrow & \gamma_{\Omega_3}\Lambda_{33}^{R\,\,T}\gamma_{\Omega_3}^{-1} \coma &
		\Lambda_{44}^R & \rightarrow & \gamma_{\Omega_4}\Lambda_{44}^{R\,\,T}\gamma_{\Omega_4}^{-1} \coma \\
		\Lambda_{55}^R & \rightarrow & \gamma_{\Omega_5}\Lambda_{55}^{R\,\,T}\gamma_{\Omega_5}^{-1} \coma &
		\Lambda_{66}^R & \rightarrow & \gamma_{\Omega_6}\Lambda_{66}^{R\,\,T}\gamma_{\Omega_6}^{-1} \coma &
		\Lambda_{77}^R & \rightarrow & \gamma_{\Omega_1}\Lambda_{11}^{R\,\,T}\gamma_{\Omega_1}^{-1} \coma &
		\Lambda_{88}^R & \rightarrow & \gamma_{\Omega_2}\Lambda_{22}^{R\,\,T}\gamma_{\Omega_2}^{-1} \fstop
	\end{array}   
	\label{eq:Q111Z2E-RFermi-invol}
\end{equation}

Using Table \ref{tab:GenQ111Z2PhaseE}, the corresponding geometric involution acting on the generators reads 
\begin{equation}
     \begin{array}{ccccccccccccccc}
	M_{1}  & \rightarrow & \bar{M}_{27} \coma &
	M_{2}  & \rightarrow & \bar{M}_{24} \coma &
	M_{3}  & \rightarrow & \bar{M}_{21} \coma &
	M_{4}  & \rightarrow & \bar{M}_{18} \coma &
	M_{5}  & \rightarrow & \bar{M}_{12} \coma \\
	M_{6}  & \rightarrow & \bar{M}_{6} \coma &
	M_{7}  & \rightarrow & \bar{M}_{26} \coma &
	M_{8}  & \rightarrow & \bar{M}_{23} \coma &
	M_{9}  & \rightarrow & \bar{M}_{20} \coma &
	M_{10} & \rightarrow & \bar{M}_{17} \coma \\
	M_{11} & \rightarrow & \bar{M}_{11} \coma &
	M_{12} & \rightarrow & \bar{M}_{5} \coma &
	M_{13} & \rightarrow & \bar{M}_{25} \coma &
	M_{14} & \rightarrow & \bar{M}_{22} \coma &
	M_{15} & \rightarrow & \bar{M}_{19} \coma \\
	M_{16} & \rightarrow & \bar{M}_{16} \coma &
	M_{17} & \rightarrow & \bar{M}_{10} \coma &
	M_{18} & \rightarrow & \bar{M}_{4} \coma &
	M_{19} & \rightarrow & \bar{M}_{15} \coma &
	M_{20} & \rightarrow & \bar{M}_{9} \coma \\ 
	M_{21} & \rightarrow & \bar{M}_{3}\coma &
	M_{22} & \rightarrow & \bar{M}_{14}\coma &
	M_{23} & \rightarrow & \bar{M}_{8}\coma &
	M_{24} & \rightarrow & \bar{M}_{2}\coma &
	M_{25} & \rightarrow & \bar{M}_{13}\coma \\
	\multicolumn{15}{c}{M_{26}   \rightarrow  \bar{M}_{7}\coma 
	M_{27}  \rightarrow  \bar{M}_{1}\fstop}
	\end{array}   
	\label{eq:Q111Z2E-Meson-invol}
    \end{equation}
This geometric involution is the same of~\eqref{eq:Q111Z2D-Meson-invol}.
 
The $\gamma_{\Omega_i}$ matrices are constrained as in \eqref{gamma_matrices_Q111/Z2}. As for phase D, we choose
\begin{equation}
\gamma_{\Omega_3}=\gamma_{\Omega_4}=\gamma_{\Omega_5}=\gamma_{\Omega_6}=\ID_N \fstop 
\end{equation}

The resulting orientifold is shown in Figure \ref{fig:orien_Q111Z2quiverE}.

\begin{figure}[H]
		\centering
		%
	\caption{Quiver diagram for the Spin(7) orientifold of phase E of $Q^{1,1,1}/\ZZ_2$ using the involution in \cref{eq:Q111Z2E-chiral-invol,eq:Q111Z2E-Fermi-invol,eq:Q111Z2E-RFermi-invol}, together with our choice of $\gamma_{\Omega_i}$ matrices.}
		\label{fig:orien_Q111Z2quiverE}
	\end{figure}

\section{Phase H}
\label{app:Q111Z2PhaseH}

The quiver for phase H is shown in Figure \ref{fig:Q111Z2quiverH}. 

\begin{figure}[H]
		\centering

	\caption{Quiver diagram for phase H of $Q^{1,1,1}/\ZZ_2$.}
	\label{fig:Q111Z2quiverH}
	\end{figure}

The $J$- and $E$-terms are
		\begin{alignat}{4}
	\renewcommand{\arraystretch}{1.1}
    & \centermathcell{J}                           &\text{\hspace{.5cm}}& \centermathcell{E                               }\nonumber \\
\Lambda_{26}^{1} \,:\,& \centermathcell{Y_{64} Z_{42}-X_{64} W_{42} }& & \centermathcell{Y_{25} X_{56}-X_{25} Z_{56} }\nonumber\\
\Lambda_{26}^{7} \,:\,& \centermathcell{Y_{64} X_{42}-Y_{42} X_{64} }& & \centermathcell{X_{25} W_{56}-Y_{25} Y_{56} }\nonumber\\
\Lambda_{27}^{1} \,:\,& \centermathcell{X_{74} W_{42}-Y_{74} Y_{42} }& & \centermathcell{Y_{25} X_{53} X_{37}-X_{25} X_{53} Y_{37} }\nonumber\\
\Lambda_{27}^{7} \,:\,& \centermathcell{X_{74} Z_{42}-Y_{74} X_{42} }& & \centermathcell{X_{25} Y_{53} Y_{37}-Y_{25} Y_{53} X_{37} }\nonumber\\
\Lambda_{68}^{1} \,:\,& \centermathcell{X_{85} W_{56}-Y_{85} Z_{56} }& & \centermathcell{X_{64} Y_{48}-Y_{64} X_{48} }\nonumber\\
\Lambda_{68}^{7} \,:\,& \centermathcell{X_{85} Y_{56}-Y_{85} X_{56} }& & \centermathcell{Y_{64} Z_{48}-X_{64} W_{48} }\nonumber\\
\Lambda_{81}^{1} \,:\,& \centermathcell{X_{14} W_{48}-Y_{14} Y_{48} }& & \centermathcell{X_{85} X_{53} Y_{31}-Y_{85} X_{53} X_{31} }\nonumber\\
\Lambda_{81}^{7} \,:\,& \centermathcell{X_{14} Z_{48}-Y_{14} X_{48} }& & \centermathcell{Y_{85} Y_{53} X_{31}-X_{85} Y_{53} Y_{31} }\nonumber\\
\Lambda_{54}^{1} \,:\,& \centermathcell{W_{48} Y_{85}-W_{42} Y_{25} }& & \centermathcell{X_{53} X_{31} X_{14}-X_{56} X_{64} }\nonumber\\
\Lambda_{54}^{7} \,:\,& \centermathcell{Z_{48} Y_{85}-Z_{42} Y_{25} }& & \centermathcell{X_{56} Y_{64}-Y_{53} X_{31} X_{14} }\label{eq:Q111Z2JEtermsH}\\
\Lambda_{54}^{3} \,:\,& \centermathcell{W_{48} X_{85}-Y_{42} Y_{25} }& & \centermathcell{Y_{56} X_{64}-X_{53} Y_{31} X_{14} }\nonumber\\
\Lambda_{54}^{4} \,:\,& \centermathcell{Z_{48} X_{85}-X_{42} Y_{25} }& & \centermathcell{Y_{53} X_{37} Y_{74}-Y_{56} Y_{64} }\nonumber\\
\Lambda_{54}^{5} \,:\,& \centermathcell{Y_{48} Y_{85}-W_{42} X_{25} }& & \centermathcell{Z_{56}X_{64}-X_{53} X_{31} Y_{14} }\nonumber\\
\Lambda_{54}^{6} \,:\,& \centermathcell{X_{48} Y_{85}-Z_{42} X_{25} }& & \centermathcell{Y_{53} Y_{37} X_{74}-Z_{56} Y_{64} }\nonumber\\
\Lambda_{54}^{2} \,:\,& \centermathcell{Y_{48} X_{85}-Y_{42} X_{25} }& & \centermathcell{X_{53} Y_{37} Y_{74}-W_{56} X_{64} }\nonumber\\
\Lambda_{54}^{8} \,:\,& \centermathcell{X_{48} X_{85}-X_{42} X_{25} }& & \centermathcell{W_{56} Y_{64}-Y_{53} Y_{37} Y_{74}  }\nonumber\\
\Lambda_{43}^{1} \,:\,& \centermathcell{Y_{31} Y_{14}-Y_{37} Y_{74} }& & \centermathcell{Y_{48} X_{85} X_{53}-X_{48} X_{85} Y_{53} }\nonumber\\
\Lambda_{43}^{7} \,:\,& \centermathcell{X_{31} Y_{14}-Y_{37} X_{74} }& & \centermathcell{X_{48} Y_{85} Y_{53}-W_{42} X_{25} X_{53} }\nonumber\\
\Lambda_{43}^{3} \,:\,& \centermathcell{Y_{31} X_{14}-X_{37} Y_{74} }& & \centermathcell{Z_{48} X_{85} Y_{53}-Y_{42} Y_{25} X_{53} }\nonumber\\
\Lambda_{43}^{4} \,:\,& \centermathcell{X_{31} X_{14}-X_{37} X_{74} }& & \centermathcell{W_{42} Y_{25} X_{53}-Z_{42} Y_{25} Y_{53} }\nonumber	
\end{alignat}
The generators of the moduli space expressed in terms of the chiral fields are listed in Table \ref{tab:GenQ111Z2PhaseH}.

\medskip

\paragraph{$\U(N)^2\times \SO(N)^4$ orientifold}\mbox{}

\medskip 

Let us consider an anti-holomorphic involution of phase H which acts on the nodes in Figure \ref{fig:Q111Z2quiverH} as $1\leftrightarrow 7$ and $2\leftrightarrow 8$ and maps all other nodes mapped to themselves. Chiral fields transform according to
\begin{equation}
\begin{array}{cccccccccccc}
		Y_{64}& \rightarrow & \gamma_{\Omega_6}\bar{X}_{64}\gamma_{\Omega_4}^{-2} \coma &
		X_{64}& \rightarrow & \gamma_{\Omega_6}\bar{Y}_{64}\gamma_{\Omega_4}^{-2} \coma &
		Z_{42}& \rightarrow & \gamma_{\Omega_4}\bar{Y}_{48}\gamma_{\Omega_8}^{-2} \coma &
		Y_{48}& \rightarrow & \gamma_{\Omega_4}\bar{Z}_{42}\gamma_{\Omega_2}^{-2} \coma \\ 
		W_{42}& \rightarrow & \gamma_{\Omega_4}\bar{X}_{48}\gamma_{\Omega_8}^{-2} \coma &
		X_{48}& \rightarrow & \gamma_{\Omega_4}\bar{W}_{42}\gamma_{\Omega_2}^{-2} \coma &
		Y_{25}& \rightarrow & \gamma_{\Omega_8}\bar{X}_{85}\gamma_{\Omega_5}^{-2} \coma &
		X_{85}& \rightarrow & \gamma_{\Omega_2}\bar{Y}_{25}\gamma_{\Omega_5}^{-2} \coma \\
		X_{56}& \rightarrow & \gamma_{\Omega_5}\bar{W}_{56}\gamma_{\Omega_6}^{-2} \coma &
		W_{56}& \rightarrow & \gamma_{\Omega_5}\bar{X}_{56}\gamma_{\Omega_6}^{-2} \coma &
		X_{25}& \rightarrow & \gamma_{\Omega_8}\bar{Y}_{85}\gamma_{\Omega_5}^{-2} \coma &
		Y_{85}& \rightarrow & \gamma_{\Omega_2}\bar{X}_{25}\gamma_{\Omega_5}^{-2} \coma \\
		Z_{56}& \rightarrow & \gamma_{\Omega_5}\bar{Z}_{56}\gamma_{\Omega_6}^{-2} \coma &
		X_{42}& \rightarrow & \gamma_{\Omega_4}\bar{W}_{48}\gamma_{\Omega_8}^{-2} \coma &
		W_{48}& \rightarrow & \gamma_{\Omega_4}\bar{X}_{42}\gamma_{\Omega_2}^{-2} \coma &
		Y_{42}& \rightarrow & \gamma_{\Omega_4}\bar{Z}_{48}\gamma_{\Omega_8}^{-2} \coma \\
		Z_{48}& \rightarrow & \gamma_{\Omega_4}\bar{Y}_{42}\gamma_{\Omega_2}^{-2} \coma &
		Y_{56}& \rightarrow & \gamma_{\Omega_5}\bar{Y}_{56}\gamma_{\Omega_6}^{-2} \coma &
		X_{74}& \rightarrow & \gamma_{\Omega_1}\bar{Y}_{14}\gamma_{\Omega_4}^{-2} \coma &
		Y_{14}& \rightarrow & \gamma_{\Omega_7}\bar{X}_{74}\gamma_{\Omega_4}^{-2} \coma \\
\end{array}  
	\label{eq:Q111Z2H-chiral-invol}
\end{equation}

Requiring the invariance of $W^{(0,1)}$, the Fermi fields transform as
\begin{equation}
	\begin{array}{cccccccccccc}
		\Lambda_{26}^{2} & \rightarrow & \gamma_{\Omega_6}\Lambda^1_{68}\gamma_{\Omega_8}^{-2} \coma &
		\Lambda_{26}^{8} & \rightarrow & -\gamma_{\Omega_6}\Lambda^2_{68}\gamma_{\Omega_8}^{-2} \coma &
		\Lambda_{27}^{2} & \rightarrow & -\gamma_{\Omega_8}\bar{\Lambda}^2_{81}\gamma_{\Omega_1}^{-2} \coma &
		\Lambda_{27}^{8} & \rightarrow & -\gamma_{\Omega_8}\bar{\Lambda}^1_{81}\gamma_{\Omega_1}^{-2} \coma \\
		\Lambda_{68}^{2} & \rightarrow & \gamma_{\Omega_2}\Lambda^1_{26}\gamma_{\Omega_6}^{-2} \coma &
		\Lambda_{68}^{8} & \rightarrow & -\gamma_{\Omega_2}\Lambda^2_{26}\gamma_{\Omega_6}^{-2} \coma &
		\Lambda_{81}^{2} & \rightarrow & -\gamma_{\Omega_2}\bar{\Lambda}^2_{27}\gamma_{\Omega_7}^{-2} \coma &
		\Lambda_{81}^{8} & \rightarrow & -\gamma_{\Omega_2}\bar{\Lambda}^1_{27}\gamma_{\Omega_7}^{-2} \coma \\
		\Lambda_{54}^{2} & \rightarrow & -\gamma_{\Omega_5}\bar{\Lambda}^8_{54}\gamma_{\Omega_4}^{-2} \coma &
		\Lambda_{54}^{8} & \rightarrow & -\gamma_{\Omega_5}\bar{\Lambda}^7_{54}\gamma_{\Omega_4}^{-2} \coma &
		\Lambda_{54}^{5} & \rightarrow & -\gamma_{\Omega_5}\bar{\Lambda}^4_{54}\gamma_{\Omega_4}^{-2} \coma &
		\Lambda_{54}^{6} & \rightarrow & -\gamma_{\Omega_5}\bar{\Lambda}^3_{54}\gamma_{\Omega_4}^{-2} \coma \\
		\Lambda_{54}^{3} & \rightarrow & -\gamma_{\Omega_5}\bar{\Lambda}^6_{54}\gamma_{\Omega_4}^{-2} \coma &
		\Lambda_{54}^{4} & \rightarrow & -\gamma_{\Omega_5}\bar{\Lambda}^5_{54}\gamma_{\Omega_4}^{-2} \coma &
		\Lambda_{54}^{1} & \rightarrow & -\gamma_{\Omega_5}\bar{\Lambda}^2_{54}\gamma_{\Omega_4}^{-2} \coma &
		\Lambda_{54}^{7} & \rightarrow & -\gamma_{\Omega_5}\bar{\Lambda}^1_{54}\gamma_{\Omega_4}^{-2} \coma \\
\end{array}   
	\label{eq:Q111Z2H-Fermi-invol}
\end{equation}
and
	\begin{equation}
	 \begin{array}{cccccccccccc}
	\Lambda_{11}^R & \rightarrow & \gamma_{\Omega_7}\Lambda_{77}^{R\,\,T}\gamma_{\Omega_7}^{-1} \coma &
    \Lambda_{22}^R & \rightarrow & \gamma_{\Omega_8}\Lambda_{88}^{R\,\,T}\gamma_{\Omega_8}^{-1} \coma &
	\Lambda_{33}^R & \rightarrow & \gamma_{\Omega_3}\Lambda_{33}^{R\,\,T}\gamma_{\Omega_3}^{-1} \coma &
	\Lambda_{44}^R & \rightarrow & \gamma_{\Omega_4}\Lambda_{44}^{R\,\,T}\gamma_{\Omega_4}^{-1} \coma \\
	\Lambda_{55}^R & \rightarrow & \gamma_{\Omega_5}\Lambda_{55}^{R\,\,T}\gamma_{\Omega_5}^{-1} \coma &
	\Lambda_{66}^R & \rightarrow & \gamma_{\Omega_6}\Lambda_{66}^{R\,\,T}\gamma_{\Omega_6}^{-1} \coma &
	\Lambda_{77}^R & \rightarrow & \gamma_{\Omega_1}\Lambda_{11}^{R\,\,T}\gamma_{\Omega_1}^{-1} \coma &
	\Lambda_{88}^R & \rightarrow & \gamma_{\Omega_2}\Lambda_{22}^{R\,\,T}\gamma_{\Omega_2}^{-1} \fstop
	\end{array}   
	\label{eq:Q111Z2H-RFermi-invol}
	\end{equation}
	
Using Table \ref{tab:GenQ111Z2PhaseH}, the corresponding geometric involution acting on the generators reads 
\begin{equation}
     \begin{array}{ccccccccccccccc}
	M_{1}  & \rightarrow & \bar{M}_{27} \coma &
	M_{2}  & \rightarrow & \bar{M}_{24} \coma &
	M_{3}  & \rightarrow & \bar{M}_{21} \coma &
	M_{4}  & \rightarrow & \bar{M}_{18} \coma &
	M_{5}  & \rightarrow & \bar{M}_{12} \coma \\
	M_{6}  & \rightarrow & \bar{M}_{6} \coma &
	M_{7}  & \rightarrow & \bar{M}_{26} \coma &
	M_{8}  & \rightarrow & \bar{M}_{23} \coma &
	M_{9}  & \rightarrow & \bar{M}_{20} \coma &
	M_{10} & \rightarrow & \bar{M}_{17} \coma \\
	M_{11} & \rightarrow & \bar{M}_{11} \coma &
	M_{12} & \rightarrow & \bar{M}_{5} \coma &
	M_{13} & \rightarrow & \bar{M}_{25} \coma &
	M_{14} & \rightarrow & \bar{M}_{22} \coma &
	M_{15} & \rightarrow & \bar{M}_{19} \coma \\
	M_{16} & \rightarrow & \bar{M}_{16} \coma &
	M_{17} & \rightarrow & \bar{M}_{10} \coma &
	M_{18} & \rightarrow & \bar{M}_{4} \coma &
	M_{19} & \rightarrow & \bar{M}_{15} \coma &
	M_{20} & \rightarrow & \bar{M}_{9} \coma \\ 
	M_{21} & \rightarrow & \bar{M}_{3}\coma &
	M_{22} & \rightarrow & \bar{M}_{14}\coma &
	M_{23} & \rightarrow & \bar{M}_{8}\coma &
	M_{24} & \rightarrow & \bar{M}_{2}\coma &
	M_{25} & \rightarrow & \bar{M}_{13}\coma \\
	\multicolumn{15}{c}{M_{26}   \rightarrow  \bar{M}_{7}\coma 
	M_{27}  \rightarrow  \bar{M}_{1}\fstop}
	\end{array}   
	\label{eq:Q111Z2H-Meson-invol}
    \end{equation}
Once again, this is the same involution of phase D in \eqref{eq:Q111Z2D-Meson-invol}. 
    
    The $\gamma_{\Omega_i}$ matrices are constrained as in \eqref{gamma_matrices_Q111/Z2}. As for phase D, we choose
\begin{equation}
\gamma_{\Omega_3}=\gamma_{\Omega_4}=\gamma_{\Omega_5}=\gamma_{\Omega_6}=\ID_N \fstop 
\end{equation}
    
Figure \ref{fig:orien_Q111Z2quiverH} shows the quiver for the resulting orientifold of phase H.
  
    \begin{figure}[H]
		\centering
		%
	\caption{Quiver diagram for the Spin(7) orientifold of phase H of $Q^{1,1,1}/\ZZ_2$ using the involution in \cref{eq:Q111Z2H-chiral-invol,eq:Q111Z2H-Fermi-invol,eq:Q111Z2H-RFermi-invol}, together with our choice of $\gamma_{\Omega_i}$ matrices.}
		\label{fig:orien_Q111Z2quiverH}
	\end{figure}

\section{Phase J}
\label{app:Q111Z2PhaseJ}

The quiver for phase J is shown in Figure \ref{fig:Q111Z2quiverJ}. 

\begin{figure}[H]
		\centering

	\caption{Quiver diagram for phase J of $Q^{1,1,1}/\ZZ_2$.}
	\label{fig:Q111Z2quiverJ}
	\end{figure}

The $J$- and $E$-terms are
		\begin{alignat}{4}
	\renewcommand{\arraystretch}{1.1}
    & \centermathcell{J}                           &\text{\hspace{.5cm}}& \centermathcell{E                               }\nonumber \\
\Lambda^1_{14} \,:\, & \centermathcell{Y_{46} Y_{61}-X_{46} W_{61} }& & \centermathcell{Y_{13} X_{34}-X_{13} Y_{34}}\nonumber \\
\Lambda^2_{14} \,:\, & \centermathcell{Y_{46} X_{61}-X_{46} Z_{61} }& & \centermathcell{X_{13} W_{34}-Y_{13} Z_{34}}\nonumber \\
\Lambda^1_{15} \,:\, & \centermathcell{Y_{56} W_{61}-W_{56} Z_{61} }& & \centermathcell{X_{18} X_{85}-X_{13} X_{35}}\nonumber \\
\Lambda^2_{15} \,:\, & \centermathcell{Z_{56} Z_{61}-X_{56} W_{61} }& & \centermathcell{X_{18} Y_{85}-Y_{13} X_{35}}\nonumber \\
\Lambda^3_{15} \,:\, & \centermathcell{Y_{56} Y_{61}-W_{56} X_{61} }& & \centermathcell{X_{13} Y_{35}-Y_{18} X_{85}}\nonumber \\
\Lambda^4_{15} \,:\, & \centermathcell{X_{56} Y_{61}-Z_{56} X_{61} }& & \centermathcell{Y_{18} Y_{85}-Y_{13} Y_{35}}\nonumber \\
\Lambda^1_{86} \,:\, & \centermathcell{Y_{61} Y_{18}-W_{61} X_{18} }& & \centermathcell{X_{85} Y_{56}-Y_{85} X_{56}}\nonumber \\
\Lambda^2_{86} \,:\, & \centermathcell{X_{61} Y_{18}-Z_{61} X_{18} }& & \centermathcell{Y_{85} Z_{56}-X_{85} W_{56}}\nonumber \\
\Lambda^1_{47} \,:\, & \centermathcell{X_{73} W_{34}-Y_{73} Y_{34} }& & \centermathcell{X_{46} Z_{67}-Y_{46} X_{67}}\nonumber \\
\Lambda^2_{47} \,:\, & \centermathcell{X_{73} Z_{34}-Y_{73} X_{34} }& & \centermathcell{Y_{46} Y_{67}-X_{46} W_{67}}\nonumber \\
\Lambda^1_{63} \,:\, & \centermathcell{Y_{35} W_{56}-W_{34} Y_{46} }& & \centermathcell{X_{61} X_{13}-X_{67} X_{73}}\nonumber \\
\Lambda^2_{63} \,:\, & \centermathcell{Y_{35} Z_{56}-Z_{34} Y_{46} }& & \centermathcell{Y_{67} X_{73}-X_{61} Y_{13}}\label{eq:Q111Z2JEtermsJ} \\
\Lambda^3_{63} \,:\, & \centermathcell{Y_{35} Y_{56}-Y_{34} Y_{46} }& & \centermathcell{X_{67} Y_{73}-Y_{61} X_{13}}\nonumber \\
\Lambda^4_{63} \,:\, & \centermathcell{Y_{35} X_{56}-X_{34} Y_{46} }& & \centermathcell{Y_{61} Y_{13}-Y_{67} Y_{73}}\nonumber \\
\Lambda^5_{63} \,:\, & \centermathcell{X_{35} W_{56}-W_{34} X_{46} }& & \centermathcell{Z_{67} X_{73}-Z_{61} X_{13}}\nonumber \\
\Lambda^6_{63} \,:\, & \centermathcell{X_{35} Z_{56}-Z_{34} X_{46} }& & \centermathcell{Z_{61} Y_{13}-W_{67} X_{73}}\nonumber \\
\Lambda^7_{63} \,:\, & \centermathcell{X_{35} Y_{56}-Y_{34} X_{46} }& & \centermathcell{W_{61} X_{13}-Z_{67} Y_{73}}\nonumber \\
\Lambda^8_{63} \,:\, & \centermathcell{X_{35} X_{56}-X_{34} X_{46} }& & \centermathcell{W_{67} Y_{73}-W_{61} Y_{13}}\nonumber \\
\Lambda^1_{62} \,:\, & \centermathcell{X_{25} W_{56}-Y_{25} Y_{56} }& & \centermathcell{X_{67} Y_{72}-Z_{67} X_{72}}\nonumber \\
\Lambda^2_{62} \,:\, & \centermathcell{X_{25} Z_{56}-Y_{25} X_{56} }& & \centermathcell{W_{67} X_{72}-Y_{67} Y_{72}}\nonumber \\
\Lambda^1_{57} \,:\, & \centermathcell{Y_{72} Y_{25}-Y_{73} Y_{35} }& & \centermathcell{Y_{56} X_{67}-X_{56} Y_{67}}\nonumber \\
\Lambda^2_{57} \,:\, & \centermathcell{Y_{72} X_{25}-X_{73} Y_{35} }& & \centermathcell{Z_{56} Y_{67}-W_{56} X_{67}}\nonumber \\
\Lambda^3_{57} \,:\, & \centermathcell{X_{72} Y_{25}-Y_{73} X_{35} }& & \centermathcell{X_{56} W_{67}-Y_{56} Z_{67}}\nonumber \\
\Lambda^4_{57} \,:\, & \centermathcell{X_{72} X_{25}-X_{73} X_{35} }& & \centermathcell{W_{56} Z_{67}-Z_{56} W_{67}}\nonumber 	
\end{alignat}
Once again, the generators can be found in Table \ref{tab:GenQ111Z2PhaseJ}. 
\medskip

\paragraph{$\U(N)^2\times \SO(N)^4$ orientifold}\mbox{}

\medskip

Let us consider an anti-holomorphic involution of phase J which acts on the nodes in Figure \ref{fig:Q111Z2quiverJ} as $1\leftrightarrow 7$ and $2\leftrightarrow 8$ and maps all other nodes mapped to themselves. Chiral fields transform according to
\begin{equation}
\begin{array}{cccccccccccc}
		Y_{46} & \rightarrow & \gamma_{\Omega_4} \bar{X}_{46}\gamma_{\Omega_6}^{-1} \coma &
		X_{46} & \rightarrow & \gamma_{\Omega_4} \bar{Y}_{46}\gamma_{\Omega_6}^{-1} \coma &
		Y_{61} & \rightarrow & \gamma_{\Omega_6} \bar{Z}_{67}\gamma_{\Omega_7}^{-1} \coma &
		Z_{67} & \rightarrow & \gamma_{\Omega_6} \bar{Y}_{61}\gamma_{\Omega_1}^{-1} \coma \\
		W_{61} & \rightarrow & \gamma_{\Omega_6} \bar{X}_{67}\gamma_{\Omega_7}^{-1} \coma &
		X_{67} & \rightarrow & \gamma_{\Omega_6} \bar{W}_{61}\gamma_{\Omega_1}^{-1} \coma &
		Y_{13} & \rightarrow & \gamma_{\Omega_7} \bar{X}_{73}\gamma_{\Omega_3}^{-1} \coma &
		X_{73} & \rightarrow & \gamma_{\Omega_1} \bar{Y}_{13}\gamma_{\Omega_3}^{-1} \coma \\
		X_{13} & \rightarrow & \gamma_{\Omega_7} \bar{Y}_{73}\gamma_{\Omega_3}^{-1} \coma &
		Y_{73} & \rightarrow & \gamma_{\Omega_1} \bar{X}_{13}\gamma_{\Omega_3}^{-1} \coma &
		X_{34} & \rightarrow & \gamma_{\Omega_3} \bar{W}_{34}\gamma_{\Omega_4}^{-1} \coma &
		W_{34} & \rightarrow & \gamma_{\Omega_3} \bar{X}_{34}\gamma_{\Omega_4}^{-1} \coma \\
		Y_{34} & \rightarrow & \gamma_{\Omega_3} \bar{Y}_{34}\gamma_{\Omega_4}^{-1} \coma &
		X_{61} & \rightarrow & \gamma_{\Omega_6} \bar{W}_{67}\gamma_{\Omega_7}^{-1} \coma &
		W_{67} & \rightarrow & \gamma_{\Omega_6} \bar{X}_{61}\gamma_{\Omega_1}^{-1} \coma &
		Z_{61} & \rightarrow & \gamma_{\Omega_6} \bar{Y}_{67}\gamma_{\Omega_7}^{-1} \coma \\ 
		Y_{67} & \rightarrow & \gamma_{\Omega_6} \bar{Z}_{61}\gamma_{\Omega_1}^{-1} \coma &
		Z_{34} & \rightarrow & \gamma_{\Omega_3} \bar{Z}_{34}\gamma_{\Omega_4}^{-1} \coma &
		Y_{56} & \rightarrow & \gamma_{\Omega_5} \bar{Y}_{56}\gamma_{\Omega_6}^{-1} \coma &
		W_{56} & \rightarrow & \gamma_{\Omega_5} \bar{X}_{56}\gamma_{\Omega_6}^{-1} \coma \\
\end{array}  
	\label{eq:Q111Z2J-chiral-invol}
\end{equation}

Requiring the invariance of $W^{(0,1)}$, the Fermi fields transform as
\begin{equation}
	\begin{array}{cccccccccccc}
		\Lambda^1_{14} & \rightarrow & \gamma_{\Omega_4} \Lambda^1_{47} \gamma_{\Omega_7}^{-1} \coma &
		\Lambda^2_{14} & \rightarrow & -\gamma_{\Omega_4} \Lambda^2_{47} \gamma_{\Omega_7}^{-1} \coma &
		\Lambda^1_{15} & \rightarrow & \gamma_{\Omega_5} \Lambda^1_{57} \gamma_{\Omega_7}^{-1} \coma &
		\Lambda^2_{15} & \rightarrow & \gamma_{\Omega_5} \Lambda^2_{57} \gamma_{\Omega_7}^{-1} \coma \\
		\Lambda^3_{15} & \rightarrow & -\gamma_{\Omega_5} \Lambda^3_{57} \gamma_{\Omega_7}^{-1} \coma &
		\Lambda^4_{15} & \rightarrow & \gamma_{\Omega_5} \Lambda^4_{57} \gamma_{\Omega_7}^{-1} \coma &
		\Lambda^1_{86} & \rightarrow & -\gamma_{\Omega_6} \Lambda^1_{62} \gamma_{\Omega_2}^{-1} \coma &
		\Lambda^2_{86} & \rightarrow & \gamma_{\Omega_6} \Lambda^2_{62} \gamma_{\Omega_2}^{-1} \coma \\
		\Lambda^1_{47} & \rightarrow & \gamma_{\Omega_1} \Lambda^1_{14} \gamma_{\Omega_4}^{-1} \coma &
		\Lambda^2_{47} & \rightarrow & -\gamma_{\Omega_4} \Lambda^2_{47} \gamma_{\Omega_7}^{-1} \coma &
		\Lambda^1_{63} & \rightarrow & \gamma_{\Omega_6} \bar{\Lambda}^8_{63} \gamma_{\Omega_3}^{-1} \coma &
		\Lambda^2_{63} & \rightarrow & \gamma_{\Omega_6} \bar{\Lambda}^6_{63} \gamma_{\Omega_3}^{-1} \coma \\
		\Lambda^3_{63} & \rightarrow & \gamma_{\Omega_6} \bar{\Lambda}^7_{63} \gamma_{\Omega_3}^{-1} \coma &
		\Lambda^4_{63} & \rightarrow & \gamma_{\Omega_6} \bar{\Lambda}^5_{63} \gamma_{\Omega_3}^{-1} \coma &
		\Lambda^5_{63} & \rightarrow & \gamma_{\Omega_6} \bar{\Lambda}^4_{63} \gamma_{\Omega_3}^{-1} \coma &
		\Lambda^6_{63} & \rightarrow & \gamma_{\Omega_6} \bar{\Lambda}^2_{63} \gamma_{\Omega_3}^{-1} \coma \\
\end{array}   
	\label{eq:Q111Z2J-Fermi-invol}
\end{equation}
and
\begin{equation}
	\begin{array}{cccccccccccc}
		\Lambda_{11}^R & \rightarrow & \gamma_{\Omega_7}\Lambda_{77}^{R\,\,T}\gamma_{\Omega_7}^{-1} \coma &
		\Lambda_{22}^R & \rightarrow & \gamma_{\Omega_8}\Lambda_{88}^{R\,\,T}\gamma_{\Omega_8}^{-1} \coma &
		\Lambda_{33}^R & \rightarrow & \gamma_{\Omega_3}\Lambda_{33}^{R\,\,T}\gamma_{\Omega_3}^{-1} \coma &
		\Lambda_{44}^R & \rightarrow & \gamma_{\Omega_4}\Lambda_{44}^{R\,\,T}\gamma_{\Omega_4}^{-1} \coma \\
		\Lambda_{55}^R & \rightarrow & \gamma_{\Omega_5}\Lambda_{55}^{R\,\,T}\gamma_{\Omega_5}^{-1} \coma &
		\Lambda_{66}^R & \rightarrow & \gamma_{\Omega_6}\Lambda_{66}^{R\,\,T}\gamma_{\Omega_6}^{-1} \coma &
		\Lambda_{77}^R & \rightarrow & \gamma_{\Omega_1}\Lambda_{11}^{R\,\,T}\gamma_{\Omega_1}^{-1} \coma &	
		\Lambda_{88}^R & \rightarrow & \gamma_{\Omega_2}\Lambda_{22}^{R\,\,T}\gamma_{\Omega_2}^{-1} \fstop
	\end{array}  
	\label{eq:Q111Z2J-RFermi-invol}
\end{equation}

Using Table \ref{tab:GenQ111Z2PhaseJ}, the corresponding geometric involution acting on the generators reads
\begin{equation}
     \begin{array}{ccccccccccccccc}
	M_{1}  & \rightarrow & \bar{M}_{27} \coma &
	M_{2}  & \rightarrow & \bar{M}_{24} \coma &
	M_{3}  & \rightarrow & \bar{M}_{21} \coma &
	M_{4}  & \rightarrow & \bar{M}_{18} \coma &
	M_{5}  & \rightarrow & \bar{M}_{12} \coma \\
	M_{6}  & \rightarrow & \bar{M}_{6} \coma &
	M_{7}  & \rightarrow & \bar{M}_{26} \coma &
	M_{8}  & \rightarrow & \bar{M}_{23} \coma &
	M_{9}  & \rightarrow & \bar{M}_{20} \coma &
	M_{10} & \rightarrow & \bar{M}_{17} \coma \\
	M_{11} & \rightarrow & \bar{M}_{11} \coma &
	M_{12} & \rightarrow & \bar{M}_{5} \coma &
	M_{13} & \rightarrow & \bar{M}_{25} \coma &
	M_{14} & \rightarrow & \bar{M}_{22} \coma &
	M_{15} & \rightarrow & \bar{M}_{19} \coma \\
	M_{16} & \rightarrow & \bar{M}_{16} \coma &
	M_{17} & \rightarrow & \bar{M}_{10} \coma &
	M_{18} & \rightarrow & \bar{M}_{4} \coma &
	M_{19} & \rightarrow & \bar{M}_{15} \coma &
	M_{20} & \rightarrow & \bar{M}_{9} \coma \\ 
	M_{21} & \rightarrow & \bar{M}_{3}\coma &
	M_{22} & \rightarrow & \bar{M}_{14}\coma &
	M_{23} & \rightarrow & \bar{M}_{8}\coma &
	M_{24} & \rightarrow & \bar{M}_{2}\coma &
	M_{25} & \rightarrow & \bar{M}_{13}\coma \\
	\multicolumn{15}{c}{M_{26}   \rightarrow  \bar{M}_{7}\coma 
	M_{27}  \rightarrow  \bar{M}_{1}\fstop}
	\end{array}   
	\label{eq:Q111Z2J-Meson-invol}
    \end{equation}
Notice, again, that this is the same geometric action that we have found for phase D in \eqref{eq:Q111Z2D-Meson-invol}. 

The $\gamma_{\Omega_i}$ matrices are constrained as in \eqref{gamma_matrices_Q111/Z2}. As for phase D, we choose
\begin{equation}
\gamma_{\Omega_3}=\gamma_{\Omega_4}=\gamma_{\Omega_5}=\gamma_{\Omega_6}=\ID_N \fstop 
\end{equation}

The resulting orientifold of phase J is shown in Figure \ref{fig:orien_Q111Z2quiverJ}.

 \begin{figure}[H]
		\centering

	\caption{Quiver diagram for the Spin(7) orientifold of phase J of $Q^{1,1,1}/\ZZ_2$ using the involution in \cref{eq:Q111Z2J-chiral-invol,eq:Q111Z2J-Fermi-invol,eq:Q111Z2J-RFermi-invol}, together with our choice of $\gamma_{\Omega_i}$ matrices.}
		\label{fig:orien_Q111Z2quiverJ}
	\end{figure}

\section{Phase L}
\label{sec:Q111Z2PhaseL}

The last $\mathcal{N}=(0,2)$ quiver we consider is phase L, shown in Figure \ref{fig:Q111Z2quiverL}. 

\begin{figure}[H]
		\centering

	\caption{Quiver diagram for phase L of $Q^{1,1,1}/\ZZ_2$.}
	\label{fig:Q111Z2quiverL}
	\end{figure}

The $J$- and $E$-terms are
		\begin{alignat}{4}
	\renewcommand{\arraystretch}{1.1}
    & \centermathcell{J}                           &\text{\hspace{.5cm}}& \centermathcell{E                               }\nonumber \\
\Lambda^1_{15} \,:\, & \centermathcell{B_{54} Y_{41}-D_{54} X_{41} }& & \centermathcell{X_{13} X_{35}-X_{18} X_{85}}\nonumber \\
\Lambda^2_{15} \,:\, & \centermathcell{C_{54} X_{41}-A_{54} Y_{41} }& & \centermathcell{X_{13} Y_{35}-Y_{18} X_{85}}\nonumber \\
\Lambda^3_{15} \,:\, & \centermathcell{W_{54} X_{41}-Y_{54} Y_{41} }& & \centermathcell{Y_{13} X_{35}-X_{18} Y_{85}}\nonumber \\
\Lambda^4_{15} \,:\, & \centermathcell{Z_{54} X_{41}-X_{54} Y_{41} }& & \centermathcell{Y_{18} Y_{85}-Y_{13} Y_{35}}\nonumber \\
\Lambda^1_{34} \,:\, & \centermathcell{Y_{41} Y_{13}-Y_{47} Y_{73} }& & \centermathcell{X_{35} Y_{54}-Y_{35} X_{54}}\nonumber \\
\Lambda^2_{34} \,:\, & \centermathcell{X_{41} Y_{13}-Y_{47} X_{73} }& & \centermathcell{Y_{35} Z_{54}-X_{35} W_{54}}\nonumber \\
\Lambda^3_{34} \,:\, & \centermathcell{Y_{41} X_{13}-X_{47} Y_{73} }& & \centermathcell{Y_{35} A_{54}-X_{35} B_{54}}\nonumber \\
\Lambda^4_{34} \,:\, & \centermathcell{X_{41} X_{13}-X_{47} X_{73} }& & \centermathcell{X_{35} D_{54}-Y_{35} C_{54}}\nonumber \\
\Lambda^1_{48} \,:\, & \centermathcell{X_{85} D_{54}-Y_{85} W_{54} }& & \centermathcell{X_{46} X_{68}-X_{41} X_{18}}\nonumber \\
\Lambda^2_{48} \,:\, & \centermathcell{X_{85} C_{54}-Y_{85} Z_{54} }& & \centermathcell{X_{41} Y_{18}-Y_{46} X_{68}}\nonumber \\
\Lambda^3_{48} \,:\, & \centermathcell{X_{85} B_{54}-Y_{85} Y_{54} }& & \centermathcell{Y_{41} X_{18}-X_{46} Y_{68}}\nonumber \\
\Lambda^4_{48} \,:\, & \centermathcell{X_{85} A_{54}-Y_{85} X_{54} }& & \centermathcell{Y_{46} Y_{68}-Y_{41} Y_{18}}\label{eq:Q111Z2JEtermsL} \\
\Lambda^1_{42} \,:\, & \centermathcell{X_{25} D_{54}-Y_{25} B_{54} }& & \centermathcell{X_{47} X_{72}-X_{46} X_{62}}\nonumber \\
\Lambda^2_{42} \,:\, & \centermathcell{X_{25} C_{54}-Y_{25} A_{54} }& & \centermathcell{Y_{46} X_{62}-X_{47} Y_{72}}\nonumber \\
\Lambda^3_{42} \,:\, & \centermathcell{X_{25} W_{54}-Y_{25} Y_{54} }& & \centermathcell{X_{46} Y_{62}-Y_{47} X_{72}}\nonumber \\
\Lambda^4_{42} \,:\, & \centermathcell{X_{25} Z_{54}-Y_{25} X_{54} }& & \centermathcell{Y_{47} Y_{72}-Y_{46} Y_{62}}\nonumber \\
\Lambda^1_{75} \,:\, & \centermathcell{W_{54} Y_{47}-D_{54} X_{47} }& & \centermathcell{X_{72} X_{25}-X_{73} X_{35}}\nonumber \\
\Lambda^2_{75} \,:\, & \centermathcell{Z_{54} Y_{47}-C_{54} X_{47} }& & \centermathcell{X_{73} Y_{35}-Y_{72} X_{25}}\nonumber \\
\Lambda^3_{75} \,:\, & \centermathcell{Y_{54} Y_{47}-B_{54} X_{47} }& & \centermathcell{Y_{73} X_{35}-X_{72} Y_{25}}\nonumber \\
\Lambda^4_{75} \,:\, & \centermathcell{X_{54} Y_{47}-A_{54} X_{47} }& & \centermathcell{Y_{72} Y_{25}-Y_{73} Y_{35}}\nonumber \\
\Lambda^1_{56} \,:\, & \centermathcell{Y_{68} Y_{85}-Y_{62} Y_{25} }& & \centermathcell{X_{54} Y_{46}-Y_{54} X_{46}}\nonumber \\
\Lambda^2_{56} \,:\, & \centermathcell{X_{68} Y_{85}-Y_{62} X_{25} }& & \centermathcell{W_{54} X_{46}-Z_{54} Y_{46}}\nonumber \\
\Lambda^3_{56} \,:\, & \centermathcell{Y_{68} X_{85}-X_{62} Y_{25} }& & \centermathcell{B_{54} X_{46}-A_{54} Y_{46}}\nonumber \\
\Lambda^4_{56} \,:\, & \centermathcell{X_{68} X_{85}-X_{62} X_{25} }& & \centermathcell{C_{54} Y_{46}-D_{54} X_{46}}\nonumber 	
\end{alignat}

\paragraph{$\U(N)^2\times \SO(N)^4$ orientifold}\mbox{}

We can, again, look for an involution that maps the nodes in Figure \ref{fig:Q111Z2quiverL} $1\leftrightarrow 7$ and $2\leftrightarrow 8$ and all the other nodes to themselves. The map on the fields is
\begin{equation}
\begin{array}{cccccccccccc}
		B_{54} & \rightarrow & \gamma_{\Omega_5}\bar{Z}_{54}\gamma_{\Omega_4}^{-1} \coma &
		Z_{54} & \rightarrow & \gamma_{\Omega_5}\bar{B}_{54}\gamma_{\Omega_4}^{-1} \coma &
		D_{54} & \rightarrow & \gamma_{\Omega_5}\bar{C}_{54}\gamma_{\Omega_4}^{-1} \coma & 
		C_{54} & \rightarrow & \gamma_{\Omega_5}\bar{D}_{54}\gamma_{\Omega_4}^{-1} \coma \\
		Y_{41} & \rightarrow & \gamma_{\Omega_4}\bar{Y}_{47}\gamma_{\Omega_7}^{-1} \coma &
		Y_{47} & \rightarrow & \gamma_{\Omega_4}\bar{Y}_{41}\gamma_{\Omega_1}^{-1} \coma &
		X_{41} & \rightarrow & \gamma_{\Omega_4}\bar{X}_{47}\gamma_{\Omega_7}^{-1} \coma & 
		X_{47} & \rightarrow & \gamma_{\Omega_4}\bar{X}_{41}\gamma_{\Omega_1}^{-1} \coma \\
		X_{13} & \rightarrow & \gamma_{\Omega_7}\bar{X}_{73}\gamma_{\Omega_3}^{-1} \coma & 
		X_{73} & \rightarrow & \gamma_{\Omega_1}\bar{X}_{13}\gamma_{\Omega_3}^{-1} \coma & 
		X_{35} & \rightarrow & \gamma_{\Omega_3}\bar{Y}_{35}\gamma_{\Omega_5}^{-1} \coma & 
		Y_{35} & \rightarrow & \gamma_{\Omega_3}\bar{X}_{35}\gamma_{\Omega_5}^{-1} \coma \\ 
		X_{18} & \rightarrow & \gamma_{\Omega_7}\bar{Y}_{72}\gamma_{\Omega_2}^{-1} \coma & 
		Y_{72} & \rightarrow & \gamma_{\Omega_1}\bar{X}_{18}\gamma_{\Omega_8}^{-1} \coma & 
		X_{85} & \rightarrow & \gamma_{\Omega_2}\bar{X}_{25}\gamma_{\Omega_5}^{-1} \coma & 
		X_{25} & \rightarrow & \gamma_{\Omega_8}\bar{X}_{85}\gamma_{\Omega_5}^{-1} \coma \\ 
		A_{54} & \rightarrow & \gamma_{\Omega_5}\bar{W}_{54}\gamma_{\Omega_4}^{-1} \coma & 
		W_{54} & \rightarrow & \gamma_{\Omega_5}\bar{A}_{54}\gamma_{\Omega_4}^{-1} \coma & 
		Y_{18} & \rightarrow & \gamma_{\Omega_7}\bar{X}_{72}\gamma_{\Omega_2}^{-1} \coma & 
		X_{72} & \rightarrow & \gamma_{\Omega_1}\bar{Y}_{18}\gamma_{\Omega_8}^{-1} \coma \\
\end{array}  
	\label{eq:Q111Z2L-chiral-invol}
\end{equation}

Requiring the invariance of the $W^{(0,1)}$ superpotential, we obtain also the following transformations for the Fermi fields:
\begin{equation}
	\begin{array}{cccccccccccc}
		\Lambda^1_{15} & \rightarrow & \gamma_{\Omega_7}\bar{\Lambda}^2_{75}\gamma_{\Omega_5}^{-1} \coma &
		\Lambda^2_{15} & \rightarrow & -\gamma_{\Omega_7}\bar{\Lambda}^1_{75}\gamma_{\Omega_5}^{-1} \coma &
		\Lambda^3_{15} & \rightarrow & -\gamma_{\Omega_7}\bar{\Lambda}^4_{75}\gamma_{\Omega_5}^{-1} \coma &
		\Lambda^4_{15} & \rightarrow & -\gamma_{\Omega_7}\bar{\Lambda}^3_{75}\gamma_{\Omega_5}^{-1} \coma \\
		\Lambda^1_{34} & \rightarrow & -\gamma_{\Omega_3}\bar{\Lambda}^1_{34}\gamma_{\Omega_4}^{-1} \coma &
		\Lambda^2_{34} & \rightarrow & -\gamma_{\Omega_3}\bar{\Lambda}^3_{34}\gamma_{\Omega_4}^{-1} \coma &
		\Lambda^3_{34} & \rightarrow & -\gamma_{\Omega_3}\bar{\Lambda}^2_{34}\gamma_{\Omega_4}^{-1} \coma &
		\Lambda^4_{34} & \rightarrow & -\gamma_{\Omega_3}\bar{\Lambda}^4_{34}\gamma_{\Omega_4}^{-1} \coma \\ 
		\Lambda^1_{48} & \rightarrow & \gamma_{\Omega_4}\bar{\Lambda}^2_{42}\gamma_{\Omega_2}^{-1} \coma &
		\Lambda^2_{48} & \rightarrow & \gamma_{\Omega_4}\bar{\Lambda}^1_{42}\gamma_{\Omega_2}^{-1} \coma &
		\Lambda^3_{48} & \rightarrow & \gamma_{\Omega_4}\bar{\Lambda}^4_{42}\gamma_{\Omega_2}^{-1} \coma &
		\Lambda^4_{48} & \rightarrow & \gamma_{\Omega_4}\bar{\Lambda}^3_{42}\gamma_{\Omega_2}^{-1} \coma \\
		\Lambda^1_{42} & \rightarrow & \gamma_{\Omega_4}\bar{\Lambda}^2_{48}\gamma_{\Omega_8}^{-1} \coma &
		\Lambda^2_{42} & \rightarrow & \gamma_{\Omega_4}\bar{\Lambda}^1_{48}\gamma_{\Omega_8}^{-1} \coma &
		\Lambda^3_{42} & \rightarrow & \gamma_{\Omega_4}\bar{\Lambda}^4_{48}\gamma_{\Omega_8}^{-1} \coma &
		\Lambda^4_{42} & \rightarrow & \gamma_{\Omega_4}\bar{\Lambda}^3_{48}\gamma_{\Omega_8}^{-1} \coma \\
\end{array}   
	\label{eq:Q111Z2L-Fermi-invol}
\end{equation}
and
\begin{equation}
	\begin{array}{cccccccccccc}
		\Lambda_{11}^R & \rightarrow & \gamma_{\Omega_7}\Lambda_{77}^{R\,\,T}\gamma_{\Omega_7}^{-1} \coma &
		\Lambda_{22}^R & \rightarrow & \gamma_{\Omega_8}\Lambda_{88}^{R\,\,T}\gamma_{\Omega_8}^{-1} \coma &
		\Lambda_{33}^R & \rightarrow & \gamma_{\Omega_3}\Lambda_{33}^{R\,\,T}\gamma_{\Omega_3}^{-1} \coma &
		\Lambda_{44}^R & \rightarrow & \gamma_{\Omega_4}\Lambda_{44}^{R\,\,T}\gamma_{\Omega_4}^{-1} \coma \\
		\Lambda_{55}^R & \rightarrow & \gamma_{\Omega_5}\Lambda_{55}^{R\,\,T}\gamma_{\Omega_5}^{-1} \coma &
		\Lambda_{66}^R & \rightarrow & \gamma_{\Omega_1}\Lambda_{11}^{R\,\,T}\gamma_{\Omega_1}^{-1} \coma &
		\Lambda_{77}^R & \rightarrow & \gamma_{\Omega_1}\Lambda_{11}^{R\,\,T}\gamma_{\Omega_1}^{-1} \coma &
		\Lambda_{88}^R & \rightarrow & \gamma_{\Omega_2}\Lambda_{22}^{R\,\,T}\gamma_{\Omega_2}^{-1} \fstop
	\end{array}   
	\label{eq:Q111Z2L-RFermi-invol}
\end{equation} 

Using Table \ref{tab:GenQ111Z2PhaseL}, the corresponding geometric involution acting on the generators reads
\begin{equation}
     %
   
	\label{eq:Q111Z2L-Meson-invol}
    \end{equation}
This is, once again, the same geometric involution. 

The $\gamma_{\Omega_i}$ matrices are constrained as in \eqref{gamma_matrices_Q111/Z2}. As for phase D, we choose
\begin{equation}
\gamma_{\Omega_3}=\gamma_{\Omega_4}=\gamma_{\Omega_5}=\gamma_{\Omega_6}=\ID_N \fstop 
\end{equation}

The resulting orientifold is shown in Figure \ref{fig:orien_Q111Z2quiverL}. 

\begin{figure}[H]
		\centering
		%
	\caption{Quiver diagram for the Spin(7) orientifold of phase L of $Q^{1,1,1}/\ZZ_2$ using the involution in \cref{eq:Q111Z2L-chiral-invol,eq:Q111Z2L-Fermi-invol,eq:Q111Z2L-RFermi-invol}, together with our choice of $\gamma_{\Omega_i}$ matrices.}
		\label{fig:orien_Q111Z2quiverL}
	\end{figure}

\section{Generators of $Q^{1,1,1}/\ZZ_2$}
\label{app:Q111Z2-gener}
	
In \cref{tab:GenQ111Z2PhaseD,tab:GenQ111Z2PhaseE,tab:GenQ111Z2PhaseH,tab:GenQ111Z2PhaseJ,tab:GenQ111Z2PhaseL} we list the generators of  $Q^{1,1,1}/\ZZ_2$ in terms of the chiral fields of phases D, E, H, J and L.

The relations among the generators are the same for all phases, and they are:
	\begin{align}
\scriptstyle \mathcal{I} = &\,\scriptstyle \left\langle M_{1} M_{3}=M_{2}^2 \coma M_{1} M_{5}=M_{2} M_{4} \coma M_{2} M_{6}=M_{3} M_{5} \coma M_{1} M_{8}=M_{2} M_{7} \coma M_{2} M_{9}=M_{3} M_{8} \coma M_{1} M_{10}=M_{4} M_{7} \coma \right.\nonumber\\ 
& \scriptstyle \left. M_{3} M_{12}=M_{6} M_{9} \coma M_{1} M_{13}=M_{7}^2 \coma M_{7} M_{14}=M_{8} M_{13} \coma M_{3} M_{15}=M_{9}^2 \coma M_{8} M_{15}=M_{9} M_{14} \coma M_{13} M_{15}=M_{14}^2 \coma \right.\nonumber\\ 
& \scriptstyle \left. M_{7} M_{16}=M_{10} M_{13} \coma M_{13} M_{17}=M_{14} M_{16} \coma M_{9} M_{18}=M_{12} M_{15} \coma M_{14} M_{18}=M_{15} M_{17} \coma M_{1} M_{19}=M_{4}^2 \coma \right.\nonumber\\ 
& \scriptstyle \left. M_{4} M_{20}=M_{5} M_{19} \coma M_{3} M_{21}=M_{6}^2 \coma M_{5} M_{21}=M_{6} M_{20} \coma M_{19} M_{21}=M_{20}^2 \coma M_{4} M_{22}=M_{10} M_{19} \coma \right.\nonumber\\ 
& \scriptstyle \left. M_{19} M_{23}=M_{20} M_{22} \coma M_{6} M_{24}=M_{12} M_{21} \coma M_{20} M_{24}=M_{21} M_{23} \coma M_{10} M_{25}=M_{16} M_{22} \coma M_{13} M_{25}=M_{16}^2 \coma \right.\nonumber\\ 
& \scriptstyle \left. M_{19} M_{25}=M_{22}^2 \coma M_{16} M_{26}=M_{17} M_{25} \coma M_{22} M_{26}=M_{23} M_{25} \coma M_{12} M_{27}=M_{18} M_{24} \coma M_{15} M_{27}=M_{18}^2 \coma \right.\nonumber\\ 
& \scriptstyle \left. M_{17} M_{27}=M_{18} M_{26} \coma M_{21} M_{27}=M_{24}^2 \coma M_{23} M_{27}=M_{24} M_{26} \coma M_{25} M_{27}=M_{26}^2 \coma M_{1} M_{6}=M_{2} M_{5}=M_{3} M_{4} \coma \right.\nonumber\\ 
& \scriptstyle \left. M_{1} M_{9}=M_{2} M_{8}=M_{3} M_{7} \coma M_{1} M_{14}=M_{2} M_{13}=M_{7} M_{8} \coma M_{2} M_{15}=M_{3} M_{14}=M_{8} M_{9} \coma \right.\nonumber\\ 
& \scriptstyle \left. M_{7} M_{15}=M_{8} M_{14}=M_{9} M_{13} \coma M_{1} M_{16}=M_{4} M_{13}=M_{7} M_{10} \coma M_{3} M_{18}=M_{6} M_{15}=M_{9} M_{12} \coma \right.\nonumber\\ 
& \scriptstyle \left. M_{13} M_{18}=M_{14} M_{17}=M_{15} M_{16} \coma M_{1} M_{20}=M_{2} M_{19}=M_{4} M_{5} \coma M_{2} M_{21}=M_{3} M_{20}=M_{5} M_{6} \coma \right.\nonumber\\ 
& \scriptstyle \left. M_{4} M_{21}=M_{5} M_{20}=M_{6} M_{19} \coma M_{1} M_{22}=M_{4} M_{10}=M_{7} M_{19} \coma M_{3} M_{24}=M_{6} M_{12}=M_{9} M_{21} \coma \right.\nonumber\\ 
& \scriptstyle \left. M_{19} M_{24}=M_{20} M_{23}=M_{21} M_{22} \coma M_{4} M_{25}=M_{10} M_{22}=M_{16} M_{19} \coma M_{7} M_{25}=M_{10} M_{16}=M_{13} M_{22} \coma \right.\nonumber\\ 
& \scriptstyle \left. M_{13} M_{26}=M_{14} M_{25}=M_{16} M_{17} \coma M_{19} M_{26}=M_{20} M_{25}=M_{22} M_{23} \coma M_{6} M_{27}=M_{12} M_{24}=M_{18} M_{21} \coma \right.\nonumber\\ 
& \scriptstyle \left. M_{9} M_{27}=M_{12} M_{18}=M_{15} M_{24} \coma M_{14} M_{27}=M_{15} M_{26}=M_{17} M_{18} \coma M_{16} M_{27}=M_{17} M_{26}=M_{18} M_{25} \coma \right.\nonumber\\ 
& \scriptstyle \left. M_{20} M_{27}=M_{21} M_{26}=M_{23} M_{24} \coma M_{22} M_{27}=M_{23} M_{26}=M_{24} M_{25} \coma M_{1} M_{11}=M_{2} M_{10}=M_{4} M_{8}=M_{5} M_{7} \coma \right.\nonumber\\ 
& \scriptstyle \left. M_{2} M_{12}=M_{3} M_{11}=M_{5} M_{9}=M_{6}M_{8} \coma M_{7} M_{17}=M_{8} M_{16}=M_{10} M_{14}=M_{11} M_{13} \coma \right.\nonumber\\ 
& \scriptstyle \left. M_{8} M_{18}=M_{9} M_{17}=M_{11} M_{15}=M_{12} M_{14} \coma M_{4} M_{23}=M_{5} M_{22}=M_{10} M_{20}=M_{11} M_{19} \coma \right.\\ 
& \scriptstyle \left. M_{5} M_{24}=M_{6} M_{23}=M_{11} M_{21}=M_{12} M_{20} \coma M_{10} M_{26}=M_{11} M_{25}=M_{16} M_{23}=M_{17} M_{22} \coma \right.\nonumber\\ 
& \scriptstyle \left. M_{11} M_{27}=M_{12} M_{26}=M_{17} M_{24}=M_{18} M_{23} \coma M_{1} M_{15}=M_{2} M_{14}=M_{3} M_{13}=M_{7} M_{9}=M_{8}^2 \coma \right.\nonumber\\ 
& \scriptstyle \left. M_{1} M_{21}=M_{2} M_{20}=M_{3} M_{19}=M_{4} M_{6}=M_{5}^2 \coma M_{1} M_{25}=M_{4} M_{16}=M_{7} M_{22}=M_{10}^2=M_{13} M_{19} \coma \right.\nonumber\\ 
& \scriptstyle \left. M_{3} M_{27}=M_{6} M_{18}=M_{9} M_{24}=M_{12}^2=M_{15} M_{21} \coma M_{13} M_{27}=M_{14} M_{26}=M_{15} M_{25}=M_{16} M_{18}=M_{17}^2 \coma \right.\nonumber\\ 
& \scriptstyle \left. M_{19} M_{27}=M_{20} M_{26}=M_{21} M_{25}=M_{22} M_{24}=M_{23}^2 \coma M_{1} M_{12}=M_{2} M_{11}=M_{3} M_{10}=M_{4} M_{9}=M_{5} M_{8}=M_{6} M_{7} \coma \right.\nonumber\\ 
& \scriptstyle \left. M_{1} M_{17}=M_{2} M_{16}=M_{4} M_{14}=M_{5} M_{13}=M_{7} M_{11}=M_{8} M_{10} \coma M_{2} M_{18}=M_{3} M_{17}=M_{5} M_{15}=M_{6} M_{14}=M_{8} M_{12}= \right.\nonumber\\ 
& \scriptstyle \left. =M_{9} M_{11} \coma M_{7} M_{18}=M_{8} M_{17}=M_{9} M_{16}=M_{10} M_{15}=M_{11} M_{14}=M_{12} M_{13} \coma M_{1} M_{23}=M_{2} M_{22}=M_{4} M_{11}=\right.\nonumber\\ 
& \scriptstyle \left. =M_{5} M_{10}=M_{7} M_{20}=M_{8} M_{19} \coma M_{2} M_{24}=M_{3} M_{23}=M_{5} M_{12}=M_{6} M_{11}=M_{8} M_{21}=M_{9} M_{20} \coma M_{4} M_{24}=M_{5} M_{23}=\right.\nonumber\\ 
& \scriptstyle \left. =M_{6} M_{22}=M_{10} M_{21}=M_{11} M_{20}=M_{12} M_{19} \coma M_{4} M_{26}=M_{5} M_{25}=M_{10} M_{23}=M_{11} M_{22}=M_{16} M_{20}=M_{17} M_{19} \coma \right.\nonumber\\ 
& \scriptstyle \left. M_{7} M_{26}=M_{8} M_{25}=M_{10} M_{17}=M_{11} M_{16}=M_{13} M_{23}=M_{14} M_{22} \coma M_{5} M_{27}=M_{6} M_{26}=M_{11} M_{24}=M_{12} M_{23}=M_{17} M_{21}=\right.\nonumber\\ 
& \scriptstyle \left. =M_{18} M_{20} \coma M_{8} M_{27}=M_{9} M_{26}=M_{11} M_{18}=M_{12} M_{17}=M_{14} M_{24}=M_{15} M_{23} \coma M_{10} M_{27}=M_{11} M_{26}=M_{12} M_{25}=\right.\nonumber\\ 
& \scriptstyle \left. =M_{16} M_{24}=M_{17} M_{23}=M_{18} M_{22} \coma M_{1} M_{18}=M_{2} M_{17}=M_{3} M_{16}=M_{4} M_{15}=M_{5} M_{14}=M_{6} M_{13}=M_{7} M_{12}=M_{8} M_{11}=\right.\nonumber\\ 
& \scriptstyle \left. =M_{9} M_{10} \coma M_{1} M_{24}=M_{2} M_{23}=M_{3} M_{22}=M_{4} M_{12}=M_{5} M_{11}=M_{6} M_{10}=M_{7} M_{21}=M_{8} M_{20}=M_{9} M_{19} \coma M_{1} M_{26}=\right.\nonumber\\ 
& \scriptstyle \left. =M_{2} M_{25}=M_{4} M_{17}=M_{5} M_{16}=M_{7} M_{23}=M_{8} M_{22}=M_{10} M_{11}=M_{13} M_{20}=M_{14} M_{19} \coma M_{2} M_{27}=M_{3} M_{26}=\right.\nonumber\\ 
& \scriptstyle \left. =M_{5} M_{18}=M_{6} M_{17}=M_{8} M_{24}=M_{9} M_{23}=M_{11} M_{12}=M_{14} M_{21}=M_{15} M_{20} \coma M_{4} M_{27}=M_{5} M_{26}=M_{6} M_{25}=M_{10}M_{24}=\right.\nonumber\\ 
& \scriptstyle \left. =M_{11} M_{23}=M_{12} M_{22}=M_{16} M_{21}=M_{17} M_{20}=M_{18} M_{19} \coma M_{7} M_{27}=M_{8} M_{26}=M_{9} M_{25}=M_{10} M_{18}=M_{11} M_{17}=\right.\nonumber\\ 
& \scriptstyle \left. =M_{12} M_{16}=M_{13} M_{24}=M_{14} M_{23}=M_{15} M_{22} \coma M_{1} M_{27}=M_{2} M_{26}=M_{3} M_{25}=M_{4} M_{18}=M_{5} M_{17}=M_{6} M_{16}=M_{7} M_{24}=\right.\nonumber\\ 
& \scriptstyle \left. =M_{8} M_{23}=M_{9} M_{22}=M_{10} M_{12}=M_{11}^2=M_{13} M_{21}=M_{14} M_{20}=M_{15} M_{19}\right\rangle\nonumber
	\end{align}

\begin{center}
		\renewcommand{\arraystretch}{1.1}

\end{center}

%% file: App/C.tex
This appendix is adapted from \cite{Heckman:2022xgu}.

\section{Other Realizations of Defects}\label{app:other}

In this appendix we have focussed on one particular realization of 4D QFTs via D3-brane probes of 
a transverse geometry. This realization is especially helpful for studying duality defects since the 
corresponding defect is realized in terms of conventional bound states of $(p,q)$ 7-branes wrapped ``at infinity'' in the 
transverse geometry. On the other hand, some structures are more manifest in other duality frames. Additionally, 
there are other ways to engineer 4D QFTs via string constructions as opposed to considering worldvolumes of probe D3-branes.

With this in mind, in this Appendix we discuss how some of the structures considered in Chapter \ref{cha:ch4} are represented in other top down constructions and how this can be used to obtain further generalizations. We begin by discussing how the defect group of $\mathcal{N} = 4 $ SYM is specified in different top down setups. After this, we discuss how duality defects are constructed in some of these alternative constructions.

\subsection{Defect Groups Revisited}

To begin, we recall that the ``defect group'' of a theory is obtained from the spectrum of heavy defects which are not screened by dynamical objects \cite{DelZotto:2015isa} (see also \cite{Tachikawa:2013hya, GarciaEtxebarria:2019caf, Albertini:2020mdx, Morrison:2020ool}). For example, in the context of a 6D SCFT engineered via F-theory on an elliptically fibered Calabi-Yau threefold $X \rightarrow B$ with base $B = \mathbb{C}^2 / \Gamma_{U(2)}$ a quotient by a finite subgroup of $U(2)$ (see \cite{Heckman:2013pva} and \cite{Heckman:2018jxk} for a review), we get extended surface defects (high tension effective strings) from D3-branes wrapped on non-compact 2-cycles of $\mathbb{C}^2 / \Gamma_{U(2)}$. The charge of these defects is screened by D3-branes wrapped on the collapsing cycles of the singularity. The corresponding quotient (as obtained from the relative homology exact sequence for $\mathbb{C}^2 / \Gamma_{U(2)}$ and its boundary geometry $S^3 / \Gamma_{U(2)}  = \partial B$) is:
\begin{equation}
0 \rightarrow H_{2}(B ) \rightarrow H_{2}(B , \partial{B}) \rightarrow H_{1}(\partial B) \rightarrow 0.
\end{equation}
In particular, the defect group for surface defects is simply given by $H_{1}(S^3 / \Gamma_{U(2)}) = \mathrm{Ab}[\Gamma]$, the abelianization of $\Gamma$.

Turning to the case of interest in Chapter \ref{cha:ch4}, observe that for the $\mathcal{N} = (2,0)$ theory obtained from compactification of IIB on the A-type singularity $\mathbb{C}^2 / \mathbb{Z}_{N}$, this leads, upon further reduction on a $T^2$, to the 4D $\mathcal{N} =  4$ SYM theory with Lie algebra $\mathfrak{su}(N)$. The surface defects of the 6D theory can be wrapped on 1-cycles of the $T^2$, and this leads to line defects of the 4D theory. In this case, the defect group for lines is just $\mathbb{Z}_N^{\mathrm{(elec)}} \times \mathbb{Z}^{\mathrm{(mag)}}_N$, and a choice of polarization serves to specify the global form of the gauge group. Similar considerations hold for theories with reduced supersymmetry, as obtained from general 6D SCFTs compactified on Riemann surfaces (see \cite{DelZotto:2015isa} for further discussion).

Returning to the case of D3-branes probing a transverse singularity, this would appear to pose a bit of a puzzle for the ``defect group picture''. To see why, observe that for D3-branes probing $\mathbb{C}^3$, the heavy line defects are obtained from F1- and D1-strings which stretch from the D3-brane out to infinity. On the other hand, the boundary topology of the $\partial \mathbb{C}^3 = S^5$ has no torsional homology.\footnote{That being said, provided one just considers the dimensional reduction to the 5D Symmetry TFT, one can still readily detect the electric and magnetic 1-form symmetries (as we did in Chapter \ref{cha:ch4}).} The physical puzzle, then, is to understand how the defect group is realized in this case.

The answer to this question relies on the fact that the precise notion of ``defect group'' is really specified by the Hilbert space of the string theory background, and quotienting heavy defects by dynamical states of the localized QFT. The general physical point is that in the boundary $S^5$, there is an asymptotic 5-form flux, and in the presence of this flux, the endpoints of the F1- and D1-strings will ``puff up'' to a finite size. The size is tracked by a single integer parameter, and the maximal value of this is precisely $N$.

To see how this flux picture works in more detail, it is helpful to work in a slightly different duality frame. Returning to the realization of $\mathcal{N} = 4$ SYM via type IIB on the background $\mathbb{R}^{3,1} \times T^2 \times \mathbb{C}^2 / \mathbb{Z}_N$, consider T-dualizing the circle fiber of the $\mathbb{C}^2 / \mathbb{Z}_N$ geometry. 
As explained in \cite{Ooguri:1995wj}, we then arrive in a type IIA background with $N$ NS5-branes filling $\mathbb{R}^{3,1} \times T^2$ and sitting at a point of the transverse $\mathbb{C}^2$. There is a dilaton gradient profile in the presence of the NS5-brane, but far away from it, the boundary geometry is simply an $S^3$ threaded by $N$ units of NSNS 3-form flux. In this realization of the 4D QFT, the line operators of interest are obtained from D2-branes which wrap a 1-cycle of the $T^2$ as well as the radial direction of the transverse geometry, ending at ``point'' of the boundary $S^3$ with flux. This appears to have the same puzzle already encountered in the case of the D3-brane realization of the QFT.

Again, the ordinary homology / K-theory for the $S^3$ is not torsional, but the \textit{twisted} K-theory is indeed torsional. Recall that the twisted K-theory for $S^3$ (see \cite{baraglia2015fourier}) involves the choice of a twist class $N \in H^3(S^3) \simeq \mathbb{Z}$, and for this choice of twist class, one gets:
\begin{equation}
K^{\ast}_{H}(S^3) \simeq \mathbb{Z}_{N}, 
\end{equation}
in the obvious notation. From a physical viewpoint, one can also see that the spectrum of ``point-like'' branes in this system is actually more involved. Indeed, the boundary $S^3$ is actually better described as an $SU(2)$ WZW model. Boundary states of the worldsheet CFT correspond to D-branes, and these are in turn characterized by fuzzy points of the geometry.\footnote{We are neglecting the additional boundary states provided by having a supersymmetric WZW model. This leads to additional extended objects / topological symmetry operators. For further discussion on these additional boundary states, see \cite{Maldacena:2001xj, Maldacena:2001ss}.} The interpretation of these boundary conditions can be visualized as ``fuzzy 2-spheres'' (see e.g., \cite{Alekseev:2000fd}), as specified by the non-commutative algebra:
\begin{equation}
[J^{a},J^{b} ] = i \varepsilon^{abc} J^c,
\end{equation}
for $a,b,c = 1,2,3$, namely a representation of $\mathfrak{su}(2)$. The size of the fuzzy 2-sphere is set by $J^{a} J^{a}$, the Casimir of the representation, and this leads to a finite list of admissible representations going from spin $j = 0,...,(N-1)/2$. Beyond this point, the stringy exclusion principle \cite{Maldacena:1998bw} is in operation, and cuts off the size of the fuzzy 2-sphere. The upshot of this is that the single point of ordinary boundary homology has now been supplemented by a whole collection of fuzzy 2-spheres, and these produce the required spectrum of heavy defects which cannot be screened by dynamical objects. Similar considerations clearly apply for topological symmetry operators generated by D4-branes wrapped on a 1-cycle of the $T^2$ and a fuzzy 2-sphere.

With this example in mind, we clearly see that similar considerations will apply in systems where the boundary geometry contains a non-trivial flux. In particular, in the D3-brane realization of $\mathcal{N} = 4$ SYM 
we can expect the F1- and D1-strings used to engineer heavy defects to also ``puff up'' to non-commutative cycles in the boundary $S^5$.

One point we wish to emphasize is that so long as we dimensionally reduce the boundary geometry to reach a lower-dimensional system (as we mainly did in Chapter \ref{cha:ch4}), then the end result of the flux can also be detected directly in the resulting 5D bulk SymTFT.

\subsection{Duality Defects Revisited}

In the previous subsection we presented a general proposal for how to identify the defect group in 
duality frames where asymptotic flux is present. Now, one of the main reasons we chose to focus on the D3-brane 
realization of our QFTs is that the top down identification of duality defects is relatively straightforward (even if the defect group computation is more subtle). Turning the discussion around, one might also ask how our top down duality defects are realized in other string backgrounds which realize the same QFT. For related discussion on this point, see the recent reference \cite{Bashmakov:2022uek}.

To illustrate, consider the IIB background $\mathbb{R}^{3,1} \times T^2 \times \mathbb{C}^2 / \mathbb{Z}_{N}$, 
in which the 4D QFT is realized purely in terms of geometry. In this case, a duality of the 4D field theory will be specified by a large diffeomorphism of the $T^2$, namely as an $SL(2,\mathbb{Z})$ transformation of the complex structure of the $T^2$.

Because the duality symmetry is now encoded purely in the geometry, the ``brane at infinity'' which implements a topological defect / interface will necessarily be a variation in the asymptotic profile of the 10D metric far from the location of the QFT. Since, however, only the topology of the configuration actually matters, it will be enough to specify how this works at the level of a holomorphic Weierstrass model.

Along these lines, we single out one of the directions $x_{\bot}$ along the $\mathbb{R}^{3,1}$ such that the duality defect / interface will be localized at $x_{\bot} = 0$ in the 4D spacetime. Combining this with the radial direction of $\mathbb{C}^2 / \mathbb{Z}_N$, we get a pair of coordinates which locally fill out a patch of the complex line $\mathbb{C}$. It is helpful to introduce the complex combination:
\begin{equation}
z = x_{\bot} + \frac{i}{r},
\end{equation}
where $r = 0$ and $r = \infty$ respectively indicate the location of the QFT and the conformal boundary, where we reach the $S^3 / \mathbb{Z}_{N}$ lens space. In terms of this local coordinate, we can now introduce a Weierstrass model with the prescribed Kodaira fiber type at $z = 0$, namely $x_{\bot} = 0$ and $r = \infty$. 
For example, a type $IV^{\ast}$ and type $III^{\ast}$ fiber would respectively be written as:
\begin{align}
\mathrm{type} \,\, III^{\ast}: \, & y^2 = x^3 + x z^3 \\
\mathrm{type} \,\, IV^{\ast}: \, & y^2 = x^3 + z^4.
\end{align}
This sort of asymptotic profile geometrizes the duality / triality defects we considered.

\section{3D TFTs from 7-branes}\label{app:minimalTFT7branes}

In the main body of Chapter \ref{cha:ch4} we showed how basic structure of 7-branes can account for duality / triality defects in 
QFTs engineered via D3-brane probes of a Calabi-Yau singularity $X$. In particular, we saw that anomaly inflow analyses constrain the resulting 3D TFT of the corresponding duality defect. Of course, given the fact that we are also claiming 
that these topological defects arise from 7-branes, it is natural to ask whether we can directly extract these terms from dimensional reduction of topological terms of the 7-brane. Our aim in this Appendix will be to show to what extent we can derive a 3D TFT living on the duality / triality defect whose 1-form anomalies match that of the appropriate minimal abelian 3D TFTs, $\mathcal{A}^{k,p}$. This is required due to in-flowing the mixed 't Hooft anomaly between the 0-form duality / triality symmetry and the 1-form symmetry of the 4D SCFT, as well as from the line operator linking arguments of Section \ref{ssec:linkingminimaltheory}. While we leave a proper match of these anomalies to future work, this appendix will highlight that, in general, that 3D TFTs on the 7-branes will differ from the minimal 3D TFTs due to the presence of a non-abelian gauge group. Additionally, we propose an 8D WZ term that allows us to determine the level of the 3D CS theory. 

We first review the case of a stack of $n$ D7-branes. The Wess-Zumino (WZ) terms are known from string perturbation theory to be \cite{Douglas:1995bn}\footnote{More generally $(p,q)$ 7-brane WZ topological actions can be inferred from $SL(2,\mathbb{Z})$ transformations of \eqref{eq:WZD7}.}:
\begin{equation}\label{eq:WZD7}
\begin{aligned}
    \mathcal{S}_{\text{WZ}} &= \int_{X_8} \left( \sum_{k}C_{2k} \mathrm{Tr} e^{ \mathcal{F}_2} \sqrt{\frac{\hat{\mathcal{A}}(R_T)}{\hat{\mathcal{A}}(R_N)}} \right)_{\text{8-form}}
\end{aligned}
\end{equation}
where 
\begin{equation}
\mathrm{Tr}\mathcal{F}_2=\mathrm{Tr}( F_2-i^{\ast}B_2)=nF^{U(1)}_2-n i^{\ast} B_2\,,
\end{equation}
with $i^{\ast}B_2$ denoting the pullback from the 10D bulk to the 8D worldvolume $X_8$ of the 7-brane, $F^{U(1)}_2$ is $U(1)$ the gauge curvature associated to factor in the numerator of the 7-brane gauge group $U(n)\simeq (U(1)\times SU(n))/\mathbb{Z}_n$. This precise combination is required because F1-strings can end on D7-branes. In particular, since F1-strings couple to the bulk 2-form, there is a gauge transformation 
$B_2 \rightarrow B_2 + d \lambda_1$ which is cancelled by introducing a compensating $U(1)$ curvature associated with the 
``center of mass'' of the 7-brane. $\mathcal{A}(R_T)$ and $\mathcal{A}(R_N)$ in \eqref{eq:WZD7} are the A-roof genera of the tangent and normal bundles which is given by the expansion
\begin{equation}
    \hat{\mathcal{A}}=1-\frac{1}{24}p_1+\frac{1}{5760}(7p^2_1-4p_2)+...
\end{equation}
where for completeness, we have included $p_i$, the $i^{\mathrm{th}}$ Pontryagin class of the tangent bundle / normal bundle. Since we are concerned with reducing the 7-brane on $S^5$, such contributions play little role in our analysis but could in principle play an important role in more intricate boundary geometries $\partial X$. Taking this into account, the only terms that concern us then are
\begin{equation}\label{eq:wzd7un}
    \int_{M_3\times S^5}\frac{1}{8\pi}C_4\wedge \mathrm{Tr}(F^2)+\frac{2\pi}{2}C_4\wedge \left(\mathrm{Tr}((F_2/2\pi)-B_2) \right)^2
\end{equation}
where we are now being careful with the overall factors of $2\pi$ and considering $\mathrm{Tr}(F)/2\pi$ as integrally quantized. Reducing \eqref{eq:wzd7un} on the $S^5$ surrounding $N$ D3s would then naively produce a level $N$ $U(n)$ 3D Chern-Simons theory living on $M_3$ with an additional coupling to the background $U(1)$ 1-form field $B_2$ that is proportional to $\int_{M_3}\mathrm{Tr}(A)\wedge B_2$. We say ``naive'' because one must first understand how the center-of-mass $U(1)$ of $n$ D7-branes is gapped out via the St\"uckelberg mechanism, i.e., how the gauge algebra reduces from $\mathfrak{u}(n)$ to $\mathfrak{su}(n)$. Indeed, observe that the coupling $C_6 \mathrm{Tr} F_2$ can gap out this $U(1)$ since integrating over $C_6$ produces the constraint\footnote{This is provided that we choose Neumann boundary conditions for $C_6$ along the D7-brane stack.}:
\begin{equation}\label{eq:Stukconstraint}
    \mathrm{Tr}(F_2)=nF^{U(1)}_2=0
\end{equation}
so $F^{U(1)}_2$ still survives as a $\mathbb{Z}_n$-valued 2-form field and is in fact equivalent to the generalized Stiefel-Whitney class \cite{Aharony:2013hda}
\begin{equation}
 F^{U(1)}_2  ~\rightarrow~ w_2\in H^2(X_8,\mathbb{Z}_n)\,.
\end{equation}
One sees this by supposing that there is a magnetic 4-brane monopole in the $U(n)$ gauge theory in the fundamental representation $\mathbf{n}_{+1}$. Then $\frac{1}{n}\int_{S^2}\mathrm{Tr}(F)$ measures a magnetic charge $+1$ with respect to the $U(1)$ in the numerator of $U(n) = (U(1) \times SU(n) ) / \mathbb{Z}_n$, where the $\mathbb{Z}_n$ embeds in the center of $SU(n)$ in the standard fashion. After gapping out the center of mass $U(1)$, we still measure a magnetic flux $1 \; \mathrm{mod}\; n$ around the monopole, which is a defining property of $w_2$. 

Similarly, when considering flat connections of the $U(n)$ gauge theory, we have that $\mathrm{Tr}(F_2)=d\mathrm{Tr}(A)$, and the constraint \eqref{eq:Stukconstraint} implies $n\mathrm{Tr}(A)=0$. Then, the integral $\int_{\gamma_1}\mathrm{Tr}A$ measures a $U(1)\subset U(n)$ monodromy around a 1-cycle $\gamma_1$ and becomes, after decoupling the center-of-mass $U(1)$, the value of $\int_{\gamma_1}w_1$ because this naturally measures the $\mathbb{Z}_n$ monodromy. A subtle distinction we should make is that while $w_2$ and $w_1$ are analogs of $\mathrm{Tr}(F)$ and $\mathrm{Tr}(A)$, the former are cohomology classes while the discrete remnants of the latter are discrete cocycles, i.e. members of $C^i(X_8,\mathbb{Z}_n)$. We therefore name these $\mathcal{J}_2$ and $a$, respectively, which satisfy $[\mathcal{J}_2]=w_2$ and $[a]=w_1$. Moreover, when $w_2=0$, we have that $\mathcal{J}_2=\delta a$ where $\delta$ is the coboundary operator. 

Generalizing to non-perturbative bound states of 7-branes,\footnote{Again, by this we mean 7-branes whose monodromy fixes $\tau$.} with some monodromy matrix $\rho$, we know from the main text that the analog of $B_2$ for the D7 case is generalized to $B^\rho_2$ which takes values in $\mathrm{ker}(\rho-1)$, it is natural then
ask whether there is an analogous discrete remnant of the ``center of mass $U(1)$'' for a general 7-brane 
of type $\mathfrak{F}$, namely the analog of the specific combination $\mathcal{F}_2 = \mathrm{Tr}(F_2 - B_2)$. From the discussion below \eqref{eq:Stukconstraint}, we can already make a reasonable guess that $\mathrm{Tr}(F_2)$ should be replaced by $\mathcal{J}_2\in C^2(M_3\times \partial X_6,\mathrm{ker}(\rho))$. In the case of perturbative IIB D7-branes, this involves the specific decomposition $U(n) = (SU(n) \times U(1)) / \mathbb{Z}_n$. In the case of constant axio-dilaton profiles, all of these cases can be obtained from the specific subgroups:\footnote{For example, we have the following subgroups of $E_8$: $(E_7 \times U(1)) / \mathbb{Z}_2$ and $(E_6 \times U(1)^2) / \mathbb{Z}_3$.} $E_8 \supset (G_{\mathfrak{F}} \times U(1)^m) / \mathbb{Z}_{k} $ with $n + m = 8$, where maximal torus of $G_{\mathfrak{F}}$ has dimension $n$, and $G_{\mathfrak{F}}$ has center $\mathbb{Z}_k$. We then see that all of the remarks below \eqref{eq:Stukconstraint} equally apply here if we start with an $E_8$ 7-brane and Higgs down to another 7-brane with constant axio-dilaton. Extending the treatment in \cite{Kapustin:2014gua} for A-type Lie groups, 
we introduce the gauge field:
\begin{equation}
    \mathbf{A}=A+\frac{1}{k}\widehat{a}
\end{equation}
and its field strength $\mathbf{F}=d\mathbf{A}+\mathbf{A}\wedge \mathbf{A}$ where the connections $A,\widehat{a}$ take values in the Lie algebras of $G_{\mathfrak{F}}$ and $U(1)^m$, respectively. As in our D7-brane discussion, the center of mass $U(1)\subset U(1)^m$ is gapped out up to a discrete $\mathbb{Z}_k$-valued gauge field $a$. The curvature on the 7-brane worldvolume therefore takes the form 
\begin{equation}
   \mathbf{F}=  F+\mathcal{J}_2
\end{equation} 
where $F$ is the traceless curvature of $A$ and $\mathcal{J}_2$ is the discrete remnant of the center of mass mode valued in $\Gamma=\mathbb{Z}_k$. Note that $\Gamma$ coincides with the center of the 7-brane gauge group (with electric polarization) which then nicely matches our guess that $\mathcal{J}_2$ should serve as our analog of $\mathrm{Tr}(F)$. Moreover, when we take electric polarization on the brane, $w_2$ is trivial for gauge bundles and therefore we have that $\mathcal{J}_2=\delta a$ with $a\in C^1(M_3\times \partial X_6, \mathrm{ker}(\rho))$

We now conjecture a non-perturbative generalization of the WZ terms in \eqref{eq:wzd7un} which applies to all types of 7-brane stacks (labelled by gauge algebra $\mathfrak{g}$) as listed in table \ref{tab:Fibs}:\footnote{A brief comment on the normalization of the instanton density: We have adopted a convention where $\mathrm{Tr} F^2 = \frac{1}{h_{G}^{\vee}} \mathrm{Tr}_{\mathrm{adj}} F^2$, with the latter a trace over the adjoint representation, and $h_{G}^{\vee}$ the dual Coxeter number of the Lie group $G$. Moreover, in our conventions, we have that for a single instanton on a compact four-manifold, $\frac{1}{4} \mathrm{Tr} F^2$ integrates to $1$. }
\begin{equation}\begin{aligned}\label{eq:WZtermsu3}
 \mathcal{S}_{WZ}^{\mathrm{(7)}} &\supset  \int_{M_3\times \partial X_6} \left(  C_4 \wedge \textnormal{tr}\exp \left(\mathbf{F}-B_2^\rho  \right)\right) \\
   &=  \int_{M_3\times \partial X_6} \left(  \frac{1}{8\pi} C_4 \wedge \mathrm{Tr}\,{F}^2 + C_4 \cup \frac{1}{2} (\mathcal{J}_2-B^\rho_2 )^2 \right)
\end{aligned}
\end{equation}
Here we have chosen normalizations in \eqref{eq:WZtermsu3} such that $\exp(i \mathcal{S}_{WZ})$ appears in the 7-brane path integral and $B_2^\rho, \mathcal{J}_2$ are $U(1)$ valued. The first term in \eqref{eq:WZtermsu3} is the D3-brane instanton density term. Namely, it can be obtained by considering a single D3-brane inside a 7-brane and viewing it as a charge-1 instanton which sets the normalization. The second term is a generalization of the term 
\begin{equation}
   \int_{D7}C_4\wedge n B_2\wedge \mathrm{Tr}F_2 = \int_{D7}C_4\wedge n B_2\wedge nF^{U(1)}_2
   \end{equation}
appearing in \eqref{eq:WZD7} by replacing $B_2$ with the more general $B_2^\rho$. The coefficient of $1$ for the second term of \eqref{eq:WZtermsu3} follows from the standard substitutions $B_2\rightarrow \frac{1}{n}B_2$ and $F^{U(1)}_2\rightarrow \frac{1}{n}F^{\mathbb{Z}_n}_2$ when one converts a $U(1)$-valued field to its  remnant field valued in $\mathbb{Z}_n\subset U(1)$.

After reducing on an $S^5$ with flux $\int F_5= N$, \eqref{eq:WZtermsu3} produces a 3D $(G_{\mathfrak{F}})_N$ CS theory\footnote{Subscript denotes the level.} along with a coupling to its electric 1-form symmetry background. In other words, our 3D action becomes:
\begin{equation}\label{eq:3dCS}
  \int_{M_3}   N \cdot CS(A) + 
\frac{N}{2 \pi}a \cup B_2^\rho+ N\cdot CS(a)\\
\end{equation}
where fields are treated as elements in $U(1)$ (suitably restricted): the background $B_2^\rho$ can be normalized to take values in $\mathbb{Z}_{\mathrm{gcd}(k,N)}$ (recall $k$ is the order of the monodromy matrix of the non-perturbative 7-brane), and similar considerations apply for $a$. Note also that a priori, the 3D TFT we get in this way need not match the minimal TFT of type $\mathcal{A}^{K,m}$, since anomaly inflow considerations do not fully fix the form of the 3D TFT. It would be interesting to carry out a complete match with the analysis presented in the main text, but we leave this for future work.

\section{D3-Brane Probe of $\mathbb{C}^3 / \mathbb{Z}_3$} \label{app:orbo}

\begin{figure}
    \centering
    \includegraphics[width=7cm]{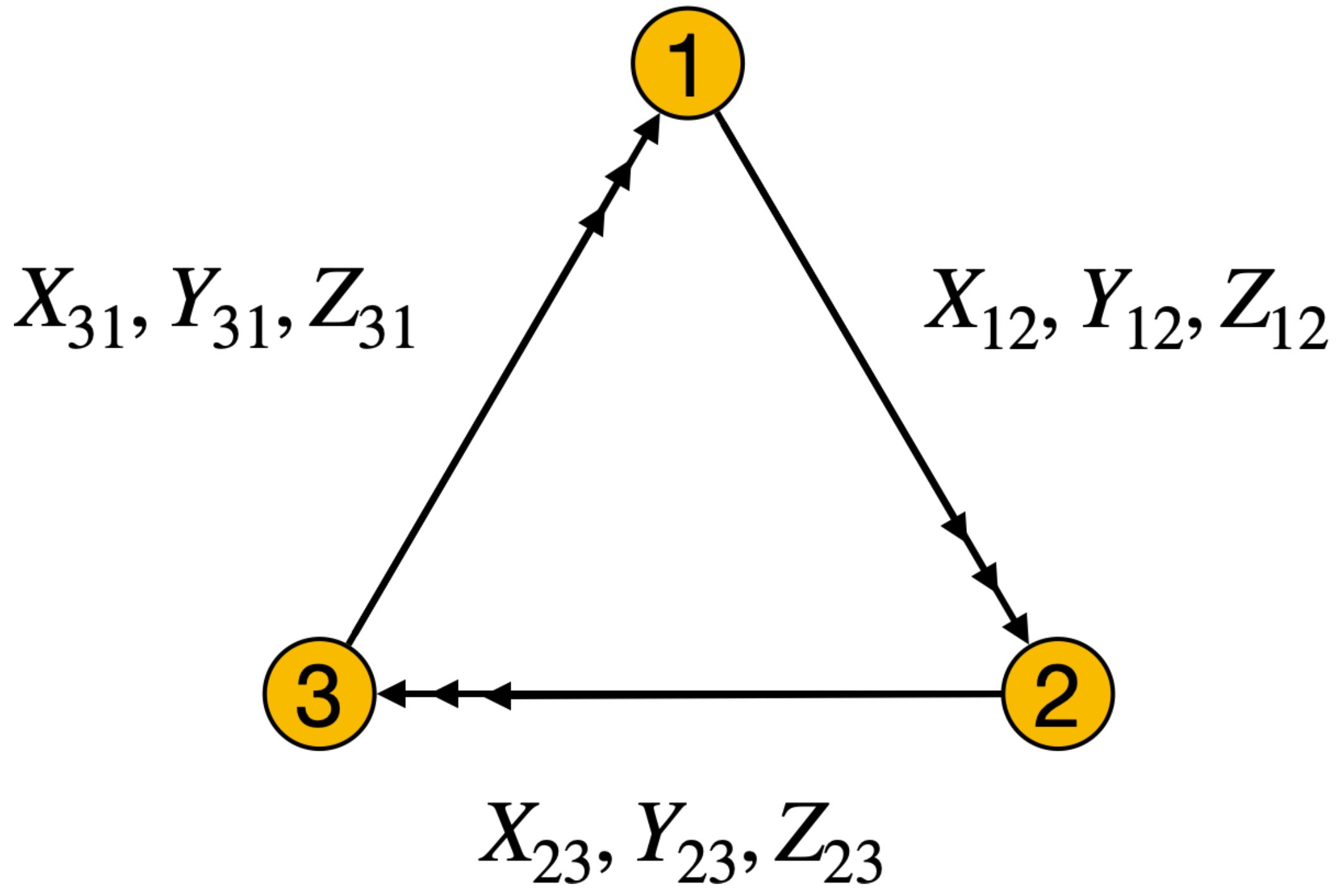}
    \caption{The quiver diagram of $\mathcal{N}=1$ theory on D3-branes probing $\mathbb{C}^3/\mathbb{Z}_3$.}
    \label{fig:quiverc3z3}
\end{figure}

In this Appendix we present further details on the case of a D3-brane probing the orbifold singularity $\mathbb{C}^3 / \mathbb{Z}_3$. The orbifold group action on $\mathbb{C}^3$ is defined by
\begin{equation}
    (z_1,z_2,z_3)\rightarrow (\zeta z_1, \zeta z_2, \zeta z_3),~  \zeta ^3=1.
\end{equation}
Following the procedure in \cite{Douglas:1996sw, Lawrence:1998ja, Kachru:1998ys}, 
the field content of the resulting 4D theory is given in Figure \ref{fig:quiverc3z3}. 
The superpotential of the theory is:
\begin{equation}
    W= \kappa \text{Tr}\big\{ X_{12}[Y_{23},Z_{31}]+X_{23}[Y_{31},Z_{13]}]+X_{31}[Y_{12},Z_{23}] \big\}.
\end{equation}

The 5D boundary $\partial \mathbb{C}^3 / \mathbb{Z}_3 = S^5/\mathbb{Z}_3$ 
of the orbifold singularity has the cohomology classes:
\begin{equation}
    H^{\ast}(S^5/\mathbb{Z}_3)=\{ \mathbb{Z},0,\mathbb{Z}_3,0,\mathbb{Z}_3,\mathbb{Z}   \},
\end{equation}
with linking numbers: 
\begin{equation}
    \int_{S^5/\mathbb{Z}_3}t_2\star t_2 \star t_2=\int_{S^5/\mathbb{Z}_3}t_2\star t_4=-\frac{E_0^3}{3\cdot 3\cdot 3}=-\frac{1}{3}.
\end{equation}
$E_0^3$ is the triple self-intersection number of the compact divisor of $O(-3)\rightarrow \mathbb{P}^2$, which can be read from the toric data.\footnote{We refer the reader to \cite{hori2003mirror} for technical details on computing intersection numbers of toric varieties.}

The 5D TFT with the generic form (\ref{eq:TFT for generic N=1}) now reduces to 
\begin{equation}\label{eq:TFT for c3z3}
    \begin{split}
        S_{\text{symTFT}}&=-\int_{M_4\times \mathbb{R}_{\ge 0}} \bigg\{ N\breve{h}_3\star \breve{g}_3 - \frac{1}{3}\breve{E}_3\star \breve{B}_1\star \breve{C}_1-\frac{1}{3}\breve{E}_{1}\star \left( \breve{B}_1\star \breve{g}_3+\breve{h}_3\star \breve{C}_1 \right) \bigg\}.
    \end{split}
\end{equation}
Let us now identify the correspondence between the symmetries in the 4D SCFT and the 5D symmetry TFT fields. The first term in (\ref{eq:TFT for c3z3}) is obvious. It is just the differential cohomology version of the familiar $N\int B_2\wedge dC_2$ term which also appears in the $\mathcal{N}=4$ SYM case. In the $\mathcal{N}=1$ quiver gauge theory, $\breve{h}_3$ (resp. $\breve{g}_3$) corresponds to $B_2^{\text{diag}}$ (resp. $C_2^{\text{diag}}$) for the diagonal $\mathbb{Z}_N$ electric (resp. magnetic) 1-form symmetry of the $\mathfrak{su}(N)^3$ theory.

Based on our previous discussion, we know $E_1, B_1$ and $C_1$ are background gauge fields for $\mathbb{Z}_3$ 0-form symmetries. In fact, as explained in \cite{Gukov:1998kn}, there are indeed three candidate $\mathbb{Z}_3$ symmetries in the 4D SCFT which act on the fields of the quiver gauge theory as follows:
\begin{equation}
\begin{split}
    &\mathbf{B}:(X_{ij},Y_{ij},Z_{ij})\rightarrow (Y_{i+1,j+1},Z_{i+1,j+1},X_{i+1,j+1})\\
    &\mathbf{C}:(X_{ij},Y_{ij},Z_{ij})\rightarrow (\zeta X_{ij}, \zeta^2 Y_{ij},Z_{ij}),\\
    &\mathbf{E}:(X_{ij},Y_{ij},Z_{ij})\rightarrow (\zeta X_{ij},\zeta Y_{ij},\zeta Z_{ij}).
\end{split}
\end{equation}
where $i$ and $j$ are mod 3 numbers denoting gauge factors. These symmetry generators transform non-trivially under IIB dualities, and their transformations are the same as those already stated in section \ref{sec:N=1}.